\documentclass[12pt]{article}
\usepackage{graphicx}
\usepackage{float}
\usepackage{subcaption}
\usepackage{dashbox}
\usepackage {arydshln}
\usepackage{amsmath,amsthm,amsfonts,amssymb,bm}
\usepackage{color}
\usepackage{enumerate}
\usepackage
[colorlinks, linkcolor=blue, anchorcolor=blue, citecolor=blue]{hyperref}
\makeatletter
\@addtoreset{equation}{section}

\makeatother

\begin{document}
\begin{titlepage}
\null

\begin{flushright}

\end{flushright}

\vskip 1.5cm
\begin{center}

 {\Large \bf On Soliton Solutions of the Anti-Self-Dual Yang-Mills}
 	
 	\vskip 0.3cm
 	
 	 {\Large \bf Equations from the Perspective of Integrable Systems}

 	  \vskip 0.7cm
 	 

\vskip 3cm
\normalsize

{\large Shan-Chi Huang  }

\vskip 1cm

{\it\large Graduate School of Mathematics, Nagoya University}

\vskip 2cm

A Dissertation in candidacy for \\
the degree of Doctor of Philosophy \\
in Mathematical Science \\
(Mathematical Physics) \\
2021

\vskip 1cm

\begin{figure*}[h]
	\centering
	\includegraphics[width=0.27\textwidth]{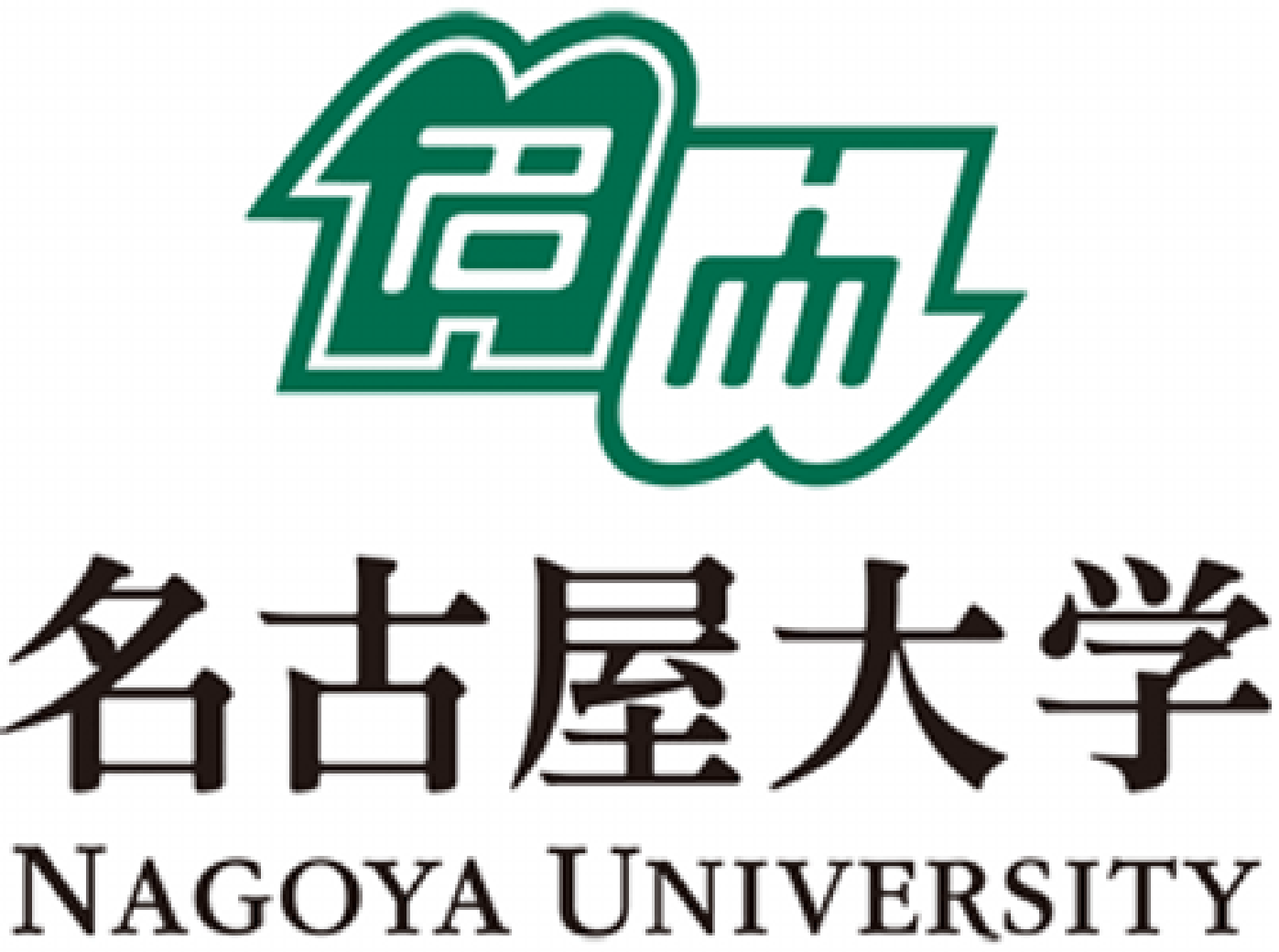}
\end{figure*}

\end{center}

\end{titlepage}

\newpage
	
\setcounter{page}{1}
\pagenumbering{roman}

\begin{center}
{\bf \large Abstract} 
\end{center}

In this thesis, we study soliton solutions of the anti-self-dual Yang-Mills (ASDYM) equations from the perspective of integrable systems. 
We construct a class of exact ASDYM 1-solitons on 4-dimensional real spaces with the Euclidean signature $(+, +, +, +)$, the Minkowski signature $(+, - , -, -)$, and the split signature ($+$, $+$, $-$, $-$) (the Ultrahyperbolic space). They are new results and  successful applications of the Darboux transformation introduced by Nimmo, Gilson, Ohta. 
In particular, the principal peak of the Lagrangian density Tr$F_{\mu\nu}F^{\mu\nu}$ is localized on a 3-dimensional hyperplane in 4 dimensional space. Therefore, we use the term "soliton walls" to distinguish them from domain walls because domain walls are described by scalar fields.
Furthermore, we propose an ansatz to obtain real-valued Lagrangian density Tr$F_{\mu\nu}F^{\mu\nu}$ even if the gauge group is non-unitary. For the case of split signature, we show that the gauge group can be $G=\mathrm{SU}(2)$ and $G=\mathrm{SU}(3)$ and hence the soliton walls could be candidates of physically interesting objects on the Ultrahyperbolic space $\mathbb{U}$. 

After $n$ iterations of the Darboux transformation, the resulting solutions can be expressed in terms of the Wronskian type quasideterminants of order $n+1$. More precisely, each element of the Wronskian type quasideterminant is a ratio of ordinary Wronskian determinants.
We call them quasi-Wronskian for short in this thesis.
On the other hand, we use the techniques of the quasideterminants to show that in the asymptotic region, a special class of the quasi-Wronskian solution possess $n$ isolated distributions of Lagrangian densities (with phase shifts). Therefore, we can interpret this kind of solution as $n$ intersecting soliton walls. For the split signature, we even show that the gauge group can be $G=\mathrm{SU}(2)$ for $n$ intersecting soliton walls, and $G=\mathrm{SU}(3)$ for each isolated soliton wall in the asymptotic region, respectively.


\clearpage
\baselineskip 6mm

\begin{center}
{\bf \large Acknowledgments}
\end{center}

When I first came to study in Nagoya university, I was almost ignorant in the field of anti-self-dual Yang-Mills equations and the related topics to integrable systems.
Even now, it is still incredible and unreal for me to realize that I indeed finish the doctoral thesis as this title. 
Of course, without the help from many peoples during my Ph.D career, without some good fortune and favor from god, the past three years of my life might be quite difficult and hard to imagine. I appreciate everyone I met during the past three years. They make me grow and become somewhat different. 

First and foremost I would like to express my sincere gratitude to my advisor Prof. Masashi Hamanaka for his continued guidance and support in my studies during the Ph.D career.
I appreciate very much for his patience, his tolerance, his encouragement, and his company in the past three years.
Especially, his positive and enthusiastic personality always make up for some lack of my personality. 
Secondly, I would like to express my sincere gratitude to my coadvisor Prof. Hiroaki Kanno for taking care of me in many aspects and supporting me behind in the past three years. 
I also thank him for very careful review on my doctoral thesis
and giving me many crucial advice that I never noticed. 
I was inspired very much during the main revision of this thesis.

I would like to thank Prof. Soichi Okada and Prof. Hidetoshi Awata for their careful review on my doctoral thesis and giving me important advice that make up for the shortcomings of this thesis. 
I would like to thank Prof. Claire R. Gilson and Prof. Jonathan J. C. Nimmo for their private notes and giving me the opportunity to do joint research with them. 
The research direction of my doctoral thesis is based very much on their pioneering work.
Finally, 
I would like to thank the Japan-Taiwan Exchange Association for funding me study in Japan that allows me to concentrate on my studies without worries.

\clearpage
\baselineskip 6mm

\tableofcontents

\newpage
\setcounter{page}{1}
\pagenumbering{arabic}

\section{Introduction}
\label{Section 1}

The Yang-Mills gauge theory \cite{YaMi} forms the foundation of the Standard Model that describes the nature law of interaction between elementary particles, and then many developments in high energy physics ensued from that beautiful idea as well. On the other hand, the classical solitons (e.g. \cite{Actor, MaSu}) in the Yang-Mills gauge theory play important roles in the study of non-perturbative aspects, duality structures, quark confinements, cosmology, condensed matter physics, nonlinear physics, and mathematical physics, etc. From the viewpoint of codimension, some well-known solitons are classified as the following, instanton of codimension 4, monopole of codimension 3, vortex (or cosmic string) of codimension 2, and domain wall of codimension 1. 
Almost all of these solitons are known for the exact solutions of the anti-self-dual Yang-Mills (ASDYM) equations (or dimensionally reduced equations of ASDYM).
For example, the instantons \cite{BePoScTy, JaNoRe} might be the most well-known classical solution of the ASDYM equations on the 4-dimensional Euclidean space. In addition, the instantons possess topological charge (instanton number) of positive integer and appear in the non-perturbative aspect of gauge theorems. Therefore, some physicists are devoted to finding new solitons for the potential  applications of them to modern physics, like the impacts of the instantons. 

In the aspect of mathematical physics, the instantons can be derived from the 't Hooft ansatz \cite{tHt3} (or known as the Corrigan-Fairlie-'t Hooft-Wilczek ansatz \cite{CoFa,Wilczek}) and studied systematically by the famous Atiyah-Drinfeld-Hitchin-Manin construction \cite{ADHM}. On the other hand, other solutions of the ASDYM equations are also discussed extensively by the Atiyah-Ward ansatz \cite{AtWa} in different branches of mathematical physics and integrable systems. 
For the Euclidean signature $(+,+,+,+)$, Yang \cite{Yang} introduce a convenient gauge with three parameters, called the $R$-gauge (Cf: \eqref{R-gauge_SL(2)_1}, \eqref{R-gauge_SL(2)_2}), and indicate that all the $G=\mathrm{SU(2)}$ (anti-)self-dual gauge fields (Cf: \eqref{R-gauge_SU(2) gauge fields_cpx_1}, \eqref{R-gauge_SU(2) gauge fields_cpx_2}, \eqref{R-gauge_SU(2) gauge fields_cpx_3}, \eqref{R-gauge_SU(2) gauge fields_cpx_4}) can be determined by solving three equations derived by him (or known as Yang's equations, Cf: \eqref{Yang's eqns for SU(2)_E}). Yang also indicate how the Corrigan-Fairlie-'t Hooft-Wilczek ansatz \cite{CoFa, tHt3, Wilczek} can be derived from these three equations. Soon after the work of Yang, the authors of \cite{CoFaYaGo1, CoFaYaGo2} apply the B\"acklund transformation to the three equations of Yang and 
construct a large class of exact solutions written by the ratios of determinants in a regular manner. The authors also indicate the correspondence between these determinant type solutions and the Atiyah-Ward ansatz \cite{AtWa}. Especially among these, the 't Hooft ansatz \cite{CoFa, tHt3, Wilczek} reappears in the form of the second simplest solutions of this class. 
In addition, these B\"acklund transformations possess an interesting property that turn $G=\mathrm{SU(2)}$ gauge fields into $G=\mathrm{SU(1,1)}$ gauge fields and vice versa. 
On the other hand, the authors of \cite{SaOhMa} indicate that the bilinear method of Hirota \cite{Hirota} provide another equivalent expression for this class of Atiyah-Ward ansatz solutions as well.
By introducing one more parameter into Yang's $R$-gauge \cite{Yang} (Cf: Mason-Woodhouse gauge \cite{MaWo}) and considering the noncommutative version of  B\"acklund transformation \cite{MaWo}, 
the authors of \cite{GiHaNi1, GiHaNi2} generalize the same class of Atiyah-Ward ansatz solutions to the noncommutative Euclidean space for $G=\mathrm{GL}(2, \mathbb{C})$. The explicit form of these solutions are expressed naturally by the quasideterminants (with noncommutative entries) \cite{GeGeReWi, GeRe} rather than the ratios of determinants (commutative entries) anymore. 

Since Yang's $R$-gauge \cite{Yang} is not applicable to $G=\mathrm{SU}(N)$ ASD gauge fields for $N > 2$, 
the authors of \cite{BrFaNuYa} indicate a gauge independent formulation for the three equations of Yang, now known as the Yang equation :
\begin{eqnarray}
\label{Yang equation_E}
\partial_{\overline{z}}(J^{-1}\partial_{z}J)+
\partial_{\overline{w}}(J^{-1}\partial_{w}J) =0.
\end{eqnarray}
Except for the advantage of gauge independence, the Yang equation is not limited to the situation of $G=\mathrm{SL(2,\mathbb{C})}$ gauge theory as well. In fact, it provide an equivalent description for all $G=\mathrm{GL}(N, \mathbb{C})$ ASD gauge fields. 
Here $J$ is in general an $N \times N$ matrix, called  Yang's $J$-matrix, and the ASD gauge fields can be formulated by the information of Yang's $J$-matrix. 
As a further study of \cite{CoFaYaGo1, CoFaYaGo2}, the authors of \cite{BrFaNuYa} generalize the B\"acklund transformations to a version that turn $G=\mathrm{SU}(N, M)$ gauge fields into $G=\mathrm{SU}(N-1, M+1)$ gauge fields and vice versa. 
Although the study of solutions of the ASDYM equations are almost discussed in the Euclidean signature, there are still some related researches based on the Minkowski signature or the split signature, and worth pondering. (e.g. \cite{ChShYe, deVega, Mason}.)
  
Under the dimensional reduction with respect to one spatial coordinate, 
the ASDYM equations are reduced to the Bogomol'nyi equations \cite{Bogomolny} which are known for possessing the BPS monopoles \cite{Bogomolny, PrSo} as exact solutions in the Yang-Mills-Higgs system. Similar to the instantons \cite{BePoScTy, JaNoRe}, the monopoles \cite{Polyakov, tHt1} can be studied systematically by the Atiyah-Drinfeld-Hitchin-Manin-Nahm construction (\cite{Hitchin, Nahm1, Nahm2}) as well. 
Under the dimensional reduction with respect to two spatial coordinates,  
the ASDYM equations can be reduced to various soliton equations in 2-dimensional integrable systems \cite{Ward}. Researches concerning these soliton equations is not only related to the study of mathematical structures or the construction of solutions, but also related to the developments in different fields of physics and applied science. For example, the KdV equation is related to fluid dynamics, the non-linear Schrodinger equation is related to nonlinear optics, the Liouville equation is related to plasma physics, and the Ernst equation (\cite{Ernst1, Ernst2, L.Witten}) (or called the axisymmetric Einstein equation) is related to axisymmetric gravitational field. 

As a bridge between mathematics and physics, systematic construction of soliton solutions in integrable systems has been one of the most attractive topic for mathematical physicist.
The term "soliton" can be found extensively in different research fields all above. 
Unfortunately, there is no consensus about the definition of soliton until now.  
In many cases, soliton is nearly equivalent to some kind of solutions of nonlinear PDEs that is stable with particle-like behavior. For some fields, researchers even use “soliton-type” to describe certain specific properties of solutions just because of a lack of better names. 

Here we mention three properties (Cf: \cite{Drazin, DrJo}) of solitons that are recognized by most researchers from the background in integrable systems, applied mathematics, and nonlinear science.
\begin{enumerate}[(1)]
\item {\bf Property 1} : 
\label{Property 1}\\
Without collision, individual soliton is of permanent form with constant velocity and amplitude over time. More precisely, the distribution of soliton is in the form of a traveling wave $u(x_j, t)=U(X)$, $X:=\kappa_j \cdot x_j - \omega t + \delta$, for some real constant $\kappa_j$, $\omega$, and $\delta$.  
\item {\bf Property 2} : 
\label{Property 2}\\
Without collision, individual soliton is localized within a specific region, so that it decays  to a constant at infinity. 
More precisely,  $U(X)$ satisfies the boundary condition
\begin{eqnarray}
\frac{d^{m}}{d X^{m}}U(X) \longrightarrow 0 ~\mbox{as}~ X \rightarrow \pm \infty, ~\mbox{for all positive integers $m$}.
\end{eqnarray}
\item {\bf Property 3} : 
\label{Property 3}\\
After soliton collisions, the velocity, amplitude, and shape of each soliton are preserved. The only difference for each soliton is a position shift, called the phase shift. 
More precisely, assume that there exists a function 
\begin{eqnarray}
u(x_j, t)=U(X_1, X_2, ..., X_n), ~X_i:=\kappa_j^{(i)} \cdot x_j - \omega^{(i)} t + \delta^{(i)}
\end{eqnarray}
describing the distribution of the $n$ soliton collisions. 
Then for any given $I \in \left\{1,2,...,n \right\}$ such that 
\begin{eqnarray}
\left\{
\begin{array}{l}
X_{I} ~\mbox{is a finite real number}  \\
X_{i, i \neq I} ~\rightarrow  \pm \infty ~~\mbox{or}~ \mp \infty
\end{array}
\right.
~~~\mbox{as} ~~ t \rightarrow \pm \infty,   \nonumber 
\end{eqnarray}
we have
\begin{eqnarray}
U(X_1, X_2, ..., X_n) ~ \stackrel{t \rightarrow \pm \infty}{\longrightarrow}~
U_{I}(X_I + \Delta_{I}^{(\pm)}),  
\end{eqnarray}
where $\Delta_{I}^{(\pm)}$ is called the phase shift of the $I$-th soliton and it depends on $2^{n-1}$ different asymptotic regions.
\end{enumerate}
Let us take (1+1)-dimensional integrable systems for example, the KdV equation
\begin{eqnarray}
-4u_{t}+6uu_{x}+u_{xxx}=0
\end{eqnarray}
is particularly notable as a typical example of the exactly solvable models. 
One kind of the exact solutions of the KdV equation is known for the 1-soliton solution (Cf: \cite{Dicky, Drazin, DrJo, Hirota, MaSa, Toda1, Toda2}) which 
is in the form of
\begin{eqnarray}
u(x,t)
:=2\frac{\partial^{2}}{\partial x^{2}}(\log\mbox{cosh}X)
=2\kappa^{2}\mbox{sech}^{2}X,~~X:=\kappa x+\kappa^{3}t + \delta .
\end{eqnarray}
First of all, $u(x,t)$ satisfies the above Property \ref{Property 1} since it is a travelling wave solution with a constant velocity $-\kappa^{2}$ and a constant amplitude $2k^2$.
Therefore it preserves its shape over time.
Secondly, $u(x,t)$ is an even function symmetric with respect to $X=0$ and $u, u_x, u_{xx}, u_{xxx} \rightarrow 0$ as $x \rightarrow \pm \infty$. Therefore, the distribution of $u(x,t)$ is localized within a region centered on $X=0$ and satisfies the above Property \ref{Property 2}. On the other hand, the KdV equation is also known for having infinitely conserved densities \cite{MiGaKr} (Cf: \cite{Dicky, Drazin, DrJo, Toda1}). More precisely, these conserved quantities are
the mass $\int u \mbox{d}x$, the momentum $\int u^{2} \mbox{d}x$, 
the energy $\int \left[ 2u^3 -(u_{x})^2 \right]\mbox{d}x$, ..., and so on. 
In particular, the energy density 
\begin{eqnarray}
\label{energy density}
2u^3 -(u_{x})^2=16\kappa^{6}\left( 2\mbox{sech}^6X-\mbox{sech}^4X \right), ~~
X:=\kappa x+\kappa^{3}t + \delta
\end{eqnarray}
possesses the behavior of 1-soliton as well since the distribution is localized within a region centered on $X=0$ and it is of permanent form over time.
By introducing one more spatial coordinate $y$, the KdV equation can be generalized to
the (2+1)-dimensional version, that is, the KP equation
\begin{eqnarray}
(-4u_{t}+6uu_{x}+u_{xxx})_x+3u_{yy}=0.
\end{eqnarray}
The KP 1-soliton solution (Cf: \cite{Kodama1, Kodama2,Toda2}) is in the form of 
\begin{eqnarray}
u(x,y,t)
&\!\!\!\!=& \!\!\!\! U(X)
:=2\frac{\partial^{2}}{\partial x^{2}}(\log\mbox{cosh}X) \nonumber \\
&\!\!\!\!=& \!\!\!\! \frac{(\kappa_{1}-\kappa_{2})^{2}}{2}\mbox{sech}^{2}X,~~
X:=\frac{\kappa_{1}-\kappa_{2}}{2} x + \frac{\kappa_{1}^{2}-\kappa_{2}^{2}}{2}y + \frac{\kappa_{1}^{3}-\kappa_{2}^{3}}{2}t + \frac{\delta_1-\delta_2}{2} ~~~~~~~~
\end{eqnarray}
which reduces to the KdV 1-soliton solution if the $y$-dependence is lost  ($i.e.$ $\kappa_{1}=-\kappa_{2}$).

The $n$-soliton solutions (Cf: \cite{Hirota, Kodama1, Kodama2, MaSa, Toda2}) of the KP equation is well-known as 
\begin{eqnarray}
u(x,y,t)
=
U(X_1, X_2, ..., X_n)
:=
2\frac{\partial^{2}}{\partial x^{2}}(\log\tau_{n}), ~~ 
\end{eqnarray}
where the function $\tau_{n}$ is defined by the Wronskian type determinant
\begin{eqnarray}
\tau_{n}
&\!\!\!\!:=& \!\!\!\!
\mbox{Wr}(f_1, f_2, ..., f_n)  
:=
\left|
\begin{array}{ccccc}
f_1^{(0)} & f_2^{(0)} & \cdots & f_n^{(0)} \\
f_1^{(1)} & f_2^{(1)} & \cdots & f_n^{(1)} \\
\vdots & \vdots & \ddots & \vdots \\
f_1^{(n-1)} & f_2^{(n-1)} & \cdots & f_n^{(n-1)}
\end{array}
\right|, ~ f_i^{(m)}:=\frac{\partial^{m}f_i}{\partial x^{m}}, ~~~~~~~~  \\
f_i&\!\!\!\!:=& \!\!\!
\mbox{cosh}X_i,~~
X_i:=\frac{\kappa_{i1}-\kappa_{i2}}{2} x + \frac{\kappa_{i1}^{2}-\kappa_{i2}^{2}}{2}y + \frac{\kappa_{i1}^{3}-\kappa_{i2}^{3}}{2}t + \frac{\delta_{i1}-\delta_{i2}}{2}.
\end{eqnarray}
The asymptotic behavior of the KP $n$-soliton ($w.r.t.$ a comoving frame related to the $I$-th KP soliton) can be proved as (Cf: Appendix \ref{Appendix A})
\begin{eqnarray}
\label{Phase shift of KP}
U(X_1, X_2, ..., X_n) ~ \stackrel{t \rightarrow \pm \infty}{\longrightarrow}~
\frac{(\kappa_{I1}-\kappa_{I2})^{2}}{2}\mbox{sech}^{2}\left( X_I + \Delta_I^{(\pm)} \right)
=U_{I}\left(X_I+\Delta_I^{(\pm)}\right),
\end{eqnarray}
where $\Delta_I^{(\pm)}$ is the phase shift of the $I$-th KP soliton and it depends on $2^{n-1}$ different asymptotic regions. 
Obviously, \eqref{Phase shift of KP} satisfies the requirement of the above Property \ref{Property 3}.
In fact, the behavior of the KP multi-soliton scattering is classified in detail by the authors of \cite{Kodama1, Kodama2, KoWi}. 
On the other hand, 
the stability of multi-solitons are closely related to the existence of infinitely many
conserved quantities which lead to an infinite dimensional symmetry of the integrable
systems. Among these, the Sato's theory \cite{Sato1, Sato2} is one of the most appealing result which reveals an infinite dimensional symmetry behind the KP equation and gives a
comprehensive viewpoint to unify the theory of lower-dimensional integrable systems.

Inspired by the KdV and KP solitons in the (1+1)-dimensional and (2+1)-dimensional integrable system respectively, it is natural to ask a question whether the ASDYM equations possess such typical solitons on 4-dimensional spaces or not. To deal with this problem, 
we simply consider the Lagrangian density Tr$F_{\mu\nu}F^{\mu\nu}$ of the ASDYM solution as an analogue of the energy density \eqref{energy density}. (Cf: conserved densities of the KP soliton \cite{Matsat}.)
If the Lagrangian density of the ASDYM solution satisfies the above Property \ref{Property 1}, \ref{Property 2} and \ref{Property 3}, we call such solution as ASDYM soliton solution. Here we give a more precise definition in the following paragraph.

Let $(x^{1}, x^{2}, x^{3}, x^{4})$ be the 4-dimensional space-time coordinate. We consider the $J$-matrices $J(x^{1}, x^{2}, x^{3}, x^{4})$, the solutions of the Yang equation, as the equivalent solutions of the ASDYM equations. Then 
\begin{enumerate}[(1)]
	\item {\bf ASDYM 1-Soliton}  \\
	We call a $J$-matrix to be an ASDYM 1-soliton solution 
	if the resulting Lagrangian density  Tr$F_{\mu\nu}F^{\mu\nu}$ satisfies the following  condition :
	\begin{itemize}
	\item  
	The Lagrangian density is in the form of a polynominal in terms of the real hyperbolic function sech$X$, more precisely, 
	\begin{eqnarray}
	U(X):=\sum\limits_{k=0}^{m}c_k~\!\mbox{sech}^{k}X \in \mathbb{R} \label{Def_ASDYM 1-Soliton},
	\end{eqnarray} 
	where $X=\ell_1x^{1}+\ell_2x^{2}+\ell_3x^{3}+\ell_4x^{4}+\ell_5$ ~is a nonhomogeneous linear function of the space-time coordinate.
	\end{itemize}  
	Here we define the principal peak of the Lagrangian density to be the peak or antipeak located on $X=0$ for convenience. Note that the principal peak need not to attain the absolute extreme value of the Lagrangian density.
	\item  {\bf ASDYM Multi-Soliton} \label{Def_ASDYM Multi-Soliton}  \\
	Let $J_i(x^{1}, x^{2}, x^{3}, x^{4})$, $i=1, 2, ...,n$ be $n$ different ASDYM 1-soliton solutions with the resulting Lagrangian densities Tr$F_{\mu\nu}F^{\mu\nu~\!(i)}$ are
	\begin{eqnarray}
	U_i(X_i):=\sum\limits_{k=0}^{m}c_k^{(i)}~\!\mbox{sech}^{k}X_{i},~~
	X_{i}=\ell_1^{(i)}x^{1}+\ell_2^{(i)}x^{2}+\ell_3^{(i)}x^{3}+\ell_4^{(i)}x^{4}+\ell_5^{(i)}.  \nonumber  
	\end{eqnarray}
	Assume that there exists a $J$ matrix $\widetilde{J}=\widetilde{J}(x^{1}, x^{2}, x^{3}, x^{4})$ such that the resulting Lagrangian density is a real function $U(X_1, X_2, ..., X_n)$ defined by $X_i$, $i=1, 2, ...n$.
	Then we call $\widetilde{J}(x^{1}, x^{2}, x^{3}, x^{4})$ to be an ASDYM $n$-soliton solution if the Lagrangian density $U(X_1, X_2, ..., X_n)$ satisfies the following asymptotic behavior :
	\begin{itemize}
		\item Let $r$ be defined by $r:=\sqrt{(x^1)^{2}+(x^2)^{2}+(x^3)^{2}+(x^4)^{2}}$.
		Then for any given $I \in \left\{1,2,...,n \right\}$ such that 
		\begin{eqnarray}
		\left\{
		\begin{array}{l}
		X_{I} ~\mbox{is a finite real number}  \\
		X_{i} ~\rightarrow  \pm \infty
		\end{array}
		\right.
		~~~\mbox{as} ~~ r \rightarrow \infty,   \nonumber 
		\end{eqnarray}
		we have
		\begin{eqnarray}
		U(X_1, X_2, ..., X_n) ~ \stackrel{r \rightarrow \infty}{\longrightarrow}~
		U_{I}(X_I + \Delta_{I}), ~~ 
		\end{eqnarray}
where $\Delta_I$ is called the phase shift of the $I$-th ASDYM 1-soliton and it depends on $2^{n-1}$ different asymptotic regions. (Cf: \eqref{Phase shift_G=SU(2)_U}, \eqref{Phase shift_G=SL(2,C)_E}, and \eqref{Phase shift_G=SU(3)_U}.) 
	\end{itemize} 
\end{enumerate}

Now let us consider the 't Hooft ansatz \cite{CoFa, tHt3, Wilczek}
\begin{eqnarray}
A_{\mu}=\frac{i}{2}\eta_{\mu\nu}^{(+)}\partial_{\nu}\log \varphi, ~\mbox{where}~
\left\{
\begin{array}{l}
\mbox{$\varphi$ ~\!satisfies the Laplace equation ~\!$\partial_{\mu}^2\varphi=0$} \\
\mbox{$\eta_{\mu \nu}^{(\pm)}$ is the 't Hooft matrices \cite{tHt2}}
\end{array},
\right.
\end{eqnarray}
and use it to construct the ASDYM 1-soliton. A simple candidate of $\varphi$ can be taken by
\begin{eqnarray}
	\label{'t Hooft ansatz_phi}
	\varphi= \frac{1}{2}(e^{X}+e^{-X}) = {\mbox{cosh}}X,~~~
	X:=k_\mu x^\mu,
\end{eqnarray}
where $k_\mu$ are real constants satisfying $k^2=k_\mu k^\mu=0$ due to  the Laplace equation.
By using some formulas on the 't Hooft symbol \cite{tHt2},
we can show that
\begin{eqnarray}
	\mbox{Tr}F_{\mu\nu}F^{\mu\nu} = -3 (k^2)^2 (4{\mbox{sech}}^4X -5{\mbox{sech}}^2X+2)
	\stackrel{k^2=0}{=}0.
\end{eqnarray}
Unfortunately, the Lagrangian density vanishes and hence it is not the interesting result we are seeking. You will see it soon in Section \ref{Section 5}, \ref{Section 6}, and \ref{Section 7}. We construct a class of ASDYM 1-solitons \cite{GiHaHuNi, HaHu1} with the resulting Lagrangian densities (Cf: \eqref{Reduced Lagrangian density_U}, \eqref{Reduced Lagrangian density_E}, \eqref{Reduced Lagrangian density_M}, and \ref{Reduced Lagrangian density_U_G=SU(3)}) are in the form of
\begin{eqnarray}
\mbox{Tr}F_{\mu\nu}F^{\mu\nu} \propto 2\mbox{sech}^2X-3\mbox{sech}^4X,
\end{eqnarray}
and the ASDYM multi-solitons \cite{HaHu2} possess quite similar features as the KP multi-solitons (Cf: Appendix \ref{Appendix A}). 
However, they are different type from the already known solitons.
\bigskip \\
{\bf Organization of this thesis}
\medskip \\
In Section \ref{Section 2}, we review some necessary knowledge of quasideterminants \cite{GeRe} which is required for the discussion in Section \ref{Section 4}, Section \ref{Section 6}, and Section \ref{Section 7}. The quasideterminants can be considered roughly as a noncommutative generalization of determinants. More than the meaning of mathematical generalization, the quasideterminants are naturally fit for the description of noncommutative integrable systems (e.g. \cite{EtGeRe, GiNi, GiNiSo}) . We introduce some elementary operation rules of quasideterminants, the noncommutative version of Jacobi identity, the homological relation, and a derivative formula of quasideterminants as useful mathematical tools in this section.    

In Section \ref{Section 3}, we review some necessary knowledge of the ASDYM theory mainly from the perspective of integrable systems. We consider the ASDYM theory in 4-dimensional complex space and introduce several equivalent descriptions of the ASDYM equation. Especially, the anti-self-dual (ASD) gauge fields can be expressed in terms of the $J$-matrices (\cite{BrFaNuYa}, Cf: \cite{MaWo}), solutions of the Yang equation. Combining the $J$-matrix formulation with the Lax representation of ASDYM, it forms almost the theoretical foundation of our solutions in this thesis.

In Section \ref{Section 4}, we introduce a Darboux transformation which is introduced firstly by Nimmo, Gilson, and Ohta \cite{NiGiOh}. After $n$ iterations of the Darboux transformation, the resulting solutions ($J$-matrix) can be expressed beautifully in terms of the Wronskian type quasideterminants of order $n+1$. More precisely, each element of the Wronskian type quasideterminant is a ratio of ordinary Wronskian determinants. We call them quasi-Wronskian for short in this thesis. This section is written mainly based on the following sub-dissertation :
\begin{itemize}
	\item 
	C.~R.~Gilson, M.~Hamanaka, \underline{S.C. Huang} and J.~J.~C.~Nimmo, \\
	"Soliton Solutions of Noncommutative Anti-Self-Dual Yang-Mills Equations," \\
	\href{https://doi.org/10.1088/1751-8121/aba72e}{\color{blue} Journal of Physics A: Mathematical and Theoretical {\bf 53}, 404002(17pp) (2020)}. \\
	~\href{https://arxiv.org/abs/2004.01718}{\color{blue}[arXiv:2004.01718]}. 
\end{itemize}
Some new results are mentioned in Appendix \ref{Appendix B} (Cf: Proposition \ref{Prop_B1} and Theorem \ref{Thm_B2}).

In Section \ref{Section 5}, we construct the ASDYM 1-solitons by applying 1 iteration of the Darboux transformation.
Firstly, we discuss general cases in 4-dimensional complex space and then impose some conditions 
to obtain the ASDYM 1-solitons on real spaces with the Euclidean signature $(+,+,+,+)$, the Minkowski signature $(+,-,-,-)$, and the split signature $(+,+,-,-)$ (Ultrahyperbolic space). 
In particular, the principal peak of the Lagrangian density is localized on a 3-dimensional hyperplane in 4-dimensional space. Therefore, we use the term "soliton walls" to distinguish them from the domain walls because the domain walls are described by scalar fields. Furthermore, we propose ansatzes on three kinds of signature to obtain real-valued Lagrangian density even if the gauge group is non-unitary. For the case of split signature, we show that the gauge group can be $G=\mathrm{SU}(2)$ and hence the soliton walls could be the candidates of physically interesting results on the Ultrahyperbolic space $\mathbb{U}$. This section is written based on the following sub-dissertation with some revisions and supplements.
\begin{itemize}
	\item 
	M.~Hamanaka and \underline{S.C. Huang}, \\
	"New Soliton Solutions of Anti-Self-Dual Yang-Mills Equations,"\\
	\href{https://doi.org/10.1007/JHEP10(2020)101}{\color{blue}Journal of High Energy Physics \textbf{10}, 101 (2020)}. 
	\href{https://arxiv.org/abs/2004.09248}{\color{blue}[arXiv:2004.09248]}.
\end{itemize} 

In Section \ref{Section 6}, we construct a special class of ASDYM $n$-solitons by applying $n$ iterations of the Darboux transformation. The resulting $n$-solitons are in the form of the quasi-Wronskian. Furthermore, we use the techniques of quasideterminants to show that in the asymptotic region, the $n$-soliton possesses $n$ isolated distribution of Lagrangian densities (with phase shifts). Therefore, we can interpret it as $n$ intersecting soliton walls. We calculate the phase shift factors explicitly and find that the Lagrangian densities can be real-valued for three kinds of signature. Especially for the split signature, we show that the gauge group can be $G=\mathrm{SU}(2)$ and hence the intersecting soliton walls could be the candidates of physically interesting results on the Ultrahyperbolic space $\mathbb{U}$. This section is written based on the following sub-dissertation (submitted to a journal for publication) and some revisions.
\begin{itemize}
	\item
	M.~Hamanaka and \underline{S.C. Huang}, \\
	"Multi-Soliton Dynamics of Anti-Self-Dual Gauge Fields",  \\	
	~\href{https://arxiv.org/abs/2106.01353}{\color{blue}[arXiv:2106.01353]}.
\end{itemize}

In Section \ref{Section 7}, we construct an example of $G=\mathrm{SU}(3)$ 1-soliton on the Ultrahyperbolic space and it can be interpreted as soliton wall as well.  After applying $n$ iterations of the Darboux transformation, we show that the resulting $n$-solitons can be interpreted as $n$ intersecting soliton walls as well. As for the gauge group, it can be $G=\mathrm{SU}(3)$ for each soliton wall in the asymptotic region. This section is written based on some unpublished results.

\newpage

\section{Quasideterminant}
\label{Section 2}

In this section, we review some basic properties of the
quasideterminant which was firstly introduced by Gelfand and Retakh \cite{GeRe}. For the purpose of this thesis, we would not pay too much attention to unnecessary mathematical structure, 
but using the quasideterminant for the benefit of studying in non-abelian integrable systems (e.g. \cite{EtGeRe, GiNi, GiNiSo}). 
As a researcher from the background of mathematical physics, we give a self-contained introduction of the quasideterminant and derive all the theorems (except for Theorem 2.9) in a more straightforward and intuitive way. It would be a readable review note for the non-professionals .
If the readers prefer using more appropriately mathematical terminology and are longing to study a more comprehensive theory of quasideterminant, further details can be referred to \cite{GeGeReWi}.
\newtheorem{defn_2.1}{Definition}[section]
\begin{defn_2.1}
	{\bf (Quasideterminant of order $n$ \cite{GeRe}) \label{Def_2.1}} \\
Let $X=(x_{i,j})_{n \times n}$ be an $n \times n$ matrix over a noncommutative ring $\mathcal{R}$.
Then the $(i,j)$-th quasideterminant of matrix $X$, denoted by
\begin{eqnarray}
\label{quasideterminant}
\left| X \right|_{ij}
=
\left|
\begin{array}{ccc:c:ccc}
x_{1,1} & \!\!\!\cdots & \!\!\!x_{1,j-1} & x_{1,j} & x_{1,j+1} & \!\!\!\cdots & \!\!\!x_{1 n} \\
\vdots &  \!\!\!\ddots & \!\!\!\vdots & \vdots & \vdots & \!\!\!\ddots & \!\!\!\vdots \\
x_{i-1,1} & \!\!\!\cdots & \!\!\!x_{i-1, j-1} & x_{i-1,j} & x_{i-1,j+1} & \!\!\!\cdots & \!\!\!x_{i-1, n} \\ \hdashline
x_{i,1} & \!\!\!\cdots & \!\!\!x_{i, j-1} & \fbox{$x_{i,j}$} & x_{i,j+1} & \!\!\!\cdots & \!\!\!x_{i, n} \\ \hdashline
x_{i+1,1} & \!\!\!\cdots & \!\!\!x_{i+1, j-1} & x_{i+1,j} & x_{i+1,j+1} & \!\!\!\cdots & \!\!\!x_{i+1, n} \\
\vdots &  \!\!\!\ddots & \!\!\!\vdots & \vdots & \vdots & \!\!\!\ddots & \!\!\!\vdots  \\
x_{n,1} & \!\!\!\cdots & \!\!\!x_{n,j-1} & x_{n,j} & x_{n,j+1} & \!\!\!\cdots & \!\!\!x_{n,n}
\end{array}\right|~,
\end{eqnarray}
is defined by 
\begin{eqnarray}
\label{quasideterminant_definition}
\left| X \right|_{ij}
:= x_{i,j} - R_{i,~\!\widehat{j}} \left(X_{\widehat{i},~\! \widehat{j}} \right)^{-1}C_{~\!\widehat{i},~\!j},
\end{eqnarray}
where $X_{\widehat{i},~\! \widehat{j}}$ denotes the $(n-1) \times (n-1)$ submatrix obtained from the matrix $X$ by deleting the $i$-th row and the $j$-th column of $X$.  $R_{i,~\!\widehat{j}}:=(x_{i,1}, \cdots, x_{i, j-1}, x_{i,j+1}, \cdots, x_{i, n})$, and $C_{~\!\widehat{i},~\!j}:=(x_{1,j}, \cdots, x_{i-1,j}, x_{i+1,j}, \cdots, x_{n,j})^{T}$ denote the submatrices obtained from the $i$-th row and the $j$-th column of $X$ by deleting the element $x_{i,j}$, respectively, and we assume that $X_{\widehat{i},~\! \widehat{j}}$ is invertible over $\mathcal{R}$. Here the symbol $\hat{i}$ means that we delete $i$ from the row or column indices of $X$.  
\end{defn_2.1}
In fact, the $i$-th row and the $j$-th column of $X$ passing through the box element $x_{ij}$ can be moved to any row and any column of $X$, and the quasideterminant remains unchanged by definition \eqref{quasideterminant_definition}. This fact implies an equivalent and convenient representation for quasideterminant   
\bigskip\\ 
{\bf The canonical form of quasideterminant} 
\begin{eqnarray}
\label{canonical form}
\left| X \right|_{ij}=\left|
\begin{array}{cccccc:c}
x_{1,1} & \!\!\!\cdots & \!\!\!x_{1,j-1} & x_{1,j+1} & \!\!\!\cdots & \!\!\!x_{1 n} & x_{1,j} \\
\vdots &  \!\!\!\ddots & \!\!\!\vdots & \vdots & \!\!\!\vdots & \!\!\!\ddots & \vdots \\
x_{i-1,1} & \!\!\!\cdots & \!\!\!x_{i-1, j-1} & x_{i-1,j+1} & \!\!\!\cdots & \!\!\!x_{i-1, n} & x_{i-1,j}  \\
x_{i+1,1} & \!\!\!\cdots & \!\!\!x_{i+1, j-1} & x_{i+1,j+1} & \!\!\!\cdots & \!\!\!x_{i+1, n} & x_{i+1,j} \\
\vdots &  \!\!\!\ddots & \!\!\!\vdots & \vdots & \!\!\!\vdots & \!\!\!\ddots & \vdots  \\
x_{n,1} & \!\!\!\cdots & \!\!\!x_{n,j-1} & x_{n,j+1} & \!\!\!\cdots & \!\!\!x_{n,n} & x_{n,j} \\ \hdashline
x_{i,1} & \!\!\!\cdots & \!\!\!x_{i, j-1} & x_{i,j+1} & \!\!\!\cdots & \!\!\!x_{i, n} & \fbox{$x_{i,j}$}
\end{array}\right|
=\left| \begin{array}{cc}
X_{\widehat{i},~\! \widehat{j}} & C_{~\!\widehat{i},~\!j} \\
R_{i,~\!\widehat{j}} & \fbox{$x_{i,j}$}
\end{array}\right|.
\end{eqnarray} 
On the other hand the above matrix, permutation of rows and columns of $X=(x_{ij})_{n \times n}$, can be redefined by a new matrix $Y=(y_{kl})_{n \times n}$. Therefore, we can rewrite \eqref{canonical form} by
\begin{eqnarray}
\label{canonical form_Y_nn}
\left| X \right|_{ij}
=
\left|
\begin{array}{ccccc:c}
y_{1,1} & \cdots & y_{1, k} & \cdots & y_{1, n-1} & y_{1, n} \\
\vdots  & \ddots & \vdots   & \ddots & \vdots     & \vdots   \\
y_{l,1} & \cdots & y_{l, k} & \cdots & y_{l, n-1} & y_{l, n} \\
\vdots  & \ddots & \vdots   & \ddots & \vdots     & \vdots   \\
\hdashline
y_{n,1} & \cdots & y_{n, k} & \cdots & y_{n, n-1} & \fbox{$y_{n, n}$}.
\end{array}
\right|
=
\left| Y \right|_{nn}
\end{eqnarray}
in which $X=(x_{ij})_{n \times n}$, $Y=(y_{kl})_{n \times n}$ satisfy the following relation
\begin{eqnarray}
\label{Y=PCP^{T}}
Y=P_{\pi}XP_{\sigma}^{T},
\end{eqnarray} 
where $P_{\pi}$, $P_{\sigma}$ denote the permutation matrices obtained by permuting the columns of the identity matrix $I_{n}$ $w.r.t.$ the following permutations :
\begin{eqnarray}
\pi&\!\!\!:=\!\!\!&
\left(
\begin{array}{cccccccc}
1 & \cdots & j-1 &  j  & j+1 & \cdots & n-1 & n \\
1 & \cdots & j-1 & j+1 & j+2 & \cdots & n   & j
\end{array}
\right),  \\
\sigma&\!\!\!:=\!\!\!&
\left(
\begin{array}{cccccccc}
1 & \cdots & i-1 &  i  & i+1 & \cdots & n-1 & n \\
1 & \cdots & i-1 & i+1 & i+2 & \cdots & n   & i
\end{array}
\right).
\end{eqnarray}
More precisely,
\begin{eqnarray}
\label{Permutation matrix}
P_{\pi}=
\left(
\begin{array}{c}
e_1 \\
\vdots \\
e_{j-1} \\
e_{j+1} \\
\vdots \\
e_{n} \\
e_{j}
\end{array}
\right),~~
P_{\sigma}=
\left(
\begin{array}{c}
e_1 \\
\vdots \\
e_{i-1} \\
e_{i+1} \\
\vdots \\
e_{n} \\
e_{i}
\end{array}
\right),
\end{eqnarray}
where $e_{k}$ denotes the row vector with 1 in the $k$-th component, and 0 otherwise.
\newtheorem{prop_2.2}[defn_2.1]{Proposition}
\begin{prop_2.2}
	{\bf \cite{GeRe}} \label{Prop_2.2} \\
Let $X=(x_{ij})_{n \times n}$ be an $n\times n$ invertible matrix 
over a noncommutative ring $\mathcal{R}$. 
Then the $(i,j)$-th quasideterminant of $X$ can be represented
as the inverse of $(j,i)$-th element of $X^{-1}$, that is, $\vert X \vert_{ij}^{-1}=(X^{-1})_{ji}$
whenever both sides make sense.
\end{prop_2.2}
\!\!\!\!{\bf (\emph{Proof})}
\smallskip \\ 
Let $Y=(y_{kl})_{n \times n}$ be the matrix defined in \eqref{canonical form_Y_nn}. 
Then by the relation \eqref{Y=PCP^{T}} and the property of permutation matrix, we have
\begin{eqnarray}
X^{-1}=P_{\sigma}^{T}Y^{-1}P_{\pi}=P_{\sigma^{-1}}Y^{-1}P_{\pi^{-1}}^{T} ~~
\Longrightarrow ~~(X^{-1})_{ji} = (Y^{-1})_{nn}.
\end{eqnarray}
Now it suffices for us to show that $(Y^{-1})_{nn}=\left| X \right|_{ij}.$
For convenience, we assign $X_{\widehat{i},\widehat{j}}$, $C_{\widehat{i},j}$, $R_{i,\widehat{j}}$, and $x_{ij}$ in \eqref{canonical form} to be $A$, $B$, $C$, and $d$, respectively. 
Then the inverse matrix formula for $2\times 2$ block matrix
\begin{eqnarray}
	\label{Inverse matrix formula}
	\left(
	\begin{array}{cc}
		A&B \\C&d
	\end{array}
	\right)^{-1}
	=\left(\begin{array}{cc}
		A^{-1}+A^{-1} B S^{-1} C  A^{-1}
		&-A^{-1} B S^{-1}\\
		-S^{-1} C A^{-1}
		&S^{-1}
	\end{array}\right),~ S:=d-C A^{-1} B ~~
\end{eqnarray}
implies that $(Y^{-1})_{nn}=S^{-1}=(d-CA^{-1}B)^{-1}=\left| X \right|_{ij}^{-1}$ immediately.
$\hfill\Box$
\medskip\\
Applying Proposition \ref{Prop_2.2} to \eqref{quasideterminant_definition}, we have
\begin{eqnarray}
\label{Quasideterminant_expansion}
\vert X \vert_{ij}=x_{ij}-\sum_{i^\prime (\neq i), j^\prime (\neq j)}
x_{ii^\prime}  (\vert X_{\widehat{i},~\! \widehat{j}}\vert_{j^\prime i^\prime })^{-1}
x_{j^\prime j}.
\end{eqnarray}
{\bf{ Examples}}
\medskip \\
{\bf (1) } For a $2\times 2$ matrix $X=(x_{ij})_{2 \times 2}$ :
\begin{eqnarray*}
	\vert X \vert_{11}=
	\begin{vmatrix}
		\fbox{$x_{11}$} &x_{12} \\x_{21}&x_{22}
	\end{vmatrix}
	=x_{11}-x_{12} x_{22}^{-1} x_{21},~~~
	\vert X \vert_{12}=
	\begin{vmatrix}
		x_{11} & \fbox{$x_{12}$} \\x_{21}&x_{22}
	\end{vmatrix}
	=x_{12}-x_{11} x_{21}^{-1} x_{22},\nonumber\\
	\vert X \vert_{21}=
	\begin{vmatrix}
		x_{11} &x_{12} \\ \fbox{$x_{21}$}&x_{22}
	\end{vmatrix}
	=x_{21}-x_{22} x_{12}^{-1} x_{11},~~~
	\vert X \vert_{22}=
	\begin{vmatrix}
		x_{11} & x_{12} \\x_{21}&\fbox{$x_{22}$}
	\end{vmatrix}
	=x_{22}-x_{21} x_{11}^{-1} x_{12}.
\end{eqnarray*}
{\bf (2) } For a $3\times 3$ matrix $X=(x_{ij})_{3 \times 3}$ :
\begin{eqnarray*}
	\vert X \vert_{11}
	&=&
	\begin{vmatrix}
		\fbox{$x_{11}$} &x_{12} &x_{13}\\ x_{21}&x_{22}&x_{23}\\x_{31}&x_{32}&x_{33}
	\end{vmatrix}
	=x_{11}-(x_{12}, x_{13}) \left(
	\begin{array}{cc}x_{22} & x_{23} \\x_{32}&x_{33}\end{array}\right)^{-1}
	\left(
	\begin{array}{c}x_{21} \\x_{31}\end{array}
	\right)
	\nonumber\\
	&=&x_{11}-x_{12}  \begin{vmatrix}
		\fbox{$x_{22}$} & x_{23} \\x_{32}&x_{33}
	\end{vmatrix}^{-1}   x_{21}
	-x_{12} \begin{vmatrix}
		x_{22} & x_{23} \\\fbox{$x_{32}$}&x_{33}
	\end{vmatrix}^{-1}  x_{31}      \nonumber\\
	&&~~~~    -x_{13} \begin{vmatrix}
		x_{22} & \fbox{$x_{23}$} \\x_{32}&x_{33}
	\end{vmatrix}^{-1}  x_{21}
	-x_{13} \begin{vmatrix}
		x_{22} & x_{23} \\x_{32}&\fbox{$x_{33}$}
	\end{vmatrix}^{-1}  x_{31},  ~~\mbox{and so on. }
\end{eqnarray*}
{\bf Commutative limit of quasideterminant}   
\medskip \\
If $n \times n$ matrix $X$ is defined over a commutative ring, we have
\begin{eqnarray}
|X|_{ij}^{-1}=(X^{-1})_{ji}=(-1)^{i+j}\frac{\mbox{det}X_{\widehat{i}, \widehat{j}}}{\mbox{det}X}
\end{eqnarray}
whenever all the terms make sense.
\\

Now let us introduce some elementary operations of quasideterminant that are quite similar to those of determinants, but not exactly the same. Without loss of generality, we can always consider the canonical form \eqref{canonical form} and relabel the indices to get a more convenient expression :
\begin{eqnarray}	
\label{quasideterminant_reindex}
|X|_{nn}
=
\left|
\begin{array}{cccc:c}
x_{1,1} & x_{1,2} & \cdots & x_{1,n-1} & x_{1,n} \\
\vdots &  \vdots & \ddots & \vdots & \vdots  \\
x_{n-1,1} & x_{n-1,2} & \cdots & x_{n-1,n-1} & x_{n-1,n} \\
\hdashline
x_{n,1} & x_{n,2} & \cdots & x_{n,n-1} & \fbox{$x_{n,n}$}
\end{array}\right|
=
\left|
\begin{array}{cc}
X_{\widehat{n}, \widehat{n}} & C_{\widehat{n}, n} \\
R_{n, \widehat{n}} & \fbox{$x_{n,n}$}
\end{array}
\right|.
\end{eqnarray}
For further simplification, we assign $X_{\widehat{n}, \widehat{n}}$, $C_{\widehat{n}, n}$, $R_{n, \widehat{n}}$, and $x_{n,n}$ to be $A, ~B, ~C$, and $d$, respectively, and use this setting to prove the following Proposition \ref{Prop_2.3}, \ref{Prop_2.4}, and \ref{Prop_2.5}.
\newtheorem{prop_2.3}[defn_2.1]{Proposition}
\begin{prop_2.3}
	{\bf (Permutation rule \cite{GeRe})}  \label{Prop_2.3} \\
The permutation of two columns (or two rows) of a quasideterminant leaves the result unchanged.
\end{prop_2.3}
{\bf (\emph{Proof})} 
\medskip \\ 
Without loss of generality, we consider the following two cases of column permutation. (The proof of row permutation is similar.)
\medskip \\
{\bf Case I : $k$-th column $\longleftrightarrow$ $l$-th column ($k \neq n, ~ l \neq n$.)}
\begin{eqnarray*}
\left|
\begin{array}{cc}
A & B \\
C & \fbox{$d$}
\end{array}
\right|_{k \leftrightarrow l}
=
\left|
\begin{array}{cc}
A P_{\pi} & B \\
C P_{\pi} & \fbox{$d$}
\end{array}
\right|
=
d - CP_{\pi}(AP_{\pi})^{-1}B
=
\left|
\begin{array}{cc}
A & B \\
C & \fbox{$d$}
\end{array}
\right|. \nonumber
\end{eqnarray*}
Here we use the informal notation $k \leftrightarrow l$ to denote the permutation of the $k$-th column and the $l$-th column of quasideterminant. 
$P_{\pi}$ denotes the permutation of columns of the $(n-1) \times (n-1)$ identity matrix $I$ $w.r.t.$ the permutation
\begin{eqnarray*}
\left(
\begin{array}{cccccccccccc}
1 & \cdots & k-1 & k &  k+1 & \cdots & l-1 & l & l+1 & \cdots & n-1 \\
1 & \cdots & k-1 & l &  k+1 & \cdots & l-1 & k & l+1 & \cdots & n-1
\end{array}
\right).   \nonumber 
\end{eqnarray*}
{\bf Case II : $k$-th column $\longleftrightarrow$ $n$-th column}~ (involving the box element $x_{nn}$)
\begin{eqnarray}
\left| X \right|_{nn}
&\!\!\!=\!\!\!&
\left|
\begin{array}{ccccccc:c}
\cdots & \!\!\!\!x_{1,k-1} & x_{1,k} & x_{1,k+1} & \!\!\!\!\cdots & \!\!\!\!x_{1,n-2} & \!\!\!\!x_{1,n-1} & x_{1,n} \\
\ddots & \!\!\!\!\vdots & \vdots & \vdots & \!\!\!\!\ddots & \!\!\!\!\vdots & \!\!\!\!\vdots & \vdots \\
\cdots & \!\!\!\!x_{n-1,k-1} & x_{n-1,k} & x_{n-1,k+1} & \!\!\!\!\cdots & \!\!\!\!x_{n-1,n-2} &\!\!x_{n-1,n-1} & x_{n-1,n} \\
\hdashline
\cdots & \!\!\!\!x_{n,k-1} & x_{n,k} & x_{n,k+1} & \!\!\!\!\cdots & \!\!\!\!x_{n,n-2} & \!\!x_{n,n-1} & \fbox{$x_{n,n}$}
\end{array}\right|  \label{X_nn_1}  \\
&\!\!\!=\!\!\!&
\left|
\begin{array}{ccccccc:c}
\cdots & \!\!\!\!x_{1,k-1} & x_{1,k+1} & \!\!\!\!x_{1,k+2} & \!\!\!\!\cdots & \!\!\!\!x_{1,n-1} & x_{1,k} & x_{1,n} \\
\ddots & \!\!\!\!\vdots &  \vdots & \vdots & \!\!\!\!\ddots & \!\!\!\!\vdots & \vdots & \vdots \\
\cdots & \!\!\!\!x_{n-1,k-1} & x_{n-1,k+1} & \!\!\!\! x_{n-1,k+2} & \!\!\!\!\cdots & \!\!\!\!x_{n-1,n-1} & x_{n-1,k}~ & x_{n-1,n} \\
\hdashline
\cdots & \!\!\!\!x_{n,k-1} & x_{n,k+1} &  x_{n,k+2} &\!\!\!\!\cdots & \!\!\!\!x_{n,n-1} & x_{n,k}~ & \fbox{$x_{n,n}$} 
\end{array}\right|  \label{X_nn_2}  \\  
&\!\!\!=\!\!\!&
\left|
\begin{array}{cc:c:ccccc}
	\cdots & \!\!\!\!x_{1,k-1} & x_{1,n} & x_{1,k+1} & \!\!\!\!\cdots & \!\!\!\!x_{1,n-2} &\!\!\!\!x_{1,n-1} & \!\!\!\!x_{1,k} \\
	\ddots & \!\!\!\!\vdots & \ddots & \vdots & \ddots & \!\!\!\!\vdots & \!\!\!\!\vdots  & \!\!\!\!\vdots\\
	\cdots & \!\!\!\!x_{n-1,k-1} & x_{n-1,n} & x_{n-1,k+1} & \!\!\!\!\cdots &  \!\!\!\!x_{n-1,n-2} & \!\!x_{n-1,n-1} & \!\!\!x_{n-1,k} \\
	\hdashline
	\cdots & \!\!\!\!x_{n,k-1} & \fbox{$x_{n,n}$} & x_{n,k+1} & \!\!\!\!\cdots & \!\!\!\!x_{n,n-2} & \!\!\!\!x_{n,n-1} & \!\!\! x_{n,k}
\end{array}\right|.     \label{X_nn_3}
\end{eqnarray}
By the result of Case I, we can permute the columns of \eqref{X_nn_1} (except for the $n$-th column) to get \eqref{X_nn_2} by the following permutation
\begin{eqnarray}
\left(
\begin{array}{ccccccccc}
1 & \cdots & k-1 & k   & k+1 & \cdots & n-2 & n-1 \\
1 & \cdots & k-1 & k+1 & k+2 & \cdots & n-1 & k
\end{array}
\right).  \nonumber 
\end{eqnarray}
Then we can adjust \eqref{X_nn_2} to \eqref{X_nn_3} by definition \eqref{quasideterminant_definition}. Finally, we find that \eqref{X_nn_3} is the permutation of \eqref{X_nn_1} $w.r.t.$ the $k$-th column and the $n$-th column.
$\hfill\Box$
\newtheorem{prop_2.4}[defn_2.1]{Proposition}
\begin{prop_2.4}
	{\bf \cite{GeRe}}  \label{Prop_2.4} \\
Let $X=(x_{ij})_{n \times n}$ be an $n \times n$ matrix over a noncommutative ring $\mathcal{R}$ and $\Lambda_{k}$ ($k=1, ~2, ..., ~n$) be invertible elements in $\mathcal{R}$. 
If $Y$ and $Z$ are matrices obtained from $X$ by multiplying common factors $\Lambda_{k}$ $(k=1,2,...,n)$ on the rightside of the same columns and the leftside of the same rows of $X$, respectively. Then the quasideterminants of $Y$ and $Z$ satisfy the following properties :
\begin{itemize}
	\item {\bf Right multiplication rule of columns}
	\begin{eqnarray}	
    \left| Y \right|_{ij}=\left| X \right|_{ij}\Lambda_{j},
	\end{eqnarray}
	or equivalently,
	\begin{eqnarray}
	\label{right multiplication law} 
	\left|
	\begin{array}{ccccc}
	x_{1,1}\Lambda_{1} & \!\! \cdots & \!\! x_{1,j}\Lambda_{j} & \!\!\cdots & \!\! x_{1,n}\Lambda_{n} \\
	\vdots &  \!\!\ddots & \!\!\vdots & \!\!\ddots & \!\!\vdots  \\
	x_{i,1}\Lambda_{1} & \!\!\cdots & \!\!\fbox{$x_{i,j}\Lambda_{j}$} & \!\!\cdots & \!\!x_{i,n}\Lambda_{n} \\
	\vdots &  \!\!\ddots & \!\!\vdots & \!\!\ddots & \!\!\vdots  \\
	x_{n,1}\Lambda_{1} & \!\!\cdots & \!\!x_{n,j}\Lambda_{j} & \!\!\cdots & \!\!x_{n,n}\Lambda_{n}
	\end{array}\right|
	=
	\left|
	\begin{array}{ccccc}
	x_{1,1} & \!\! \cdots & \!\! x_{1,j} & \!\! \cdots & \!\! x_{1,n} \\
	\vdots &  \!\! \ddots & \!\! \vdots & \!\! \ddots & \!\! \vdots  \\
	x_{i,1} & \!\! \cdots & \!\! \fbox{$x_{i,j}$} & \!\! \cdots & \!\! x_{i,n} \\
	\vdots &  \!\! \ddots & \!\! \vdots & \!\! \ddots & \!\! \vdots  \\
	x_{n,1} & \!\! \cdots & \!\! x_{n,j} & \!\! \cdots & \!\! x_{n,n}
	\end{array}\right|\Lambda_{j}. 
	\end{eqnarray}
	\item {\bf Left multiplication rule of rows}
	\begin{eqnarray}
	\left| Z \right|_{ij}=\Lambda_{i}\left| X \right|_{ij},
	\end{eqnarray}
	or equivalently,
	\begin{eqnarray}
	\label{left multiplication law}
	\left|
	\begin{array}{ccccc}
	\Lambda_{1}x_{1,1} & \!\!\cdots & \!\!\Lambda_{1}x_{1,j} & \!\!\cdots & \!\!\Lambda_{1} x_{1,n} \\
	\vdots &  \!\!\ddots & \!\!\vdots & \!\!\ddots & \!\!\vdots  \\
	\Lambda_{i}x_{i,1} & \!\!\cdots & \!\!\fbox{$\Lambda_{i}x_{i,j}$} & \!\!\cdots & \!\!\Lambda_{i}x_{i,n} \\
	\vdots &  \!\!\ddots & \!\!\vdots & \!\!\ddots & \!\!\vdots  \\
	\Lambda_{n}x_{n,1} & \!\!\cdots & \!\!\Lambda_{n}x_{n,j} & \!\!\cdots & \!\!\Lambda_{n}x_{n,n}
	\end{array}\right|	
	=
	\Lambda_{i} \left|
	\begin{array}{ccccc}
	x_{1,1} & \!\! \cdots & \!\! x_{1,j} & \!\!\cdots & \!\!x_{1,n} \\
	\vdots &  \!\! \ddots & \!\! \vdots & \!\!\ddots & \!\!\vdots  \\
	x_{i,1} & \!\!\cdots & \!\!\fbox{$x_{i,j}$} & \!\!\cdots & \!\!x_{i,n} \\
	\vdots &  \!\!\ddots & \!\!\vdots & \!\!\ddots & \!\!\vdots  \\
	x_{n,1} & \!\!\cdots & \!\!x_{n,j} & \!\!\cdots & \!\!x_{n,n}
	\end{array}\right|. ~~~~
	\end{eqnarray}
\end{itemize}
\end{prop_2.4}
{\bf (\emph{Proof})} 
\smallskip \\
Without loss of generality, we just consider the canonical form of \eqref{right multiplication law} as following. (The proof of \eqref{left multiplication law} is similar.)
\begin{eqnarray*}	
&&\left|
\begin{array}{cccc:c}
x_{1,1}\Lambda_{1} & x_{1,2}\Lambda_{2} & \cdots & x_{1,n-1}\Lambda_{n-1} & x_{1,n}\Lambda_{n} \\
\vdots &  \vdots & \ddots & \vdots & \vdots  \\
x_{n-1,1}\Lambda_{1} & x_{n-1,2}\Lambda_{2} & \cdots & x_{n-1,n-1}\Lambda_{n-1} & x_{n-1,n}\Lambda_{n} \\
\hdashline
x_{n,1}\Lambda_{1} & x_{n,2}\Lambda_{2} & \cdots & x_{n,n-1}\Lambda_{n-1} & \fbox{$x_{n,n}\Lambda_{n}$}
\end{array}\right|  \nonumber \\
&=& 
\left| \begin{array}{cc}
AD & B \Lambda_{n}  \\
CD & \fbox{$d\Lambda_{n}$}
\end{array}\right| 
=
d\Lambda_{n} - CD(AD)^{-1} B\Lambda_{n}
=\left(d - CA^{-1}B \right)\Lambda_{n} 
= |X|_{nn}\Lambda_{n},   \nonumber 
\end{eqnarray*}
where $D$:=diag($\Lambda_{1}$,~$\Lambda_{2}$,~$\cdots$,~$\Lambda_{n-1}$) is a $(n-1) \times (n-1)$ diagonal matrix.
$\hfill\Box$
\newtheorem{prop_2.5}[defn_2.1]{Prposition}
\begin{prop_2.5}
	{\bf (Row (Column) operation \cite{GeRe})} \label{Prop_2.5} \\
Suppose that the $l$-th row (column) of a quasideterminant does not involve the box element. 
Then the operation of adding the $l$-th row (column) to the $k$-th row (column) leaves the quasideterminant unchanged.
\end{prop_2.5}
{\bf (\emph{Proof})} 
\smallskip \\ 
Without loss of generality, we just consider the row operations of the canonical from. (The proof of the column operations are similar.) We also introduce an informal notation $k \rightarrow k + l$ to denote the operation of adding the $l$-th row to the $k$-th row in quasideterminant.\\
{\bf Case I : $k$-th row $\longrightarrow$ $k$-th row + $l$-th row ~($k \neq n, ~ l \neq n$)}
\begin{eqnarray*}
\left|
\begin{array}{cc}
A & B \\
C & \fbox{$d$}
\end{array}
\right|_{k \rightarrow k + l}
=
\left|
\begin{array}{cc}
EA & EB \\
C & \fbox{$d$}
\end{array}
\right|
=
d - C(EA)^{-1}EB
=
\left|
\begin{array}{cc}
A & B \\
C & \fbox{$d$}
\end{array}
\right|,   \nonumber 
\end{eqnarray*}
where $E:= I + M $, $I$ is the $(n-1) \times (n-1)$ identity matrix, and $M$ denotes the $(n-1) \times (n-1)$ matrix with 1 in the $(k, l)$-th component and 0 elsewhere.
\smallskip \\
{\bf Case II : $n$-th row $\longrightarrow$ $n$-th row + $l$-th row}  ~(involving the box element $x_{nn}$)
\begin{eqnarray*}
\left|
\begin{array}{cc}
A & B \\
C & \fbox{$d$}
\end{array}
\right|_{n \rightarrow n + l}  
=
\left|
\begin{array}{cc}
A & B \\
FA + C & \fbox{$ FB + d$}
\end{array}
\right|  
=
FB + d - (FA + C)A^{-1}B
=
\left|
\begin{array}{cc}
A & B \\
C & \fbox{$d$}
\end{array}
\right|,   \nonumber 
\end{eqnarray*}
where $F$ denotes the row vector with 1 in the $l$-th component and 0 elsewhere.
\medskip \\
{\bf Case III : $k$-th row $\longrightarrow$ $k$-th row + $n$-th row} ~ (involving the box element $x_{nn}$)
\smallskip \\
For convenience, we define $e_{k}$ to be the column vector with 1 in the $k$-th component and 0 elsewhere and use the informal notation
$\fbox{$X$}_{~\!k}$ to denote a column vector with box element in the $k$-th component of $X$, that is, $\fbox{$X$}_{~\!k}:=(x_1, ..., \fbox{$x_k$} ~\!, ..., x_{n-1})^{T}$.  
Now we have
\begin{eqnarray*}
\left|
\begin{array}{cc}
A & B  \\
C & \fbox{$d$}
\end{array}
\right|_{k \rightarrow k + n}  \nonumber 
&\!\!\!\!\!\!\!\!=&\!\!\!\!
\left|
\begin{array}{cc}
A + e_{k}C & B + e_{k}d \\
C & \fbox{$d$}
\end{array}
\right|   
=
\left|
\begin{array}{cc}
A + e_{k}C & e_{k} \\
C & \fbox{$0$}
\end{array}
\right|
\left|
\begin{array}{cc}
A + e_{k}C & \fbox{$B + e_{k}d$}_{~\!k} \\
C & d
\end{array}
\right|  \nonumber  \\
&& ~~~~~~~~~~~~~~~~~~~~~~~~~~~\mbox{By Proposition \ref{Prop_2.7} (Homological relation)}  \nonumber \\
&\!\!\!\!\!\!\!\!=&\!\!\!\!
\left|
\begin{array}{cc}
A + e_{k}C & e_{k} \\
C & \fbox{$0$}
\end{array}
\right|
\left|
\begin{array}{cc}
A & \fbox{$B$}_{~\!k} \\
C & d
\end{array}
\right| 
=
\left|
\begin{array}{cc}
A + e_{k}C & e_{k} \\
C & \fbox{$0$}
\end{array}
\right|
\left|
\begin{array}{cc}
A & \fbox{$\mathbf{0}$}_{~\!k} \\
C &  1
\end{array}
\right|
\left|
\begin{array}{cc}
A & B \\
C & \fbox{$d$}
\end{array}
\right|   \nonumber  \\
&& ~~~~~~~~~~~~~~~~~~~~~~~~~~~~~~~~~~~ \mbox{By Proposition \ref{Prop_2.7} (Homological relation)} \\
&\!\!\!\!\!\!\!\!=&\!\!\!\!
\left\{
\left|
\begin{array}{cc}
A + e_kC & e_{k} \\
C & \fbox{$0$}
\end{array}
\right|
\left|
\begin{array}{cc}
A & e_{k} \\
C & \fbox{0}
\end{array}
\right|^{-1}
\right\}
\left|
\begin{array}{cc}
A & B \\
C & \fbox{$d$}
\end{array}
\right|  \nonumber 
\neq
\left|
\begin{array}{cc}
A & B \\
C & \fbox{$d$}
\end{array}
\right|.   \nonumber \\
&& ~~~~~~~~~ \mbox{By Corollary \ref{Cor_2.8} (Inverse relation)}
\end{eqnarray*}
Therefore, the quasideterminant is not invariant in this case. $\hfill\Box$
\smallskip \\

A general rule for reducing the quasideterminant of higher order to lower order is called the noncommutative version of Sylvester's Theorem \cite{GeRe}. The simplest version of this theorem is given by the following noncommutative version of Jacobi identity. 
The readers will see it soon that this identity plays a crucial role throughout this thesis.
\newtheorem{thm_2.6}[defn_2.1]{Theorem}
\begin{thm_2.6}
	{\bf (Noncommutative version of Jacobi Identity \cite{GeRe})} \label{Thm_2.6}\\
Let $X=(x_{i,j})_{n \times n}$ be an $n \times n$ matrix over a noncommutative ring $\mathcal{R}$.
Then the $(i,j)$-th quasideterminant of $X$ can be  expressed explicitly as four quasideterminants 
of the
$(n-1) \times (n-1) $ submatrices $X_{\widehat{l},~\!\widehat{m}}$, $X_{\widehat{l},~\!\widehat{j}}$, $X_{\widehat{i},~\!\widehat{j}}$, $X_{\widehat{i},~\!\widehat{m}}$ of $X$ :
\begin{eqnarray}
\label{Jacobi Identity}
\left| X \right|_{i,j}
=
\left|
\begin{array}{cc:c:c:c:cc}
\!\!x_{1,1} & \!\!\!\! \cdots & \! x_{1,m} & \! \cdots & \! x_{1,j} & \! \cdots & \!\!\!\!\! x_{1,n} \\
\!\!\vdots & \!\!\!\! \ddots & \! \vdots & \! \ddots & \! \vdots & \! \ddots & \!\!\!\!\! \vdots \\
\hdashline
\!\!x_{l,1} & \!\!\!\! \cdots & \! x_{l,m} & \! \cdots & \! x_{l,j} & \! \cdots & \!\!\!\!\! x_{l,n} \\
\hdashline
\!\!\vdots & \!\!\!\! \ddots & \! \vdots & \! \ddots & \! \vdots & \! \ddots & \!\!\!\!\! \vdots \\
\hdashline
\!\!x_{i,1} & \!\!\!\! \cdots & \! x_{i,m} & \! \cdots & \! \fbox{$x_{i,j}$} & \! \cdots & \!\!\!\!\! x_{i,n} \\
\hdashline
\!\!\vdots & \!\!\!\! \ddots & \! \vdots & \! \ddots & \! \vdots & \! \ddots & \!\!\!\!\! \vdots \\
\!\!x_{n,1} & \!\!\!\! \cdots & \! x_{n,m} & \! \cdots & \! x_{n,j} & \! \cdots & \!\!\!\!\! x_{n,n} 
\end{array}
\right| 
= 
\left|
X_{\widehat{l},~\!\widehat{m}}
\right|_{i,j}
\!\! -
\left|
X_{\widehat{l},~\!\widehat{j}}
\right|_{i,m}
\left|
X_{\widehat{i},~\!\widehat{j}}
\right|_{l,m}^{-1}
\left|
X_{\widehat{i},~\!\widehat{m}}
\right|_{l,j}. ~~
\end{eqnarray}
\end{thm_2.6}
 {\bf (\emph{Proof})} 
 \smallskip \\
 By using Proposition \ref{Prop_2.3} (Permutation rule), we can adjust the $l$-th and $i$-th row, the $m$-th and $j$-th column to the lower right corner, and then obtain the following simplified form \cite{GiNi, Nimmo} :
\begin{eqnarray}
\label{Jacobi Identity 2}
&&\left|
\begin{array}{ccc}
A & B & C \\
D & f & g  \\
E & h & \fbox{$i$}
\end{array}
\right| 
=
\left|
\begin{array}{cc}
A & C \\
E & \fbox{$i$}
\end{array}
\right|
-
\left|
\begin{array}{cc}
A & B \\
E & \fbox{$h$}
\end{array}
\right|
\left|
\begin{array}{cc}
A & B \\
D & \fbox{$f$}
\end{array}
\right|^{-1}
\left|
\begin{array}{cc}
A & C \\
D & \fbox{$g$}
\end{array}
\right|, \\
&&\left(
\begin{array}{cc}
f & g \\
h & i
\end{array}
\right)
:= 
\left(
\begin{array}{cc}
x_{l,m} & x_{l,j} \\
x_{i,m} & x_{i,j}
\end{array}
\right), \nonumber
\end{eqnarray}
where the matrix sizes are $A: (n-2) \times (n-2)$; $B, C: (n-2)  \times 1$; $D, E : 1 \times (n-2)$.
By applying the inverse matrix formula \eqref{Inverse matrix formula}, we can complete this proof : 
\begin{eqnarray*}
\left|
\begin{array}{ccc}
A & B & C \\
D & f & g  \\
E & h & \fbox{$i$}
\end{array}
\right|   
&\!\!\!\!=&
\!\!\! i-(E,h)
\left(
\begin{array}{cc}
A & B \\
D & f
\end{array}
\right)^{-1} 
\left(
\begin{array}{c}
C  \\
g 
\end{array}
\right) \nonumber \\
&\!\!\!\!=&
\!\!\! i-(E,h)
\left(
\begin{array}{cc}
A^{-1}+A^{-1}BS^{-1}DA^{-1}  & -A^{-1}BS^{-1} \\
-S^{-1}DA^{-1} & S^{-1}
\end{array}
\right) 
\left(
\begin{array}{c}
C  \\
g 
\end{array}
\right) \nonumber \\
&\!\!\!\!=&
\!\!\! i-\!EA^{-1}C-\!EA^{-1}BS^{-1}DA^{-1}C+\!hS^{-1}DA^{-1}C+\!EA^{-1}BS^{-1}g-\!hS^{-1}g  \nonumber  \\
&\!\!\!\!=&
\!\!\! (i-EA^{-1}C)-(h-EA^{-1}B)S^{-1}(g-DA^{-1}C)  \nonumber \\
&\!\!\!\!=&
\!\!\! \left|
\begin{array}{cc}
A & C \\
E & \fbox{$i$}
\end{array}
\right|
-
\left|
\begin{array}{cc}
A & B \\
E & \fbox{$h$}
\end{array}
\right|
\left|
\begin{array}{cc}
A & B \\
D & \fbox{$f$}
\end{array}
\right|^{-1}
\left|
\begin{array}{cc}
A & C \\
D & \fbox{$g$}
\end{array}
\right|. \nonumber
\end{eqnarray*}
$\hfill\Box$
\medskip

The following homological relation between two quasideterminants is a direct application of the noncommutative version of Jacobi identity. It is a very powerful tool when the box element need to be displaced horizontally or vertically in some hard calculations of quasideterminants.
\newtheorem{prop_2.7}[defn_2.1]{Proposition}
\begin{prop_2.7}
	{\bf (Homological relation \cite{GiNiSo,Nimmo}, Cf: \cite{GeRe})}  \label{Prop_2.7}
\begin{itemize}
\item  {\bf Vertical transition of the box element : }
\begin{eqnarray}
\label{Homological relation}
\left|
\begin{array}{ccccc}
	\!\! x_{1,1} & \!\!\! \cdots & \!\!\!\! x_{1,j} & \!\!\! \cdots & \!\!\!\! x_{1,n} \\
	\!\! \vdots  & \!\!\! \ddots & \!\!\!\! \vdots  & \!\!\! \ddots & \!\!\!\! \vdots \\
	\!\! x_{m,1} & \!\!\! \cdots & \!\!\!\! x_{m,j} & \!\!\! \cdots & \!\!\!\! x_{m,n} \\
	\!\! \vdots  & \!\!\! \ddots & \!\!\!\! \vdots  & \!\!\! \ddots &  \!\!\!\! \vdots  \\
	\!\! x_{l,1} & \!\!\! \cdots & \!\!\!\! \fbox{$x_{l,j}$}  &  \!\!\! \cdots & \!\!\!\! x_{l,n} \\
	\!\! \vdots  & \!\!\! \ddots & \!\!\!\! \vdots  & \!\!\! \ddots &  \!\!\!\! \vdots  \\
	\!\! x_{n,1} & \!\!\! \cdots & \!\!\!\! x_{n,j} & \!\!\! \cdots & \!\!\!\! x_{n,n}
\end{array}\right|
=
\left|
\begin{array}{ccccc}
\!\! x_{1,1} & \!\!\! \cdots & \!\!\!\! 0        & \!\!\! \cdots & \!\!\!\! x_{1,n} \\ 
\!\! \vdots  & \!\!\! \ddots & \!\!\!\! \vdots   & \!\!\! \ddots & \!\!\!\! \vdots \\
\!\! x_{m,1} & \!\!\! \cdots & \!\!\!\! 1        & \!\!\! \cdots & \!\!\!\! x_{m,n} \\
\!\! \vdots  & \!\!\! \ddots & \!\!\!\! \vdots   & \!\!\! \ddots & \!\!\!\! \vdots  \\
\!\! x_{l,1} & \!\!\! \cdots & \!\!\!\! \fbox{$0$} & \!\!\! \cdots & \!\!\!\! x_{l,n} \\
\!\! \vdots  & \!\!\! \ddots & \!\!\!\! \vdots   & \!\!\! \ddots & \!\!\!\! \vdots  \\
\!\! x_{n,1} & \!\!\! \cdots & \!\!\!\! 0        & \!\!\! \cdots & \!\!\!\! x_{n,n}
\end{array}\right|
\left|
\begin{array}{ccccc}
\!\! x_{1,1} & \!\!\! \cdots & \!\!\!\! x_{1,j} & \!\!\! \cdots & \!\!\!\! x_{1,n} \\
\!\! \vdots  & \!\!\! \ddots & \!\!\!\! \vdots  & \!\!\! \ddots & \!\!\!\! \vdots \\
\!\! x_{m,1} & \!\!\! \cdots & \!\!\!\! \fbox{$x_{m,j}$} &  \!\!\! \cdots & \!\!\!\! x_{m,n} \\
\!\! \vdots  & \!\!\! \ddots & \!\!\!\! \vdots  & \!\!\! \ddots & \!\!\!\! \vdots  \\
\!\! x_{l,1} & \!\!\! \cdots & \!\!\!\! x_{l,j} & \!\!\! \cdots & \!\!\!\! x_{l,n} \\
\!\! \vdots  & \!\!\! \ddots & \!\!\!\! \vdots  & \!\!\! \ddots & \!\!\!\! \vdots  \\
\!\! x_{n,1} & \!\!\! \cdots & \!\!\!\! x_{n,j} & \!\!\! \cdots & \!\!\!\! x_{n,n}
\end{array}\right|~~ 
\end{eqnarray}
\item  {\bf Horizontal transition of the box element :}
\begin{eqnarray}
\label{Homological relation 2}
&&\left| 
\begin{array}{ccccccc}
\!\! x_{1,1} & \!\!\! \cdots & \!\!\!\! x_{1,m} & \!\!\! \cdots & \!\!\!\! x_{1,l} & \!\!\! \cdots & \!\!\!\! x_{1,n} \\
\!\! \vdots  & \!\!\! \ddots & \!\!\!\! \vdots  & \!\!\! \ddots & \!\!\!\! \vdots & \!\!\! \ddots & \!\!\!\! \vdots \\
\!\! x_{i,1} & \!\!\! \cdots & \!\!\!\! x_{i,m} & \!\!\! \cdots & \!\!\!\! \fbox{$x_{i,l}$} & \!\!\! \cdots & \!\!\!\! x_{i,n} \\
\!\! \vdots  & \!\!\! \ddots & \!\!\!\! \vdots  & \!\!\! \ddots & \!\!\!\! \vdots & \!\!\! \ddots & \!\!\!\! \vdots \\
\!\! x_{n,1} & \!\!\! \cdots & \!\!\!\! x_{n,m} & \!\!\! \cdots & \!\!\!\! x_{n,l} & \!\!\! \cdots & \!\!\!\! x_{n,n}
\end{array}
\right|  \nonumber \\
&=&
\left| 
\begin{array}{ccccccc}
\!\! x_{1,1} & \!\!\! \cdots & \!\!\!\! x_{1,m} & \!\!\! \cdots & \!\!\!\! x_{1,l} & \!\!\! \cdots & \!\!\!\! x_{1,n} \\
\!\! \vdots  & \!\!\! \ddots & \!\!\!\! \vdots  & \!\!\! \ddots & \!\!\!\! \vdots  & \!\!\! \ddots & \!\!\!\! \vdots \\
\!\! x_{i,1} & \!\!\! \cdots & \!\!\!\! \fbox{$x_{i,m}$} & \!\!\! \cdots & \!\!\!\! x_{i,l} & \!\!\! \cdots & \!\!\!\! x_{i,n} \\
\!\! \vdots  & \!\!\! \ddots & \!\!\!\! \vdots  & \!\!\! \ddots & \!\!\!\! \vdots  & \!\!\! \ddots & \!\!\!\! \vdots \\
\!\! x_{n,1} & \!\!\! \cdots & \!\!\!\! x_{n,m} & \!\!\! \cdots & \!\!\!\! x_{n,l} & \!\!\! \cdots & \!\!\!\! x_{n,n}
\end{array}
\right|
\left| 
\begin{array}{ccccccc}
\!\! x_{1,1} & \!\!\! \cdots & \!\!\!\! x_{1,m} & \!\!\! \cdots & \!\!\!\! x_{1,l} & \!\!\! \cdots & \!\!\!\! x_{1,n} \\
\!\! \vdots  & \!\!\! \ddots & \!\!\!\! \vdots  & \!\!\! \ddots & \!\!\!\! \vdots  & \!\!\! \ddots & \!\!\!\! \vdots \\
\!\! 0       & \!\!\! \cdots & \!\!\!\! 1       & \!\!\! \cdots & \!\!\!\! \fbox{$0$}& \!\!\! \cdots & \!\!\!\! 0 \\
\!\! \vdots  & \!\!\! \ddots & \!\!\!\! \vdots  & \!\!\! \ddots & \!\!\!\! \vdots  & \!\!\! \ddots & \!\!\!\! \vdots \\
\!\! x_{n,1} & \!\!\! \cdots & \!\!\!\! x_{n,m} & \!\!\! \cdots & \!\!\!\! x_{n,l} & \!\!\! \cdots & \!\!\!\! x_{n,n}
\end{array}
\right|~~~~~~~
\end{eqnarray}
\end{itemize}
\end{prop_2.7}
{\bf(\emph{Proof})} 
\smallskip \\
By adjusting the first two quasideterminants in \eqref{Homological relation} to the canonical form  and  using Proposition \ref{Prop_2.3} (Permutation rule) to adjust the last quasideterminant, \eqref{Homological relation} can be simplified as an equivalent representation \cite{GiNiSo,Nimmo} :
\begin{eqnarray*}
\left|
\begin{array}{ccc}
A & B & C \\
D & f & g  \\
E & h & \fbox{$i$}
\end{array}
\right|
=
\left|
\begin{array}{ccc}
A & B & \boldsymbol{0} \\
D & f & 1  \\
E & h & \fbox{0}
\end{array}
\right|
\left|
\begin{array}{ccc}
A & B & C \\
D & f & \fbox{$g$}  \\
E & h & i
\end{array}
\right|, ~
\left(
\begin{array}{cc}
f & g \\
h & i
\end{array}
\right)
:= 
\left(
\begin{array}{cc}
x_{m,n} & x_{m,j} \\
x_{l,n} & x_{l,j}
\end{array}
\right). \nonumber
\end{eqnarray*}
By using the Jacobi Identity \eqref{Jacobi Identity 2},  we can check
\begin{eqnarray*}
&&\mbox{RHS}  \nonumber \\
&=& \left(\left|
\begin{array}{cc}
A & \boldsymbol{0} \\
E & \fbox{0}
\end{array}
\right|
-
\left|
\begin{array}{cc}
A & B \\
E & \fbox{$h$}
\end{array}
\right|
\left|
\begin{array}{cc}
A & B \\
D & \fbox{$f$}
\end{array}
\right|^{-1}
\left|
\begin{array}{cc}
A & \boldsymbol{0} \nonumber \\
D & \fbox{1}
\end{array}
\right|\right)\\
&&\times \left(\left|
\begin{array}{cc}
A & C \\
D & \fbox{$g$}
\end{array}
\right|
-
\left|
\begin{array}{cc}
A & B \\
D & \fbox{$f$}
\end{array}
\right|
\left|
\begin{array}{cc}
A & B \\
E & \fbox{$h$}
\end{array}
\right|^{-1}
\left|
\begin{array}{cc}
A & C \\
E & \fbox{$i$}
\end{array}
\right|\right) \nonumber \\
&=&
\left|
\begin{array}{cc}
A & C \\
E & \fbox{$i$}
\end{array}
\right|
-
\left|
\begin{array}{cc}
A & B \\
E & \fbox{$h$}
\end{array}
\right|
\left|
\begin{array}{cc}
A & B \\
D & \fbox{$f$}
\end{array}
\right|^{-1}
\left|
\begin{array}{cc}
A & C \\
D & \fbox{$g$}
\end{array}
\right|  \nonumber 
= \mbox{LHS}.  \nonumber
\end{eqnarray*}
Through a very similar process, \eqref{Homological relation 2} can be simplified as
\begin{eqnarray*}
\left|
\begin{array}{ccc}
A & B & C \\
D & f & g  \\
E & h & \fbox{$i$}
\end{array}
\right|
=
\left|
\begin{array}{ccc}
A & B & C \\
D & f & g  \\
E & \fbox{$h$} & i
\end{array}
\right|
\left|
\begin{array}{ccc}
A & B & C \\
D & f & g  \\
E & 1 & \fbox{0}
\end{array}
\right|, \nonumber
\end{eqnarray*}
and the proof is similar to the above.
$\hfill\Box$
\newpage 
Note that if we apply Proposition \ref{Prop_2.7} (Homological relation) to the RHS of \eqref{Homological relation} and \eqref{Homological relation 2} again, we can obtain the following results immediately.
\newtheorem{cor_2.8}[defn_2.1]{Corollary}
\begin{cor_2.8}
	{\bf (Inverse relation)}  \label{Cor_2.8}
\begin{eqnarray}
\label{Inverse relation}
&&\left|
\begin{array}{ccccc}
x_{1,1} & \cdots & 0 & \cdots & x_{1,n} \\
\vdots & \ddots & \vdots & \ddots & \vdots \\
x_{m,1} & \cdots & 1 & \cdots & x_{m,n} \\
\vdots &  \ddots & \vdots & \ddots & \vdots  \\
x_{l,1} & \cdots & \fbox{$0$} & \cdots & x_{l,n} \\
\vdots &  \ddots & \vdots & \ddots & \vdots  \\
x_{n,1} & \cdots & 0 & \cdots & x_{n,n}
\end{array}\right|^{-1}
=\left|
\begin{array}{ccccc}
x_{1,1} & \cdots & 0 & \cdots & x_{1,n} \\
\vdots & \ddots & \vdots & \ddots & \vdots \\
x_{m,1} & \cdots & \fbox{$0$} & \cdots & x_{m,n} \\
\vdots &  \ddots & \vdots & \ddots & \vdots  \\
x_{l,1} & \cdots & 1 & \cdots & x_{l,n} \\
\vdots &  \ddots & \vdots & \ddots & \vdots  \\
x_{n,1} & \cdots & 0 & \cdots & x_{n,n}
\end{array}\right|, \\
&&\left| 
\begin{array}{ccccccc}
\!\! x_{1,1} & \!\!\! \cdots & \!\!\!\! x_{1,m} & \!\!\! \cdots & \!\!\!\! x_{1,l} & \!\!\! \cdots & \!\!\!\! x_{1,n} \\
\!\! \vdots  & \!\!\! \ddots & \!\!\!\! \vdots  & \!\!\! \ddots & \!\!\!\! \vdots  & \!\!\! \ddots & \!\!\!\! \vdots \\
\!\! 0       & \!\!\! \cdots & \!\!\!\! 1       & \!\!\! \cdots & \!\!\!\! \fbox{$0$}& \!\!\! \cdots & \!\!\!\! 0 \\
\!\! \vdots  & \!\!\! \ddots & \!\!\!\! \vdots  & \!\!\! \ddots & \!\!\!\! \vdots  & \!\!\! \ddots & \!\!\!\! \vdots \\
\!\! x_{n,1} & \!\!\! \cdots & \!\!\!\! x_{n,m} & \!\!\! \cdots & \!\!\!\! x_{n,l} & \!\!\! \cdots & \!\!\!\! x_{n,n}
\end{array}
\right|^{-1}
=
\left| 
\begin{array}{ccccccc}
\!\! x_{1,1} & \!\!\! \cdots & \!\!\!\! x_{1,m}  & \!\!\! \cdots & \!\!\!\! x_{1,l} & \!\!\! \cdots & \!\!\!\! x_{1,n} \\
\!\! \vdots  & \!\!\! \ddots & \!\!\!\! \vdots   & \!\!\! \ddots & \!\!\!\! \vdots & \!\!\! \ddots & \!\!\!\! \vdots \\
\!\! 0       & \!\!\! \cdots & \!\!\!\! \fbox{$0$} & \!\!\! \cdots & \!\!\!\! 1 & \!\!\! \cdots & \!\!\!\! 0 \\
\!\! \vdots  & \!\!\! \ddots & \!\!\!\! \vdots   & \!\!\! \ddots & \!\!\!\! \vdots & \!\!\! \ddots & \!\!\!\! \vdots \\
\!\! x_{n,1} & \!\!\! \cdots & \!\!\!\! x_{n,m}  & \!\!\! \cdots & \!\!\!\! x_{n,l} & \!\!\! \cdots & \!\!\!\! x_{n,n}
\end{array}
\right|.
\end{eqnarray}
\end{cor_2.8}

Finally, we introduce a rather appealing and useful formula for calculating the derivative of quasideterminant as the ending of this review section. More detailed discussions and applications of the derivative of quasideterminant can be found in \cite{GiNi}.    
\newtheorem{thm_2.9}[defn_2.1]{Theorem}
\begin{thm_2.9}
	{\bf (Derivative formula of quasideterminant \cite{GiNi,Nimmo}) } \label{Thm_2.9} \\
Let $A$ be an $n \times n$ matrix, $B$ be an $n \times 1$ column matrix, $C$ be an $1 \times n$ row matrix, and $d$ be an $1 \times 1$ matrix with noncommutative elements. Then we have the following derivative formula of quasideterminant :
\begin{eqnarray}
\left|
\begin{array}{cc}
A & B \\
C & \fbox{$d$}
\end{array}
\right|^{~\prime}
=
\left\{ 
\begin{array}{c}
\left|
\begin{array}{cc}
A & B \\
C^{\prime} & \fbox{$d^{\prime}$}
\end{array}
\right| 
+\sum\limits_{i=1}^{n}
\left|
\begin{array}{cc}
A & e_{i} \\
C & \fbox{$0$}
\end{array}
\right| 
\left|
\begin{array}{cc}
A & B \\
(A^{i})^{\prime} & \fbox{$(B^{i})^{\prime}$}
\end{array}
\right| 
\\
\left|
\begin{array}{cc}
A & B^{\prime} \\
C & \fbox{$d^{\prime}$}
\end{array}
\right|
+\sum\limits_{i=1}^{n}
\left|
\begin{array}{cc}
A & (A_{i})^{\prime} \\
C & \fbox{$(C_{i})^{\prime}$}
\end{array}
\right| 
\left|
\begin{array}{cc}
A & B \\
e^{i} & \fbox{$0$}
\end{array}
\right| 
\end{array}
\right. 
\end{eqnarray}
Here we use the upper index notation $A^{i}$, $B^{i}$, $e^{i}$ to denote the $i$-th row of matrix $A$, $B$ and identity matrix $I$, the lower index notation $A_{i}$, $C_{i}$, $e_{i}$ to denote the $i$-th column of matrix $A$, $C$ and identity $I$. 
\end{thm_2.9}
{\bf (\emph{Proof})} 
\smallskip \\
(The following proof is excerpted form \cite{GiNi,Nimmo}.)
\begin{eqnarray}
&&\left|
\begin{array}{cc}
A & B \\
C & \fbox{$d$}
\end{array}
\right|^{~\prime} \nonumber \\
&=& (d-CA^{-1}B)^{\prime}
=d^{\prime}-C^{\prime}A^{-1}B-C(A^{-1})^{\prime}B-CA^{-1}B^{\prime} \nonumber \\
&=& d^{\prime}-C^{\prime}A^{-1}B-CA^{-1}B^{\prime}+CA^{-1}A^{\prime}A^{-1}B \nonumber \\
&=& 
\left\{ 
\begin{array}{c}
\left|
\begin{array}{cc}
A & B \\
C^{\prime} & \fbox{$d^{\prime}$}
\end{array}
\right|-CA^{-1}\sum_{i=1}^{n}e_{i}e^{i}(B^{\prime}-A^{\prime}A^{-1}B) \\
\left|
\begin{array}{cc}
A & B^{\prime} \\
C & \fbox{$d^{\prime}$}
\end{array}
\right|+(C^{\prime}-CA^{-1}A^{\prime})\sum_{i=1}^{n}e_{i}e^{i}(-A^{-1}B)
\end{array}
\right. \nonumber  \\
&=&
\left\{ 
\begin{array}{c}
\left|
\begin{array}{cc}
A & B \\
C^{\prime} & \fbox{$d^{\prime}$}
\end{array}
\right|
+\sum_{i=1}^{n}(-CA^{-1}e_{i})\left[(B^{i})^{\prime}-(A^{i})^{\prime}A^{-1}B)\right] \\
\left|
\begin{array}{cc}
A & B^{\prime} \\
C & \fbox{$d^{\prime}$}
\end{array}
\right|
+\sum_{i=1}^{n}\left[(C_{i}^{\prime}-CA^{-1}(A_{i})^{\prime}\right](-e^{i}A^{-1}B)
\end{array}
\right. \nonumber \\
&=&
\left\{ 
\begin{array}{c}
\left|
\begin{array}{cc}
A & B \\
C^{\prime} & \fbox{$d^{\prime}$}
\end{array}
\right| 
+\sum_{i=1}^{n}
\left|
\begin{array}{cc}
A & e_{i} \\
C & \fbox{0}
\end{array}
\right| 
\left|
\begin{array}{cc}
A & B \\
(A^{i})^{\prime} & \fbox{$(B^{i})^{\prime}$}
\end{array}
\right| 
\\
\left|
\begin{array}{cc}
A & B^{\prime} \\
C & \fbox{$d^{\prime}$}
\end{array}
\right|
+\sum_{i=1}^{n}
\left|
\begin{array}{cc}
A & (A_{i})^{\prime} \\
C & \fbox{$(C_{i})^{\prime}$}
\end{array}
\right| 
\left|
\begin{array}{cc}
A & B \\
e^{i} & \fbox{0}
\end{array}
\right| 
\end{array}.
\right. \nonumber 
\end{eqnarray}
$\hfill\Box$

\newpage

\section{Yang's form of the Anti-Self-Dual Yang-Mills (ASDYM) equations}
\label{Section 3}

In this section, we review some necessary knowledge of the ASDYM theory mainly from the perspective of integrable systems. In Subsection \ref{Section 3.1}, we firstly review Yang-Mills theory on 4-dimensional real spaces briefly. In Subsection \ref{Section 3.2}, we then review the general theory of ASDYM on 4-dimensional complex space. In Subsection \ref{Section 3.3}, we introduce several equivalent descriptions of the ASDYM equation. Especially, the anti-self-dual (ASD) gauge fields can be formulated in terms of the so-called $J$-matrix \cite{BrFaNuYa} (Cf: \cite{MaWo}) which is the solution of Yang equation (Cf:~\cite{Yang} for $G=\mathrm{SU}(2)$, and \cite{BrFaNuYa} for $G=\mathrm{SU}(N)$). In Subsection \ref{Section 3.4}, we give a comprehensive review of \cite{Yang} in which Yang gives the condition for finding all $G=\mathrm{SU}(2)$ ASD gauge fields on the 4-dimensional Euclidean space, and he also point out a special class of ASD gauge fields which is in fact equivalent to the 't Hooft ansatz \cite{tHt3} (or known as the Corrigan-Fairlie-'t Hooft-Wilczek ansatz \cite{CoFa,Wilczek}). As a supplement of \cite{Yang}, we discuss the condition of $G=\mathrm{U}(N)$ ASD gauge fields for the Euclidean and the split signature, respectively.

\subsection{Brief review of Yang-Mills theory on 4D real spaces}
\label{Section 3.1}

In this subsection, we give a brief review of non-abelian gauge theory, some more details can be referred to \cite{MaSu}.  
In theoretical or mathematical physics, a gauge theory is a type of field theory in which the Lagrangian is invariant under local transformations from certain Lie groups, called gauge groups in these field theories. 
For our purpose in this thesis, we assume the gauge group $G$ to be $\mathrm{GL}(N,\mathbb{C})$ or subgroup of it, and the Lie algebra of $G$ is defined by
$\mathfrak{g}:=
\left\{
X \in \mathrm{Mat}(N,\mathbb{C}) \big| ~ \mbox{exp}(X) \in G
\right\}$.
Now we can consider a field theory with complex scalar field $\Phi$ in the fundamental representation of $G$,
\begin{eqnarray}
\label{Gauge transf of Phi}
\Phi  \longmapsto  g\Phi, ~ g \in G.
\end{eqnarray}
To construct a gauge invariant Lagrangian, we need to introduce additional fields $A_{\mu}$, called gauge fields, to define the covariant derivatives of $\Phi$ as :
\begin{eqnarray}
\label{Covariant derivatives_real expression}
D_{\mu}\Phi = \partial_{\mu}\Phi + A_{\mu}\Phi, ~~ \mu = 0, 1, 2, 3
\end{eqnarray}
and define the gauge transformation of $A_{\mu}$ by :
\begin{equation}
\label{Gauge transf of A_mu}
A_{\mu} \longmapsto 
gA_{\mu}g^{-1} - (\partial_{\mu}g)g^{-1}, ~~ g \in G,
\end{equation}
where $A_{\mu}$ take values in the Lie algebra $\mathfrak{g}$. Under the gauge transformation \eqref{Gauge transf of Phi} and \eqref{Gauge transf of A_mu}, we can conclude that
\begin{eqnarray}
D_{\mu}\Phi \longmapsto  gD_{\mu}\Phi  ~~~~~~ \mbox{($i.e$ gauge covariant)}.
\end{eqnarray}
The final ingredient of the Lagrangian is the Yang-Mills field strength $F_{\mu \nu}$ defined by the commutator of covariant derivatives $D_{\mu}$ and $D_{\nu}$, $\mu$, $\nu$ = 0, 1, 2, 3 as follows : 
\begin{eqnarray}
\label{Field strengths_real expression}
F_{\mu \nu} = [D_{\mu}, D_{\nu}]
=\partial_{\mu}A_{\nu} - \partial_{\nu}A_{\mu} + [A_{\mu}, A_{\nu}].
\end{eqnarray} 
Under the gauge transformation \eqref{Gauge transf of A_mu}, we can conclude that
\begin{eqnarray}
\label{Gauge covariant_F}
F_{\mu \nu}  \longmapsto  gF_{\mu \nu}g^{-1} ~~~~~~ \mbox{($i.e$ ~gauge covariant)}.
\end{eqnarray}
Now we find that 
\begin{eqnarray}
\mbox{Tr}(F_{\mu \nu}F^{\mu \nu})  \longmapsto  
\mbox{Tr}(gF_{\mu \nu}g^{-1}gF^{\mu \nu}g^{-1})
=
\mbox{Tr}(F_{\mu \nu}F^{\mu \nu})  
\end{eqnarray}
is a gauge invariant quantity due to the cyclic property of the trace.
Furthermore, if restricting the gauge group to be unitary, we also find that $(D_{\mu}\Phi)^{\dagger}(D^{\mu}\Phi)$ and $\Phi^{\dagger}\Phi$ are other gauge invariant quantities. 
As a result, a gauge invariant Lagrangian density, called the Yang-Mills Lagrangian density, can be written as :
\begin{eqnarray}
\mathcal{L} = 
\frac{-1}{2}\mbox{Tr}(F_{\mu \nu}F^{\mu \nu}) 
+ (D_{\mu}\Phi)^{\dagger}(D^{\mu}\Phi)
-U(\Phi^{\dagger}\Phi).
\end{eqnarray}
If we just consider the kinetic term of the Yang-Mills Lagrangian density 
\begin{eqnarray}
\label{Pure YM Lagrangian density}
\mathcal{L} = 
\frac{-1}{2}\mbox{Tr}(F_{\mu \nu}F^{\mu \nu}), 
\end{eqnarray}
and define the Yang-Mills action on 4-dimensional real spaces as :
\begin{eqnarray}
\label{Pure YM action}
S = \frac{-1}{2} \int d^{4}x \mbox{Tr}(F_{\mu \nu}F^{\mu \nu}),
\end{eqnarray}
the gauge theory is called pure Yang-Mills theory because the only contribution of the Lagrangian density is from the gauge fields $A_{\mu}$ and its field strengths $F_{\mu \nu}$. 
The condition of unitary gauge group here is no longer necessary for mathematical purpose because \eqref{Pure YM Lagrangian density} and \eqref{Pure YM action} are invariant under gauge transformation \eqref{Gauge transf of A_mu} which is independent of gauge groups. 
By using the method of variation, the condition $\delta S =0$ implies the so-called Yang-Mills field equation :
\begin{eqnarray}
\label{YM field eq}
D_{\mu}F_{\mu\nu} := [D_{\mu}, F_{\mu\nu}]=
\partial_{\mu}F_{\mu \nu} + [A_{\mu}, F_{\mu\nu}] = 0.
\end{eqnarray}  
In general, to find the exact solutions of the Yang-Mills field equation \eqref{YM field eq} without setting any constraint is very hard and almost impossible. One popular approach to simplify this difficulty is to impose the condition of the anti-self-duality or the self-duality on the Yang-Mills field strengths in the sense of Hodge dual. More precisely, we define the Hodge star operator $\star$ here by 
\begin{eqnarray}
\label{Hodge star operator}
\star F_{\mu \nu} = \frac{1}{2} \epsilon_{\mu \nu \rho \sigma} F^{\rho \sigma},  ~~\epsilon_{\mu \nu \rho \sigma} : \mbox{anti-symmetric tensor},
\end{eqnarray}
and $\star F_{\mu \nu}$ is called the Hodge dual of $F^{\mu\nu}$.
For different signature of 4-dimensional real spaces, one can check immediately that the anti-self-duality (or self-duality) of \eqref{Hodge star operator} is equivalent to the so-called anti-self-dual (or self-dual) Yang-Mills equations (Cf:~\cite{MaWo}):
\begin{itemize}
	\item On the Euclidean real space $\mathbb{E}$ with signature $(+, +, +, +)$ :
	\begin{eqnarray}
	\label{ASDYM_E}
	&&\star F_{\mu \nu} = \pm F_{\mu \nu} ~~~~
	(+ : \mbox{self-duality}, ~- : \mbox{anti-self-duality}) 
	\nonumber \\
	\Longleftrightarrow  
	&&F_{12} \pm F_{34}=0, ~ F_{13} \mp F_{24}=0, ~ F_{14} \pm F_{23}=0. ~~~~~~~~~~~~~~~~~~~~~~~~~~~~~~~~~~~~~~~~~
	\end{eqnarray}
	\item On the Minkowski real space $\mathbb{M}$ with signature $(+, -, -, -)$ :
	\begin{eqnarray}
	\label{ASDYM_M}
	&&\star F_{\mu \nu} = \pm i F_{\mu \nu} ~~~~
	(+ : \mbox{self-duality}, ~- : \mbox{anti-self-duality}) 
	\nonumber \\
	\Longleftrightarrow
	&&F_{01} \pm iF_{23}=0, ~ F_{02} \mp iF_{13}=0, ~ F_{03} \pm iF_{12}=0. ~~~~~~~~~~~~~~~~~~~~~~~~~~~~~~~~~~~~~
	\end{eqnarray}
	\item On the Ultrahyperbolic real space $\mathbb{U}$ with split signature $(+, +, -, -)$:
	\begin{eqnarray}
	\label{ASDYM_U}
	&&\star F_{\mu \nu} = \pm F_{\mu \nu} ~~~~
	(+ : \mbox{self-duality}, ~- : \mbox{anti-self-duality}) 
	\nonumber \\
	\Longleftrightarrow  
	&&F_{12} \pm F_{34}=0, ~ F_{13} \pm F_{24}=0, ~ F_{14} \mp F_{23}=0. ~~~~~~~~~~~~~~~~~~~~~~~~~~~~~~~~~~~~~~~~~
	\end{eqnarray}
\end{itemize}
Note that ASD (or SD \footnote{Without loss of generality, we just consider the anti-self-duality part of \eqref{ASDYM_E}, \eqref{ASDYM_M} and \eqref{ASDYM_U} in this thesis.}) Yang-Mills equations \eqref{ASDYM_E}, \eqref{ASDYM_M}, \eqref{ASDYM_U} are all invariant under gauge transformation \eqref{Gauge transf of A_mu} due to the fact of \eqref{Gauge covariant_F}.

\subsection{Brief review of ASDYM theory on 4D complex space}
\label{Section 3.2}
\smallskip
\subsubsection{Complex representation of the ASDYM equations}  
In this subsection, we give a brief review of the ASDYM theory on 4-dimensional complex space, some more details can be referred to \cite{MaWo}. 
Firstly, the ASDYM equations \eqref{ASDYM_E}, \eqref{ASDYM_M}, and \eqref{ASDYM_U} can be unified into a more general theory mathematically by considering 4-dimensional complex coordinates.
Let us define a  4-dimensional complex flat space with coordinates $(z, \widetilde{z}, w, \widetilde{w})$, in which the metric is defined by
\begin{eqnarray}
\label{complex metric}
ds^2&\!\!\!:=\!\!\!&g_{mn}dz^{m}dz^{n}
=2(dz d\widetilde z -dw d\widetilde w),
~~~m,n=1,2,3,4, ~\\
&&{\mbox{where}}~~
\label{N=2_com}
g_{mn}:=\left(
\begin{array}{cccc}
0&1&0&0 \\
1&0&0&0 \\ 
0&0&0&-1 \\
0&0&-1&0 
\end{array}\right),~
(z^{1},z^{2},z^{3},z^{4}):=(z,\widetilde z, w, \widetilde w).\nonumber
\end{eqnarray}
Then 4-dimensional real flat spaces with three different signatures can be embedded successfully into the 4-dimensional complex space and recovered from    
\eqref{complex metric} by imposing some suitable reality conditions 
on $z,\widetilde z, w, \widetilde w$ as follows.
\begin{itemize}
	
	\item On the Euclidean real space $\mathbb{E}$ with signature $(+, +, +, +)$, \\ 
	the reality condition is
	$\widetilde z=\overline z,~
	\widetilde w= -\overline w$, or without loss of generality, we can take
	\begin{eqnarray}
	\label{Reality condition_E}
	\left(\begin{array}{cc}
		z & w \\
		\widetilde{w} & \widetilde{z}
	\end{array}\right)
	=
	\frac{1}{\sqrt{2}}
	\left(\begin{array}{cc}
	x^{1}+ix^{2} & x^{3}+ix^{4} \\
	-(x^{3}-ix^{4}) & x^{1}-ix^{2}
	\end{array}\right).	
	\end{eqnarray}	
	
	\item On the Minkowski real space $\mathbb{M}$ with signature $(+, -, -, -)$, \\ 
	the reality condition is
	$z,\widetilde z \in \mathbb{R},~\widetilde w= \overline w$, or without loss of generality, we can take
	\begin{eqnarray}
	\label{Reality condition_M}
	\left(\begin{array}{cc}
	z & w \\
	\widetilde{w} & \widetilde{z}
	\end{array}\right)
	=
	\frac{1}{\sqrt{2}}
	\left(\begin{array}{cc}
	x^{0}+x^{1} & x^{2}+ix^{3} \\
	x^{2}-ix^{3} & x^{0}-x^{1}
	\end{array}\right).	
	\end{eqnarray}	
	
	\item On the Ultrahyperbolic real space $\mathbb{U}$ with split signature $(+, +, -, -)$, \\ 
	the reality condition is
	$z, \widetilde z,w, \widetilde w\in \mathbb{R}$, or without loss of generality, we can take
	\begin{eqnarray}
	\label{Reality condition_U}
	\left(\begin{array}{cc}
	z & w \\
	\widetilde{w} & \widetilde{z}
	\end{array}\right)
	=
	\frac{1}{\sqrt{2}}
	\left(\begin{array}{cc}
	x^{1}+x^{3} & (x^{2}+x^{4}) \\
	-(x^{2}-x^{4}) & x^{1}-x^{3} 
	\end{array}\right).	
	\end{eqnarray}	
\end{itemize}
Now let us consider a gauge theory on this complex space and assume the gauge group to be $G=\mathrm{GL}(N,\mathbb{C})$. 
A natural way to define the complex representation of covariant derivatives and field strengths can be done by the replacement of the real coordinates $x^{\mu}$ with the complex coordinates $z, \widetilde{z}, w, \widetilde{w}$ as follows :
\begin{eqnarray}
\label{Field strength_complex}
F_{mn} := [D_{m}, D_{n}]
= \partial_{m} A_{n}- \partial_{n} A_{m}+[A_m, A_n], ~~~~
D_{m}:=\partial_{m} +A_{m},
\end{eqnarray}
where $m, n$ denote the complex coordinates $z, \widetilde{z}, w, \widetilde{w}$ and $A_m, A_n$ denote the gauge fields 
taking values in the Lie algebra $\mathfrak{gl}(N,\mathbb{C})$. 
To construct a unified theory for the real representations of ASDYM equations \eqref{ASDYM_E}, \eqref{ASDYM_M}, and \eqref{ASDYM_U}, 
we can define the complex representation of ASDYM equations in the following way: 
\begin{eqnarray}
\label{ASDYM_complex}
F_{zw}
&\!\!\!=&\!\!\! \partial_{z} A_{w}- \partial_{w} A_{z}+[A_z, A_w]=0, 
\nonumber \\
F_{\widetilde z \widetilde w}
&\!\!\!=&\!\!\! \partial_{\widetilde z} A_{\widetilde w}- \partial_{\widetilde w} A_{\widetilde z}+[A_{\widetilde z}, A_{\widetilde w}]=0, 
 \\
F_{z \widetilde z}-F_{w \widetilde w}
&\!\!\!=&\!\!\! \partial_{z}A_{\widetilde{z}} - \partial_{\widetilde z}A_{z}
+\partial_{\widetilde w}A_{w} - \partial_{w}A_{\widetilde{w}}
+[A_{z}, A_{\widetilde{z}}]-[A_{w}, A_{\widetilde{w}}]
=0,  \nonumber
\end{eqnarray}
or equivalently,
\begin{eqnarray}
\label{ASDYM_complex2}
[D_{z}, D_{w}]=0, ~
\left[D_{\widetilde z}, D_{\widetilde w}\right]=0, ~
\left[D_{z}, D_{\widetilde{z}}\right] - \left[D_{w}, ~ D_{\widetilde{w}}\right] =0.
\end{eqnarray}
Now we can easily check that \eqref{ASDYM_complex} reduces to \eqref{ASDYM_E}, \eqref{ASDYM_M}, and \eqref{ASDYM_U} 
under the reality conditions \eqref{Reality condition_E}, \eqref{Reality condition_M}, and \eqref{Reality condition_U}, respectively.
\medskip \\
\\
\subsubsection{Complex representation of the Lagrangian density}
Although the Lagrangian density of complex variables is lack of physical interpretation, we define it mathematically in the sub-dissertation \cite{HaHu1} as a complex analogue of the ordinary Lagrangian density Tr$F_{\mu\nu}F^{\mu\nu}$. The advantage is that the Lagrangian density Tr$F_{\mu\nu}F^{\mu\nu}$ of three kinds of signatures can be unified into one complex space. More precisely, we use the convention \eqref{complex metric} and \eqref{Field strength_complex} to define the complex representation of Lagrangian density in the following way:
\begin{eqnarray}
{\mbox{Tr}}F^2&\!\!\!:=\!\!\!&{\mbox{Tr}}F_{mn}F^{mn}  ~~~(F^{mn}:=g^{mk}g^{nl}F_{kl}) \nonumber \\
&\!\!\!=& \!\!\!
-2{\mbox{Tr}}(
{{F}_{w{\widetilde{w}}}^2}+{{F}_{z{\widetilde{z}}}^2}
+2F_{{\widetilde{z}}w} F_{z{\widetilde{w}}}
+2F_{zw} F_{{\widetilde{z}}{\widetilde{w}}}
).
\end{eqnarray}
Here we ignore the $\frac{-1}{2}$ factor for simplicity (~\!Cf:~\eqref{Pure YM Lagrangian density}~\!).
For the anti-self-dual gauge fields, the Lagrangian density becomes simpler :
\begin{equation}
\label{Complex Lagrangian density_ASD}
{\mbox{Tr}}F^2
=
2{\mbox{Tr}}(F_{w{\widetilde{z}}} F_{z{\widetilde{w}}}
-{{F}_{z{\widetilde{z}}}^2}-{{F}_{w{\widetilde{w}}}^2})
=
4{\mbox{Tr}}(F_{w{\widetilde{z}}} F_{z{\widetilde{w}}}
-{{F}_{w{\widetilde{w}}}^2})
=
4{\mbox{Tr}}(F_{w{\widetilde{z}}} F_{z{\widetilde{w}}}
-{{F}_{z{\widetilde{z}}}^2}). 
\end{equation}
By imposing reality conditions \eqref{Reality condition_E}, \eqref{Reality condition_M}, and \eqref{Reality condition_U} on \eqref{Complex Lagrangian density_ASD} respectively, we can obtain the ordinary Lagrangian density ($w.r.t.$ the ASD gauge fields) on each real space:
\begin{itemize}
	\item On the Euclidean space $\mathbb{E}$ :  $\mbox{Tr}F_{\mu\nu}F^{\mu\nu}
	=4\mbox{Tr}(F_{12}^{2}+F_{13}^{2}+F_{14}^{2})$.  
	\item On the Minkowski space $\mathbb{M}$ : $\mbox{Tr}F_{\mu\nu}F^{\mu\nu}
	=-4\mbox{Tr}(F_{01}^{2}+F_{02}^{2}+F_{03}^{2})$.
	\item On the Ultrahyperbolic space $\mathbb{U}$ : 
	$\mbox{Tr}F_{\mu\nu}F^{\mu\nu}
	=4\mbox{Tr}(F_{12}^{2}-F_{13}^{2}-F_{14}^{2})$.
\end{itemize}   

\subsection{The equivalent descriptions of ASDYM equations on 4D complex space }
\label{Section 3.3}

\subsubsection{The $R$-gauge formulation of ASDYM ($G=\mathrm{SL}(2,\mathbb{C})$)}
In this subsection, we review the work of Yang in \cite{Yang} and give some detailed derivation as a supplementary note.
From the first two equations of ASDYM \eqref{ASDYM_complex2}, we find that there exist $N \times N$ invertible matrices $h$ and $\widetilde{h}$ such that 
\begin{eqnarray}
\label{Dh=0}
D_z h= 0, ~ D_{w}h=0, ~ D_{\widetilde{z}}\widetilde{h}=0, ~ D_{\widetilde{w}}\widetilde{h}=0, 
\end{eqnarray}
or equivalently,
\begin{eqnarray}
\label{Gauge fields in terms of h}
A_{z}=-(\partial_{z}h)h^{-1},~
A_{w}=-(\partial_{w}h)h^{-1},~
A_{\widetilde{z}}=-(\partial_{\widetilde{z}}\widetilde{h})\widetilde{h}^{-1},~
A_{\widetilde{w}}=-(\partial_{\widetilde{w}}\widetilde{h})\widetilde{h}^{-1}.~ 
\end{eqnarray}   
Conversely, we can use \eqref{Gauge fields in terms of h} 
and the fact that $\partial_{m}h^{-1}=-h^{-1}(\partial_{m}h)h^{-1}$
to check
\begin{eqnarray*}
&&F_{zw}  \nonumber  \\
&=& \partial_{z}\left(-(\partial_{w}h)h^{-1} \right)
-\partial_{w}\left(-(\partial_{z}h)h^{-1} \right)
+[-(\partial_{z}h)h^{-1}, -(\partial_{w}h)h^{-1}]   \nonumber  \\
&=& 
-(\partial_{zw}h)h^{-1}-(\partial_{w}h)(\partial_{z}h^{-1})
+(\partial_{wz}h)h^{-1}+(\partial_{z}h)(\partial_{w}h^{-1}) \nonumber \\
&&+(\partial_{z}h)h^{-1}(\partial_{w}h)h^{-1}  \nonumber
-(\partial_{w}h)h^{-1}(\partial_{z}h)h^{-1}  \nonumber   \\
&=& 0 ~~~~ (~\mbox{Similarly}, ~F_{\widetilde{z}\widetilde{w}} =0).  \nonumber
\end{eqnarray*}
Therefore, the first two equations of ASDYM \eqref{ASDYM_complex} are equivalent to the existence of \eqref{Gauge fields in terms of h}. Now let us assume the gauge group to be $G= \mathrm{GL}(2, \mathbb{C})$ and introduce a convenient gauge, called Yang's $R$-gauge \cite{Yang}, to find an equivalent description of the ASDYM equations. The $R$-gauge is defined by
\begin{eqnarray}
h_{R}
&\!\!\!\!:=&\!\!\!\! \frac{1}{2\sqrt{\varphi}}\left[
(1+\varphi)I-\rho(\sigma_{1}-i\sigma_{2})-(1-\varphi)\sigma_{3}
\right]
=\frac{1}{\sqrt{\varphi}}\left(
\begin{array}{cc}
\varphi & 0  \\
-\rho & 1
\end{array} \right),  \label{R-gauge_SL(2)_1} \\
\widetilde{h}_{R}
&\!\!\!\!:=&\!\!\!\! \frac{1}{2\sqrt{\varphi}}\left[
(1+\varphi)I+\widetilde\rho(\sigma_{1}+i\sigma_{2})+(1-\varphi)\sigma_{3}
\right]
=\frac{1}{\sqrt{\varphi}}\left(
\begin{array}{cc}
1 & \widetilde\rho  \\
0 & \varphi
\end{array} \right),  \label{R-gauge_SL(2)_2} 
\end{eqnarray}
where $\sigma_{i}$ are Pauli matrices, $\varphi$ is any real function, $\rho, \widetilde{\rho}$ are any complex functions, and the ASD gauge fields can be formulated as
\begin{eqnarray}
A_{z}&\!\!\! =& \!\!\! -(\partial_{z}h_R)h_R^{-1}
=
\frac{1}{2\varphi}
\left(
\begin{array}{ll}
-\partial_z\varphi & 0 \\
2\partial_z\rho & \partial_z\varphi
\end{array}
\right), ~~~~ \label{ASD fields_R-gauge_SL(2)_1} \\
A_{w} &\!\!\! =& \!\!\! -(\partial_{w}h_R)h_R^{-1}
=
\frac{1}{2\varphi}
\left(
\begin{array}{ll}
-\partial_w \varphi & 0 \\
2\partial_w\rho & \partial_w\varphi
\end{array}
\right), ~~~~~~~~~~ \label{ASD fields_R-gauge_SL(2)_2}    \\
A_{\widetilde{z}}&\!\!\!=& \!\!\!-(\partial_{\widetilde z}{\widetilde{h}_R}){\widetilde{h}_R}^{-1}
=
\frac{1}{2\varphi}
\left(
\begin{array}{ll}
\partial_{\widetilde z}\varphi & -2\partial_{\widetilde{z}}\widetilde \rho \\
0 & -\partial_{\widetilde{z}}\varphi
\end{array}
\right), ~ \label{ASD fields_R-gauge_SL(2)_3}\\
A_{\widetilde{w}}
&\!\!\!\! =& \!\!\!\!-(\partial_{\widetilde w}\widetilde{h}_R){\widetilde{h}_R}^{-1}
=
\frac{1}{2\varphi}
\left(
\begin{array}{ll}
\partial_{\widetilde{w}}\varphi & -2\partial_{\widetilde{w}}\widetilde \rho   \\
0 & -\partial_{\widetilde{w}}\varphi   \label{ASD fields_R-gauge_SL(2)_4} 
\end{array}
\right).
\end{eqnarray}
Note that the gauge fields are all traceless, therefore, the gauge group is in fact $G$=SL(2, $\mathbb{C}$). By using \eqref{ASD fields_R-gauge_SL(2)_1}, \eqref{ASD fields_R-gauge_SL(2)_2}, \eqref{ASD fields_R-gauge_SL(2)_3}, and \eqref{ASD fields_R-gauge_SL(2)_4}, we can calculate the field strengths
\begin{eqnarray*}
F_{z\widetilde{z}}
&\!\!\!\!=&\!\!\!\!
\frac{1}{\varphi^2}
\!\left(
\begin{array}{cc}
\!\varphi(\partial_{z}\partial_{\widetilde{z}}\varphi)
-(\partial_z\varphi) (\partial_{\widetilde{z}}\varphi)
+(\partial_z\rho) (\partial_{\widetilde{z}}\widetilde\rho) & 
-\varphi(\partial_{z}\partial_{\widetilde{z}}\widetilde\rho) 
+2(\partial_z\varphi)(\partial_{\widetilde{z}}\widetilde \rho)  \\
-\varphi(\partial_{z}\partial_{\widetilde{z}}\rho) 
+ 2(\partial_z\rho)(\partial_{\widetilde{z}}\varphi)  & 
\!\!\!-\varphi(\partial_{z}\partial_{\widetilde{z}}\varphi)
+(\partial_z\varphi) (\partial_{\widetilde{z}}\varphi)
-(\partial_z\rho) (\partial_{\widetilde{z}}\widetilde\rho)
\end{array}
\!\!\!\right),  \nonumber  \\
F_{w\widetilde{w}}
&\!\!\!\!=& \!\!\!\!
\frac{1}{\varphi^2}
\!\left(\!\!\!
\begin{array}{cc}
\varphi(\partial_{w}\partial_{\widetilde{w}}\varphi)
\!-\!(\partial_w\varphi) (\partial_{\widetilde{w}}\varphi)
\!+\!(\partial_w\rho) (\partial_{\widetilde{w}}\widetilde\rho) & 
-\varphi(\partial_{w}\partial_{\widetilde{w}}\widetilde\rho) 
+2(\partial_w\varphi)(\partial_{\widetilde{w}}\widetilde \rho)  \\
-\varphi(\partial_{w}\partial_{\widetilde{w}}\rho) 
+ 2(\partial_w\rho)(\partial_{\widetilde{w}}\varphi)  & 
\!\!\!\!-\varphi(\partial_{w}\partial_{\widetilde{w}}\varphi)
\!+\!(\partial_w\varphi) (\partial_{\widetilde{w}}\varphi)
\!-\!(\partial_w\rho) (\partial_{\widetilde{w}}\widetilde\rho)
\end{array}
\!\!\!\right),  \nonumber 
\end{eqnarray*}
and conclude that the third equation of ASDYM \eqref{ASDYM_complex} ($i.e.$ $F_{z\widetilde{z}} - F_{z\widetilde{z}} =0$) is equivalent to the following three equations of Yang.
\newtheorem{thm_3.1}{Theorem}[section]
\begin{thm_3.1}
{\bf ($R$-gauge Formulation for $G=\mathrm{SL}(2,\mathbb{C})$, {Cf: \cite{Yang}} for $G=\mathrm{SU}(2)$)}  \label{Thm_3.1} \\ 	
Under Yang's $R$-gauge \eqref{R-gauge_SL(2)_1} and \eqref{R-gauge_SL(2)_2}, the ASDYM equations \eqref{ASDYM_complex} can be formulated as the following three equations of Yang : 
\begin{eqnarray}
\label{Yang's eqns_SL(2,C)}
\begin{array}{l}
\varphi({\partial_{z}}{\partial_{\widetilde{z}}} \varphi
-{\partial_{w}}{\partial_{\widetilde{w}}}\varphi)
-\left[
(\partial_{z}\varphi) (\partial_{\widetilde{z}}\varphi)-(\partial_{w}\varphi)(\partial_{\widetilde{w}}\varphi)
\right]
+\left[
(\partial_{z}\rho) (\partial_{\widetilde{z}} \widetilde\rho)
-(\partial_{w}\rho) (\partial_{\widetilde{w}}\widetilde\rho)
\right]=0, ~~\\
\varphi({\partial_{z}}{\partial_{\widetilde{z}}}\rho
-{\partial_{w}}{\partial_{\widetilde{w}}}\rho)
-2\left[
(\partial_{z}\rho) (\partial_{\widetilde{z}}\varphi)
-(\partial_{w}\rho) (\partial_{\widetilde{w}}\varphi)
\right]=0, \\
\varphi({\partial_{z}}{\partial_{\widetilde{z}}}\widetilde\rho
-{\partial_{w}}{\partial_{\widetilde{w}}}\widetilde \rho)
-2\left[
(\partial_{z}\varphi) (\partial_{\widetilde{z}}\widetilde \rho)
-(\partial_{w}\varphi) (\partial_{\widetilde{w}}\widetilde \rho)
\right]=0.
\end{array}
\end{eqnarray}
\end{thm_3.1}

\subsubsection{$J$-matrix formulation of ASDYM ($G=\mathrm{GL}(N,\mathbb{C})$)}
In this subsection, we follow \cite{MaWo} and review the $J$-matrix formulation of ASDYM equations for gauge group $G=\mathrm{GL}(N,\mathbb{C})$.
Recall that in \eqref{Gauge fields in terms of h}, the ASD gauge fields can be expressed in terms of $N \times N$ invertible matrices $h$ and $\widetilde{h}$. 
In \cite{BrFaNuYa}, the authors point out that the three equations \eqref{Yang's eqns_SL(2,C)} of Yang  can be cast in a gauge independent formulation, called the Yang equation
\begin{eqnarray}
\label{Yang equation}
\partial_{\widetilde{z}}(J^{-1}\partial_{z}J)-
\partial_{\widetilde{w}}(J^{-1}\partial_{w}J) =0,
\end{eqnarray}
by introducing the matrix $J:=h^{-1}\widetilde{h}$ (Cf: ~\cite{Yang} for $G=\mathrm{SU}(2)$).
Here the $N \times N$ matrix $J$ is called Yang's $J$-matrix and invariant under the gauge transformation (Cf: \eqref{Dh=0})
\begin{eqnarray}
h\longmapsto gh,~~ \widetilde h\longmapsto g\widetilde h,~~ g \in \mathrm{G}.
\end{eqnarray}
Therefore, the Yang equation \eqref{Yang equation} is gauge invariant.
\newtheorem{thm_3.2}[thm_3.1]{Theorem}
\begin{thm_3.2}
{\bf ($J$-matrix formulation for $G=\mathrm{GL}(N,\mathbb{C})$ \cite{BrFaNuYa}, Cf: \cite{MaWo})}
\label{Thm_3.2} \\
The ASDYM equations \eqref{ASDYM_complex} can be formulated as the Yang equation :
\begin{eqnarray}
\partial_{\widetilde{z}}(J^{-1}\partial_{z}J)-
\partial_{\widetilde{w}}(J^{-1}\partial_{w}J) =0, ~ J:=h^{-1}\widetilde{h},  \nonumber 
\end{eqnarray}
where $h$ and $\widetilde{h}$ satisfy
\begin{eqnarray}
A_{z}=-(\partial_{z}h)h^{-1},~
A_{w}=-(\partial_{w}h)h^{-1},~
A_{\widetilde{z}}=-(\partial_{\widetilde{z}}\widetilde{h})\widetilde{h}^{-1},~
A_{\widetilde{w}}=-(\partial_{\widetilde{w}}\widetilde{h})\widetilde{h}^{-1}.~ \nonumber
\end{eqnarray}
\end{thm_3.2}
{\bf (\emph{Proof})} \\
In the original paper \cite{BrFaNuYa} the authors didn't mention their derivation of \eqref{Yang equation}, we give the following proof as a supplementary note here.
By direct calculation, we have
\begin{eqnarray}
&&\!J^{-1}\partial_{w}J  \nonumber \\
&=\!\!\!&
\widetilde{h}^{-1}h
\left[(\partial_{w}h^{-1})\widetilde{h}+h^{-1}(\partial_{w}\widetilde{h})\right]
=
-\widetilde{h}
\left[(\partial_{w}h)h^{-1}\right]
\widetilde{h}+\widetilde{h}^{-1}(\partial_{w}\widetilde{h}) \nonumber \\
&=\!\!\!&  \widetilde{h}^{-1}A_{w}\widetilde{h}+\widetilde{h}^{-1}(\partial_{w}\widetilde{h}) \nonumber 
\end{eqnarray}
\begin{eqnarray}
&&\!\partial_{\widetilde{w}}(J^{-1}\partial_{w}J)  \nonumber  \\
&=\!\!\!& 
(\partial_{\widetilde{w}}\widetilde{h}^{-1})A_{w}\widetilde{h}
+\widetilde{h}^{-1}\left[ (\partial_{\widetilde w}A_{w})\widetilde{h}+A_{w}(\partial_{\widetilde{w}}\widetilde{h}) \right]
+(\partial_{\widetilde{w}}\widetilde{h}^{-1})(\partial_{w}\widetilde{h})+\widetilde{h}^{-1}(\partial_{\widetilde{w}w}\widetilde{h}) 
\nonumber   \\
&=\!\!\!&
\widetilde{h}^{-1}
\left[
-(\partial_{\widetilde w}\widetilde{h})\widetilde{h}^{-1}A_{w}
+\partial_{\widetilde w}A_{w}
+A_{w}(\partial_{\widetilde w}\widetilde{h})\widetilde{h}^{-1}
\underbrace{-(\partial_{\widetilde{w}}\widetilde{h})\widetilde{h}^{-1}(\partial_{w}\widetilde{h})\widetilde{h}^{-1}
+(\partial_{\widetilde{w}w}\widetilde{h})\widetilde{h}^{-1}}
\right]\widetilde{h} \nonumber \\
&& ~~~~~~~~~~~~~~~~~~~~~~~~~~~~~~~~~~~~~~~~~~~~~~~~~~~~~~~~~~
=\partial_{w}
\left[(\partial_{\widetilde{w}}\widetilde{h})\widetilde{h}^{-1}\right]
=
-\partial_{w}A_{\widetilde{w}}
\nonumber \\
&=\!\!\!&
\widetilde{h}^{-1}
(\partial_{\widetilde{w}}A_{w}
-\partial_{w}A_{\widetilde{w}}
+A_{w}A_{\widetilde{w}}
-A_{\widetilde{w}}A_{w})\widetilde{h} \nonumber  \\
&=\!\!\!&
-\widetilde{h}^{-1}F_{w\widetilde{w}}\widetilde{h}.~~~~  
(~\mbox{Similarly},~
\partial_{\widetilde{z}}(J^{-1}\partial_{z}J)
=-\widetilde{h}^{-1}F_{z\widetilde{z}} \widetilde{h}.)
~~~~~~~~~~~~~~~~~~~~~~~~~~~~~~~~\nonumber
\end{eqnarray}
We can conclude a beautiful relation 
\begin{eqnarray}
\partial_{\widetilde{z}}(J^{-1}\partial_{z}J)-
\partial_{\widetilde{w}}(J^{-1}\partial_{w}J)
=\widetilde{h}^{-1}(F_{w\widetilde{w}}-F_{z\widetilde{z}})\widetilde{h}
\end{eqnarray}
which implies that 
\begin{eqnarray}
F_{z\widetilde{z}}-F_{w\widetilde{w}}=0  
\iff  
\partial_{\widetilde{z}}(J^{-1}\partial_{z}J)-
\partial_{\widetilde{w}}(J^{-1}\partial_{w}J) = 0.    
\end{eqnarray}
In other words, the third equation of ASDYM \eqref{ASDYM_complex} is equivalent to the Yang equation.
(In \cite{MaWo}, the authors give another proof which is made by taking a special gauge transformation.) $\hfill\Box$
\medskip  \\
For gauge group $G=\mathrm{SL}(2,\mathbb{C})$, a parametrization of Yang's $J$-matrix can be obtained immediately from the $R$-gauge \eqref{R-gauge_SL(2)_1} and \eqref{R-gauge_SL(2)_2} :
\begin{eqnarray}
\label{J_R}
J=
h_R^{-1}\widetilde{h}_{R}
=
\frac{1}{\varphi}\left(
\begin{array}{cc}
1 & \widetilde{\rho} \\
\rho & \varphi^2+\rho\widetilde{\rho}
\end{array}
\right),
\end{eqnarray}
which implies the following result directly. 
\newtheorem{cor_3.3}[thm_3.1]{Corollary}
\begin{cor_3.3}
	{\color{white}\bf Corollary 3.3}  \label{Cor_3.3}  \\
Under the parametrization \eqref{J_R} of $J$-matrix, the three equations \eqref{Yang's eqns_SL(2,C)} of Yang is equivalent to the Yang equation \eqref{Yang equation} for $G=\mathrm{SL}(2,\mathbb{C})$. 
\end{cor_3.3}

\subsubsection{$K$-matrix formulation of ASDYM ($G=\mathrm{GL}(N,\mathbb{C})$)}
In this subsection, we follow \cite{MaWo} and review the $K$-matrix formulation of ASDYM equations for $G=\mathrm{GL}(N,\mathbb{C})$. Firstly, let us consider the following gauge, and call it $S$-gauge for convenience.
\begin{eqnarray}
\label{S-gauge}
h=J^{-1},~ \widetilde{h}=I.
\end{eqnarray}
Under the $S$-gauge, we can get two vanishing gauge fields and the remaining two are expressed in terms of $J$-matrix :
\begin{eqnarray}
\label{Gauge fields_S-gauge_general}
A_{\widetilde z}=A_{\widetilde w}=0,~~ A_{z}=J^{-1}\partial_{z}J,~~ A_{w}=J^{-1}\partial_{w}J.
\end{eqnarray}
By the first two equations of \eqref{Gauge fields_S-gauge_general}, the ASDYM equations \eqref{ASDYM_complex} reduces to
\begin{eqnarray}
\label{ASDYM_complex_reduced}
\partial_{z} A_{w}- \partial_{w} A_{z}+[A_z, A_w]=0,~~ 
\partial_{\widetilde z}A_{z} - \partial_{\widetilde w}A_{w} = 0.
\end{eqnarray}
We find that the second equation of \eqref{ASDYM_complex_reduced} is equivalent to the existence of a potential $K$ such that 
\begin{eqnarray}
\label{Potential form of gauge fields}
A_{z}=\partial_{\widetilde{w}}K, ~~ 
A_{w}=\partial_{\widetilde{z}}K.
\end{eqnarray}
Substituting \eqref{Potential form of gauge fields} into the first equation of \eqref{ASDYM_complex_reduced}, we have 
\begin{eqnarray}
\label{K equation}
\partial_{z}\partial_{\widetilde z}K-
\partial_{w}\partial_{\widetilde w}K+
[\partial_{\widetilde w}K, ~\partial_{\widetilde z}K] = 0. 
\end{eqnarray} 
Comparing  \eqref{Potential form of gauge fields} with the last two equations of \eqref{Gauge fields_S-gauge_general}, we have
\begin{eqnarray}
\label{Potential form of gauge fields_J}
\partial_{\widetilde{z}}K=J^{-1}\partial_{w}J, ~~
\partial_{\widetilde{w}}K=J^{-1}\partial_{z}J.  
\end{eqnarray}
Now we can conclude the following theorem.
\newtheorem{thm_3.4}[thm_3.1]{Theorem}
\begin{thm_3.4}
	{\bf ($K$-matrix formulation for $G=\mathrm{GL}(N,\mathbb{C})$, Cf: \cite{MaWo})} \label{Thm_3.4} \\
Under the $S$-gauge \eqref{S-gauge},
the ASDYM equations \eqref{ASDYM_complex} can be formulated as the equation \eqref{K equation}, where the relationship between the $K$-matrix and the $J$-matrix is presented as \eqref{Potential form of gauge fields_J}.
\end{thm_3.4}

\subsubsection{Lax representation of ASDYM ($G=\mathrm{GL}(N,\mathbb{C})$)}
The Lax representation of soliton equations is a very typical technique in solving exact solutions in the field of integrable system.  The details of further discussion can be referred to any typical textbook (e.g. \cite{Dicky, DrJo, Hirota, MaSa, Toda1}). For our purpose in this thesis, we just mention the Lax representation of the ASDYM equations as the following Theorem \ref{Thm_3.5} and excerpt the proof from \cite{MaWo}. 
\newtheorem{thm_3.5}[thm_3.1]{Theorem}
\begin{thm_3.5}
	{\bf (Lax representation, Cf: \cite{MaWo}) } \label{Thm_3.5} \\
	Let $(L, M)$ be the Lax operators formulated by  
	\begin{eqnarray}
	\label{Lax representation of ASDYM}
	\left\{
	\begin{array}{c}
	L:= D_{w}-\zeta D_{\widetilde z}=(\partial_{w}+A_{w})-\zeta(
	\partial_{\widetilde{z}}+A_{\widetilde z}) \\
	M:= D_{z}-\zeta D_{\widetilde w}=(\partial_{z}+A_{z})-\zeta(
	\partial_{\widetilde{w}}+A_{\widetilde w})
	\end{array}	
	\right.,
	\end{eqnarray} 
	where $\zeta$ is an nonzero complex number, called spectral parameter. Then $[L, M]$ =0, for all $\zeta$ if and only if the ASDYM equations hold. 
\end{thm_3.5}
	{\bf (\emph{Proof})}
	\begin{eqnarray}
	[L,M]       
	&=& 
	[D_{w}-\zeta D_{\widetilde z}, ~D_{z}-\zeta D_{\widetilde w}]  
	=
	[D_{w}, D_{z}] + \zeta([ D_{z}, D_{\widetilde{z}},]+[D_{w}, D_{\widetilde{w}}]) + \zeta^{2}[D_{\widetilde{z}}, D_{\widetilde{w}}]  \nonumber  \\ 
	&=&
	F_{wz}+\zeta(F_{z\widetilde{z}}-F_{w \widetilde{w}}) + \zeta^{2}F_{\widetilde{z}\widetilde{w}}
	=0, ~\mbox{for all~} \zeta.  \nonumber \\
	&&\Longleftrightarrow ~~
	F_{zw}=0, ~F_{\widetilde{z}\widetilde{w}}=0, ~F_{z\widetilde{z}}-F_{w \widetilde{w}}=0.   \nonumber
	\end{eqnarray}$\hfill\Box$
	\\
Note that the linear system $L\phi=M\phi=0$ is invariant under the gauge transformation 
\begin{eqnarray}
L \longmapsto gLg^{-1}, ~~ M \longmapsto gMg^{-1},~~ \phi \longmapsto g\phi. 
\end{eqnarray}

Now let us introduce a novel idea of Nimmo-Gilson-Ohta \cite{NiGiOh} that generalize the spectral parameter $\zeta$ (scalar quantity) in \eqref{Lax representation of ASDYM} to $N \times N$ constant matrix and consider "Lax operators" as follows :
\begin{eqnarray}
\label{Lax representation of ASDYM_NGO}
\left\{
\begin{array}{l}
L(\phi):= D_{w}\phi-(D_{\widetilde z}\phi)\zeta=(\partial_{w}+A_{w})\phi-\left((
\partial_{\widetilde{z}}+A_{\widetilde z})\phi\right)\zeta  \\
M(\phi):= D_{z}\phi-(D_{\widetilde w}\phi)\zeta=(\partial_{z}+A_{z})\phi-\left((
\partial_{\widetilde{w}}+A_{\widetilde w})\phi\right)\zeta
\end{array}	
\right.,
\end{eqnarray} 
where $\phi$ and $\zeta$ are $N \times N$ matrices. The crucial difference between \eqref{Lax representation of ASDYM_NGO} and \eqref{Lax representation of ASDYM} is that the constant matrix $\zeta$ must follow an additional operation rule. More precisely, $\zeta$ must act on the matrix function $\phi$ from the rightside. As a result, $L$, $M$ are not conventional operators anymore. Nevertheless, we can consider \eqref{Lax representation of ASDYM_NGO} as a generalization of the Lax operators because \eqref{Lax representation of ASDYM_NGO} actually return to 
\begin{eqnarray}
\left\{
\begin{array}{l}
L\phi= D_{w}\phi-\zeta D_{\widetilde z}\phi=(\partial_{w}+A_{w})\phi-\zeta (
\partial_{\widetilde{z}}+A_{\widetilde z})\phi \\
M\phi= D_{z}\phi-\zeta D_{\widetilde w}\phi=(\partial_{z}+A_{z})\phi-\zeta (
\partial_{\widetilde{w}}+A_{\widetilde w})\phi
\end{array}	
\right.
\end{eqnarray} 
if $\zeta$ is an $N \times N$ scalar matrix.
Now we can derive the ASDYM equations from \eqref{Lax representation of ASDYM_NGO} under the condition that $L(M(\phi))=M(L(\phi))$ holds for all constant matrix $\zeta$ :     
\begin{eqnarray}
L(M(\phi))-M(L(\phi))  
&\!\!\!=&\!\!\!
D_{w}\left(D_{z}\phi-(D_{\widetilde{w}}\phi)\zeta\right)
-D_{\widetilde{z}}\left(D_{z}\phi-(D_{\widetilde{w}}\phi)\zeta\right)\zeta  \nonumber  \\
&\!\!\!=&\!\!\!
[D_{w}, D_{z}]\phi + ([D_{z}, D_{\widetilde{z}}]-[D_{w}, D_{\widetilde{w}}])\phi\zeta + [D_{\widetilde{z}}, D_{\widetilde{w}}]\phi\zeta^2  \nonumber  \\
&\!\!\!=&\!\!\!
F_{wz}\phi 
+ (F_{z\widetilde{z}}-F_{w\widetilde{w}})\phi\zeta
+ F_{\widetilde{z}\widetilde{w}}\phi\zeta^2
=0 , ~~\mbox{for all}~ \zeta   \nonumber  \\
&&\Longleftrightarrow ~~
F_{zw}=0, ~F_{\widetilde{z}\widetilde{w}}=0, ~F_{z\widetilde{z}}-F_{w \widetilde{w}}=0. \nonumber
\end{eqnarray}
Note that the linear system $L(\phi)=M(\phi)=0$ is invariant under the gauge transformation 
\begin{eqnarray}
A_m \longmapsto gA_mg^{-1}-(\partial_mg)g^{-1}, ~~\phi \longmapsto g\phi ~~(i.e. ~ D_m\phi \longmapsto gD_m\phi),~ m=z, \widetilde{z}, w, \widetilde{w}.~~~ 
\end{eqnarray}

\subsection{Conditions of $G=\mathrm{SU}(N)$ ASD gauge fields on 4D real spaces}
\label{Section 3.4}

\subsubsection{Property of gauge fields in $G=\mathrm{SU}(N)$ gauge theory} 
According to the requirement of $G=\mathrm{SU}(N)$ gauge theory, the gauge fields $A_{\mu}$ must belong to the Lie algebra $\mathfrak{su}(N)$. Conversely, if a gauge field $A_{\mu}$ belongs to the Lie algebra $\mathfrak{su}(N)$, then we can always make sure of the existence of $\mathrm{SU}(N)$ gauge theory. More precisely, if a gauge field $A_{\mu} \in \mathfrak{su}(N)$, then it preserves the properties of 
\begin{eqnarray}
\mathfrak{su}(N)
=
\left\{
X \in \mathrm{Mat}(N,\mathbb{C}) \big| X^{\dagger}=-X,~\det(X)=0
\right\}  \nonumber 
\end{eqnarray}
under the $G=\mathrm{SU}(N)$ gauge transformation  
\begin{equation}
\label{Gauge transf of A_mu G=SU(N)}
A_{\mu} \longmapsto 
A_{\mu}^{\prime} := 
gA_{\mu}g^{-1} - (\partial_{\mu}g)g^{-1}, ~~ g \in \mathrm{SU}(N). 
\end{equation} 
This property is not hard to verify because 
$A_{\mu} \in \mathfrak{su}(N)$ and $g \in \mathrm{SU}(N)$ directly imply that
\begin{eqnarray*}
(A_{\mu}^{\prime})^{\dagger}
&\!\!\!=&\!\!\!
(g^{-1})^{\dagger}A_{\mu}^{\dagger}g^{\dagger}
-(g^{-1})^{\dagger}\partial_{\mu}g^{\dagger}
=
g(-A_{\mu})g^{-1} - g(\partial_{\mu}g^{-1}) 
=
-g^{-1}A_{\mu}g + (\partial_{\mu}g)g^{-1}  \nonumber \\
&\!\!\!=&\!\!\!
-A_{\mu}^{\prime}, \nonumber  \\
\mbox{Tr}(A_{\mu}^{\prime})
&\!\!\!=&\!\!\!
\mbox{Tr}(gA_{\mu}g^{-1}) - \mbox{Tr}\left[(\partial_{\mu}g)g^{-1}\right]
=\mbox{Tr}A_{\mu} - \partial_{\mu}\log \left[\mbox{det}(g)\right]
=0.  \nonumber 
\end{eqnarray*} 
Here we use the cyclic property of trace and the Jacobi's formula : 
\begin{eqnarray}
\label{Jacobi's formula}
\frac{d}{dt}\mbox{det}A(t) = \mbox{Tr}\left[\mbox{adj}(A(t)) \frac{dA(t)}{dt}\right]
= \mbox{det}A(t)\cdot\mbox{Tr}(A^{-1}\frac{dA(t)}{dt}). 
\end{eqnarray}
$\hfill\Box$
\\
{\bf Remark} \\
In fact, the gauge fields $A_{\mu}$ preserve the hermiticity of $\mathfrak{su}\mathrm{(N)}$ even under a gauge transformation of $G=\mathrm{GL}(N,\mathbb{C})$
\begin{equation*}
A_{\mu} \longmapsto
gA_{\mu}g^{-1} - (\partial_{\mu}g)g^{-1}, ~~ g \in \mathrm{G}  \nonumber
\end{equation*}
if there exist $C \in \mathrm{Mat}(N, \mathbb{C})$ which is independent of all $x^{\mu}$, commutes with $A_{\mu}$ such that $gg^{\dagger}=C$. 
In particular, we can take $C=I$ and hence $g$ is unitary in this case.  
\subsubsection{$G=\mathrm{U}(N)$ ASD gauge fields on the Euclidean space $\mathbb{E}$}
\label{Section 3.4.2}

In this subsection, we review the work of Yang \cite{Yang} and give some detailed derivation as a supplementary note.
The real representation of ASD gauge fields $A_{\mu}~ (\mu =1, 2, 3, 4)$ on the Euclidean space $\mathbb{E}$ can be reduced from  the complex representation \eqref{Gauge fields in terms of h}. For instance, we can use the reality condition \eqref{Reality condition_E} to obtain:
\begin{eqnarray}
\label{Derivative_E_cpx as real}
\partial_{z}&\!\!\!\! =& \!\!\!\! \frac{1}{\sqrt{2}}(\partial_{1}-i\partial_{2}),~ 
\partial_{\overline{z}}=\frac{1}{\sqrt{2}}(\partial_{1}+i\partial_{2}),~
\partial_{w}=\frac{1}{\sqrt{2}}(\partial_{3}-i\partial_{4}),~
\partial_{\overline{w}}=\frac{1}{\sqrt{2}}(\partial_{3}+i\partial_{4}), ~~~~~~~~~~~~ \\
A_{z}&\!\!\!\! = & \!\!\!\! \frac{1}{\sqrt{2}}(A_{1} \!- iA_{2}), 
A_{\overline{z}}\!=\!\frac{1}{\sqrt{2}}(A_{1}+iA_{2}),
A_{w}\!=\!\frac{1}{\sqrt{2}}(A_{3}\!-iA_{4}),
A_{\overline{w}}\!=\!\frac{1}{\sqrt{2}}(A_{3}+iA_{4}) \label{Gauge fields_E_cpx as real},
\end{eqnarray}
or equivalently, 
\begin{eqnarray}
\label{Derivative_E_real as cpx}
\partial_{1}&\!\!\!\! =& \!\!\!\!\frac{1}{\sqrt{2}}(\partial_{z}+\partial_{\overline{z}}),~
\partial_{2}=\frac{i}{\sqrt{2}}(\partial_{z}-\partial_{\overline{z}}),
\partial_{3}=\frac{1}{\sqrt{2}}(\partial_{w}+\partial_{\overline{w}}),
\partial_{4}=\frac{i}{\sqrt{2}}(\partial_{w}-\partial_{\overline{w}}), ~~~~~~~~~~~~~~~~~ \\
A_{1}&\!\!\!\! =& \!\!\!\!\frac{1}{\sqrt{2}}(A_{z}+A_{\overline{z}}),
A_{2}=\frac{i}{\sqrt{2}}(A_{z}\!-\!A_{\overline{z}}),~
A_{3}=\frac{1}{\sqrt{2}}(A_{w}+A_{\overline{w}}),~
A_{4}=\frac{i}{\sqrt{2}}(A_{w}\!-\!A_{\overline{w}}).   \label{Gauge fields_E_real as cpx}
\end{eqnarray}
Then the complex representation of ASD gauge fields \eqref{Gauge fields in terms of h} reduce to
\begin{eqnarray*}
A_{z}
&\!\!\!\!=& \!\!\!\!
-(\partial_{z}h)h^{-1}
=
\frac{-1}{\sqrt{2}}(\partial_{1}h-i\partial_{2}h)h^{-1},~~~
A_{\overline{z}}
=
-(\partial_{\overline{z}}\widetilde{h})\widetilde{h}^{-1}
=
\frac{-1}{\sqrt{2}}(\partial_{1}\widetilde{h}+i\partial_{2}\widetilde{h})\widetilde{h}^{-1},  \nonumber \\
A_{w}
&\!\!\!\!=& \!\!\!\!
-(\partial_{w}h)h^{-1}
=
\frac{-1}{\sqrt{2}}(\partial_{3}h-i\partial_{4}h)h^{-1},~~~
A_{\overline{w}}
=
-(\partial_{\overline{w}}\widetilde{h})\widetilde{h}^{-1}
=
\frac{-1}{\sqrt{2}}(\partial_{3}\widetilde{h}+i\partial_{4}\widetilde{h})\widetilde{h}^{-1}.   \nonumber
\end{eqnarray*}
By direct substitution, we obtain
\begin{eqnarray}
\label{Real representation of gauge fields_E}
&&A_{1}
=\frac{-1}{2}
\left\{ 
\left[(\partial_{1}\widetilde{h})\widetilde{h}^{-1}+(\partial_{1}h)h^{-1} \right]+
i\left[(\partial_{2}\widetilde{h})\widetilde{h}^{-1}-(\partial_{2}h)h^{-1}
\right]
\right\},  \nonumber  \\
&&A_{2}
=\frac{-1}{2}
\left\{ 
\left[(\partial_{2}\widetilde{h})\widetilde{h}^{-1}+(\partial_{2}h)h^{-1} \right]
-i\left[(\partial_{1}\widetilde{h})\widetilde{h}^{-1}-(\partial_{1}h)h^{-1}
\right]
\right\},   \\
&&A_{3}
=\frac{-1}{2}
\left\{ 
\left[(\partial_{3}\widetilde{h})\widetilde{h}^{-1}+(\partial_{3}h)h^{-1} \right]
+i\left[(\partial_{4}\widetilde{h})\widetilde{h}^{-1}-(\partial_{4}h)h^{-1}
\right]
\right\},  \nonumber  \\
&&A_{4}
=\frac{-1}{2}
\left\{ 
\left[(\partial_{4}\widetilde{h})\widetilde{h}^{-1}+(\partial_{4}h)h^{-1} \right]
-i\left[(\partial_{3}\widetilde{h})\widetilde{h}^{-1}-(\partial_{3}h)h^{-1}
\right]
\right\}.  \nonumber
\end{eqnarray}
If we impose the condition
\begin{eqnarray} 
\label{h^{dagger}tilde{h}=C}
h^{\dagger}\widetilde{h}=C ~~~ ( \mbox{$C$: independent of $x^{\mu}$} ) 
\end{eqnarray}
on \eqref{Real representation of gauge fields_E}, it implies  
\begin{eqnarray}
\label{Condition of U(N) gauge group_E}
\begin{array}{c}
\left[ (\partial_{\mu}h)h^{-1}\right]^{\dagger}
=
h^{-\dagger}(\partial_{\mu}h^{\dagger})
=
\widetilde{h}C^{-1}\partial_{\mu}(C\widetilde{h}^{-1})=-(\partial_{\mu}\widetilde{h})\widetilde{h}^{-1}, \\
\left[ (\partial_{\mu}\widetilde{h})\widetilde{h}^{-1}\right]^{\dagger}
=
\widetilde{h}^{-\dagger}(\partial_{\mu}\widetilde{h}^{\dagger})
=
hC^{-\dagger}\partial_{\mu}(C^{\dagger}h^{-1})
=
-(\partial_{\mu}h)h^{-1},
\end{array}
\end{eqnarray}
and therefore $A_{\mu}^{\dagger}=-A_{\mu}$, for all $\mu=1, 2, 3, 4$. That is, the gauge group can be $G=\mathrm{U}(N)$ under the condition \eqref{h^{dagger}tilde{h}=C}.
\newtheorem{prop_3.6}[thm_3.1]{Theorem}
\begin{prop_3.6}
	{\bf (Condition of unitary gauge group on $\mathbb{E}$, Cf: \cite{Yang})} \label{Prop_3.6} \\ 
Let $A_{\mu} ~ (\mu=1,2,3,4)$ be gauge fields defined on the Euclidean space $\mathbb{E}$ as \eqref{Real representation of gauge fields_E}. If $h,~ \widetilde{h}$ satisfy the relation $h^{\dagger}\widetilde{h}=C$, for some $N \times N$ matrix $C$ ($C$: independent of $x^{\mu}$), then the gauge group can be $G=\mathrm{U}(N)$. In particular, if we take $C = I$, then the condition for $G=\mathrm{U}(N)$ is $h^{\dagger}=\widetilde{h}^{-1}$ in this case.   
\end{prop_3.6}

Especially for $N=2$ case we can impose the condition $h_R^{\dagger}=\widetilde{h}_R^{-1}$ on the $R$-gauge \eqref{R-gauge_SL(2)_1} and \eqref{R-gauge_SL(2)_2}, or equivalently the condition $\widetilde{\rho}=\overline{\rho}$ (complex conjugate of $\rho$), to get $G=\mathrm{SU}(2)$ ASD gauge fields \cite{Yang} :
\begin{eqnarray}
A_{z}&\!\!\!\! =& \!\!\!\! -(\partial_{z}h_R)h_R^{-1}
=
\frac{1}{2\varphi}
\left(
\begin{array}{cc}
-\partial_{z}\varphi & 0 \\
2\partial_{z}\rho & \partial_{z}\varphi
\end{array}
\right), ~  \label{R-gauge_SU(2) gauge fields_cpx_1} \\
A_{w}&\!\!\!\!=&\!\!\!\! -(\partial_{w}h_R)h_R^{-1}
=
\frac{1}{2\varphi}
\left(
\begin{array}{cc}
-\partial_{w}\varphi & 0 \\
2\partial_{w}\rho & \partial_{w}\varphi
\end{array}
\right), ~~~~~~~~~~  \label{R-gauge_SU(2) gauge fields_cpx_2}
\end{eqnarray}
\begin{eqnarray}
A_{\overline{z}}&\!\!\!\!=& \!\!\!\!-(\partial_{\overline z}{\widetilde{h}_R}){\widetilde{h}_R}^{-1}
=
\frac{1}{2\varphi}
\left(
\begin{array}{cc}
\partial_{\overline{z}}\varphi & -2\partial_{\overline{z}}\overline{\rho} \\
0 & -\partial_{\overline{z}}\varphi
\end{array}
\right), ~  \label{R-gauge_SU(2) gauge fields_cpx_3} \\
A_{\overline{w}}
&\!\!\!\!=& \!\!\!\!-(\partial_{\overline w}\widetilde{h}_R){\widetilde{h}_R}^{-1}
=
\frac{1}{2\varphi}
\left(
\begin{array}{cc}
\partial_{\overline{w}}\varphi & -2\partial_{\overline{w}}\overline{\rho}  \label{R-gauge_SU(2) gauge fields_cpx_4}    \\
0 & -\partial_{\overline{w}}\varphi
\end{array}
\right).  ~~~~~~~~~
\end{eqnarray}
Rewriting them in real coordinates $x^{\mu}$, we get 
\begin{eqnarray}
A_{1}
&\!\!\!=&
\!\!\! \frac{1}{2\sqrt{2}\varphi}
\left(
\begin{array}{cc}
\!-(\partial_{z} - \partial_{\overline{z}})\varphi & -2\partial_{\overline{z}}\overline{\rho}  \\
2\partial_{z}\rho & (\partial_{z} - \partial_{\overline{z}})\varphi
\end{array}
\!\right)
=
\frac{i}{2\varphi}
\left(
\begin{array}{cc}
\partial_{2}\varphi & -(\partial_{2}-i\partial_{1})\overline{\rho}  \\
\!\!\!-(\partial_{2}+i\partial_{1})\rho & -\partial_{2}\varphi       
\end{array}
\!\right),  ~~~~~~~~~~  \label{Gauge fields_R-gauge_SU(2)_E_1} \\
A_{2}
&\!\!\!=&
\!\!\! \frac{i}{2\sqrt{2}\varphi}
\left(
\begin{array}{cc}
\!-(\partial_{z} + \partial_{\overline{z}})\varphi & 2\partial_{\overline{z}}\overline{\rho}   \\
2\partial_{z}\rho & (\partial_{z} + \partial_{\overline{z}})\varphi
\end{array}
\!\right)
=
\frac{i}{2\varphi}
\left(
\begin{array}{cc}
-\partial_{1}\varphi & (\partial_{1}+i\partial_{2})\overline{\rho}  \\
\!(\partial_{1}-i\partial_{2})\rho & \partial_{1}\varphi       
\end{array}
\!\right),   \label{Gauge fields_R-gauge_SU(2)_E_2}  \\
A_{3}
&\!\!\!=&
\!\!\! \frac{1}{2\sqrt{2}\varphi}
\left(
\begin{array}{cc}
\!-(\partial_{w} - \partial_{\overline{w}})\varphi & -2\partial_{\overline{w}}\overline{\rho}   \\
2\partial_{w}\rho & \!\!(\partial_{w} - \partial_{\overline{w}})\varphi
\end{array}
\!\right)
=
\frac{i}{2\varphi}
\left(
\begin{array}{cc}
\partial_{4}\varphi & -(\partial_{4}-i\partial_{3})\overline{\rho}  \\
\!\!\!\!-(\partial_{4}+i\partial_{3})\rho & -\partial_{4}\varphi       
\end{array}
\!\right),   \label{Gauge fields_R-gauge_SU(2)_E_3}  \\
A_{4}
&\!\!\!=&
\!\!\! \frac{i}{2\sqrt{2}\varphi}
\left(
\begin{array}{cc}
\!-(\partial_{w} + \partial_{\overline{w}})\varphi & 2\partial_{\overline{w}}\overline{\rho}   \\
2\partial_{w}\rho & \!(\partial_{w} + \partial_{\overline{w}})\varphi
\end{array}
\!\right)
=
\frac{i}{2\varphi}
\left(
\begin{array}{cc}
-\partial_{3}\varphi & (\partial_{3}+i\partial_{4})\overline{\rho}  \\
\!(\partial_{3}-i\partial_{4})\rho & \partial_{3}\varphi       
\end{array}
\!\right).   \label{Gauge fields_R-gauge_SU(2)_E_4}
\end{eqnarray}
On the other hand the three equations \eqref{Yang's eqns_SL(2,C)} of Yang for $G=\mathrm{SL(2,\mathbb{C})}$ reduce to the following three equations, mentioned by Yang  in \cite{Yang},  for $G=\mathrm{SU(2)}$ and the Euclidean signature:
\begin{eqnarray}
\label{Yang's eqns for SU(2)_E}
\begin{array}{l}
\varphi(\partial_{z}\partial_{\overline{z}}\varphi + \partial_{w}\partial_{\overline{w}}\varphi)
-\left[
(\partial_{z}\varphi) (\partial_{\overline{z}}\varphi) + (\partial_{w}\varphi)(\partial_{\overline{w}}\varphi)
\right]
+\left[
(\partial_{z}\rho) (\partial_{\overline{z}} \overline{\rho})
+ (\partial_{w}\rho) (\partial_{\overline{w}}\overline{\rho})
\right]=0, ~~\\
\varphi(\partial_{z}\partial_{\overline{z}}\rho + \partial_{w}\partial_{\overline{w}}\rho)
-2\left[
(\partial_{z}\rho) (\partial_{\overline{z}}\varphi)
+(\partial_{w}\rho) (\partial_{\overline{w}}\varphi)
\right]=0,~~ \\
\varphi(\partial_{z}\partial_{\overline{z}}\overline{\rho} + \partial_{w}\partial_{\overline{w}}\overline{\rho})
-2\left[
(\partial_{z}\varphi) (\partial_{\overline{z}}\overline{\rho})
+(\partial_{w}\varphi) (\partial_{\overline{w}}\overline{\rho})
\right]=0,~~
\end{array}
\end{eqnarray}
which is equivalent to the Yang equation \eqref{Yang equation} under the following parametrization of $J$-matrix :
\begin{eqnarray}
J=(h_R^{\dagger}h_R)^{-1}=\frac{1}{\varphi}
\left(
\begin{array}{cc}
1 & \overline{\rho}  \\
\rho & \varphi^2+|\rho|^2
\end{array}
\right).
\end{eqnarray}
A special set of solutions of the Yang equations \eqref{Yang's eqns for SU(2)_E} can be solved by imposing the following conditions on the scalar functions $\varphi$, $\rho$, $\overline{\rho}$ as follows:
\begin{eqnarray}
\label{Special solution of Yang_cpx coordinate}
&&\partial_{z}\rho = \partial_{\overline{w}}\varphi, ~~
\partial_{\overline{z}}\overline{\rho} = \partial_{w}\varphi, ~~
\partial_{w}\rho = -\partial_{\overline{z}}\varphi, ~~
\partial_{\overline{w}}\overline{\rho} = -\partial_{z}\varphi \\
&& \!\!\!\!\!\!\!\!\!\!\!\!\!\!\!\! \Longleftrightarrow 
~\left\{
\begin{array}{l}
(\partial_{1}-i\partial_{2})\rho = (\partial_{3}+i\partial_{4})\varphi,~
(\partial_{1}+i\partial_{2})\overline{\rho} = (\partial_{3}-i\partial_{4})\varphi,~ \\
(\partial_{3}-i\partial_{4})\rho = -(\partial_{1}+i\partial_{2})\varphi,~
(\partial_{3}+i\partial_{4})\overline{\rho} = -(\partial_{1}-i\partial_{2})\varphi.
\end{array}
\right.  \label{Special solution of Yang_real coordinate}
\end{eqnarray}
In fact, these relations implies that $\varphi$ satisfies the Laplace equation $(\partial_{z}\partial_{\overline{z}} + \partial_{w}\partial_{\overline{w}})\varphi=0$ 
($\Longleftrightarrow$ $\partial_{\mu}^2\varphi=0$) directly.
Substituting \eqref{Special solution of Yang_real coordinate} into \eqref{Gauge fields_R-gauge_SU(2)_E_1}, \eqref{Gauge fields_R-gauge_SU(2)_E_2}, \eqref{Gauge fields_R-gauge_SU(2)_E_3} and \eqref{Gauge fields_R-gauge_SU(2)_E_4}, we find that the ASD gauge fields are now expressed in terms of a single scalar function $\varphi$, rather than three scalar functions $\varphi$, $\rho$, $\overline{\rho}$ at all :
\begin{eqnarray}
A_{1}
&\!\!\!=&\!\!\! 
\frac{i}{2\varphi}
\left(
\begin{array}{cc}
\partial_{2} & \partial_{4} + i\partial_{3})  \\
\partial_{4} - i\partial_{3}  & -\partial_{2}       
\end{array}
\right)\varphi
=
\frac{i}{2\varphi}\left[
\eta_{12}^{(+)}\partial_{2}+
\eta_{13}^{(+)}\partial_{3}+
\eta_{14}^{(+)}\partial_{4}
\right]\varphi,~~~~     \label{ASD gauge fields_tHft matrices_A1} \\
A_{2}
&\!\!\!=&\!\!\! 
\frac{i}{2\varphi}
\left(
\begin{array}{cc}
-\partial_{1} & \partial_{3} - i\partial_{4}  \\
\partial_{3} + i\partial_{4} & \partial_{1}       
\end{array}
\right)\varphi
=
\frac{i}{2\varphi}\left[
\eta_{21}^{(+)}\partial_{1}+
\eta_{23}^{(+)}\partial_{3}+
\eta_{24}^{(+)}\partial_{4}
\right]\varphi,~~~~     \label{ASD gauge fields_tHft matrices_A2}   \\
A_{3}
&\!\!\!=& \!\!\! 
\frac{i}{2\varphi}
\left(
\begin{array}{cc}
\partial_{4} & -(\partial_{2} + i\partial_{1})  \\
-(\partial_{2} - i\partial_{1}) & -\partial_{4}       
\end{array}
\right)\varphi  
=
\frac{i}{2\varphi}\left[
\eta_{31}^{(+)}\partial_{1}+
\eta_{32}^{(+)}\partial_{2}+
\eta_{34}^{(+)}\partial_{4}
\right]\varphi,~~~~   \label{ASD gauge fields_tHft matrices_A3}  \\
A_{4}
&\!\!\!=&\!\!\! 
\frac{i}{2\varphi}
\left(
\begin{array}{cc}
-\partial_{3} & -(\partial_{1} - i\partial_{2})  \\
-(\partial_{1} + i\partial_{2}) & \partial_{3}       
\end{array}
\right)\varphi
=
\frac{i}{2\varphi}\left[
\eta_{41}^{(+)}\partial_{1}+
\eta_{42}^{(+)}\partial_{2}+
\eta_{43}^{(+)}\partial_{3}
\right]\varphi, ~~~~  \label{ASD gauge fields_tHft matrices_A4} 
\end{eqnarray} 
where $\eta_{\mu \nu}^{(\pm)}$ is the 't Hooft matrices defined by $\eta_{\mu \nu}^{(\pm)}:=\eta_{\mu\nu}^{i(\pm)}\sigma_i,~\mu,~ \nu= 1, 2, 3, 4,~ i=1,2,3.$ $\sigma_i$ is the Pauli matrices. 
$\eta_{\mu \nu}^{i(\pm)}:=\epsilon_{i\mu\nu4}\pm(\delta_{i\mu}\delta_{\nu4}-\delta_{i\nu}\delta_{\mu4})$ is called the 't Hooft symbol \cite{tHt2}. 
Substituting the 't Hooft matrices
\begin{eqnarray}
&&\!\!\!\!\!\!\!\!\!\!\! \eta_{12}^{(\pm)}
=-\eta_{21}^{(\pm)}
=\!\left(
\begin{array}{cc}
1 & 0 \\
0 & -1
\end{array}
\right),~~
\eta_{13}^{(\pm)}
=-\eta_{31}^{(\pm)}
=\!\left(
\begin{array}{cc}
0 & i \\
-i & 0
\end{array}
\right),~~
\eta_{14}^{(\pm)}
=-\eta_{41}^{(\pm)}
=\pm\left(
\begin{array}{cc}
0 & 1 \\
1 & 0
\end{array}
\right),~~   \nonumber  \\
&&\!\!\!\!\!\!\!\!\!\!\!  \eta_{23}^{(\pm)}
=-\eta_{32}^{(\pm)}
=\!\left(
\begin{array}{cc}
0 & 1 \\
1 & 0
\end{array}
\right),~~
\eta_{24}^{(\pm)}
=-\eta_{42}^{(\pm)}
=\pm\left(
\begin{array}{cc}
0 & -i \\
i & 0
\end{array}
\!\right),~~
\eta_{34}^{(\pm)}
=-\eta_{43}^{(\pm)}
=\pm\left(
\begin{array}{cc}
\!1 & \!0 \\
\!0 & \!\!-1
\end{array}
\!\right).  \nonumber 
\end{eqnarray}
into \eqref{ASD gauge fields_tHft matrices_A1}, \eqref{ASD gauge fields_tHft matrices_A2}, \eqref{ASD gauge fields_tHft matrices_A3} and \eqref{ASD gauge fields_tHft matrices_A4}, we find that the ASD fields can be expressed exactly in the form of the 't Hooft ansatz \cite{tHt3} (Cf: \cite{CoFa,Wilczek}) :
\begin{eqnarray}
A_{\mu}=\frac{i}{2}\eta_{\mu\nu}^{(+)}\partial_{\nu}\log \varphi
\end{eqnarray}
which belongs to a special class of solutions of the Yang equations \eqref{Yang's eqns for SU(2)_E}.
This important relationship was pointed out firstly by Yang in \cite{Yang} and we summarize it as the following:
\newtheorem{thm_3.7}[thm_3.1]{Theorem}
\begin{thm_3.7}
	{\bf (Relationship between the 't Hooft ansatz and the Yang equations~~\!\cite{Yang})} \label{Thm_3.7} \\
If we impose the condition \eqref{Special solution of Yang_cpx coordinate} on the Yang equations \eqref{Yang's eqns for SU(2)_E}, the solution set is equivalent to the 't Hooft ansatz :
\begin{eqnarray}
A_{\mu}=\frac{i}{2}\eta_{\mu\nu}^{(+)}\partial_{\nu}\log \varphi,~ \mbox{where} ~\varphi~ \mbox{satisfies the Laplace equation}~ \partial_{\mu}^2\varphi=0. \nonumber 
\end{eqnarray}
\end{thm_3.7}

So far, we have shown that on the Euclidean space $\mathbb{E}$, Yang's $R$-gauge gives $G=\mathrm{SU}(2)$ ASD gauge fields \eqref{Gauge fields_R-gauge_SU(2)_E_1} to \eqref{Gauge fields_R-gauge_SU(2)_E_4} which contain the 't Hooft ansatz as a special subset. On the other hand, if taking the $S$-gauge \eqref{S-gauge}, we get $A_{\widetilde z} = A_{\widetilde{w}}=0$. Comparing with \eqref{Gauge fields_E_cpx as real}, we find that $A_{\mu}$ cannot be all anti-hermitian because $A_1 = -iA_2$ and $A_3 = -iA_4$. 
That is to say, the $G=\mathrm{U}(N)$ symmetry is lost if we take the $S$-gauge on the Euclidean space $\mathbb{E}$.
\newtheorem{rem_3.8}[thm_3.1]{Remark}
\begin{rem_3.8}
	{\bf ($S$-gauge on $\mathbb{E}$)} \label{Rem_3.8} \\ 
There is no $G=\mathrm{U}(N)$ gauge theory on the Euclidean space $\mathbb{E}$ under the $S$-gauge.
\end{rem_3.8}
Now a natural question is that can we find a $G=\mathrm{U}(N)$ gauge theory on the Ultrahyperbolic space $\mathbb{U}$ under the $S$-gauge ?

\subsubsection{$G=\mathrm{U}(N)$ ASD gauge fields on the Ultrahyperbolic space $\mathbb{U}$}

In this subsection, we discuss the condition of $G=\mathrm{U}(N)$ gauge theory on the Ultrahyperbolic space $\mathbb{U}$, and show that the $G=\mathrm{U}(N)$ symmetry is preserved on the Ultrahyperbolic space $\mathbb{U}$ even when taking the $S$-gauge. In fact, this subsection is written as a supplement for the sub-dissertation \cite{HaHu1, HaHu2}.

The real representation of ASD gauge fields $A_{\mu}~ (\mu =1, 2, 3, 4)$ on the Ultrahyperbolic space $\mathbb{U}$ can be reduced from  the complex representation \eqref{Gauge fields in terms of h}. For instance, we can use the reality condition \eqref{Reality condition_U} to obtain:
\begin{eqnarray}
\label{Derivative_U_cpx as real}
\partial_{z}&\!\!\!\!=&\!\!\!\!\frac{1}{\sqrt{2}}(\partial_{1} + \partial_{3}),~ 
\partial_{\widetilde{z}}=\frac{1}{\sqrt{2}}(\partial_{1} - \partial_{3}),~
\partial_{w}=\frac{1}{\sqrt{2}}(\partial_{2} + \partial_{4}),~ 
\partial_{\widetilde{w}}=\frac{-1}{\sqrt{2}}(\partial_{2} - \partial_{4}), ~~~~~~~~~~~~~~~~ \\
A_{z}&\!\!\!\!=&\!\!\!\!\frac{1}{\sqrt{2}}(A_{1} + \!A_{3}),~ 
A_{\widetilde{z}}=\frac{1}{\sqrt{2}}(A_{1} - \!A_{3}),~
A_{w}=\frac{1}{\sqrt{2}}(A_{2} + \!A_{4}),~ 
A_{\widetilde{w}}=\frac{-1}{\sqrt{2}}(A_{2} -\! A_{4})  \label{Gauge fields_U_cpx as real},
\end{eqnarray}
or equivalently, 
\begin{eqnarray}
\label{Derivative_U_real as cpx}
\partial_{1}&\!\!\!\!=&\!\!\!\! \frac{1}{\sqrt{2}}(\partial_{z} + \partial_{\widetilde{z}}),~
\partial_{3}=\frac{1}{\sqrt{2}}(\partial_{z} - \partial_{\widetilde{z}}),~
\partial_{2}=\frac{1}{\sqrt{2}}(\partial_{w} - \partial_{\widetilde{w}}),~
\partial_{4}=\frac{1}{\sqrt{2}}(\partial_{w} + \partial_{\widetilde{w}}),  ~~~~~~~~~~~~~~~\\
A_{1}&\!\!\!\!=&\!\!\!\! \frac{1}{\sqrt{2}}(A_{z} + \!A_{\widetilde{z}}),~
A_{3}=\frac{1}{\sqrt{2}}(A_{z} - \!A_{\widetilde{z}}),~
A_{2}=\frac{1}{\sqrt{2}}(A_{w} - \!A_{\widetilde{w}}),~
A_{4}=\frac{1}{\sqrt{2}}(A_{w} +\! A_{\widetilde{w}}).  \label{Gauge fields_U_real as cpx}
\end{eqnarray}
Then the complex representation of ASD gauge fields \eqref{Gauge fields in terms of h} reduce to
\begin{eqnarray*}
A_{z}
=
-(\partial_{z}h)h^{-1}
=
\frac{-1}{\sqrt{2}}(\partial_{1}h + \partial_{3}h)h^{-1},~~~
A_{\widetilde{z}}
=
-(\partial_{\widetilde{z}}\widetilde{h})\widetilde{h}^{-1}
=
\frac{-1}{\sqrt{2}}(\partial_{1}\widetilde{h} - \partial_{3}\widetilde{h})\widetilde{h}^{-1},  \nonumber \\
A_{w}
=
-(\partial_{w}h)h^{-1}
=
\frac{-1}{\sqrt{2}}(\partial_{2}h + \partial_{4}h)h^{-1},~~~
A_{\widetilde{w}}
=
-(\partial_{\widetilde{w}}\widetilde{h})\widetilde{h}^{-1}
=
\frac{1}{\sqrt{2}}(\partial_{2}\widetilde{h} - \partial_{4}\widetilde{h})\widetilde{h}^{-1}.   \nonumber
\end{eqnarray*}
By direct substitution, we have
\begin{eqnarray}
\label{Real representation of gauge fields_U} 
&&A_{1}
=\frac{-1}{2}
\left\{ 
\left[(\partial_{1}h)h^{-1}+(\partial_{3}h)h^{-1} \right]+
\left[(\partial_{1}\widetilde{h})\widetilde{h}^{-1}-(\partial_{3}\widetilde{h})\widetilde{h}^{-1}
\right]
\right\},  \nonumber  \\
&&A_{3}
=\frac{-1}{2}
\left\{ 
\left[(\partial_{1}h)h^{-1} + (\partial_{3}h)h^{-1} \right]
-\left[(\partial_{1}\widetilde{h})\widetilde{h}^{-1}-(\partial_{3}\widetilde{h})\widetilde{h}^{-1}
\right]
\right\},  \\
&&A_{2}
=\frac{-1}{2}
\left\{ 
\left[(\partial_{2}h)h^{-1} + (\partial_{4}h)h^{-1} \right]
+\left[(\partial_{2}\widetilde{h})\widetilde{h}^{-1}- (\partial_{4}\widetilde{h})\widetilde{h}^{-1}
\right]
\right\},  \nonumber  \\
&&A_{4}
=\frac{-1}{2}
\left\{ 
\left[(\partial_{2}h)h^{-1} + (\partial_{4}h)h^{-1}\right]
-\left[(\partial_{2}\widetilde{h})\widetilde{h}^{-1} - (\partial_{4}\widetilde{h})\widetilde{h}^{-1}
\right]
\right\}.  \nonumber
\end{eqnarray}
If we impose the condition
\begin{eqnarray}
\label{h^{dagger}h=C}
h^{\dagger}h=C,~~  \widetilde{h}^{\dagger}\widetilde{h}=\widetilde{C} ~~ ( C, \widetilde{C}: \mbox{independent of}~ x^{\mu} ~)
\end{eqnarray}
on \eqref{Real representation of gauge fields_U}, it implies
\begin{eqnarray}
\begin{array}{c}
\label{Condition of U(N) gauge group_U}
\left[ (\partial_{\mu}h)h^{-1} \right]^{\dagger}
=
h^{-\dagger}(\partial_{\mu}h^{\dagger})
=
hC^{-1}\partial_{\mu}(Ch^{-1})
=
-(\partial_{\mu}h)h^{-1},  \\
\left[ (\partial_{\mu}\widetilde{h})\widetilde{h}^{-1} \right]^{\dagger}
=
\widetilde{h}^{-\dagger}(\partial_{\mu}\widetilde{h}^{\dagger})
=
\widetilde{h}\widetilde{C}^{-1}\partial_{\mu}(\widetilde{C}\widetilde{h}^{-1})
=
-(\partial_{\mu}\widetilde{h})\widetilde{h}^{-1},
\end{array}
\end{eqnarray}
and therefore $A_{\mu}^{\dagger}=-A_{\mu}$, for all $\mu=1, 2, 3, 4$. That is, the gauge group can be $G=\mathrm{U}(N)$ under the condition \eqref{h^{dagger}h=C}.
\newtheorem{prop_3.9}[thm_3.1]{Proposition}
\begin{prop_3.9}
{\bf (Condition of unitary gauge group on $\mathbb{U}$, Cf: \cite{HaHu1,HaHu2})} \label{Prop_3.9}\\
Let $A_{\mu} ~ (\mu=1,2,3,4)$ be gauge fields defined on the Ultrahyperbolic space $\mathbb{U}$ as \eqref{Real representation of gauge fields_U}. If $h,~ \widetilde{h}$ satisfy the relation $h^{\dagger}h = C, ~\widetilde{h}^{\dagger}\widetilde{h} = \widetilde{C}$, for some $N \times N$ matrix $C$ and $\widetilde{C}$ (independent of $x^{\mu}$), then the gauge group can be $G=\mathrm{U}(N)$. In particular, if we take $C = \widetilde{C} = I$, then the condition for $G=\mathrm{U}(N)$ is $h, \widetilde{h} \in \mathrm{U}(N)$ in this case. 
\end{prop_3.9}  
On the other hand, if we substitute the $S$-gauge \eqref{S-gauge} into \eqref{Real representation of gauge fields_U}, the ASD gauge fields are simplified to
\begin{eqnarray}
\label{Real representation of gauge fields_U_S-gauge} 
A_{1} =  A_{3} 
=\frac{1}{2} 
(J^{-1}\partial_{1}J + J^{-1}\partial_{3}J), ~~
A_{2} = A_{4}
=\frac{1}{2}
(J^{-1}\partial_{2}J + J^{-1}\partial_{4}J).  
\end{eqnarray}
We find that $A_{\mu}~ (\mu=1, 2, 3, 4)$ can be all anti-hermitian if $JJ^{\dagger}=C$ for some $N \times N$ matrix $C$ (independent of $x^{\mu}$). In particular, we can set $J \in \mathrm{U}(N)$ to satisfy this condition. 
Now we can conclude that the $G=\mathrm{U}(N)$ symmetry is still preserved even when we take the $S$-gauge on the Ultrahyperbolic space $\mathbb{U}$.
\\

Next, let us check whether the $R$-gauge \eqref{R-gauge_SL(2)_1} and \eqref{R-gauge_SL(2)_2} can give $G=\mathrm{SU}(2)$ ASD gauge fields on the Ultrahyperbolic space $\mathbb{U}$ or not. Clearly, if $h_R, \widetilde{h}_R \in \mathrm{U}(N)$, the only result is $h_R =\widetilde{h}_R = I$ which cannot give nontrivial gauge fields. In general, we can substitute the $R$-gauge \eqref{R-gauge_SL(2)_1} and \eqref{R-gauge_SL(2)_2} into \eqref{Real representation of gauge fields_U}, and then get
\begin{eqnarray}
A_1
&\!\!\!=\!\!\!&
\frac{1}{2\varphi}
\left(
\begin{array}{cc}
-\partial_{3}\varphi & -(\partial_{1}-\partial_{3})\widetilde{\rho} \\
(\partial_{1}+\partial_{3})\rho & \partial_{3}\varphi
\end{array}
\right),~ 
A_3
=
\frac{1}{2\varphi}
\left(
\begin{array}{cc}
-\partial_{1}\varphi & (\partial_{1}-\partial_{3})\widetilde{\rho} \\
(\partial_{1}+\partial_{3})\rho & \partial_{1}\varphi
\end{array}
\right),  \nonumber  \\
A_2
&\!\!\!=\!\!\!&
\frac{1}{2\varphi}
\left(
\begin{array}{cc}
-\partial_{4}\varphi & -(\partial_{2}-\partial_{4})\widetilde{\rho} \\
(\partial_{2}+\partial_{4})\rho & \partial_{4}\varphi
\end{array}
\right),~
A_4=
\frac{1}{2\varphi}
\left(
\begin{array}{cc}
-\partial_{2}\varphi & (\partial_{2}-\partial_{4})\widetilde{\rho} \\
(\partial_{2}+\partial_{4})\rho & \partial_{2}\varphi
\end{array}
\right).   \nonumber 
\end{eqnarray}
One can easily check that the only anti-hermitian gauge fields are  
\begin{eqnarray}
A_1\!=\!
\frac{1}{2\varphi}
\!\left(
\begin{array}{cc}
\!\!-\partial_{3} & \!\!0 \\
0 & \!\!\partial_{3}
\end{array}
\!\!\right)\!\varphi,~
A_3\!=\!
\frac{1}{2\varphi}
\!\left(
\begin{array}{cc}
\!\!-\partial_{1} & \!\!0 \\
0 & \!\!\partial_{1}
\end{array}
\!\!\right)\!\varphi,~  
A_2\!=\!
\frac{1}{2\varphi}
\!\left(
\begin{array}{cc}
\!\!-\partial_{4} & \!\!0 \\
0 & \!\!\partial_{4}
\end{array}
\!\!\right)\!\varphi,~
A_4\!=\!
\frac{1}{2\varphi}
\!\left(
\begin{array}{cc}
\!\!-\partial_{2} & \!\!0 \\
0 & \!\!\partial_{2}
\end{array}
\!\!\right)\!\varphi  \nonumber 
\end{eqnarray}
which are abelian fields with the field strengths $F_{\mu \nu} =0~ (\mu,~\nu = 1,2,3,4)$.
\newtheorem{rem_3.10}[thm_3.1]{Remark}
\begin{rem_3.10}
We can conclude that on the Ultrahyperbolic space $\mathbb{U}$ : \label{Rem_3.10}
\begin{itemize}
	\item  There is no nontrivial $G=\mathrm{SU}(2)$ gauge theory under the $R$-gauge.
	\item  $G=\mathrm{U}(N)$ gauge theory can be realized under the $S$-gauge.
\end{itemize}
\end{rem_3.10}






\newpage

\section{Darboux transformation}
\label{Section 4}

Systematic construction for exact solutions of soliton equations has been a crucial subject in the research of integrable systems. A typical technique developed in this field is to construct the Lax representation, and then use the form invariance of the linear system under the Darboux transformation to generate more and more exact solutions . In this section, we introduce a Darboux transformation for "Lax representation" that considered by Nimmo, Gilson, and Ohta \cite{NiGiOh}. After applying the Darboux transformation, we obtain a class of exact solutions which can be expressed in terms of quasideterminants \cite{GiHaHuNi}. 
We call this kind of solutions quasi-Wronskian type solutions.

\subsection{Linear system of the ASDYM equations}
\label{Section 4.1}

In this subsection, we review a Darboux transformation mentioned in \cite{NiGiOh} (Cf: the sub-dissertation \cite{GiHaHuNi}).
Recall that in Subsection \ref{Section 3.4.2}, the $R$-gauge formulation gives a great successful description for $G=\mathrm{SU}(2)$ ASD gauge fields on the Euclidean space $\mathbb{E}$, however, it unavoidably loses the $\mathrm{SU(2)}$ gauge symmetry on the Ultrahyperbolic space $\mathbb{U}$. Therefore, it is natural to adopt the $S$-gauge for seeking unitary ASD gauge fields on the Ultrahyperbolic space $\mathbb{U}$.
In \cite{NiGiOh}, the authors take the $S$-gauge and consider the following linear system (Cf: \eqref{Lax representation of ASDYM_NGO}) :
\begin{eqnarray}
\label{Linear system of ASDYM_NGO}
\left\{
\begin{array}{c}
L(\phi):= \left(\partial_{w}-(\partial_{w}J)J^{-1}\right)\phi-(\partial_{\widetilde{z}}\phi)\zeta = 0 \\
M(\phi):=	
\left(\partial_{z}-(\partial_{z}J)J^{-1}\right)\phi-(\partial_{\widetilde{w}}\phi)\zeta = 0   
\end{array}.
\right.
\end{eqnarray}
The corresponding Yang equation and ASD gauge fields are
\begin{eqnarray}
\label{Yang equation_2}
&&\partial_{\widetilde{z}}\left((\partial_{z}J)J^{-1}\right)-
\partial_{\widetilde{w}}\left((\partial_{w}J)J^{-1}\right) =0,  \\
&&A_{z}=-(\partial_{z}J)J^{-1}, ~~
A_{w}=-(\partial_{w}J)J^{-1}, ~~
A_{\widetilde{z}}=A_{\widetilde{w}}=0. \label{A_z, A_w in terms of J}
\end{eqnarray}
In fact, the authors of \cite{NiGiOh} adjust the $J$-matrix $J$ in \eqref{S-gauge} and \eqref{Gauge fields_S-gauge_general} to $J^{-1}$, but it would not change the essence of $J$-matrix eventually.
\newtheorem{lem_4.1}{Lemma}[section]
\begin{lem_4.1}
{\bf (Darboux transformation \cite{NiGiOh}, Cf: \cite{GiHaHuNi})} \label{Lemma_4.1} \\ 
Let $J$ be a given $J$-matrix and $\phi(\zeta)$ be the general solution of the linear system \eqref{Linear system of ASDYM_NGO} $w.r.t.$ the unspecified spectral parameter $\zeta$.
Assume that $\psi(\Lambda)$ be a specified solution of \eqref{Linear system of ASDYM_NGO} for a specific value of the spectral parameter $\zeta=\Lambda$.
Then the linear system \eqref{Linear system of ASDYM_NGO} is form invariant under the following Darboux transformation :
\begin{eqnarray}
\label{Darboux transf}
\widetilde{J}=-\psi\Lambda \psi^{-1}J
=
\left|
\begin{array}{cc}
\psi & 1 \\
\psi\Lambda & \fbox{$0$} 
\end{array}
\right|J,~~
\widetilde{\phi}=\phi\zeta-\psi\Lambda \psi^{-1}\phi
=
	\left|
	\begin{array}{cc}
	\psi & \phi \\
	\psi\Lambda & \fbox{$\phi \zeta$} 
	\end{array}
	\right|.
\end{eqnarray}
That is, ($\widetilde{J}$, $\widetilde{\phi}$) satisfies
\begin{eqnarray}
\left\{
\begin{array}{c}
\widetilde{L}(\widetilde{\phi}):= \left(\partial_{w}-(\partial_{w}\widetilde{J})\widetilde{J}^{-1}\right)\widetilde{\phi}-(\partial_{\widetilde{z}}\widetilde{\phi})\zeta = 0 \\
\widetilde{M}(\widetilde{\phi}):=	
\left(\partial_{z}-(\partial_{z}\widetilde{J})\widetilde{J}^{-1}\right)\widetilde{\phi}-(\partial_{\widetilde{w}}\widetilde{\phi})\zeta = 0
\end{array}.
\right.
\end{eqnarray}
\end{lem_4.1}
{\bf (\emph{Proof})} 
\\
As a supplement for \cite{GiHaHuNi, NiGiOh}, we give a more straightforward and detailed derivation here.
For convenience, we set $H=-\psi\Lambda \psi^{-1}$, 
then 
\begin{eqnarray*} 
\widetilde{L}(\widetilde{\phi})
&\!\!\!=\!\!\!&\left(
\partial_{w}-\left((\partial_{w}H)J+H(\partial_{w}{J})\right){J}^{-1}H^{-1}
\right)(\phi\zeta+H\phi)
-
\left(\partial_{\widetilde z}(\phi\zeta+H\phi)\right)\zeta  \nonumber \\ 
&\!\!\!=\!\!\!&
(\partial_{w}\phi)\zeta-(\partial_{w}H)H^{-1}\phi\zeta-H(\partial_{w}J)J^{-1}H^{-1}\phi\zeta  \nonumber \\
&& +(\partial_{w}H)\phi+H(\partial_{w}\phi)-(\partial_{w}H)\phi-H(\partial_{w}J)J^{-1}\phi   \nonumber  \\
&&
-(\partial_{\widetilde{z}}\phi)\zeta^{2}  
-(\partial_{\widetilde{z}}H)\phi\zeta-H(\partial_{\widetilde{z}}\phi)\zeta  \nonumber  \\
&\!\!\!=\!\!\!&
H\left(\partial_{w}\phi-(\partial_{w}J)J^{-1}\phi-(\partial_{\widetilde{z}}\phi)\zeta
\right)  \nonumber \\
&&+
\left(
\partial_{w}\phi
-\left(
(\partial_{w}H)H^{-1}\phi+H(\partial_{w}J)J^{-1}H^{-1}\phi
+\partial_{\widetilde{z}}H\phi 
\right)
-(\partial_{\widetilde{z}}\phi)\zeta
\right)\zeta    \nonumber   \\
&\!\!\!=\!\!\!&
L(\phi)\zeta -\psi\Lambda \psi^{-1}L(\phi)  \nonumber  \\
&\!\!\!=\!\!\!& 
\left|
\begin{array}{cc}
\psi & L(\phi) \\
\psi\Lambda & \fbox{$L(\phi)\zeta$}
\end{array}
\right|
= 0 ~~  \mbox{by the fact that} ~ L(\phi)=0.  \nonumber 
\end{eqnarray*}
Similarly, we have
\begin{eqnarray*}
\widetilde{M}(\widetilde{\phi})
=
\left|
\begin{array}{cc}
\psi & M(\phi) \\
\psi\Lambda & \fbox{$M(\phi)\zeta$}
\end{array}
\right|
=0.  \nonumber
\end{eqnarray*}
It remains for us to show that
$(\partial_{w}J)J^{-1}
=
(\partial_{w}H)H^{-1}+H(\partial_{w}J)J^{-1}H^{-1}
+\partial_{\widetilde{z}}H$ 
in the third equality above. This can be done by calculating $\partial_{\widetilde z}H$ explicitly :
\begin{eqnarray*}
\partial_{\widetilde z}H
&\!\!\!=\!\!\!&
-\partial_{\widetilde z}(\psi\Lambda \psi^{-1}) =
-(\partial_{\widetilde z}\psi)\Lambda \psi^{-1}+\psi\Lambda \psi^{-1}(\partial_{\widetilde z}\psi)\psi^{-1}  \nonumber  \\
&\!\!\!=\!\!\!& 
-(\partial_{w}\psi)\psi^{-1}+(\partial_{w}J)J^{-1} 
-\psi\Lambda \psi^{-1}(\partial_{w}J)J^{-1}\psi\Lambda^{-1}\psi^{-1}
+\psi\Lambda \psi^{-1}(\partial_{w}\psi)\Lambda^{-1}\psi^{-1}
\nonumber  \\
&&(~\mbox{Here we use the fact that~~} 
\partial_{w}\psi-(\partial_{w}J)J^{-1}\psi-(\partial_{\widetilde{z}}\psi)\Lambda=0~)  \nonumber  \\
&\!\!\!=\!\!\!&
(\partial_{w}J)J^{-1}-H(\partial_{w}J)J^{-1}H^{-1}
+\left(
(\partial_{w}\psi)\Lambda \psi^{-1}-\psi\Lambda \psi^{-1}(\partial_{w}\psi)\psi^{-1}
\right)
(-\psi\Lambda \psi^{-1})^{-1}  \nonumber  \\
&\!\!\!=\!\!\!&
(\partial_{w}J)J^{-1}-H(\partial_{w}J)J^{-1}H^{-1}
-(\partial_{w}H)H^{-1}.    \nonumber
\end{eqnarray*}
$\hfill\Box$ 
\\
Now we can apply the Darboux transformation \eqref{Darboux transf} to generate more solutions by using the specified solutions $\psi_{i}(\Lambda_{i})$, $i=1, 2, 3, ...$ of the linear system \eqref{Linear system of ASDYM_NGO}. The solution generated by $i$ iterations of the Darboux transformation can be expressed in terms of the quasideterminant of order $i-1$. The explicit procedure of applying the Darboux transformation is mentioned in the next subsection.
\newpage

\subsection{Quasi-Wronskian type solutions of the ASDYM equations}
\label{Section 4.2}

In this subsection, we review the quasi-Wronskian type solution mentioned in the sub-dissertation \cite{GiHaHuNi} and the private note \cite{Nimmo}. 
(Cf: \cite{NiGiOh} for ordinary Wronskian version.) First of all, let us explain the procedure of applying the Darboux transformation explicitly:
\begin{itemize} 
\item {\bf Preparation} 
\\
Let us set $J_{1}$ to be an initial $J$-matrix for the linear system \eqref{Linear system of ASDYM_NGO}. In this case, we call $J_1$ as seed solution, and call the following \eqref{Initial linear system} as initial linear system : 
\begin{eqnarray} \label{Initial linear system}
\left\{
\begin{array}{c}
L_{1}(\phi)= [\partial_{w}-(\partial_{w}J_{1})J_{1}^{-1}]\phi-(\partial_{\widetilde{z}}\phi)\zeta = 0 \\
M_{1}(\phi)=	
[\partial_{z}-(\partial_{z}J_{1})J_{1}^{-1}]\phi-(\partial_{\widetilde{w}}\phi)\zeta = 0   
\end{array},
\right.   
\end{eqnarray}
where $\phi(\zeta)$ denotes the general solution of \eqref{Initial linear system} $w.r.t.$ the unspecified spectral parameter $\zeta$. Now we can solve \eqref{Initial linear system}, choose the solutions whatever we prefer, and specify them as $\psi_{i}(\Lambda_{i}),~i=1, 2, 3, ...$ and so on.
To prevent the reader from confusing about the terminology, let us remind you the following remarks before going ahead to the next step : 
	\begin{itemize}
		\item The lowercase letters $\phi$ and $\psi_{i}$ denote the solutions of the initial linear system \eqref{Initial linear system} ($w.r.t.$ the seed solution $J_{1}$).
		\item The capital letters $\Phi_{i}$ and $\Psi_{i}$ denote the solutions generated by $i-1$ iterations of the Darboux transformation \eqref{Darboux transf} with $\Phi_{1}:=\phi,~ \Psi_{1}:=\psi$. 
		Here $\Phi_{i}$ is the general (unspecified) solution of the linear system \eqref{Linear system of ASDYM_NGO} ($w.r.t.$ $J_{i}$ : generated by $i-1$ iterations of the Darboux transformation) and $\Psi_{i}$ is the specified solution defined by
		$\Psi_{i}:=\Phi_{i}\Big|_{(\phi, \zeta) \rightarrow (\psi_{i}, \Lambda_{i})}$, 
		here the rightarrow means the replacement of $(\phi, \zeta)$ by $(\psi_{i}, \Lambda_{i})$.
\end{itemize}
\item {\bf After 1 iteration of the Darboux transformation}
\medskip \\
Let us choose a specified solution $\psi_1(\Lambda_{1})$ of the initial linear system \eqref{Initial linear system} $w.r.t.$ a specific $\Lambda_{1}$. Then by applying the Darboux transformation \eqref{Darboux transf} to \eqref{Initial linear system}, we have
\begin{eqnarray}
\label{J_2}
J_{2} := \widetilde{J_{1}} &\!\!\!\!=& \!\!\!\! -\Psi_{1}\Lambda_{1}\Psi_{1}^{-1}J_{1}~~~(\Psi_{1}:=\psi_{1})  \nonumber \\ 
&\!\!\!\!=& \!\!\!\! 
\left|
\begin{array}{cc}
\psi_{1} & 1 \\
\psi_{1}\Lambda_{1} & \fbox{0}
\end{array}
\right| J_{1}.  
\\
\Phi_{2} := \widetilde{\Phi}_{1} &\!\!\!\!=& \!\!\!\! \Phi_{1}\zeta - \Psi_{1}\Lambda_{1}\Psi_{1}^{-1}\Phi_{1} ~~~(\Phi_{1}:= \phi   \label{Phi_2})  \nonumber \\
&\!\!\!\!=& \!\!\!\!
\left|
\begin{array}{cc}
\psi_{1} & \phi   
\\
\psi_{1}\Lambda_{1} & \fbox{$\phi\zeta$}
\end{array}
\right|.
\\
&&\!\!\!\!\!\!\!\!\!\!\!\!\!\!\!\!\!\!\!\!\!\!\!\!\!\!\!\!\!\!\!\!\!\!(J_{2}, \Phi_{2}) ~~ {\mbox{satisfies}} ~~ \left\{
\begin{array}{c}
L_{2}(\Phi_{2})= [\partial_{w}-(\partial_{w}J_{2})J_{2}^{-1}]\Phi_{2}-(\partial_{\widetilde{z}}\Phi_{2})\zeta = 0 \\
M_{2}(\Phi_{2})=	
[\partial_{z}-(\partial_{z}J_{2})J_{2}^{-1}]\Phi_{2}-(\partial_{\widetilde{w}}\Phi_{2})\zeta = 0   
\end{array},
\right.   \label{Linear system_J_2}
\end{eqnarray}
where $\Phi_{2}(\zeta)$ denotes the general (unspecified) solution of the  above linear system \eqref{Linear system_J_2} $w.r.t.$ $J_{2}$.
\item {\bf{After 2 iterations of the Darboux transformation}}
\medskip \\
Before applying the Darboux transformation \eqref{Darboux transf} to the linear system \eqref{Linear system_J_2}, we need to find a specified solution of \eqref{Linear system_J_2}. In fact, it can be done by choosing another specified solution $\psi_2(\Lambda_2)$ (differs form $\psi_{1}(\Lambda_{1})$) of the initial linear system \eqref{Initial linear system} and use $\psi_2(\Lambda_2)$ to specify the general solution $\Phi_{2}(\zeta)$ of \eqref{Linear system_J_2} to be $\Psi_{2}(\Lambda_2)$, where
\begin{eqnarray} 
	\Psi_{2}:=\Phi_{2}\Big|_{(\phi, \zeta) \rightarrow (\psi_{2}, \Lambda_{2})}
	=
	\left|
	\begin{array}{cc}
		\psi_{1} & \psi_{2} \\
		\psi_{1}\Lambda_{1} & \fbox{$\psi_{2}\Lambda_{2}$}
	\end{array}
	\right|
\end{eqnarray}
is a specified solution of \eqref{Linear system_J_2} now. 
Then by applying the Darboux transformation \eqref{Darboux transf} to \eqref{Linear system_J_2}, we have
\begin{eqnarray}
J_{3} &\!\!\! :=& \!\!\! \widetilde{J_2}
= - \Psi_{2}\Lambda_{2}\Psi_{2}^{-1}J_{2} ~~~~ \nonumber \\
&\!\!\! =& \!\!\!
-\left|
\begin{array}{cc}
\psi_{1} & \psi_{2} \\
\psi_{1}\Lambda_{1} & \fbox{$\psi_{2}\Lambda_{2}$}
\end{array}
\right|\Lambda_{2}
\left|
\begin{array}{cc}
\psi_{1} & \psi_{2} \\
\psi_{1}\Lambda_{1} & \fbox{$\psi_{2}\Lambda_{2}$}
\end{array}
\right|^{-1}\!\!\!\!J_{2}
=
\left|
\begin{array}{ccc}
\psi_{1} & \psi_{2} & 1 \\
\psi_{1}\Lambda_{1} & \psi_{2}\Lambda_{2} & 0 \\
\psi_{1}\Lambda_{1}^{2} & \psi_{2}\Lambda_{2}^{2} & \fbox{0}
\end{array}
\right|J_{1}, 
~~~~~~~~~~\\
\Phi_{3} &\!\!\! :=& \!\!\! \widetilde{\Phi}_2
= \Phi_{2}\zeta - \Psi_{2}\Lambda_{2}\Psi_{2}^{-1}\Phi_{2}  \nonumber 
\\
&\!\!\! =& \!\!\!
\left|
\begin{array}{cc}
\psi_{1} & \!\! \phi  \\
\psi_{1}\Lambda_{1} & \! \fbox{$\phi\zeta$}
\end{array}
\right|\zeta \!\!- 
\left|
\begin{array}{cc}
\psi_{1} & \!\! \psi_{2} \\
\psi_{1}\Lambda_{1} & \!\! \fbox{$\psi_{2}\Lambda_{2}$}
\end{array}
\right|\Lambda_{2}
\left|
\begin{array}{cc}
\psi_{1} & \!\! \psi_{2} \\
\psi_{1}\Lambda_{1} & \!\! \fbox{$\psi_{2}\Lambda_{2}$}
\end{array}
\right|^{-1}
\left|
\begin{array}{cc}
\psi_{1} & \!\! \phi  \\
\psi_{1}\Lambda_{1} & \!\! \fbox{$\phi\zeta$}
\end{array}
\right|   \nonumber
\\
&\!\!\! =& \!\!\!
\left|
\begin{array}{ccc}
\psi_{1} & \psi_{2} & \phi \\
\psi_{1}\Lambda_{1} & \psi_{2}\Lambda_{2} & \phi\zeta \\
\psi_{1}\Lambda_{1}^{2} & \psi_{2}\Lambda_{2}^{2} & \fbox{$\phi\zeta^{2}$}
\end{array}
\right|.      
\\
&&\!\!\!\!\!\!\!\!\!\!\!\!\!\!\!\!\!\!\!\!(J_{3}, \Phi_{3}) ~~ {\mbox{satisfies}} ~~ 
\left\{
\begin{array}{c}
L_{3}(\Phi_{3})= [\partial_{w}-(\partial_{w}J_{3})J_{3}^{-1}]\Phi_{3}-(\partial_{\widetilde{z}}\Phi_{3})\zeta = 0 \\
M_{3}(\Phi_{3})=	
[\partial_{z}-(\partial_{z}J_{3})J_{3}^{-1}]\Phi_{3}-(\partial_{\widetilde{w}}\Phi_{3})\zeta = 0   
\end{array},
\right.   \label{Linear system_J_3}
\end{eqnarray}
where $\Phi_{3}(\zeta)$ denotes the general (unspecified) solution of the  above linear system \eqref{Linear system_J_3} $w.r.t.$ $J_{3}$.
\item {\bf After 3 iterations of the Darboux transformation}
\medskip \\
Following the same process as the previous step, 
we can choose $\psi_3(\Lambda_3)$, a specified solution of the initial linear system \eqref{Initial linear system} differs from $\psi_2(\Lambda_2)$ and $\psi_1(\Lambda_1)$, and use $\psi_3(\Lambda_3)$ to specify the general solution $\Phi_{3}(\zeta)$ of \eqref{Linear system_J_3} to be $\Psi_{3}(\Lambda_3)$, where
\begin{eqnarray}
	\Psi_{3}:=\Phi_{3}\Big|_{(\phi, \zeta) \rightarrow (\psi_{3}, \Lambda_{3})}
	=
	\left|
	\begin{array}{ccc}
	\psi_{1} & \psi_{2} & \psi_{3} \\
	\psi_{1}\Lambda_{1} & \psi_{2}\Lambda_{2} & \psi_{3}\Lambda_{3} \\
	\psi_{1}\Lambda_{1}^{2} & \psi_{2}\Lambda_{2}^{2} & \fbox{$\psi_{3}\Lambda_{3}^{2}$}
	\end{array}
	\right|~
\end{eqnarray}
is a specified solution of \eqref{Linear system_J_3} now. 
Then by applying the Darboux transformation \eqref{Darboux transf} to \eqref{Linear system_J_3}, we have
\begin{eqnarray}
J_{4} := {\widetilde{J_{3}}= -\Psi_{3}\Lambda_{3}\Psi_{3}^{-1}J_3
=
\left|
\begin{array}{cccc}
\psi_{1} & \psi_{2} & \psi_{3} & 1 \\
\psi_{1}\Lambda_{1} & \psi_{2}\Lambda_{2} & \psi_{3}\Lambda_{3} & 0 \\
\psi_{1}\Lambda_{1}^{2} & \psi_{2}\Lambda_{2}^{2} & \psi_{3}\Lambda_{3}^{2} & 0 \\
\psi_{1}\Lambda_{1}^{3} & \psi_{2}\Lambda_{2}^{3} & \psi_{3}\Lambda_{3}^{3} & \fbox{0}
\end{array}
\right|J_1},
\end{eqnarray}
and so on.
\end{itemize}
After $n$ iterations of the Darboux transformation \eqref{Darboux transf}, we can conclude the following theorem.
\newtheorem{thm_4.2}[lem_4.1]{Theorem}
\begin{thm_4.2}
{\bf (Quasi-Wronskian solutions ~\cite{GiHaHuNi,Nimmo})} \label{Thm_4.2} \\
Let $J_1$ be a given seed solution. Assume that $\psi_i(\Lambda_i)$, $1\leq i \leq n$ be $n$ specified solutions of the initial linear system \eqref{Initial linear system} $w.r.t.$ the specific spectral parameters $\Lambda_{i}$. Then under the following Darboux transformation
\begin{eqnarray}
\label{Darboux transf_(J_k+1, Phi_k+1)}
\left\{
\begin{array}{l}
J_{k+1}:=-\Psi_{k}\Lambda_{k}\Psi_{k}^{-1}J_{k}
=
\left|
\begin{array}{cc}
\Psi_{k} & 1 \\
\Psi_{k}\Lambda_{k} & \fbox{$0$}
\end{array}
\right|J_{k} \\
\Phi_{k+1}:=\Phi_{k}\zeta - \Psi_{k}\Lambda_{k}\Psi_{k}^{-1}\Phi_{k}
=
\left|
\begin{array}{cc}
\Psi_{k} & \Phi_{k} \\
\Psi_{k}\Lambda_{k} & \fbox{$\Phi_{k}\zeta$}
\end{array}
\right| \\
\Phi_{1}:=\phi,~~ \Psi_{k}:=\Phi_{k}\Big|_{(\phi, \zeta) \rightarrow (\psi_{k}, \Lambda_{k})}
\end{array}
\right., ~k=1,2,...,n,
\end{eqnarray}
we have 
\begin{eqnarray} 
J_{n+1} 
&\!\!\!=\!\!\!&\left|
\begin{array}{ccccc}
\psi_{1} & \psi_{2} & \!\! \cdots & \!\! \psi_{n} & 1 \\
\psi_{1}\Lambda_{1} & \psi_{2}\Lambda_{2} & \!\! \cdots & \!\! \psi_{n}\Lambda_{n} & 0 \\
\vdots &  \vdots & \!\! \ddots & \!\! \vdots & \vdots  \\
\psi_{1}\Lambda_{1}^{n} & \psi_{2}\Lambda_{2}^{n} & \!\! \cdots & \!\! \psi_{n}\Lambda_{n}^{n} & \fbox{$0$}
\end{array} \right| J_{1},  \label{J_n}  \\
\Phi_{n+1}  
&\!\!\!=\!\!\!&\left|
\begin{array}{ccccc}
\psi_{1} & \psi_{2} & \!\!\cdots & \!\! \psi_{n} & \phi \\
\psi_{1}\Lambda_{1} & \psi_{2}\Lambda_{2} & \!\! \cdots & \!\! \psi_{n}\Lambda_{n} & \phi \zeta \\
\vdots & \vdots & \!\! \ddots & \!\! \vdots & \!\! \vdots  \\
\psi_{1}\Lambda_{1}^{n} & \psi_{2}\Lambda_{2}^{n} & \!\! \cdots & \!\! \psi_{n}\Lambda_{n}^{n} & \fbox{$\phi\zeta^{n}$}
\end{array}\right|. \label{Phi_n}
\end{eqnarray}
\end{thm_4.2}
{\bf (\emph{Proof})~~Mathematical induction}  
\smallskip \\
(The following proof was given firstly in the private note \cite{Nimmo} (Cf: \cite{GiHaHuNi}).)
\smallskip 
\\
Due to \eqref{J_2} and \eqref{Phi_2}, the statement clearly holds for $n=1$. It remains for us to show that the statement is correct for $n=k+1$ if it holds for $1 \leq n \leq k$. 
For convenience, we use the bold letters ${\bm{\psi}}$ and ${\bm{\Lambda}}$ to define $1 \times k$ matrix ${\bm{\psi}}:=(\psi_{1}, \psi_{2}, ..., \psi_{k})$, and $k \times k$ diagonal matrix ${\bm{\Lambda}}$:=diag($\Lambda_{1}, \Lambda_{2}, ..., \Lambda_{k}$), then $J_{k+1}$ and $\Phi_{k+1}$ can be written by
\begin{eqnarray}
J_{k+1}
=
\left|
\begin{array}{cc}
{\bm{\psi}} & 1 \\
{\bm{\psi}} {\bm{\Lambda}} & 0  \\
\vdots & \vdots \\
{\bm{\psi}} {\bm{\Lambda}}^{k} & \fbox{0}
\end{array}
\right|,~~
\Phi_{k+1}
=
\left|
\begin{array}{cc}
{\bm{\psi}} & \phi \\
{\bm{\psi}} {\bm{\Lambda}} & \phi\zeta  \\
\vdots & \vdots \\
{\bm{\psi}} {\bm{\Lambda}}^{k} & \fbox{$\phi\zeta$}
\end{array}
\right|.
\end{eqnarray}
Here we just show the proof of \eqref{Phi_n} because the proof of \eqref{J_n} is the same.
\\
\\
{\bf \emph{Proof} of \eqref{Phi_n}} 
\medskip \\
By the right multiplication rule \eqref{right multiplication law}, we have
\begin{eqnarray*}	
&&\Phi_{k+1}\zeta
	=\left|
	\begin{array}{ccccc}
	{\bm{\psi}} & \phi \\
	{\bm{\psi}} {\bm{\Lambda}} & \phi\zeta \\
	\vdots & \vdots \\
	{\bm{\psi}} {\bm{\Lambda}^{k}} & \fbox{$\phi\zeta^{k}$}
	\end{array}\right| \zeta
	=\left|
	\begin{array}{ccccc}
	{\bm{\psi}} {\bm{\Lambda}} & \phi\zeta \\
	{\bm{\psi}} {\bm{\Lambda}}^{2} & \phi\zeta^{2} \\
	\vdots & \vdots \\
	{\bm{\psi}} {\bm{\Lambda}}^{k+1} & \fbox{$\phi\zeta^{k+1}$}
	\end{array}\right|, \nonumber  \\
&&\Psi_{k+1}\Lambda_{k+1}
	=
	\left|
	\begin{array}{cccc}
	{\bm{\psi}} & \psi_{k+1}  \\
	{\bm{\psi}} {\bm{\Lambda}} & \psi_{k+1}\Lambda_{k+1} \\
	\vdots & \vdots \\
	{\bm{\psi}} {\bm{\Lambda}}^{k} & \fbox{$\psi_{k+1}\Lambda_{k+1}^{k}$}
	\end{array}\right|\Lambda_{k+1} \nonumber 	
	=
	\left|
	\begin{array}{cccc}
	{\bm{\psi}} {\bm{\Lambda}} & \psi_{k+1}\Lambda_{k+1} \\
	{\bm{\psi}} {\bm{\Lambda}}^{2} & \psi_{k+1}\Lambda_{k+1}^{2} \\
	\vdots & \vdots \\
	{\bm{\psi}} {\bm{\Lambda}}^{k+1} & \fbox{$\psi_{k+1}\Lambda_{k+1}^{k+1}$}
\end{array}\right|. \nonumber ~~~~~
    \end{eqnarray*}
By the homological relation \eqref{Homological relation}, we have
\begin{eqnarray*}
\Psi_{k+1}^{-1}\Phi_{k+1} 
&=&
{\left|
\begin{array}{cccc}
{\bm{\psi}} & \psi_{k+1} \\
{\bm{\psi}} {\bm{\Lambda}} & \psi_{k+1}\Lambda_{k+1} \\
\vdots & \vdots \\
{\bm{\psi}} {\bm{\Lambda}}^{k} & \fbox{$\psi_{k+1}\Lambda_{k+1}^{k}$}
\end{array}\right|}^{-1}
\left|
\begin{array}{ccccc}
{\bm{\psi}} & \phi \\
{\bm{\psi}} {\bm{\Lambda}} & \phi\zeta \\
\vdots & \vdots \\
{\bm{\psi}} {\bm{\Lambda}}^{k} & \fbox{$\phi\zeta^{k}$}
\end{array}\right| ~~ \nonumber \\
&=&
\left({\left|
	\begin{array}{cccc}
	{\bm{\psi}} & \fbox{$\psi_{k+1}$} \\
	{\bm{\psi}} {\bm{\Lambda}} & \psi_{k+1}\Lambda_{k+1}  \\
	\vdots & \vdots \\
	{\bm{\psi}} {\bm{\Lambda}}^{k} & \psi_{k+1}\Lambda_{k+1}^{k} 
	\end{array}\right|}^{-1}
{\left|
	\begin{array}{ccccc}
	{\bm{\psi}} & 1 \\
	{\bm{\psi}} {\bm{\Lambda}} & 0 \\
	\vdots & \vdots \\
	{\bm{\psi}} {\bm{\Lambda}}^{k} & \fbox{0}
	\end{array}\right|}^{-1} \right) \!\times  
\!\left( \left|
\begin{array}{ccccc}
{\bm{\psi}} & 1 \\
{\bm{\psi}} {\bm{\Lambda}} & 0 \\
\vdots & \vdots \\
{\bm{\psi}} {\bm{\Lambda}}^{k} & \fbox{0}
\end{array}\right|
\left|
\begin{array}{ccccc}
{\bm{\psi}} & \fbox{$\phi$} \\
{\bm{\psi}} {\bm{\Lambda}} & \phi\zeta \\
\vdots & \vdots \\
{\bm{\psi}} {\bm{\Lambda}}^{k} & \phi\zeta^{k}
\end{array}\right| \right) \nonumber  \\
&=&
{\left|
\begin{array}{cccc}
{\bm{\psi}} & \fbox{$\psi_{k+1}$} \\
{\bm{\psi}} {\bm{\Lambda}} & \psi_{k+1}\Lambda_{k+1} \\
\vdots & \vdots \\
{\bm{\psi}} {\bm{\Lambda}}^{k} & \psi_{k+1}\Lambda_{k+1}^{k} 
\end{array}\right|}^{-1}
\left|
\begin{array}{ccccc}
{\bm{\psi}} & \fbox{$\phi$} \\
{\bm{\psi}} {\bm{\Lambda}} & \phi \zeta \\
\vdots & \vdots \\
{\bm{\psi}} {\bm{\Lambda}}^{k} & \phi\zeta^{k}
\end{array}\right|. \nonumber
\end{eqnarray*}
Then by the Jacobi Identity \eqref{Jacobi Identity}, we can conclude that
\begin{eqnarray*}
\Phi_{k+2}
&=& \Phi_{k+1}\zeta -\Psi_{k+1}\Lambda_{k+1} \Psi_{k+1}^{-1} \Phi_{k+1}  \nonumber \\
&=&
\left|
\begin{array}{c:c:c}
{\bm{\psi}} & \psi_{k+1} & \phi \\
\hdashline
{\bm{\psi}} {\bm{\Lambda}} & \psi_{k+1}\Lambda_{k+1} &\phi \zeta \\
\vdots & \vdots & \vdots \\
{\bm{\psi}} {\bm{\Lambda}}^{k} & \psi_{k+1}\Lambda_{k+1}^{k} & \phi\zeta^{k} \\
\hdashline
{\bm{\psi}} {\bm{\Lambda}}^{k+1} & \psi_{k+1}\Lambda_{k+1}^{k+1} & \fbox{$\phi\zeta^{k+1}$}
\end{array}\right|
=
\left|
\begin{array}{ccc:c:c}
\psi_{1} & \cdots & \psi_{k} & \psi_{k+1} & \phi \\
\hdashline
\psi_{1}\Lambda_{1} & \cdots & \psi_{k}\Lambda_{k} & \psi_{k+1}\Lambda_{k+1} &\phi \zeta \\
\vdots & \ddots & \vdots & \vdots & \vdots \\
\psi_{1}\Lambda_{1}^{k} & \cdots & \psi_{k}\Lambda_{k}^{k} & \psi_{k+1}\Lambda_{k+1}^{k} & \phi\zeta^{k} \\
\hdashline
\psi_{1}\Lambda_{1}^{k+1} & \cdots & \psi_{k}\Lambda_{k}^{k+1} & 
\psi_{k+1}\Lambda_{k+1}^{k+1} & \fbox{$\phi\zeta^{k+1}$}
\end{array}\right|. \nonumber
\end{eqnarray*}
Here we use the dashed line to indicate the four box elements $\phi\zeta^{k+1}$, $\psi_{k+1}\Lambda_{k+1}^{k+1}$, $\psi_{k+1}$ and $\phi$ in the RHS of \eqref{Jacobi Identity}.
$\hfill\Box$ 

\subsection{The potential form of gauge fields and $K$-matrix }
\label{Section 4.3}

In this subsection, we review the potential form of the following gauge fields 
\begin{eqnarray}
\label{Gauge fields_n+1}
A_{w}^{(n+1)}:=-(\partial_{w}J_{n+1})J_{n+1}^{-1},~~
A_{z}^{(n+1)}:=-(\partial_{z}J_{n+1})J_{n+1}^{-1}
\end{eqnarray}
which was mentioned firstly in  the private note \cite{Nimmo} (Cf: the sub-dissertation \cite{GiHaHuNi}). We mention the explicit form of \eqref{Gauge fields_n+1} in Theorem \ref{Thm_4.6} which can be considered as the quasi-Wronskian version of \eqref{Potential form of gauge fields} because the corresponding potential ($i.e.~ K$-matrix) is determined by the specified solutions $\psi_{i}(\Lambda_i),~ i=1,2,...,n$ of the initial linear system \eqref{Initial linear system}.
As a supplementary note of \cite{GiHaHuNi,Nimmo}, we separate the derivation into Lemma \ref{Lem_4.3} and Lemma \ref{Lem_4.4} for more detailed discussion. In fact, $k=0$ case of Lemma \ref{Lem_4.4} ($i.e.$ Corollary \ref{Cor_4.5}) is enough for the proof of Theorem \ref{Thm_4.6}. Nevertheless, we apply the same idea of the proof mentioned in \cite{GiHaHuNi,Nimmo} to obtain the formula for $k > 0$ case as well. Much more than the purpose of a supplement for \cite{GiHaHuNi,Nimmo}, we also apply Lemma \ref{Lem_4.4} to obtain an independent new result and mention it in Appendix \ref{Appendix B} as Theorem \ref{Thm_B2} in which the formula of $k>0$ is necessary for this proof.
For convenience, we set ${\bm{\psi}}:= (\psi_1, \psi_2, ...,\psi_{n})$, and ${\bm{\Lambda}}$:=diag$(\Lambda_{1}, \Lambda_{2}, ..., \Lambda_{n})$ to simplify the following $n \times (n+1)$ matrix   
\begin{eqnarray}
\label{Nimmo notation}
\left(
\begin{array}{ccccc}
\psi_{1} & \!\!\cdots & \!\! \psi_{n}  \\
\vdots & \!\! \ddots & \!\! \vdots \\
\psi_{1}\Lambda_{1}^{k} & \!\! \cdots & \!\! \psi_{n}\Lambda_{n}^{k} \\
\vdots & \!\! \ddots & \!\! \vdots  \\
\psi_{1}\Lambda_{1}^{n} & \!\! \cdots & \!\! \psi_{n}\Lambda_{n}^{n}
\end{array}
\right)
=
\left(
\begin{array}{cc}
{\bm{\psi}} \\
\vdots \\
{\bm{\psi}}{\bm{\Lambda}}^{k} \\
\vdots \\
{\bm{\psi}}{\bm{\Lambda}}^{n}
\end{array}
\right)
\end{eqnarray}
and begin with our discussion.
\newtheorem{lem_4.3}[lem_4.1]{Lemma}
\begin{lem_4.3}
{\bf (Cf: \cite{GiHaHuNi,Nimmo} for $k=0$ case)} \label{Lem_4.3}
\begin{eqnarray*}
&&\left|
\begin{array}{cc}
\! {\bm{\psi}} & 0 \\ 
\! \vdots & \vdots \\
\! {\bm{\psi}}{\bm{\Lambda}}^{k} & 1 \\
\! \vdots & \vdots \\
\! {\bm{\psi}}{\bm{\Lambda}}^{n-1} & 0 \\
\! {(\partial_{w}\bm{\psi}}){\bm{\Lambda}}^{m} & \fbox{$0$}
\end{array}
\right| ~~ (~\!0 \leq k \leq n-1, ~ 0 \leq m \leq n~\!)   \nonumber  \\
&=&
\!\!\! -
\left|
\begin{array}{cc}
\! {\bm{\psi}} & 0 \\ 
\! \vdots & \vdots \\
\! {\bm{\psi}}{\bm{\Lambda}}^{k} & 0 \\
\! \vdots & \vdots \\
\! {\bm{\psi}}{\bm{\Lambda}}^{n-1} & 1 \\ 
\! (\partial_{\widetilde{z}}\bm{\psi}){\bm{\Lambda}}^{m} & \fbox{$0$}
\end{array}
\right|
\left|
\begin{array}{cc}
\! {\bm{\psi}} & 0 \\ 
\! \vdots & \vdots \\
\! {\bm{\psi}}{\bm{\Lambda}}^{k} & 1 \\
\! \vdots & \vdots \\
\! {\bm{\psi}}{\bm{\Lambda}}^{n-1} & 0 \\
\! {\bm{\psi}}{\bm{\Lambda}}^{n} & \fbox{$0$}
\end{array}
\right|
+
\left|
\begin{array}{cc}
\! {\bm{\psi}} & 0 \\ 
\! \vdots & \vdots \\
\! {\bm{\psi}}{\bm{\Lambda}}^{k-1} & 1 \\
\! \vdots & \vdots \\
\! {\bm{\psi}}{\bm{\Lambda}}^{n-1} & 0 \\
\! {(\partial_{\widetilde{z}}\bm{\psi}}){\bm{\Lambda}}^{m} & \fbox{$0$}
\end{array}
\right|
+
(\partial_{w}J_{1})J_{1}^{-1}
\left|
\begin{array}{cc}
\! {\bm{\psi}} & 0 \\ 
\! \vdots & \vdots \\
\! {\bm{\psi}}{\bm{\Lambda}}^{k} & 1 \\
\! \vdots & \vdots \\
\! {\bm{\psi}}{\bm{\Lambda}}^{n-1} & 0 \\
\! {\bm{\psi}}{\bm{\Lambda}}^{m} & \fbox{$0$}
\end{array}
\right|
\nonumber
\end{eqnarray*}
provided that the power of \bm{$\Lambda$} must be non-negative integers. In addition, this result is preserved if we replace the variable pair $(w,~ \widetilde{z})$ by $(z,~ \widetilde{w})$.
\end{lem_4.3}
{\bf (\emph{Proof})}
\begin{eqnarray*}
&&\left|
\begin{array}{cc}
{\bm{\psi}} & 0 \\ 
\vdots & \vdots \\
{\bm{\psi}}{\bm{\Lambda}}^{k} & 1 \\
\vdots & \vdots \\
{\bm{\psi}}{\bm{\Lambda}}^{n-1} & 0 \\
{(\partial_{w}\bm{\psi})}{\bm{\Lambda}}^{m} & \fbox{0}
\end{array}
\right|
=
-(\partial_{w}{\bm{\psi}}){\bm{\Lambda}}^{m}
\left(
\begin{array}{c}
{\bm{\psi}} \\ 
\vdots \\
{\bm{\psi}}{\bm{\Lambda}}^{k} \\
\vdots \\
{\bm{\psi}}{\bm{\Lambda}}^{n-1}
\end{array}
\right)^{-1}
\left(
\begin{array}{c}
0 \\
\vdots \\
1 \\
\vdots \\
0
\end{array}
\right)  \nonumber
\end{eqnarray*}
\begin{eqnarray*}
&\!\!\!=\!\!\!&
-(\partial_{w}{\bm{\psi}}){\bm{\Lambda}}^{m-1}
\left(
\begin{array}{c}
{\bm{\psi}} \\ 
\vdots \\
{\bm{\psi}}{\bm{\Lambda}}^{k} \\
\vdots \\
{\bm{\psi}}{\bm{\Lambda}}^{n-1}
\end{array}
\right)^{-1}\sum_{l=1}^{n}e_{l}^{T}e_{l}
\left(
\begin{array}{c}
{\bm{\psi}} \\ 
\vdots \\
{\bm{\psi}}{\bm{\Lambda}}^{k} \\
\vdots \\
{\bm{\psi}}{\bm{\Lambda}}^{n-1}
\end{array}
\right){\bm{\Lambda}}
\left(
\begin{array}{c}
{\bm{\psi}} \\ 
\vdots \\
{\bm{\psi}}{\bm{\Lambda}}^{k} \\
\vdots \\
{\bm{\psi}}{\bm{\Lambda}}^{n-1}
\end{array}
\right)^{-1}
\left(
\begin{array}{c}
0 \\
\vdots \\
1 \\
\vdots \\
0
\end{array}
\right)  \nonumber  \\
&& {\mbox{(where $e_{l}:= (0,...,0,1,0,...,0)$ with $l$-th component is 1, and 0 elsewhere.)}} \nonumber \\
&\!\!\!=&
\!\!\!\! -\sum_{l=0}^{n-1}
\left[
-(\partial_{w}{\bm{\psi}}){\bm{\Lambda}}^{m-1}
\left(
\begin{array}{c}
{\bm{\psi}} \\ 
\vdots \\
{\bm{\psi}}{\bm{\Lambda}}^{k} \\
\vdots \\
{\bm{\psi}}{\bm{\Lambda}}^{n-1}
\end{array}
\right)^{-1}
\!\!\!\!\!\!e_{l+1}^{T}
\right]
\!\! \left[
-e_{l+1}
\left(
\begin{array}{c}
{\bm{\psi}}{\bm{\Lambda}} \\ 
\vdots \\
{\bm{\psi}}{\bm{\Lambda}}^{k+1} \\
\vdots \\
{\bm{\psi}}{\bm{\Lambda}}^{n}
\end{array}
\right)
\left(
\begin{array}{c}
{\bm{\psi}} \\ 
\vdots \\
{\bm{\psi}}{\bm{\Lambda}}^{k} \\
\vdots \\
{\bm{\psi}}{\bm{\Lambda}}^{n-1}
\end{array}
\right)^{-1}
\!\!\!\! \left(
\begin{array}{c}
0 \\
\vdots \\
1 \\
\vdots \\
0
\end{array}
\right)
\right]  \nonumber 
\end{eqnarray*}
~~~~~~~(By \eqref{Initial linear system}, we can replace $\partial_{w}{\bm{\psi}}$  by $(\partial_{\widetilde{z}}{\bm{\psi}}){\bm{\Lambda}} + (\partial_{w}J_{1})J_{1}^{-1}{\bm{\psi}}$.)
\begin{eqnarray}
\label{*}
&\!\!\!\!=&\!\!\!\!
-\sum_{l=0}^{n-1}
\left(
\left|
\begin{array}{cc}
{\bm{\psi}} & 0 \\
\vdots & \vdots \\
{\bm{\psi}}{\bm{\Lambda}}^{l} & 1 \\
\vdots & \vdots \\
{\bm{\psi}}{\bm{\Lambda}}^{n-1} &  0 \\
(\partial_{\widetilde{z}}{\bm{\psi}}){\bm{\Lambda}}^{m} & \fbox{$0$}
\end{array}
\right|
+
(\partial_{w}J_{1})J_{1}^{-1}
\left|
\begin{array}{cc}
{\bm{\psi}} & 0 \\
\vdots & \vdots \\
{\bm{\psi}}{\bm{\Lambda}}^{l} & 1 \\
\vdots & \vdots \\
{\bm{\psi}}{\bm{\Lambda}}^{n-1} &  0 \\
{\bm{\psi}}{\bm{\Lambda}^{m-1}} & \fbox{$0$}
\end{array}
\right| 
\right)
\left|
\begin{array}{cc}
{\bm{\psi}} & 0 \\
\vdots & \vdots \\
{\bm{\psi}}{\bm{\Lambda}}^{k} & 1 \\
\vdots & \vdots \\
{\bm{\psi}}{\bm{\Lambda}}^{n-1} & 0 \\
{\bm{\psi}}{\bm{\Lambda}}^{l+1} & \fbox{$0$}
\end{array}
\right| ~~~~~~~~~~~~~~~~~~  
\end{eqnarray}
By using the row operation of quasideterminant (Proposition \ref{Prop_2.5}), we can obtain
\begin{eqnarray}
\label{**}
&& \!\!\!\!\!\!
\left|
\begin{array}{cc}
{\bm{\psi}} & 0 \\
\vdots & \vdots \\
{\bm{\psi}}{\bm{\Lambda}}^{k} & 1 \\
\vdots & \vdots \\
{\bm{\psi}}{\bm{\Lambda}}^{n-1} & 0 \\
{\bm{\psi}}{\bm{\Lambda}}^{l+1} & \fbox{0}
\end{array}
\right|   
=
\left\{
\begin{array}{c}
\!\!\!\!\!\!\!\!\!\!\!\!\!\!\!\!\!\!\!\!\!\!\!\!\!\!\!\!\!\!\!\!\!\!\!\!\!\!\!\!\!\!\!\!\!\!\!\!\!\!\!\!\!\!\!\!\!\!\! 
\left|
\begin{array}{cc}
{\bm{\psi}} & 0 \\
\vdots & \vdots \\
{\bm{\psi}}{\bm{\Lambda}}^{k} & 1 \\
\vdots & \vdots \\
{\bm{\psi}}{\bm{\Lambda}}^{n} & \fbox{0}
\end{array}
\right|,~~ \mbox{if~~} l=n-1   
\\
\left|
\begin{array}{cc}
{\bm{\psi}} & 0 \\
\vdots & \vdots \\
{\bm{\psi}}{\bm{\Lambda}}^{k} & 1 \\
\vdots & \vdots \\
{\bm{\psi}}{\bm{\Lambda}}^{k} & \fbox{0}
\end{array}
\right|
=
\left|
\begin{array}{cc}
{\bm{\psi}} & 0 \\
\vdots & \vdots \\
{\bm{\psi}}{\bm{\Lambda}}^{k} & 1 \\
\vdots & \vdots \\
0 & \fbox{$-1$}
\end{array}
\right|
=
-1,~~ {\mbox{if}~~} l=k-1
\\
\!\!\!\!\!\! \left|
\begin{array}{cc}
{\bm{\psi}} & 0 \\
\vdots & \vdots \\
{\bm{\psi}}{\bm{\Lambda}}^{k} & 1 \\
\vdots & \vdots \\
{\bm{\psi}}{\bm{\Lambda}}^{l+1} & 0 \\
\vdots & \vdots \\
{\bm{\psi}}{\bm{\Lambda}}^{l+1} & \fbox{0}
\end{array}
\right|
=
\left|
\begin{array}{cc}
{\bm{\psi}} & 0 \\
\vdots & \vdots \\
{\bm{\psi}}{\bm{\Lambda}}^{k} & 1 \\
\vdots & \vdots \\
{\bm{\psi}}{\bm{\Lambda}}^{l+1} & 0 \\
\vdots & \vdots \\
0 & \fbox{$0$}
\end{array}
\right|
=
0 ~\mbox{,~~otherwise} 
\end{array}
\right. ~~~~ 
\end{eqnarray}
Similarly, we can obtain
\begin{eqnarray}
\label{***}
&& \!\!\!\!\!\!
\left|
\begin{array}{cc}
{\bm{\psi}} & 0 \\
\vdots & \vdots \\
{\bm{\psi}}{\bm{\Lambda}}^{l} & 1 \\
\vdots & \vdots \\
{\bm{\psi}}{\bm{\Lambda}}^{m-1} & \fbox{0}  
\end{array}
\right|
=\left\{
\begin{array}{c}
-1~, ~ \mbox{if}~~ l=m-1 \\
\\
\!\!\!\! 0~, ~ \mbox{otherwise}  
\end{array}
\right.  ~~   
\end{eqnarray}
Substituting \eqref{**} and \eqref{***} into the summation \eqref{*}, it remains three terms. That is, 
\begin{eqnarray*}
\eqref{*} =
-\left|
\begin{array}{cc}
\!\! {\bm{\psi}} & \!\!0 \\ 
\!\! \vdots & \!\!\vdots \\
\!\! {\bm{\psi}}{\bm{\Lambda}}^{k} & \!\!0 \\
\!\! \vdots & \!\!\vdots \\
\!\! {\bm{\psi}}{\bm{\Lambda}}^{n-1} & \!\!1 \\ 
\!\! (\partial_{\widetilde{z}}\bm{\psi}){\bm{\Lambda}}^{m} & \!\!\fbox{$0$}
\end{array}
\right|
\left|
\begin{array}{cc}
\! {\bm{\psi}} & \!\!0 \\ 
\! \vdots & \!\!\vdots \\
\! {\bm{\psi}}{\bm{\Lambda}}^{k} & \!\!1 \\
\! \vdots & \!\!\vdots \\
\! {\bm{\psi}}{\bm{\Lambda}}^{n-1} & \!\!0 \\
\! {\bm{\psi}}{\bm{\Lambda}}^{n} & \!\!\fbox{$0$}
\end{array}
\right|
+
\left|
\begin{array}{cc}
\!\! {\bm{\psi}} & \!\!0 \\ 
\!\! \vdots & \!\!\vdots \\
\!\! {\bm{\psi}}{\bm{\Lambda}}^{k-1} & \!\!1 \\
\!\! \vdots & \!\!\vdots \\
\!\! {\bm{\psi}}{\bm{\Lambda}}^{n-1} & \!\!0 \\
\!\! {(\partial_{\widetilde{z}}\bm{\psi}}){\bm{\Lambda}}^{m} & \!\!\fbox{$0$}
\end{array}
\right|
+
(\partial_{w}J_{1})J_{1}^{-1}
\left|
\begin{array}{cc}
\! {\bm{\psi}} & \!\!0 \\ 
\! \vdots & \!\!\vdots \\
\! {\bm{\psi}}{\bm{\Lambda}}^{k} & \!\!1 \\
\! \vdots & \!\!\vdots \\
\! {\bm{\psi}}{\bm{\Lambda}}^{n-1} & \!\!0 \\
\! {\bm{\psi}}{\bm{\Lambda}}^{m} & \!\!\fbox{$0$}
\end{array}
\right|. \nonumber 
\end{eqnarray*}
$\hfill\Box$
\newtheorem{lem_4.4}[lem_4.1]{Lemma}
\begin{lem_4.4}
{\bf (Cf: \cite{GiHaHuNi,Nimmo} for $k=0$ case)} \label{Lem_4.4} \\
For $0 \leq k \leq n-1$,
\begin{eqnarray}
\partial_{w}
\left|
\begin{array}{cc}
{\bm{\psi}} & 0 \\
\vdots & \vdots  \\
{\bm{\psi}}{\bm{\Lambda}}^{k} & 1 \\
\vdots & \vdots  \\
{\bm{\psi}}{\bm{\Lambda}}^{n} & \fbox{$0$}
\end{array}
\right|  
&=&
-\left(
\partial_{\widetilde{z}}\left|
\begin{array}{cc}
{\bm{\psi}} & 0 \\
{\bm{\psi}}{\bm{\Lambda}} & 0 \\
\vdots & \vdots  \\
{\bm{\psi}}{\bm{\Lambda}}^{n-1} & 1 \\
{\bm{\psi}}{\bm{\Lambda}}^{n} & \fbox{$0$}
\end{array}
\right|
\right)
\left|
\begin{array}{ccccc}
{\bm{\psi}} & 0 \\
\vdots & \vdots  \\
{\bm{\psi}}{\bm{\Lambda}}^{k} & 1 \\
\vdots & \vdots  \\
{\bm{\psi}}{\bm{\Lambda}}^{n} & \fbox{$0$}
\end{array}
\right|  
+
\partial_{\widetilde{z}}
\left|
\begin{array}{cc}
{\bm{\psi}} & 0 \\
\vdots & \vdots  \\
{\bm{\psi}}{\bm{\Lambda}}^{k-1} & 1 \\
\vdots & \vdots  \\
{\bm{\psi}}{\bm{\Lambda}}^{n} & \fbox{$0$}
\end{array}
\right|  \nonumber  \\
&&+~
[~
(\partial_{w}J_{1})J_{1}^{-1}~,~
\left|
\begin{array}{cc}
{\bm{\psi}} & 0 \\
\vdots & \vdots  \\
{\bm{\psi}}{\bm{\Lambda}}^{k} & 1 \\
\vdots & \vdots  \\
{\bm{\psi}}{\bm{\Lambda}}^{n} & \fbox{$0$}
\end{array}
\right| 
~]   \nonumber 
\end{eqnarray}
provided that the power of \bm{$\Lambda$} must be non-negative integers. In addition, this result is preserved if we replace variable pair $(w,~ \widetilde{z})$ by $(z,~ \widetilde{w})$.
\end{lem_4.4}
{\bf (\emph{Proof})}
\\
By using the derivative formula of quasideterminant (Theorem \ref{Thm_2.9}), we have
\begin{eqnarray*}
\partial_{w}
\left|
\begin{array}{cc}
{\bm{\psi}} & 0 \\
\vdots & \vdots  \\
{\bm{\psi}}{\bm{\Lambda}}^{k} & 1 \\
\vdots & \vdots  \\
{\bm{\psi}}{\bm{\Lambda}}^{n} & \fbox{0}
\end{array}
\right|
=
\left|
\begin{array}{cc}
{\bm{\psi}} & 0 \\
\vdots & \vdots  \\
{\bm{\psi}}{\bm{\Lambda}}^{k} & 1 \\
\vdots & \vdots  \\
{(\partial_{w}\bm{\psi}}){\bm{\Lambda}}^{n} & \fbox{$0$}
\end{array}
\right|
+
\sum_{m=0}^{n-1}
\left|
\begin{array}{cc}
{\bm{\psi}} & 0 \\
\vdots & \vdots  \\
{\bm{\psi}}{\bm{\Lambda}}^{m} & 1 \\
\vdots & \vdots  \\
{\bm{\psi}}{\bm{\Lambda}}^{n} & \fbox{$0$}
\end{array}
\right| 
\left|
\begin{array}{cc}
{\bm{\psi}} & 0 \\
\vdots & \vdots  \\
{\bm{\psi}}{\bm{\Lambda}}^{k} & 1 \\
\vdots & \vdots  \\
{(\partial_{w}\bm{\psi}}){\bm{\Lambda}}^{m} & \fbox{$0$}
\end{array}
\right|   \nonumber
\end{eqnarray*}
~~~~~~~~~~~~~~(By Lemma \ref{Lem_4.3}, we get the following equality.)
\begin{eqnarray*}
&=&
\!\!\! - \left\{
\left|
\begin{array}{cc}
\! {\bm{\psi}} & 0 \\ 
\! \vdots & \vdots \\
\! {\bm{\psi}}{\bm{\Lambda}}^{k} & 0 \\
\! \vdots & \vdots \\
\! {\bm{\psi}}{\bm{\Lambda}}^{n-1} & 1 \\ 
\! (\partial_{\widetilde{z}}\bm{\psi}){\bm{\Lambda}}^{n} & \fbox{0}
\end{array}
\right|
+
\sum_{m=0}^{n-1}
\left|
\begin{array}{cc}
{\bm{\psi}} & 0 \\
\vdots & \vdots  \\
{\bm{\psi}}{\bm{\Lambda}}^{m} & 1 \\
\vdots & \vdots  \\
\! {\bm{\psi}}{\bm{\Lambda}}^{n-1} & 0 \\ 
{\bm{\psi}}{\bm{\Lambda}}^{n} & \fbox{0}
\end{array}
\right| 
\left|
\begin{array}{cc}
\! {\bm{\psi}} & 0 \\ 
\! \vdots & \vdots \\
\! {\bm{\psi}}{\bm{\Lambda}}^{k} & 0 \\
\! \vdots & \vdots \\
\! {\bm{\psi}}{\bm{\Lambda}}^{n-1} & 1 \\ 
\! (\partial_{\widetilde{z}}\bm{\psi}){\bm{\Lambda}}^{m} & \fbox{0}
\end{array}
\right|
\right\}
\left|
\begin{array}{cc}
\! {\bm{\psi}} & 0 \\ 
\! \vdots & \vdots \\
\! {\bm{\psi}}{\bm{\Lambda}}^{k} & 1 \\
\! \vdots & \vdots \\
\! {\bm{\psi}}{\bm{\Lambda}}^{n-1} & 0 \\
\! {\bm{\psi}}{\bm{\Lambda}}^{n} & \fbox{0}
\end{array}
\right|  \nonumber  \\
&&\!\!\!+
\left\{
\left|
\begin{array}{cc}
\! {\bm{\psi}} & 0 \\ 
\! \vdots & \vdots \\
\! {\bm{\psi}}{\bm{\Lambda}}^{k-1} & 1 \\
\! \vdots & \vdots \\
\! {\bm{\psi}}{\bm{\Lambda}}^{n-1} & 0 \\
\! {(\partial_{\widetilde{z}}\bm{\psi}}){\bm{\Lambda}}^{n} & \fbox{0}
\end{array}
\right|
+
\sum_{m=0}^{n-1}
\left|
\begin{array}{cc}
{\bm{\psi}} & 0 \\
\vdots & \vdots  \\
{\bm{\psi}}{\bm{\Lambda}}^{m} & 1 \\
\vdots & \vdots  \\
\! {\bm{\psi}}{\bm{\Lambda}}^{n-1} & 0 \\ 
{\bm{\psi}}{\bm{\Lambda}}^{n} & \fbox{0}
\end{array}
\right|
\left|
\begin{array}{cc}
\! {\bm{\psi}} & 0 \\ 
\! \vdots & \vdots \\
\! {\bm{\psi}}{\bm{\Lambda}}^{k-1} & 1 \\
\! \vdots & \vdots \\
\! {\bm{\psi}}{\bm{\Lambda}}^{n-1} & 0 \\
\! {(\partial_{\widetilde{z}}\bm{\psi}}){\bm{\Lambda}}^{m} & \fbox{0}
\end{array}
\right| 
\right\}  \nonumber  \\
&&\!\!\!+ \left\{
(\partial_{w}J_{1})J_{1}^{-1}
\left|
\begin{array}{cc}
\! {\bm{\psi}} & 0 \\ 
\! \vdots & \vdots \\
\! {\bm{\psi}}{\bm{\Lambda}}^{k} & 1 \\
\! \vdots & \vdots \\
\! {\bm{\psi}}{\bm{\Lambda}}^{n-1} & 0 \\
\! {\bm{\psi}}{\bm{\Lambda}}^{n} & \fbox{0}
\end{array}
\right|
+
\sum_{m=0}^{n-1}
\left|
\begin{array}{cc}
{\bm{\psi}} & 0 \\
\vdots & \vdots  \\
{\bm{\psi}}{\bm{\Lambda}}^{m} & 1 \\
\vdots & \vdots  \\
\! {\bm{\psi}}{\bm{\Lambda}}^{n-1} & 0 \\ 
{\bm{\psi}}{\bm{\Lambda}}^{n} & \fbox{0}
\end{array}
\right| 
(\partial_{w}J_{1})J_{1}^{-1}
\left|
\begin{array}{cc}
\! {\bm{\psi}} & 0 \\ 
\! \vdots & \vdots \\
\! {\bm{\psi}}{\bm{\Lambda}}^{k} & 1 \\
\! \vdots & \vdots \\
\! {\bm{\psi}}{\bm{\Lambda}}^{n-1} & 0 \\
\! {\bm{\psi}}{\bm{\Lambda}}^{m} & \fbox{0}
\end{array}
\right|
\right\}   \nonumber \\  \nonumber \\
&& \!\!{(\mbox{By using the derivative formula (Theorem \ref{Thm_2.9}) on the first two bracket, and}} 
\nonumber \\
&& {\mbox{the last bracket only takes value at $m=k$, we get the following equality.)}} \nonumber \\ \nonumber \\
&=&
\!\!\!
-\left(
\partial_{\widetilde{z}}
\left|
\begin{array}{cc}
\! {\bm{\psi}} & 0 \\ 
\! \vdots & \vdots \\
\! {\bm{\psi}}{\bm{\Lambda}}^{k} & 0 \\
\! \vdots & \vdots \\
\! {\bm{\psi}}{\bm{\Lambda}}^{n-1} & 1 \\
\! {\bm{\psi}}{\bm{\Lambda}}^{n} & \fbox{0}
\end{array}
\right|
\right)
\left|
\begin{array}{cc}
\! {\bm{\psi}} & 0 \\ 
\! \vdots & \vdots \\
\! {\bm{\psi}}{\bm{\Lambda}}^{k} & 1 \\
\! \vdots & \vdots \\
\! {\bm{\psi}}{\bm{\Lambda}}^{n-1} & 0 \\
\! {\bm{\psi}}{\bm{\Lambda}}^{n} & \fbox{0}
\end{array}
\right| 
+
\partial_{\widetilde{z}}
\left|
\begin{array}{cc}
\! {\bm{\psi}} & 0 \\ 
\! \vdots & \vdots \\
\! {\bm{\psi}}{\bm{\Lambda}}^{k-1} & 1 \\
\! \vdots & \vdots \\
\! {\bm{\psi}}{\bm{\Lambda}}^{n-1} & 0 \\
\! {\bm{\psi}}{\bm{\Lambda}}^{n} & \fbox{0}
\end{array}
\right|    \nonumber \\
&&\!\!\!+
(\partial_{w}J_{1})J_{1}^{-1}
\left|
\begin{array}{cc}
\! {\bm{\psi}} & 0 \\ 
\! \vdots & \vdots \\
\! {\bm{\psi}}{\bm{\Lambda}}^{k} & 1 \\
\! \vdots & \vdots \\
\! {\bm{\psi}}{\bm{\Lambda}}^{n-1} & 0 \\
\! {\bm{\psi}}{\bm{\Lambda}}^{n} & \fbox{0}
\end{array}
\right| 
-
\left|
\begin{array}{cc}
{\bm{\psi}} & 0 \\
\vdots & \vdots  \\
{\bm{\psi}}{\bm{\Lambda}}^{k} & 1 \\
\vdots & \vdots  \\
\! {\bm{\psi}}{\bm{\Lambda}}^{n-1} & 0 \\ 
{\bm{\psi}}{\bm{\Lambda}}^{n} & \fbox{0}
\end{array}
\right| 
(\partial_{w}J_{1})J_{1}^{-1}.   \nonumber
\end{eqnarray*}
$\hfill\Box$
\\
In particular, when $k=0$ we have the following result.
\newtheorem{cor_4.5}[lem_4.1]{Corollary}
\begin{cor_4.5}
{\bf (\cite{Nimmo}, Cf: \cite{GiHaHuNi} for $J_1=I$ case )} \label{Cor_4.5}
\begin{eqnarray}
\partial_{w}
\left|
\begin{array}{cc}
{\bm{\psi}} & \!\! 1 \\
{\bm{\psi}}{\bm{\Lambda}} & \!\! 0  \\
\vdots & \!\! \vdots  \\
{\bm{\psi}}{\bm{\Lambda}}^{n-1} & \!\! 0 \\ 
{\bm{\psi}}{\bm{\Lambda}}^{n} & \!\! \fbox{$0$}
\end{array}
\right|  
=
\!
-\left(
\partial_{\widetilde{z}}\left|
\begin{array}{cc}
{\bm{\psi}} & \!\! 0 \\
{\bm{\psi}}{\bm{\Lambda}} & \!\! 0 \\
\vdots & \!\! \vdots  \\
{\bm{\psi}}{\bm{\Lambda}}^{n-1} & \!\! 1 \\
{\bm{\psi}}{\bm{\Lambda}}^{n} & \!\! \fbox{$0$}
\end{array}
\right|
\right)
\!\left|
\begin{array}{ccccc}
{\bm{\psi}} & \!\! 1 \\
{\bm{\psi}}{\bm{\Lambda}} & \!\! 0  \\
\vdots & \!\! \vdots  \\
{\bm{\psi}}{\bm{\Lambda}}^{n-1} & \!\! 0 \\ 
{\bm{\psi}}{\bm{\Lambda}}^{n} & \!\! \fbox{$0$}
\end{array}
\right|  
\!+
[~\!
(\partial_{w}J_{1})J_{1}^{-1}~,~
\left|
\begin{array}{cc}
{\bm{\psi}} & \!\! 0 \\
{\bm{\psi}}{\bm{\Lambda}} & \!\! 0 \\
\vdots & \!\! \vdots  \\
{\bm{\psi}}{\bm{\Lambda}}^{n-1} & \!\! 1 \\
{\bm{\psi}}{\bm{\Lambda}}^{n} & \!\! \fbox{$0$}
\end{array}
\right| 
~\!\!].   \nonumber 
\end{eqnarray}
\end{cor_4.5}
{\color{white} blank}
\newtheorem{thm_4.6}[lem_4.1]{Theorem}
\begin{thm_4.6}
{\bf (Gauge fields \cite{Nimmo}, Cf: \cite{GiHaHuNi} for $J_1=I$ case)} \label{Thm_4.6} \\
The gauge fields $A_{w}^{(n+1)}
:= -(\partial_{w}J_{n+1})J_{n+1}^{-1}$ and $A_{z}^{(n+1)}
:= -(\partial_{z}J_{n+1})J_{n+1}^{-1}$ can be expressed as the potential form with the potential $K_1+K_{n+1}$, more precisely, 
\begin{eqnarray}
\begin{array}{l}
A_{w}^{(n+1)} = \partial_{\widetilde z}(K_{1} + K_{n+1}) \\
A_{z}^{(n+1)} = \partial_{\widetilde w}(K_{1} + K_{n+1}) 
\end{array}
~~ \mbox{where} ~~
\left\{
\begin{array}{l}
K_{n+1}
=
-
\left|
\begin{array}{ccccc}
\psi_{1} & \!\! \cdots & \!\! \psi_{n} & 0 \\
\psi_{1}\Lambda_{1} & \!\! \cdots & \!\! \psi_{n}\Lambda_{n} & 0 \\
\vdots & \!\! \ddots & \!\! \vdots & \vdots  \\
\psi_{1}\Lambda_{1}^{n-1} & \!\! \cdots & \!\! \psi_{n}\Lambda_{n}^{n-1} & 1 \\
\psi_{1}\Lambda_{1}^{n} & \!\! \cdots & \!\! \psi_{n}\Lambda_{n}^{n} & \fbox{$0$}
\end{array} \right|  \\
\partial_{\widetilde{w}}K_{1} = -(\partial_{z}J_{1})J_{1}^{-1}  \\
\partial_{\widetilde z}K_{1} = -(\partial_{w}J_{1})J_{1}^{-1} 
\end{array}.  \nonumber 
\right.
\end{eqnarray}
\end{thm_4.6}
{\bf (\emph{Proof})}
\\
\begin{eqnarray*}
\partial_{w}J_{n+1}
&=&
\partial_{w}
\left(
\left|
\begin{array}{cc}
{\bm{\psi}} & 1 \\
{\bm{\psi}}{\bm{\Lambda}} & 0 \\
\vdots & \vdots \\
{\bm{\psi}}{\bm{\Lambda}}^{n-1} & 0 \\ 
{\bm{\psi}}{\bm{\Lambda}}^{n} & \fbox{0}
\end{array}
\right|J_{1}
\right)
=
\left(
\partial_{w}
\left|
\begin{array}{cc}
{\bm{\psi}} & 1 \\
{\bm{\psi}}{\bm{\Lambda}} & 0 \\
\vdots & \vdots \\
{\bm{\psi}}{\bm{\Lambda}}^{n-1} & 0 \\ 
{\bm{\psi}}{\bm{\Lambda}}^{n} & \fbox{0}
\end{array}
\right|
\right)
J_{1}
+
\left|
\begin{array}{cc}
{\bm{\psi}} & 1 \\
{\bm{\psi}}{\bm{\Lambda}} & 0 \\
\vdots & \vdots \\
{\bm{\psi}}{\bm{\Lambda}}^{n-1} & 0 \\ 
{\bm{\psi}}{\bm{\Lambda}}^{n} & \fbox{0}
\end{array}
\right|(\partial_{w}J_{1})  \nonumber  \\
&=&
\left(
(\partial_{w}J_{1})J_{1}^{-1}-
\partial_{\widetilde z}
\left|
\begin{array}{cc}
{\bm{\psi}} & 0 \\
{\bm{\psi}}{\bm{\Lambda}} & 0 \\
\vdots & \vdots \\
{\bm{\psi}}{\bm{\Lambda}}^{n-1} & 1 \\ 
{\bm{\psi}}{\bm{\Lambda}}^{n} & \fbox{0}
\end{array}
\right|
\right)
\left|
\begin{array}{cc}
{\bm{\psi}} & 1 \\
{\bm{\psi}}{\bm{\Lambda}} & 0 \\
\vdots & \vdots \\
{\bm{\psi}}{\bm{\Lambda}}^{n-1} & 0 \\
{\bm{\psi}}{\bm{\Lambda}}^{n} & \fbox{0}
\end{array}
\right|J_{1}   \nonumber  \\
&=&
(K_{1} + K_{n+1})J_{n+1}.   \nonumber 
\end{eqnarray*}
Here we use Corollary \ref{Cor_4.5} in the first term of the second equality.
$\hfill\Box$
\\

\newpage


\section{ASDYM 1-Solitons ($G=\mathrm{SU}(2)$ or $\mathrm{SL}(2,\mathbb{C})$)}
\label{Section 5}

In this section, we construct the ASDYM 1-solitons successfully by applying 1 iteration of the Darboux transformation. The results are written based on the sub-dissertation \cite{HaHu1} with some revisions and supplements.
In Subsection \ref{Section 5.1}, we firstly gives a candidate of soliton solution on the 4-dimensional complex space $\mathbb{C}^{4}$ (Cf: Subsection \ref{Section 3.2}) and calculate the corresponding complex-valued Lagrangian density Tr$F^2$ (Cf: \eqref{Complex Lagrangian density_ASD}) explicitly. To avoid encountering the singularity problem on the complex space, we consider a special slice of $\mathbb{C}^4$ and find that the resulting Lagrangian density is the origin of the real-valued Lagrangian densities mentioned later in Subsection \ref{Section 5.2} and \ref{Section 5.3}.
 
In Subsection \ref{Section 5.2} and \ref{Section 5.3}, we firstly impose the reality conditions \eqref{Reality condition_E}, \eqref{Reality condition_M}, and \eqref{Reality condition_U} on the complex space $\mathbb{C}^4$ to realize the Euclidean signature $(+,+,+,+)$, the Minkowski signature $(+,-,-,-)$, and the split signature $(+,+,-,-)$ (Ultrahyperbolic space) respectively.
We then propose ansatz \eqref{Ansatz_U}, \eqref{Ansatz_E}, and \eqref{Ansatz_M} for obtaining real-valued Lagrangian densities on each real spaces. Especially for the split signature, we show that the gauge group can be $G=\mathrm{SU}(2)$ and therefore our results could be some physical objects expectantly in open $\mathrm{N=2}$ string theories \cite{OoVa1, OoVa2}. Subsection \ref{Section 5.4} is a summary of our results.

\subsection{ASDYM 1-Soliton Solutions on 4D complex space}
\label{Section 5.1}
Let us consider 1 iteration of the Darboux transformation and test the solution \eqref{J_2} if it can induce soliton solution or not. We set the gauge group to be $G=\mathrm{GL}(2,\mathbb{C})$, and the seed solution to be $J_1=I$. Under these setup, the initial linear system \eqref{Initial linear system} reduces to the following two by two linear system :
\begin{eqnarray}
\label{Initial linear system_J_1=I}
\partial_z \psi=(\partial_{\widetilde{w}}\psi) \Lambda,~
\partial_w \psi=(\partial_{\widetilde{z}}\psi) \Lambda.
\end{eqnarray}
We can solve it immediately and find a candidate of soliton solution as following :
\medskip \\
{\bf{Candidate of Soliton Solution}}
\begin{eqnarray}
\label{Soliton solution_cpx}
J\!\!\!\!&=\!\!\!\!&
\left|
\begin{array}{cc}
\!\psi & \!\!1 \\
\!\psi\Lambda & \!\!\fbox{0}
\end{array}
\right|
=
-\psi\Lambda \psi^{-1},~ \nonumber \\
&&\psi:=
\left(
\begin{array}{cc}
\!\!A & \!\!C \\
\!\!B & \!\!D
\end{array}
\!\right)
:=\left(
\begin{array}{cc}
\!a_1e^{L}+a_2e^{-L} & \!\!c_1e^{M}+c_2e^{-M} \\
\!b_1e^{L}+b_2e^{-L} & \!\!d_1e^{M}+d_2e^{-M}
\end{array}\!\!\right),~
\Lambda:=
\left(
\begin{array}{cc}
\!\lambda & \!0 \\
\!0 & \!\mu
\end{array}
\!\!\right),~~~~~~~~~ \\
\label{LM_cpx}
&&L:=(\lambda \alpha) z
+\beta \widetilde{z}
+(\lambda \beta) w
+\alpha\widetilde{w},~~
M:=(\mu\gamma) z
+\delta \widetilde{z}
+(\mu\delta) w
+\gamma \widetilde{w},~ \\
&&\mbox{where~ $a_{1}, a_{2},
b_{1}, b_{2},
c_{1}, c_{2},
d_{1}, d_{2},
\alpha, \beta, \gamma, \delta, \lambda, \mu$
are complex constants.} \nonumber 
\end{eqnarray}
Then we can use the above $J$-matrix to calculate the gauge fields \eqref{A_z, A_w in terms of J}, the field strength \eqref{Field strength_complex}, and the Lagrangian density \eqref{Complex Lagrangian density_ASD}. The detailed calculations can be referred to Appendix \ref{Appendix C}. Here we just mention the main results as the following :
\medskip \\
{\bf{Gauge Fields}}
\begin{eqnarray}
\label{A_m}
{A_{m}}&\!\!\!\!:=\!\!\!\!& -(\partial_{m}J)J^{-1}
=
\frac{2(\lambda - \mu)}{(AD-BC)^2}\left(
\begin{array}{cc}
\widetilde{p}_mAB - p_mCD  & -\widetilde{p}_mA^2 + p_mC^2   \\ 
\widetilde{p}_mB^2 - p_mD^2  & -\widetilde{p}_mAB + p_mCD
\end{array}\right),
\\
&&
(p_m,~\widetilde{p}_m):=
\left\{
\begin{array}{l}
(\alpha(a_{2}b_{1}-a_{1}b_{2}),~\gamma(c_{2}d_{1}-c_{1}d_{2}))~~
{\mbox{if}} ~~ m=z,~~ \\
(\beta(a_{2}b_{1}-a_{1}b_{2}),~\delta(c_{2}d_{1}-c_{1}d_{2}))~~
{\mbox{if}} ~~ m=w \nonumber \\
(0, 0)~~ {\mbox{if}} ~~ m= \widetilde{z}, \widetilde{w}
\end{array}.
\right.
\end{eqnarray}
\\
{\bf Lagrangian Density}
\begin{eqnarray}
\label{Lagrangian density_cpx}
{\mbox{Tr}} F^2\!\!\!\!&=\!\!\!\!&
8(\lambda-\mu)^2(\alpha\delta-\beta\gamma)^2
\varepsilon_0\widetilde{\varepsilon}_0 
\left\{\frac{2\varepsilon_1\widetilde{\varepsilon}_1
	\sinh^2 X_1
	-2\varepsilon_2\widetilde{\varepsilon}_2
	\sinh^2 X_2 
	-\varepsilon_0\widetilde{\varepsilon}_0}
{\left[(\varepsilon_1\widetilde{\varepsilon}_1)^{\frac12}
	\cosh X_1 
	+(\varepsilon_2\widetilde{\varepsilon}_2)^{\frac12}
	\cosh X_2\right]^4}
\right\},	 \\
&&X_1:=M+L+\dfrac12 \log(\varepsilon_1/\widetilde{\varepsilon}_1),~~~ 
X_2:=M-L+\dfrac12 \log(\varepsilon_2/\widetilde{\varepsilon}_2), \nonumber \\
&&
\left\{
\begin{array}{l}
\varepsilon_0:=a_2b_1-a_1b_2,~~~
\widetilde{\varepsilon_0}:=c_2d_1-c_1d_2,~ \\ 
\varepsilon_1:=a_1d_1-b_1c_1,~~~
\widetilde{\varepsilon}_1:=a_2d_2-b_2c_2,~ \\
\varepsilon_2:=a_1d_2-b_1c_2,~~~
\widetilde{\varepsilon}_2:=a_2d_1-b_2c_1.   \\
\mbox{$\varepsilon_{0},~\varepsilon_{1},~\varepsilon_{2}$~ satisfy the relation~
$\varepsilon_{1}\widetilde{\varepsilon_{1}}
=
\varepsilon_{0}\widetilde{\varepsilon_{0}} + \varepsilon_{2}\widetilde{\varepsilon_{2}}$}
\label{Relation of epsilons_C^4}
\end{array}.
\right.
\end{eqnarray}
Note that the Lagrangian density is nonzero if $\lambda \neq \mu$, $\alpha\delta \neq \beta\gamma$, and
$\varepsilon_{0}\widetilde{\epsilon_{0}} \neq 0$.
Now the remaining task is to deal with the problem of singularities. 
By the relation \eqref{Relation of epsilons_C^4}, we have 
$\varepsilon_{0}\widetilde{\epsilon_{0}} \neq 0 \iff \varepsilon_{1}\widetilde{\epsilon_{1}} \neq \varepsilon_{2}\widetilde{\epsilon_{2}}$.
Therefore, all the singularities appear on
\begin{eqnarray}
\label{D}
D:=\left\{(z, \widetilde{z}, w,\widetilde{w})\in\mathbb{C}^4~|~\
(\varepsilon_1\widetilde{\varepsilon}_1)^{\frac12}
\cosh X_1 
+(\varepsilon_2\widetilde{\varepsilon}_2)^{\frac12}
\cosh X_2=0,~
\varepsilon_1\widetilde{\varepsilon}_1 \neq \varepsilon_2\widetilde{\varepsilon}_2
\right\}.
\end{eqnarray}
To get a well-defined Lagrangian density, we need to exclude all singularities from $D$. However, the classification of all singularities over $\mathbb{C}^{4}$ is almost impossible job and not our aim here. We can simply consider some subspaces of the complement set of $D$ and restrict the domain of the "soliton solutions" \eqref{Soliton solution_cpx} to these subspaces.
\subsubsection{Periodic 1-Soliton type distribution of Lagrangian density}
Let us consider a subspace of $\mathbb{C}^{4}$ defined by
\begin{eqnarray}
\label{widetilde{D}_1}
\widetilde{D}_{1}:=
\left\{ {(z, \widetilde{z}, w,\widetilde{w})\in\mathbb{C}^4~|~X_1 = \mbox{Re}}X_1 + in \pi, n \in \mathbb{Z},~ X_2 = i\mbox{Im}X_2
\right\}.
\end{eqnarray}
By the argument formulas
\begin{eqnarray}
\cosh X_1 &\!\!\!=&\!\!\! \cosh({\mbox{Re}}X_1+i n_1 \pi)  
= (-1)^{n_1}\cosh({\mbox{Re}}X_1) \in \mathbb{R}, ~~\label{Argument formula_coshX_1} \\
\cosh X_2 &\!\!\!=&\!\!\! \cosh(i\mbox{Im}X_2)=\cos (\mbox{Im}X_2), ~~\label{Argument formula_coshX_2}    \\
\sinh X_1 &\!\!\!=&\!\!\! \sinh({\mbox{Re}}X_1+i n_1 \pi) 
= (-1)^{n_1}\sinh({\mbox{Re}}X_1) \in \mathbb{R}, ~~\label{Argument formula_sinhX_1} \\
\sinh X_2 &\!\!\!=&\!\!\! \sinh(i\mbox{Im}X_2)= i \sin (\mbox{Im}X_2). ~~\label{Argument formula_sinhX_2}
\end{eqnarray}
and the properties of hyperbolic functions, we have
\begin{eqnarray}
\label{Bound of coshX}
\left| \cosh X_1 \right| = \left| \cosh (\mbox{Re}X_1) \right| \geq 1,~ 
\left| \cosh X_2 \right| = \left| \cosh(i\mbox{Im}X_2) \right| = \left| \cos(\mbox{Im}X_2) \right| \leq 1. ~~~~~
\end{eqnarray}
If we impose the condition 
$\left|\varepsilon_1\widetilde{\varepsilon}_1 \right| \textgreater \left|\varepsilon_2\widetilde{\varepsilon}_2 \right|$ on \eqref{D}, then
\begin{eqnarray*}
&&\widetilde{D}_1 \cap D  \nonumber \\
&\!\!\!\!=& \!\!\!\! \left\{(z, \widetilde{z}, w,\widetilde{w})\in\mathbb{C}^4~|~\
(\varepsilon_1\widetilde{\varepsilon}_1)^{\frac12}
\cosh (\mbox{Re}X_1) 
+
(\varepsilon_2\widetilde{\varepsilon}_2)^{\frac12}
\cos (\mbox{Im}X_2)=0, ~
\left|\varepsilon_1\widetilde{\varepsilon}_1 \right| \textgreater \left|\varepsilon_2\widetilde{\varepsilon}_2 \right|
\right\} ~~~~~~~~ \nonumber \\
&\!\!\!\!=& \!\!\!\! \varnothing  ~~\mbox{because} ~~  \nonumber  \\
&& \!\!\! \left| (\varepsilon_1\widetilde{\varepsilon}_1)^{\frac12}
\cosh (\mbox{Re}X_1) 
\!+
\!(\varepsilon_2\widetilde{\varepsilon}_2)^{\frac12}
\cos (\mbox{Im}X_2)
\right|  \nonumber  \\
&\geq&
\!\!\! \left| \varepsilon_1\widetilde{\varepsilon}_1 \right|^{\frac12}
\left| \cosh (\mbox{Re}X_1) \right| 
-
\left| \varepsilon_2\widetilde{\varepsilon}_2 \right|^{\frac12}
\left| \cos (\mbox{Im}X_2) \right|
\geq
\left| \varepsilon_1\widetilde{\varepsilon}_1 \right|^{\frac12}
-
\left| \varepsilon_2\widetilde{\varepsilon}_2 \right|^{\frac12}
\textgreater ~
0  ~~ \mbox{by}~~ \eqref{Bound of coshX}. \nonumber
\end{eqnarray*}
In other words, we can choose some suitable $a_1, a_2, b_1, b_2, c_1, c_2, d_1, d_2$ in \eqref{Soliton solution_cpx} to fit the condition $\left|\varepsilon_1\widetilde{\varepsilon}_1 \right| \textgreater \left|\varepsilon_2\widetilde{\varepsilon}_2 \right|$ 
and then obtain a well-defined Lagrangian density defined on the complex slice $\widetilde{D}_1 \subset \mathbb{C}^4$ :
\begin{eqnarray}
\label{Lagrangian density_cpx_Periodic 1-Soliton}
{\mbox{Tr}} F^2\!\!\!\!&=\!\!\!\!&
8(\lambda-\mu)^2(\alpha\delta-\beta\gamma)^2
\varepsilon_0\widetilde{\varepsilon}_0  \nonumber  \\
&& \times \left\{\frac{2\varepsilon_1\widetilde{\varepsilon}_1
	\sinh^2 (\mbox{Re}X_1)
	+2\varepsilon_2\widetilde{\varepsilon}_2
	\sin^2 (\mbox{Im}X_2) 
	-\varepsilon_0\widetilde{\varepsilon}_0}
{\left[(\varepsilon_1\widetilde{\varepsilon}_1)^{\frac12}
	(-1)^{n}\cosh (\mbox{Re}X_1) 
	+(\varepsilon_2\widetilde{\varepsilon}_2)^{\frac12}
	\cos (\mbox{Im}X_2) \right]^4}
\right\}.	~~~~~~~~
\end{eqnarray}
Here we use the argument formulas \eqref{Argument formula_coshX_1}, \eqref{Argument formula_coshX_2}, \eqref{Argument formula_sinhX_1}, and \eqref{Argument formula_sinhX_2}.
Note that the distribution of Lagrangian density has bounded and periodic behavior in the ${\mbox{Im}}X_2$-direction. We will convince you in the next subsection that 
\eqref{Lagrangian density_cpx_Periodic 1-Soliton} exactly possesses the behavior of 1-soliton in the Re$X_1$-direction with principal peak localized on Re$X_1$=0 if the Im$X_2$ part vanishes.
Therefore, we call it periodic 1-soliton type distribution. Due to the periodicity, some people also use the term "breather solution" to call this type of soliton.

\subsubsection{Pure 1-Soliton type distribution of Lagrangian density}
If we impose $a_1, a_2, b_1, b_2, c_1, c_2, d_1, d_2$ on \eqref{Soliton solution_cpx} to fit the condition $\varepsilon_{2}\widetilde{\varepsilon_{2}}=0$, we can remove the $X_2$-part from the Lagrangian density \eqref{Lagrangian density_cpx} completely. In addition,
 $\varepsilon_{2}\widetilde{\varepsilon_{2}}=0$ also implies 
$\varepsilon_{0}\widetilde{\varepsilon_{0}} = \varepsilon_{1}\widetilde{\varepsilon_{1}}$
by the relation \eqref{Relation of epsilons_C^4}.
This fact allows us to remove all the epsilon factors in the numerator and denominator of \eqref{Lagrangian density_cpx} and hence we can obtain a more concise form as the following :
\begin{eqnarray}
\label{Reduced Lagrangian density_cpx}
{\mbox{Tr}} F^2=
8(\lambda-\mu)^2(\alpha\delta-\beta\gamma)^2
\left(
2{\mbox{sech}}^2 X-3{\mbox{sech}}^4 X
\right).	
\end{eqnarray} 
Now let us deal with the annoying problem of singularity and periodicity.
By the following argument formula :
\begin{eqnarray}
\label{sech^2X}
\mbox{sech}^2(x+iy)
=2\left[
\frac{\mbox{cosh}2x ~\mbox{cos}2y +1}
{(\mbox{cosh}2x + \mbox{cos}2y)^2}
-i\frac{\mbox{sinh}2x~\mbox{sin}2y}{(\mbox{cosh}2x + \mbox{cos}2y)^2}
\right],
\end{eqnarray} 
we find that the Lagrangian density \eqref{Reduced Lagrangian density_cpx} has 
periodicity on the subspace $X= r + i~{\mbox{Im}}X$, for any given real number $r$. 
Especially for $r=0$ case, the 
singularities appear periodically because ${\mbox{sech}}(i\mbox{Im}X)={\mbox{sec}}({\mbox{Im}}X)$. Therefore, the Lagrangian density \eqref{Reduced Lagrangian density_cpx} has no solitonic behavior on the subspace spaces $X= r + i~{\mbox{Im}}X$, for fixed $r \in \mathbb{R}$. 
Next, let us fix the imaginary part of $X$. For simplicity, we consider 
\begin{eqnarray}
\widetilde{D}_2 :=
\left\{
(z, \widetilde{z}, w, \widetilde{w}) \in \mathbb{C}^{4}~ \big|~
X={\mbox{Re}}X + in\pi, ~n \in \mathbb{Z}
\right\}. \label{widetilde_D2}
\end{eqnarray}
By the fact that ${\mbox{sech}}(\mbox{Re}X + in\pi)={\mbox{sech}}({\mbox{Re}}X)$ (No singularity and periodicity),~
the Lagrangian density is clearly well-defined on the complex slice $\widetilde{D}_2$ $\subset \mathbb{C}^4$ :
\begin{eqnarray}
\label{Reduced Lagrangian density_cpx_D2}
{\mbox{Tr}} F^2=
8(\lambda-\mu)^2(\alpha\delta-\beta\gamma)^2
\left[
2{\mbox{sech}}^2 ({\mbox{Re}}X)-3{\mbox{sech}}^4 ({\mbox{Re}}X)
\right].
\end{eqnarray}
Note that if we ignore the complex constant $8(\lambda-\mu)^2(\alpha\delta-\beta\gamma)^2$,  the distribution of Lagrangian density exactly possesses the solitonic behavior due to our definition of the ASDYM 1-soliton (Cf: \eqref{Def_ASDYM 1-Soliton}),
and the principal peak is localized on Re$X$=0. Therefore, we call it pure 1-soliton type distribution.
In fact, one of the simplest choice for fitting $\varepsilon_{2}\widetilde{\varepsilon_{2}}=0$ can be done by setting $a_2=b_1=c_1=d_2=0 ~(a_1=a, b_2=b, c_2=c, d_1=d)$ in \eqref{Soliton solution_cpx}, or equivalently,
\begin{eqnarray}
\label{Reduced 1-soliton_cpx}
J = -\psi\Lambda\psi^{-1},~~ 
\psi=\left(
\begin{array}{ll}
ae^{L} & ce^{-M} \\ 
be^{-L}& de^{M} 
\end{array}\right), ~~
\Lambda
=
\left(
\begin{array}{cc}
\lambda & 0 \\
0 & \mu
\end{array}
\right).
\end{eqnarray}
In this case, 
$X=M+L+\dfrac12 \log(-ad/bc)$.


Since the physical quantities must be real value for the purpose of physical observation, 
we should discuss the ASDYM solitons on the real coordinate spaces. 
Moreover, it would be better to request the gauge group to be unitary because the unitary gauge group guarantees that the Lagrangian density Tr$F_{\mu\nu}F^{\mu\nu}$ is real-valued. As a result,
we propose ansatzes in the remaining subsections to construct the real-valued Lagrangian densities \eqref{Reduced Lagrangian density_U}, \eqref{Reduced Lagrangian density_E}, and \eqref{Reduced Lagrangian density_M} which are defined on the real subspaces of  $\widetilde{D}_2$ (\eqref{widetilde_D2} for $n=0$) and 
belong to the same class of the Lagrangian density \eqref{Reduced Lagrangian density_cpx_D2}.

\subsection{ASDYM 1-Solitons on the 4D Ultrahyperbolic space $\mathbb{U}$, $G=\mathrm{SU}(2)$}
\label{Section 5.2}
In this subsection, we take the reality condition \eqref{Reality condition_U} to realize the split signature $ds^2=(dx^1)^2 + (dx^2)^2 - (dx^3)^2 - (dx^4)^2$ and then consider the ASDYM solitons on the Ultrahyperbolic space $\mathbb{U}$. Our aim is to impose further conditions on the "soliton solution" \eqref{Soliton solution_cpx} to seek the anti-hermitian gauge fields $A_{\mu}$. If the gauge fields $A_{\mu}$ can be all anti-hermitian, it guarantees the unitarity of gauge group and hence the Lagrangian density is real-valued.


\subsubsection{Periodic 1-Soliton type distribution of Lagrangian density}
Let us make the following attempt : 
\begin{eqnarray}
\label{Ansatz_U}
\left\{
\begin{array}{l}
b_i \rightarrow -b_i,~ c_i \rightarrow \overline{b_i},~ d_i \rightarrow \overline{a_i},~ i=1,2. \\
M=\overline{L} ~~\mbox{under reality condition \eqref{Reality condition_U}} ~ \Longleftrightarrow~ \gamma=\overline{\alpha},~
\delta=\overline{\beta},~\mu=\overline{\lambda}.
\end{array}
\right.
\end{eqnarray}
to adjust the parameters in the "soliton solution" \eqref{Soliton solution_cpx}.
Under this substitution, we get a candidate of 1-soliton solution on the Ultrahyperbolic space $\mathbb{U}$ as follows.
\medskip \\
{\bf{1-Soliton Solution}}
\begin{eqnarray}
\label{Soliton solution_U}
J\!\!\!\!&=\!\!\!\!&
\left|
\begin{array}{cc}
\!\psi & \!\!1 \\
\!\psi\Lambda & \!\!\fbox{0}
\end{array}
\right|
=
-\psi\Lambda \psi^{-1} ~~
\left\{
\begin{array}{l}
\psi
=\left(
\begin{array}{cc}
a_1e^{L}+a_2e^{-L}
& 
\overline{b}_1e^{\overline{L}} + \overline{b}_2e^{-\overline{L}}
\\ 
-b_1e^{L} - b_2e^{-L}
& 
\overline{a}_1e^{\overline{L}}
+\overline{a}_2e^{-\overline{L}} 
\end{array}\right)  \\
\Lambda=
\left(
\begin{array}{cc}
\lambda & 0 \\
0 & \overline{\lambda}
\end{array}
\right)
\end{array}
\right.  \\
&&\!\!\!\!\mbox{where}~
L=(\lambda \alpha) z
+\beta\widetilde{z}
+(\lambda \beta) w
+\alpha \widetilde{w},~~ z,\widetilde{z}, w, \widetilde{w} \in \mathbb{R}   \nonumber \\
&&~~~~~~~~~\!=\frac{1}{\sqrt{2}}
\left[
(\lambda\alpha+\beta)x^1
+(\lambda\beta-\alpha)x^2
+(\lambda\alpha-\beta)x^3
+(\lambda\beta+\alpha)x^4 
\right]
=:\ell_\mu x^{\mu}  \label{l_mu_U},  ~~~~~~~  \\
&&~~~~~~~~~~~~~~ a_1, a_2, b_1, b_2, \alpha, \beta, \lambda \in \mathbb{C}. \nonumber 
\end{eqnarray}
The expansion of $J$-matrix is in the form of
\begin{eqnarray}
\label{J_2_U}
J=\frac{-1}{|A|^{2}+|B|^{2}}
\left(
\begin{array}{cc}
\lambda|A|^2 + \overline{\lambda}|B|^2  & (\overline{\lambda}-\lambda)A\overline{B} \\
(\overline{\lambda}-\lambda)\overline{A}B & \overline{\lambda}|A|^2 + \lambda|B|^2
\end{array}
\right),~
\begin{array}{l}
A:= a_1e^{L}+a_2e^{-L}  \\
B:= b_1e^{L}+b_2e^{-L} 
\end{array}, 
\end{eqnarray}
and it satisfies the relation $JJ^{\dagger}=|\lambda|^2I$. This fact guarantees that all the gauge fields $A_{\mu} \in G=\mathrm{U}(N)$ by Proposition \ref{Prop_3.9} (for $h=J,~\widetilde{h}=I$ case). Next, we need to check whether the soliton solution \eqref{Soliton solution_U} induces nontrivial gauge fields or not. 
\medskip\\
{\bf{Gauge Fields}}
\medskip \\
By taking the $S$-gauge ($h=J,~\widetilde{h}=I$) on \eqref{Real representation of gauge fields_U}, we get the ASD gauge fields
\begin{eqnarray}
\label{Gauge Fields_U_A_1=A_3, A_2=A_4}
A_1=A_3
=\frac{-1}{2}\left[
(\partial_{1}J)J^{-1} + (\partial_{3}J)J^{-1}
\right],~
A_2=A_4
=\frac{-1}{2}\left[
(\partial_{2}J)J^{-1} + (\partial_{4}J)J^{-1} 
\right]. ~~
\end{eqnarray}
After further substitutions of $J$-matrix \eqref{Soliton solution_U} (or by imposing the ansatz \eqref{Ansatz_U} directly on \eqref{A_m}), the explicit form of the ASD gauge fields can be calculated as :
\begin{eqnarray}
\label{Gauge fields_U_soliton}
{A_{\mu}}&\!\!\!\!:=\!\!\!\!&
\frac{\sqrt{2}(\overline{\lambda}-\lambda)}{(|A|^{2}+|B|^{2})^2}\left(
\begin{array}{cc}
\overline{p}_{\mu}AB + p_{\mu}\overline{A}\overline{B} &  \overline{p}_{\mu}A^2- p_{\mu}\overline{B}^2
\\ 
-\overline{p}_{\mu}B^2 + p_{\mu}\overline{A}^2 & - \overline{p}_{\mu}AB- p_{\mu}\overline{A}\overline{B}
\end{array}\right),
\\
&&
p_{\mu}:=
\left\{
\begin{array}{l}
\alpha\varepsilon_0 ~~~~~
{\mbox{if}} ~~ \mu=1,3 \\
\beta\varepsilon_0 ~~~~~
{\mbox{if}} ~~ \mu=2,4 \nonumber \\
\varepsilon_0:=a_1b_2 - a_2b_1
\end{array}.
\right.
\end{eqnarray}
Note that the gauge fields $A_{\mu}$ are nonzero if $\lambda \notin \mathbb{R}$. Furthermore, all of them are not only anti-hermitian but traceless, therefore, the group group even can be $G=\mathrm{SU}(2)$. Now the remaining task is to check whether the gauge fields induce nontrivial Lagrangian density Tr$F_{\mu\nu}F^{\mu\nu}$ or not. 
By imposing the ansatz \eqref{Ansatz_U} on \eqref{Lagrangian density_cpx} directly, we obtain the following Lagrangian density.
\medskip \\
{\bf Real-valued Lagrangian Density}
\begin{eqnarray}
\label{Lagrangian density_U}
&&{\mbox{Tr}} F_{\mu \nu}F^{\mu \nu} \nonumber \\
\!\!\!&=&\!\!\!
8\left[(\alpha\overline{\beta} \!-\! \overline{\alpha}\beta)
(\lambda \!-\!\overline{\lambda}) \left|\varepsilon_0 \right| \right]^2
\!
\left\{\frac{2\varepsilon_1\widetilde{\varepsilon}_1
	\sinh^2 X_1
	-2\left|\varepsilon_2 \right|^2
	\sinh^2 X_2 
	-\left|\varepsilon_0 \right|^2}
{\left[(\varepsilon_1\widetilde{\varepsilon}_1)^{\frac12}
	\cosh X_1 
	+\left|\varepsilon_2 \right|
	\cosh X_2\right]^4} 
\right\},~  \varepsilon_{0} \neq 0,~~~~~~~~~~	\\
&&X_1=\overline{L} + L+\dfrac12 \log(\varepsilon_1/\widetilde{\varepsilon}_1),~
X_2=\overline{L} - L+\dfrac12 \log(\varepsilon_2/\overline{\varepsilon}_2),  \label{X_1, X_2}\\
&&
\left\{
\begin{array}{l}
\varepsilon_0=a_1 b_2 - a_2b_1,~   \\
\varepsilon_1=\left|a_1\right|^2 + \left|b_1 \right|^2,~
\widetilde{\varepsilon}_1=\left|{a}_2 \right|^2 + \left|{b_2} \right|^2 \in \mathbb{R},  \\
\varepsilon_2=a_1\overline{a}_2 + b_1\overline{b}_2, \\
\mbox{$\varepsilon_{0},~ \varepsilon_{1},~ \varepsilon_{2}$~~satisfy the relation : $\varepsilon_1\widetilde{\varepsilon}_1 = \vert \varepsilon_2\vert^2 + \vert \varepsilon_0\vert^2$}. \label{Relation of epsilons_U}
\end{array}
\right.
\end{eqnarray}
Note that the Lagrangian density is nonzero if $\alpha\overline{\beta} \notin \mathbb{R}$ and $\lambda \notin \mathbb{R}$.
Since the nontrivial Lagrangian density is defined for $\varepsilon_{0} \neq 0$ and
by relation \eqref{Relation of epsilons_U}, we get a nonnegative bound for the epsilon factors :
$\varepsilon_1\widetilde{\varepsilon}_1 \textgreater \vert \varepsilon_2\vert^2 \geq 0$ if $\varepsilon_{0} \neq 0$.
By taking complex conjugate on \eqref{X_1, X_2}, we can find that $X_1$ is real and $X_2$ is pure imaginary immediately. This fact restricts $\cosh X_1$ and $\cosh X_2$ to have the following real bounds : $\cosh X_1\geq 1$, and $|\cosh X_2|=|\cos 
\left({\mbox{Im}} X_2\right)| \leq 1$, respectively. 
As a result, all singularities must appear on the subspace
\begin{eqnarray}
\widetilde{\mathbb{U}}:=\left\{x^\mu\in\mathbb{R}^4~|~\cosh X_1= 1,~ 
\cosh X_2=\cos \left({\mbox{Im}} X_2\right)= -1,~
\varepsilon_1\widetilde{\varepsilon}_1=\vert \varepsilon_2\vert^2
\right\} \subset \mathbb{U} ~~~~
\end{eqnarray} 
which is nonempty only if $\varepsilon_{0}=0$.
Therefore, we can conclude that the Lagrangian density \eqref{Lagrangian density_U} is well-defined on the whole Ultrahyperbolic space $\mathbb{U}$. 

Furthermore, by the properties of hyperbolic functions
\begin{eqnarray}
\label{Properties of hyperbolic functions}
\left\{
\begin{array}{l}
\cosh X_1 \in \mathbb{R},~ \sinh X_1 \in \mathbb{R}~ \\
\cosh X_2 = \cos \left({\mbox{Im}} X_2\right) \in \mathbb{R}~   \\
\sinh X_2 = i\sin \left({\mbox{Im}} X_2\right) \in i\mathbb{R} ~\Rightarrow~ \sinh^2 X_2 = -\sin^2 \left({\mbox{Im}} X_2\right) \in \mathbb{R}
\end{array},
\right.
\end{eqnarray}
we find that the distribution of the Lagrangian density has solitonic behavior in $X_1$-direction with principal peak localized on $X_1=0$, and has periodic behavior in $X_2$-direction. It is the periodic 1-soliton type distribution (or breather solution) and belongs to a subclass of \eqref{Lagrangian density_cpx_Periodic 1-Soliton}. 
On the other hand, we also can check that the Lagrangian density is exactly real-valued by taking complex conjugate on the coefficient 
$\left[(\alpha\overline{\beta} - \overline{\alpha}\beta)
(\lambda - \overline{\lambda}) \left|\varepsilon_0 \right| \right]^2$ and using \eqref{Properties of hyperbolic functions}. This property can be guaranteed due to $G=\mathrm{SU}(2)$, however, it is a highly nontrivial result for the Euclidean and the Minkowski signature because the gauge group is $G=\mathrm{SL}(2,\mathbb{C})$ in these cases. 

\subsubsection{Pure 1-Soliton type distribution of Lagrangian density}

If we choose suitable parameters $a_1, a_2, b_1, b_2$ in \eqref{Soliton solution_U} to fit the condition $\varepsilon_{2}=0$ (Cf: \eqref{Relation of epsilons_U}), the periodic part of the Lagrangian density \eqref{Lagrangian density_U} can be removed completely. Moreover, the relation \eqref{Relation of epsilons_U} implies that
$|\varepsilon_{0}|^2 = \varepsilon_{1}\widetilde{\varepsilon_{1}}$ if $\varepsilon_{2}=0$.
This fact allows us to cancel out all the epsilon factors in the numerator and denominator of \eqref{Lagrangian density_U}, and then get a more concise form as the following :
\begin{eqnarray}
\label{Reduced Lagrangian density_U}
{\mbox{Tr}} F_{\mu \nu}F^{\mu \nu}&\!\!\!\!=& \!\!\!\!
8\left[(\alpha\overline{\beta}-\overline{\alpha}\beta)
(\lambda -\overline{\lambda})  \right]^2
\left(2{\mbox{sech}}^2 X-3{\mbox{sech}}^4 X\right),  \\
X&\!\!\!\!=& \!\!\!\! L + \overline{L} + \frac{1}{2}\log\left[(\left|a_1 \right|^2 + \left|b_2 \right|^2) / (\left|a_2 \right|^2+\left|b_2 \right|^2)\right].  
\end{eqnarray}
Now we find that the distribution of the Lagrangian density satisfies our definition of the ASDYM 1-soliton (Cf: \eqref{Def_ASDYM 1-Soliton}) and hence we call it the pure 1-soliton type distribution (Cf: \eqref{Reduced Lagrangian density_cpx_D2}).
On the other hand, 
the principal peak of the Lagrangian density is localized on a 3-dimensional hyperplane $X=0$ (with normal vector $\ell_\mu+\overline{\ell}_\mu$, Cf: \eqref{l_mu_U}) in 4-dimensional space.
Therefore, we can interpret it as the codimension 1 type soliton on the Ultrahyperbolic space $\mathbb{U}$ and use the term "soliton wall" to distinguish it from the domain wall.
Furthermore, since the gauge group is $G=\mathrm{SU}(2)$ and the ASDYM equations for the split signature are equations of motion of effective action for open $\mathrm{N=2}$ string theories \cite{OoVa1, OoVa2}, therefore the soliton walls could be some physical objects expectantly in open $\mathrm{N=2}$ string theories.

More interesting, the integration of the Lagrangian density over the whole space is zero. 
This result suggests that the soliton solution \eqref{Soliton solution_U} or \eqref{J_2_U} 
belongs to the sector of instanton number zero.
To verify it, let us introduce three independent axes $X^1, X^2, X^3$
in the directions orthogonal to the $X$-axis 
(normal direction of the soliton wall (SW)). 
Then the integration of the Lagrangian density can be calculated as
\begin{eqnarray}
\label{Integration of Lagrangian density_U}
\int_{\mathbb{U}}{\mbox{Tr}} F_{\mu \nu}F^{\mu \nu}d^4x
&\propto& \lim_{R\rightarrow\infty}
\int_{-R}^{R} 
\int_{-R}^{R} 
\int_{-R}^{R} 
dX^1dX^2dX^3
\int_{-R}^{R} 
(2{\mbox{sech}}^2 X-3{\mbox{sech}}^4 X) dX\nonumber\\
&=&
\int_{\scriptsize{\mbox{SW}}} dX^1dX^2dX^3
\int_{-\infty}^{\infty} 
(2{\mbox{sech}}^2 X-3{\mbox{sech}}^4 X) dX \nonumber\\.
&=& \int_{\scriptsize{\mbox{SW}}} dX^1dX^2dX^3 \left.({\mbox{tanh}}X \cdot {\mbox{sech}^2X}) \right|_{-\infty}^{~\infty} = 0.
\end{eqnarray}

\subsection{ASDYM 1-Solitons on the 4D Euclidean space $\mathbb{E}$ and Minkowski space $\mathbb{M}$, $G=\mathrm{SL}(2,\mathbb{C})$}

\label{Section 5.3}
In this subsection, we take the reality conditions \eqref{Reality condition_E}, \eqref{Reality condition_M} to realize the Euclidean metric $ds^{2}=(dx^{1})^2+(dx^{2})^2+(dx^{3})^2+(dx^{4})^2$ and the Minkowski metric $ds^{2}=(dx^{0})^2-(dx^{1})^2-(dx^{2})^2-(dx^{3})^2$, respectively.
Then we follow quite similar steps as Subsection \ref{Section 5.2} to discuss the ASDYM 1-solitons on the above two spaces. As mentioned before, the unitary gauge symmetry is lost on the Euclidean space $\mathbb{E}$ if we take the $S$-gauge \eqref{S-gauge}. On the other hand, the existence of the unitary ASD gauge fields are not allowed on the Minkowski space $\mathbb{M}$ due to the ASDYM equations \eqref{ASDYM_M}. Nevertheless, we also can impose further conditions on the "soliton solution" \eqref{Soliton solution_cpx} to seek nonzero and real-valued Lagrangian density on the two spaces. They are highly nontrivial results for $G=\mathrm{SL}(2,\mathbb{C})$. 


\subsubsection{1-Soliton type distribution of Lagrangian density on $\mathbb{E}$}

Let us impose the following conditions 
\begin{eqnarray}
\label{Ansatz_E}
\left\{
\begin{array}{l}
b_i \rightarrow -b_i,~ c_i \rightarrow \overline{b_i},~ d_i \rightarrow \overline{a_i},~ i=1,2. \\
M=\overline{L} ~~\mbox{under reality condition \eqref{Reality condition_E}} ~ \Longleftrightarrow~ \gamma=-\overline{\lambda}\overline{\beta},~
\delta=\overline{\lambda}\overline{\alpha},~\mu=-1/\overline{\lambda}. ~~~~
\end{array}
\right.
\end{eqnarray}
on the "soliton solution" \eqref{Soliton solution_cpx} to obtain the ASDYM 1-solitons of the Euclidean space version. The result is the following :
\medskip \\
{\bf{1-Soliton Solution}}
\begin{eqnarray}
\label{Soliton solution_E}
J\!\!\!\!&=\!\!\!\!&
\left|
\begin{array}{cc}
\!\psi & \!\!1 \\
\!\psi\Lambda & \!\!\fbox{0}
\end{array}
\right|
=
-\psi\Lambda \psi^{-1} ~~
\left\{
\begin{array}{l}
\psi
=\left(
\begin{array}{cc}
a_1e^{L}+a_2e^{-L}
& 
\overline{b}_1e^{\overline{L}} + \overline{b}_2e^{-\overline{L}}
\\ 
-b_1e^{L} - b_2e^{-L}
& 
\overline{a}_1e^{\overline{L}}
+\overline{a}_2e^{-\overline{L}} 
\end{array}\right)  \\
\Lambda=
\left(
\begin{array}{cc}
\lambda & 0 \\
0 & -1/\overline{\lambda}
\end{array}
\right)
\end{array},
\right.  \\
&&
\mbox{where}~
L=(\lambda \alpha) z
+\beta\overline{z}
+(\lambda \beta) w
-\alpha \overline{w}  \nonumber \\
&&~~~~~~~~~~~ =\frac{1}{\sqrt{2}}
\left[
(\lambda\alpha+\beta)x^1
+i(\lambda\alpha-\beta)x^2
+(\lambda\beta-\alpha)x^3
+i(\lambda\beta+\alpha)x^4
\right], ~~~~~~   \nonumber  \\
&&~~~~~~~~~~~~~~~ a_1, a_2, b_1, b_2, \alpha, \beta, \lambda \in \mathbb{C}.   \nonumber 
\end{eqnarray}
{\bf Gauge Fields}
\medskip \\
By taking the $S$-gauge ($h=J,~\widetilde{h}=I$) on \eqref{Real representation of gauge fields_E}, we get the ASD gauge fields
\begin{eqnarray}
\label{Gauge Fields_E_A_2=iA_1}
A_1
&\!\!\!\!=& \!\!\!\! \frac{-1}{2}\left[
(\partial_{1}J)J^{-1} -i (\partial_{2}J)J^{-1}
\right],~~ A_2=iA_1,~  \\
A_3
&\!\!\!\! =& \!\!\!\! \frac{-1}{2}\left[
(\partial_{3}J)J^{-1} -i (\partial_{4}J)J^{-1} 
\right],~~ A_4=iA_3.   \label{Gauge Fields_E_A_4=iA_3}
\end{eqnarray}
After further substitutions of the $J$-matrix \eqref{Soliton solution_E} (or by imposing the ansatz \eqref{Ansatz_E} on \eqref{A_m} directly), the explicit form of the ASD gauge fields can be calculated as :
\begin{eqnarray}
\label{Gauge fields_E_soliton}
{A_{\mu}}&\!\!\!\!:=\!\!\!\!&
\frac{-\sqrt{2}(|\lambda|^2+1)}{(|A|^{2}+|B|^{2})^2}\left(
\begin{array}{cc}
\widetilde{p}_{\mu}AB + p_{\mu}\overline{A}\overline{B} & \widetilde{p}_{\mu}A^2-  p_{\mu}\overline{B}^2
\\ 
-\widetilde{p}_{\mu}B^2 + p_{\mu}\overline{A}^2 & - \widetilde{p}_{\mu}AB- p_{\mu}\overline{A}\overline{B}
\end{array}\right),
\\
&&A:=a_{1}e^{L}+a_{2}e^{-L},~B:=b_{1}e^{L}+b_{2}e^{-L},  \nonumber 
\\
&&
(p_{\mu},~\widetilde{p}_{\mu}):=
\left\{
\begin{array}{l}
\left((\alpha/\overline{\lambda})\varepsilon_0,~\!\! -\overline{\beta} ~\! \overline{\varepsilon_0} ~\right) ~~~~~
{\mbox{if}} ~~ \mu=1 \\
\left((i\alpha/\overline{\lambda})\varepsilon_0,~\!\! -i\overline{\beta} ~\! \overline{\varepsilon_0} ~\right) ~~~
{\mbox{if}} ~~ \mu=2 \\
\left( (\beta/\overline{\lambda})\varepsilon_0,~ \overline{\alpha} ~\! \overline{\varepsilon_0} ~\right) ~~~~~~~~
{\mbox{if}} ~~ \mu=3 \nonumber \\
\left( (i\beta/\overline{\lambda})\varepsilon_0,~ i\overline{\alpha} ~\! \overline{\varepsilon_0} ~\right) ~~~~~~
{\mbox{if}} ~~ \mu=4 \nonumber \\
\varepsilon_0:=a_1b_2 -a_2b_1
\end{array}.
\right.
\end{eqnarray}
Note that $A_1,~A_3$ are anti-hermitian ($A_2,~A_4$ are hermitian) if $\widetilde{p}_{\mu}=-\overline{p}_{\mu}$. In this case, $\alpha=\overline{\lambda}\beta,~\beta=-\overline{\lambda}\alpha$ which holds only if $\lambda=\pm i$. On the other hand, we find that the gauge fields are all traceless and hence the gauge group is $G=\mathrm{SL}(2,\mathbb{C})$.
Now the remaining task is to check whether the gauge fields induce nontrivial and real-valued Lagrangian density Tr$F_{\mu\nu}F^{\mu\nu}$. 
By imposing the ansatz \eqref{Ansatz_E} on \eqref{Lagrangian density_cpx} directly, we obtain the following Lagrangian density.
\medskip \\
{\bf Periodic 1-Soliton type distribution of Lagrangian Density}
\begin{eqnarray}
\label{Lagrangian density_E}
&&{\mbox{Tr}} F_{\mu \nu}F^{\mu \nu}  \nonumber \\
&\!\!\!\!=&\!\!\!\!
8\left[(\left|\alpha\right|^2\!+\left|\beta\right|^2)
(\left|\lambda \right|^2\!+1) \left|\varepsilon_0 \right| \right]^2\!\!
\left\{\!
\frac{2\varepsilon_1\widetilde{\varepsilon}_1
	\sinh^2 X_1
	-2\left|\varepsilon_2 \right|^2
	\sinh^2 X_2 
	-\left|\varepsilon_0 \right|^2}
{\left[(\varepsilon_1\widetilde{\varepsilon}_1)^{\frac12}
	\cosh X_1 
	+\left|\varepsilon_2 \right|
	\cosh X_2\right]^4}\!
\right\},~ \varepsilon_{0} \neq 0, ~~~~~~~~  \\
&&X_1=\overline {L} + L+\dfrac12 \log(\varepsilon_1/\widetilde{\varepsilon}_1),~
X_2=\overline{L} - L+\dfrac12 \log(\varepsilon_2/\overline{\varepsilon}_2),  \nonumber \\
&&
\left\{
\begin{array}{l}
\varepsilon_0=a_1b_2 - a_2b_1,~   \\
\varepsilon_1=\left|a_1\right|^2 + \left|b_1 \right|^2,~
\widetilde{\varepsilon}_1=\left|{a}_2 \right|^2 + \left|{b_2} \right|^2 \in \mathbb{R},  \\
\varepsilon_2=a_1\overline{a}_2 + b_1\overline{b}_2.  \label{Relation of epsilons_E}  \\
\end{array}
\right. 
\end{eqnarray}
Note that the Lagrangian density is nonzero if $(\alpha,\beta) \neq (0,0)$.
By the properties of hyperbolic functions \eqref{Properties of hyperbolic functions} and the coefficient $8\left[(\left|\alpha\right|^2\!+\left|\beta\right|^2)
(\left|\lambda \right|^2\!+1) \left|\varepsilon_0 \right| \right]^2  \in \mathbb{R}$,
the Lagrangian density is clearly real-valued and belongs to a subclass of \eqref{Lagrangian density_cpx_Periodic 1-Soliton}. 
\medskip 
\\
{\bf Pure 1-Soliton type distribution of Lagrangian density}
\medskip \\
If we choose suitable parameters $a_1, a_2, b_1, b_2$ in \eqref{Soliton solution_E} to fit the condition $\varepsilon_{2}=0$ (Cf: \eqref{Relation of epsilons_E}), we can get the pure 1-soliton type distribution (Cf: \eqref{Reduced Lagrangian density_cpx_D2}) :
\begin{eqnarray}
\label{Reduced Lagrangian density_E}
{\mbox{Tr}} F_{\mu \nu}F^{\mu \nu}&\!\!\!\!=& \!\!\!\!
8\left[(\left|\alpha\right|^2\!+\left|\beta\right|^2)
(\left|\lambda \right|^2\!+1) \right]^2
\left(2{\mbox{sech}}^2 X-3{\mbox{sech}}^4 X\right),  \\
&& \!\!\!\!\! X= L + \overline{L} + \frac{1}{2}\log\left[(\left|a_1 \right|^2 + \left|b_2 \right|^2) / (\left|a_2 \right|^2+\left|b_2 \right|^2)\right].
\end{eqnarray}


\subsubsection{1-Soliton type distribution of Lagrangian density on $\mathbb{M}$}

Let us impose the following conditions 
\begin{eqnarray}
\label{Ansatz_M}
\left\{
\begin{array}{l}
b_i \rightarrow -b_i,~ c_i \rightarrow \overline{b_i},~ d_i \rightarrow \overline{a_i},~ i=1,2. \\
M=\overline{L}~~\mbox{under reality condition \eqref{Reality condition_M}} ~ \Longleftrightarrow~ \alpha=\overline{\mu}\beta,~ \gamma=\overline{\lambda}\overline{\beta},~ \delta=\overline{\beta}. ~~
\end{array}
\right.
\end{eqnarray}
on the "soliton solution" \eqref{Soliton solution_cpx} to obtain the ASDYM 1-solitons of the Minkowski space version. The result is the following :
\medskip \\
{\bf{1-Soliton Solution}}
\begin{eqnarray}
\label{Soliton solution_M}
J\!\!\!\!&=\!\!\!\!&
\left|
\begin{array}{cc}
\!\psi & \!\!1 \\
\!\psi\Lambda & \!\!\fbox{0}
\end{array}
\right|
=
-\psi\Lambda \psi^{-1} ~~
\left\{
\begin{array}{l}
\psi
=\left(
\begin{array}{cc}
a_1e^{L}+a_2e^{-L}
& 
\overline{b}_1e^{\overline{L}} + \overline{b}_2e^{-\overline{L}}
\\ 
-b_1e^{L} - b_2e^{-L}
& 
\overline{a}_1e^{\overline{L}}
+\overline{a}_2e^{-\overline{L}} 
\end{array}\right)  \\
\Lambda=
\left(
\begin{array}{cc}
\lambda & 0 \\
0 & \mu
\end{array}
\right)
\end{array},
\right.  \\
&&
\mbox{where}~
L=
\beta\left[(\lambda\overline{\mu}) z
+\widetilde{z}
+\lambda w
+\overline{\mu}~\!\overline{w}
\right] ~~ (z, \widetilde{z} \in \mathbb{R})  \nonumber \\
&&~~~~~~~~~~~=\frac{\beta}{\sqrt{2}}
\left[
(\lambda\overline{\mu}+1)x^0
+(\lambda\overline{\mu}-1)x^1
+(\lambda+\overline{\mu})x^2
+i(\lambda-\overline{\mu})x^3
\right], ~~~~~~   \nonumber  \\
&&~~~~~~~~~~~~~~~ a_1, a_2, b_1, b_2, \beta, \lambda, \mu \in \mathbb{C}.    \nonumber 
\end{eqnarray}
{\bf Gauge Fields}
\medskip \\
Following a quite similar procedure as we derived \eqref{Real representation of gauge fields_E}, \eqref{Real representation of gauge fields_U}, and taking the $S$-gauge, we can obtain the ASD gauge fields of the Minkowski space version :
\begin{eqnarray}
\label{Gauge Fields_M_A_0=A_1}
A_0&\!\!\!\!=& \!\!\!\! A_1=\frac{-1}{2}\left[
(\partial_{0}J)J^{-1} + (\partial_{1}J)J^{-1}
\right], \\
A_2&\!\!\!\!=& \!\!\!\! \frac{-1}{2}\left[
(\partial_{2}J)J^{-1} - i(\partial_{3}J)J^{-1}
\right],~~ A_3=iA_2.    \label{Gauge Fields_M_A_3 = iA_2}
\end{eqnarray}
After further substitutions of the $J$-matrix \eqref{Soliton solution_M} (or by imposing the ansatz \eqref{Ansatz_M} on \eqref{A_m}), the explicit form of the ASD gauge fields can be calculated as :
\begin{eqnarray}
\label{Gauge fields_M_soliton}
{A_{\mu}}&\!\!\!\!:=\!\!\!\!&
\frac{\sqrt{2}(\mu-\lambda)}{(|A|^{2}+|B|^{2})^2}\left(
\begin{array}{cc}
\widetilde{p}_{\mu}AB + p_{\mu}\overline{A}\overline{B} & {\widetilde{p}_{\mu}}A^2- p_{\mu}\overline{B}^2
\\ 
-\widetilde{p}_{\mu}B^2 + p_{\mu}\overline{A}^2 & - \widetilde{p}_{\mu}AB- p_{\mu}\overline{A}\overline{B}
\end{array}\right),
\\
&&A:=a_{1}e^{L}+a_{2}e^{-L},~B:=b_{1}e^{L}+b_{2}e^{-L},  \nonumber 
\\
&&
(p_{\mu},~\widetilde{p}_{\mu}):=
\left\{
\begin{array}{l}
\left( \overline{\mu}\beta\varepsilon_0,~ \overline{\lambda}~\!\overline{\beta} ~\! \overline{\varepsilon_0} ~\right) ~~~~~
{\mbox{if}} ~~ \mu=0,~1 \\
\left( \beta\varepsilon_0,~\overline{\beta} ~\! \overline{\varepsilon_0} ~\right) ~~~~~~~~~
{\mbox{if}} ~~ \mu=2 \\
\left( i\beta\varepsilon_0,~ i\overline{\beta} ~\! \overline{\varepsilon_0} ~\right) ~~~~~~~
{\mbox{if}} ~~ \mu=3  \\
\varepsilon_0:=a_1b_2 -a_2b_1
\end{array}.
\right.
\end{eqnarray}
Note that $A_0,~A_1,~A_2$ are anti-hermitian ($A_3$ is hermitian) if  $\widetilde{p}_{\mu}=\overline{p}_{\mu}$ ($i.e.$ $\mu=\overline{\lambda}$). In this case, 
$L=
\frac{\beta}{\sqrt{2}}\left[
(\lambda^2 + 1)x^0
+(\lambda^2 - 1)x^1
+2\lambda x^2
\right] $ and hence the gauge fields are independent of $x^3$-direction. On the other hand, we find that the gauge group is $G=\mathrm{SL}(2,\mathbb{C})$ because all the gauge fields are traceless.
Now the remaining task is to check whether the gauge fields induce nontrivial and real-valued Lagrangian density Tr$F_{\mu\nu}F^{\mu\nu}$. 
By imposing the ansatz \eqref{Ansatz_M} on \eqref{Lagrangian density_cpx} directly, we obtain the following Lagrangian density.
\medskip \\
{\bf Periodic 1-Soliton type distribution of Lagrangian Density}
\begin{eqnarray}
\label{Lagrangian density_M}
&&{\mbox{Tr}} F_{\mu \nu}F^{\mu \nu} \nonumber \\
&\!\!\!\!=&\!\!\!\!
8\left| \beta(\lambda-\mu) \right|^4 \left|\varepsilon_0 \right|^2
\left\{
\frac{2\varepsilon_1\widetilde{\varepsilon}_1
	\sinh^2 X_1
	-2\left|\varepsilon_2 \right|^2
	\sinh^2 X_2 
	-\left|\varepsilon_0 \right|^2}
{\left[(\varepsilon_1\widetilde{\varepsilon}_1)^{\frac12}
	\cosh X_1 
	+\left|\varepsilon_2 \right|
	\cosh X_2\right]^4}
\right\},~  \varepsilon_{0} \neq 0, ~~~~~~  \\
&&X_1=\overline {L} + L+\dfrac12 \log(\varepsilon_1/\widetilde{\varepsilon}_1),~
X_2=\overline{L} - L+\dfrac12 \log(\varepsilon_2/\overline{\varepsilon}_2),  \nonumber \\
&&
\left\{
\begin{array}{l}
\varepsilon_0=a_1b_2 - a_2b_1,~   \\
\varepsilon_1=\left|a_1\right|^2 + \left|b_1 \right|^2,~
\widetilde{\varepsilon}_1=\left|{a}_2 \right|^2 + \left|{b_2} \right|^2 \in \mathbb{R},  \\
\varepsilon_2=a_1\overline{a}_2 + b_1\overline{b}_2.  \label{Relation of epsilons_M} \\
\end{array}
\right. 
\end{eqnarray}
Note that the Lagrangian density is nonzero if $\lambda \neq \mu$ and $\beta \neq 0$. 
By the properties of hyperbolic functions \eqref{Properties of hyperbolic functions} and the coefficient $8\left| \beta(\lambda-\mu) \right|^4 \left|\varepsilon_0 \right|^2 \in \mathbb{R}$, the Lagrangian density is clearly real-valued and belongs to a subclass of \eqref{Lagrangian density_cpx_Periodic 1-Soliton}.
\medskip \\
{\bf Pure 1-Soliton type distribution of Lagrangian density}
\medskip \\
If we choose suitable parameters $a_1, a_2, b_1, b_2$ in \eqref{Soliton solution_M} to fit the condition $\varepsilon_{2}=0$ (Cf: \eqref{Relation of epsilons_M}), we can get the pure 1-soliton type distribution (Cf: \eqref{Reduced Lagrangian density_cpx_D2}) :
\begin{eqnarray}
\label{Reduced Lagrangian density_M}
{\mbox{Tr}} F_{\mu \nu}F^{\mu \nu}&\!\!\!\!=& \!\!\!\!
8\left| \beta(\lambda-\mu) \right|^4
\left(2{\mbox{sech}}^2 X-3{\mbox{sech}}^4 X\right),  \\
&& \!\!\!\!\! X= L + \overline{L} + \frac{1}{2}\log\left[(\left|a_1 \right|^2 + \left|b_2 \right|^2) / (\left|a_2 \right|^2+\left|b_2 \right|^2)\right].
\end{eqnarray}

\subsection{Summary of the ASDYM 1-Soliton on 4D real spaces}

\label{Section 5.4}
So far we found a class of ASDYM 1-solitons (and breather solutions) on three kinds of real spaces successfully by the ansatz \eqref{Ansatz_U}, \eqref{Ansatz_E}, and \eqref{Ansatz_M} respectively. The resulting Lagrangian density is of 1-soliton (or periodic 1-soliton) type distribution and well-defined on each real space. Our results are in fact split-new for the ASDYM equations because the 1-soliton solutions \eqref{Soliton solution_U}, \eqref{Soliton solution_E}, and \eqref{Soliton solution_M} cannot be reproduced in the ordinary Lax formulation.
More precisely, if the spectral parameter $\Lambda$ is 
a scalar matrix (i.e. $\lambda=\mu$), the "Lax operators" \eqref{Lax representation of ASDYM_NGO} 
is equivalent to the ordinary Lax operators \eqref{Lax representation of ASDYM}. 
In this case, $J=-\psi\Lambda\psi^{-1} = -\lambda I$ and hence it is not the nontrivial solution we seek.
Under the condition $\varepsilon_{2}=0$, we can remove the periodic part of the Lagrangian density \eqref{Lagrangian density_U}, \eqref{Lagrangian density_E}, and \eqref{Lagrangian density_M} and obtain the pure 1-soliton type distribution (Cf: \eqref{Reduced Lagrangian density_U}, \eqref{Reduced Lagrangian density_E}, \eqref{Reduced Lagrangian density_M}) :
\begin{eqnarray}
\label{Reduced Lagrangian density_3 real spaces}
{\mbox{Tr}} F_{\mu\nu}F^{\mu\nu}
=C\left(
2{\mbox{sech}}^2 X-3{\mbox{sech}}^4 X
\right),
\end{eqnarray}
where 
$X = L + \overline{L} + \frac{1}{2}\log\left[(\left|a_1 \right|^2 + \left|b_2 \right|^2) / (\left|a_2 \right|^2+\left|b_2 \right|^2)\right]$ and
$C$ is a real constant depending on the signature 
of different real spaces. 
Note that the principal peak of the Lagrangian density \eqref{Reduced Lagrangian density_3 real spaces} 
is localized on a three dimensional hyperplane $X=0$ and hence it is the codimension 1 type soliton on 4-dimensional space.
We call it soliton wall for convenience in this thesis.
In fact, one of the simplest type soliton wall can be captured by the following reduced 1-soliton solution : (A simple choice is $a_2 = b_1 =0$ which implies that $\varepsilon_{2}=0$.)
\begin{eqnarray}
\label{Reduced 1-soliton_3 real spaces}
J = -\psi\Lambda\psi^{-1},~~ 
\psi=
\left(
\begin{array}{cc}
ae^{L} & \overline{b}e^{-\overline{L}}
\\ 
-be^{-L} & \overline{a}e^{\overline{L}} 
\end{array}\right), ~~
\Lambda
=
\left(
\begin{array}{cc}
\lambda & 0 \\
0 & \mu
\end{array}
\right).
\end{eqnarray}
In this case, 
\begin{eqnarray}
\label{Reduced Lagrangian density_special_3 real spaces}
{\mbox{Tr}} F_{\mu \nu}F^{\mu \nu}
=
C\left(2{\mbox{sech}}^2 X\!-3{\mbox{sech}}^4 X\right),~
X\!=\!L + \overline{L} +\log\left|a/b\right|. 
\end{eqnarray}
Here $C$ and $\Lambda$ are dependent on the signature of real spaces (Cf: Table \ref{table_2}).
We will discuss the multi-soliton dynamics of this special type soliton wall later in Subsection \ref{Section 6.2}.
\medskip \\
\noindent
\begin{table}[h]
	\!\!\!\!\!\!\caption{Summary of the 1-Soliton Solutions on 4D Real Spaces}
	\label{table_2}
	\begin{tabular}{|c|c|c|c|}
		\hline
		signature & $\mathbb{U}$ & $\mathbb{E}$  & $\mathbb{M}$   \\\hline\hline
		reality condition 
		& $z, \widetilde z,w, \widetilde w\in \mathbb{R}$ & $\widetilde z=\overline z,~
		\widetilde w= -\overline w$  & 
		$z, \widetilde z \in \mathbb{R},~\widetilde w= \overline w$ 
		\\
		$\displaystyle
		\sqrt{2}
		\left(\begin{array}{cc}\!\!z & w\!\!\\\!\!\widetilde{w} & \widetilde{z}\!\!\end{array}\right)
		\!\!=\!\!$
		&
		$\displaystyle \!\!
		\left(\begin{array}{cc}\!\!\!\!x^1\!+x^3 & \!\!x^2\!-x^4 \!\!\\
		\!\!\!\!-x^2\!+x^4& \!\! x^1\!-x^3\!\!
		\end{array}\right)\!\!$
		&
		$\!\!\displaystyle
		\left(\begin{array}{cc}\!\!x^1+ix^2\!\!& x^3+ix^4\!\!\\\!\!-x^3+ix^4\!\!&
		x^1-ix^2\!\!
		\end{array}\right)\!\!$
		&
		$\displaystyle
		\left(\begin{array}{cc}\!\!x^0+x^1&x^2+ix^3\!\!\\\!\!x^2-ix^3&x^0-x^1\!\!
		\end{array}\right)$
		\\\hline
		& & &  \\
		$\!\!$ansatz$\!\!$ &$\mu=\overline{\lambda}$ & $\mu=-1/\overline{\lambda}$ & None
		\\
		$L=\overline{M}$
		& 
		$L=(\lambda \alpha) z
		+\beta\widetilde{z}$ 
		&
		$L=(\lambda \alpha) z
		+\beta\overline{z}$  & 
		$L=(\lambda \overline{\mu}\beta) z
		+\beta\widetilde{z}$
		\\
		&
		$+(\lambda \beta) w
		+\alpha \widetilde{w}$
		&
		$+(\lambda \beta) w
		-\alpha \overline{w}$
		&$+(\lambda \beta) w
		+(\overline{\mu}\beta)\overline{w} $
		\\\hline
		$L=l_\mu x^\mu$ & 
		$l_\mu\!\!=\!\!\displaystyle\frac{1}{\sqrt{2}}\left(\begin{array}{c}
		\!\!\lambda\alpha+\beta\!\! \\ \!\! \lambda\beta-\alpha\!\! \\
		\!\!\lambda\alpha-\beta\!\! \\ \!\! \lambda\beta+\alpha\!\!
		\end{array}\right)$ 
		&
		$l_\mu\!\!=\!\!\displaystyle\frac{1}{\sqrt{2}}\left(\begin{array}{c}
		\!\!\lambda\alpha+\beta\!\!\\\!\!i(\lambda\alpha-\beta)\!\!  \\
		\!\!\lambda\beta-\alpha\!\! \\\!\!i(\lambda\beta+\alpha)\!\!
		\end{array}\right)$ & 
		$l_\mu\!\!=\!\!\displaystyle\frac{\beta}{\sqrt{2}}\left(\begin{array}{c}
		\lambda \overline{\mu}+1 \\ \lambda \overline{\mu}-1  \\
		\lambda + \overline{\mu} \\ i(\lambda - \overline{\mu})
		\end{array}\right)$    \\\hline
		constant $C$
		&
		$\!\!8\left[(\alpha\overline{\beta}\!-\!\overline{\alpha}\beta)
		(\lambda \!-\!\overline{\lambda})  \right]^2\!\!$
		&$\!\!8(\left|\alpha\right|^2\!\!+\!\left|\beta\right|^2)^2
		(\left|\lambda \right|^2+1)^2\!\!$
		&
		$8\left|{\beta}(\lambda-\mu)\right|^4$
		\\\hline
		hermiticity
		&
		$A_1,A_2,A_3,A_4$:
		& 
		$A_1,A_3$:anti-hermitian
		&
		$\!\!A_0,A_1,A_2$:anti-hermitian$\!\!$
		\\
		of $A_\mu$
		&anti-hermitian
		& $A_2,A_4$:hermitian
		&$A_3$:hermitian
		\\
		& (None)
		&(when $\lambda=\pm i$)
		&(when $\lambda=\overline{\mu}$) 
		\\\hline
		gauge group
		&$G=\mathrm{SU}(2)$
		&$G=\mathrm{SL}(2,\mathbb{C})$
		&$G=\mathrm{SL}(2,\mathbb{C})$
		\\\hline
	\end{tabular} 
\end{table}

\newpage

\section{ASDYM Multi-Solitons ($G=\mathrm{SU}(2)$ or $\mathrm{SL}(2,\mathbb{C})$)}
\label{Section 6}

As mentioned in Theorem \ref{Thm_4.2},
we can obtain  $n$ different specified solutions
$\psi_i(\Lambda_i)$ by solving the initial linear system \eqref{Initial linear system}  and 
use them to form the exact solutions $J_2, J_3, \cdots J_{n+1}$  of the Yang equation. 
For gauge group $G=\mathrm{SL}(2,\mathbb{C})$, a class of $\psi_i(\Lambda_i)$ can be solved by the initial linear system \eqref{Initial linear system_J_1=I} as the following form :
\begin{eqnarray}
\label{1-Soliton Solutions_psi_i}
\psi_i=\left(
\begin{array}{cc}
a_{i1}e^{L_i} + a_{i2}e^{-L_i}
& 
\overline{b}_{i1}~\!e^{\overline{L}_i} + \overline{b}_{i2}~\!e^{-\overline{L}_i}
\\ 
-b_{i1}e^{L_i} - b_{i2}e^{-L_i}
& 
\overline{a}_{i1}~\!e^{\overline{L}_i} + \overline{a}_{i2}~\!e^{-\overline{L}_i}
\end{array}\right)
,~
\Lambda_i=
\left(
\begin{array}{cc}
\lambda_i & 0 \\
0 & \mu_i
\end{array}
\right),~i=1,2,\cdots,n~~
\end{eqnarray}
where  
$L_i
=\lambda_i \alpha_i z
+\beta_i\widetilde{z}
+\lambda_i \beta_i w
+\alpha_i\widetilde{w}$, 
and 
$\lambda_i, \mu_i, a_i, b_i, \alpha_i,\beta_i$ are complex constants. 

We also showed in previous section that the solutions $\psi_i(\Lambda_i)$ exactly give the ASDYM 1-solitons, respectively. 
Now we can use the $n$ different solutions $\psi_i(\Lambda_i)$ to form a series of new "soliton solutions" $ J_{3}, J_{4}, \cdots, J_{n+1}$ of the Yang equation. A natural question is whether such kind of $J$-matrix $J_{k+1}$ can give the ASDYM $k$-soliton or not. Another nature question is whether the $J_{k+1}$ can give a $G=\mathrm{SU}(2)$ gauge theory on the Ultrahyperbolic space $\mathbb{U}$ (like $J_2$ does) or not. 

\subsection{Hermiticity of the gauge fields given by "Multi-Solitons"}
\label{Section 6.1}

Let us consider the following candidate of multi-soliton solutions :
\begin{eqnarray}
\label{Multi-Soliton Solutions_J_n+1_Candidate}
J_{n+1}
&\!\!\!\!:=& \!\!\!\!
\left| 
\begin{array}{ccccc}
\!\psi_{1} & \!\! \psi_{2} & \!\!\!\!\cdots & \!\!\!\!\psi_{n}  & \!\!1 \\
\!\psi_{1}\Lambda_{1} & \!\! \psi_{2}\Lambda_{2} & \!\!\!\!\cdots & \!\!\!\! \psi_{n}\Lambda_{n}  & \!\!0 \\
\!\vdots & \!\! \vdots & \!\!\!\! \ddots & \!\!\!\!\vdots & \!\!\vdots \\
\!\psi_{1}\Lambda_{1}^{n-1} & \!\! \psi_{2}\Lambda_{2}^{n-1} & \!\!\!\! \cdots & \!\!\!\!\psi_{n}\Lambda_{n}^{n-1}  & \!\!0 \\
\!\psi_{1}\Lambda_{1}^{n} & \!\! \psi_{2}\Lambda_{2}^{n} & \!\!\!\! \cdots & \!\!\!\!\psi_{n}\Lambda_{n}^{n}  & \!\!\fbox{0}
\end{array} \!
\right|,~ 
\begin{array}{l}
\psi_i
=\left(
\begin{array}{cc}
\!\!A_i & \!\overline{B}_i \\
\!\!-B_i & \!\overline{A}_i
\end{array}
\right),~
\Lambda_i=
\left(
\begin{array}{cc}
\!\lambda_i & \!0 \\
\!0 & \!\mu_i
\end{array}
\right),  ~~~~~~ \\
A_i:=a_{i1}e^{L_i}+a_{i2}e^{-L_i},   \\
B_i:=b_{i1}e^{L_i} + b_{i2}e^{-L_i}, \\
L_i
:=
(\lambda_i \alpha_i) z +\beta_i\widetilde{z} +(\lambda_i \beta_i) w +\alpha_i \widetilde{w},  
\end{array} ~~ \\
&&{\mbox{where~ $a_{i1}, a_{i2}, b_{i1}, b_{i2}, \alpha_i, \beta_i, \lambda_i, \mu_i$~ are complex constants}},~ i=1,2,...,n.  \nonumber
\end{eqnarray}
Recall in the Table \ref{table_2} that $\mu_i$ relies on the metric of real spaces.
Under the reality conditions \eqref{Reality condition_E}, \eqref{Reality condition_M}, and \eqref{Reality condition_U}, we can obtain
\medskip \\
{\bf (i) Ultrahyperbolic space} ~ ($\mu_i=\overline{\lambda}_i$)
\begin{eqnarray}
L_i\!\!\!&=& \!\!\!
\frac{1}{\sqrt{2}}
\left[
(\lambda_i\alpha_i+\beta_i)x^1
+(\lambda_i\beta_i-\alpha_i)x^2
+(\lambda_i\alpha_i-\beta_i)x^3
+(\lambda_i\beta_i+\alpha_i)x^4 
\right].  \label{l_imu_U}  
\end{eqnarray}
{\bf (ii) Euclidean space} ~ ($\mu_i= -1/\overline{\lambda}_i$)
\begin{eqnarray}
L_i\!\!\!&=& \!\!\!
\frac{1}{\sqrt{2}}
\left[
(\lambda_i\alpha_i+\beta_i)x^1
+i(\lambda_i\alpha_i-\beta_i)x^2
+(\lambda_i\beta_i-\alpha_i)x^3
+i(\lambda_i\beta_i+\alpha_i)x^4 
\right].  \label{l_imu_E} 
\end{eqnarray}
{\bf (iii) Minkowski space}
\begin{eqnarray}
L_i\!\!\!&=& \!\!\!
\frac{\beta}{\sqrt{2}}
\left[
(\lambda_i\overline{\mu_i}+1)x^0
+(\lambda_i\overline{\mu_i}-1)x^1
+(\lambda_i+\mu_i)x^2
+i(\lambda_i-\overline{\mu_i})x^3 
\right].  \label{l_imu_M}  
\end{eqnarray}
\newtheorem{prop_6.1}{Proposition}[section] \label{Prop_6.1}
\begin{prop_6.1}
\begin{eqnarray}
\label{det(J_n+1)}
\det(J_{n+1})=\prod_{i=1}^{n}\lambda_i\mu_i.
\end{eqnarray}
\end{prop_6.1}
{\bf ({\emph{Proof}})}
\\
For $n=1$,
\begin{eqnarray*}
\det(J_{2})
=
\det(\left|
\begin{array}{cc}
\psi_{1} & 1 \\
\psi_{1}\Lambda_{1} & \fbox{0}
\end{array}
\right|)
=
\det(-\psi_{1}\Lambda_{1}\psi_{1}^{-1})
=
(-1)^{2}\det(\Lambda_{1})
=
\lambda_{1}\mu_{1} ~~ \mbox{holds.}   \nonumber 
\end{eqnarray*} 
Suppose that the statement holds for $1 \leq n \leq k-1$,~that is,~ 
$\det(J_{k})=\prod\limits_{i=1}^{k-1}\lambda_i\mu_i$. Then for $n=k$,~ we have
\begin{eqnarray*}
J_{k+1}
&\!\!\!\!=& \!\!\!\! 
-\Psi_{k}\Lambda_{k}\Psi_{k}^{-1}J_{k} ~~ \mbox{by the Darboux transformation \eqref{Darboux transf_(J_k+1, Phi_k+1)}} \nonumber  \\
&& \!\!\!\! \Longrightarrow ~
\det(J_{k+1})=\det(-\Lambda_{k})\det(J_{k})
=(-1)^2\lambda_{k}\mu_{k}\prod_{i=1}^{k-1}\lambda_{i}\mu_{i}
=\prod_{i=1}^{k}\lambda_{i}\mu_{i}.   \nonumber  
\end{eqnarray*}
\hfill$\Box$
\\
A further result can be obtained by applying Jacobi's formula \eqref{Jacobi's formula} to \eqref{det(J_n+1)} directly :
\begin{eqnarray}
\label{Tr(A_mu)=0}
&&\underbrace{\partial_{\mu}\det(J_{n+1}^{-1}) }
= \det(J_{n+1}^{-1})\cdot\underbrace{\mbox{Tr}\left[J_{n+1}(\partial_{\mu}J_{n+1}^{-1})\right]} ~~  \Longrightarrow ~~ \mbox{Tr}\left[(\partial_{\mu}J_{n+1})J_{n+1}^{-1}\right]=0.  ~~~~~~~~  \\
&&=0~~  \mbox{by} ~ \eqref{det(J_n+1)}. ~~~~~~~~~~
=-\mbox{Tr}\left[(\partial_{\mu}J_{n+1})J_{n+1}^{-1}\right] \nonumber  
\end{eqnarray}
Comparing \eqref{Tr(A_mu)=0} with \eqref{Gauge Fields_U_A_1=A_3, A_2=A_4}, \eqref{Gauge Fields_E_A_2=iA_1}, \eqref{Gauge Fields_E_A_4=iA_3}, \eqref{Gauge Fields_M_A_0=A_1}, and \eqref{Gauge Fields_M_A_3 = iA_2} (under the replacement of $J \rightarrow J_{n+1}$), we can conclude immediately that the gauge fields $A_{\mu}^{(n+1)}$ given by the "multi-soliton solutions" \eqref{Multi-Soliton Solutions_J_n+1_Candidate} are all traceless on the three kinds of real spaces $\mathbb{E}$, $\mathbb{M}$, and $\mathbb{U}$. Therefore, the gauge group is $G=\mathrm{SL}(2,\mathbb{C})$ for the three kinds of signature. 

The remaining problem is if the gauge group can be $G=\mathrm{SU}(2)$ on the Ultrahyperbolic space $\mathbb{U}$ or not. Before verifying this, we need some properties of the $J$-matrix $J_{n+1}$ to check the hermiticity of the gauge fields $A_{\mu}^{(n+1)}$ (Cf: Theorem \ref{Thm_4.6}). Let us begin with an observation of a special class of quasideterminant. 
If we consider the quasideterminant of order $n$ whose components are all in the same form as the following $2 \times 2$ matrices $x_{pq}$.
\begin{eqnarray}
\label{x_pq_special form}
\left| X^{(n)} \right|_{ij}
=\left|
\begin{array}{cccccc}
x_{11} & \cdots & x_{1j} & \cdots & x_{1n}  \\
\vdots & \ddots & \vdots & \ddots & \vdots  \\
x_{i1} & \cdots & \fbox{$x_{ij}$} & \cdots & x_{in}  \\
\vdots & \ddots & \vdots & \ddots & \vdots  \\
x_{n1} & \cdots & x_{nj} & \cdots & x_{nn} \\
\end{array}   
\right|,~~
x_{pq}
:=
\left(
\begin{array}{cc}
r_{pq} & \overline{s}_{pq}  \\
-s_{pq} & \overline{r}_{pq}
\end{array}
\right),~ 1 \leq i,j,p,q \leq n. ~~~ 
\end{eqnarray}
Then we can show that $\left|X^{(n)}\right|_{ij}$ is in the same form as its components, that is, 
\begin{eqnarray}
\label{Quasideterminant_special form}
\left| X^{(n)} \right|_{ij}=
\left(
\begin{array}{cc}
R_{ij}^{(n)} & \overline{S}_{ij}^{(n)} \\
-S_{ij}^{(n)} & \overline{R}_{ij}^{(n)}
\end{array}
\right). ~~
\end{eqnarray}
This fact can be verified immediately by mathematical induction.
For $n=1$, we have $\left|X^{(1)}\right|_{ij}=x_{11}$ and hence the statement is clearly true.
Suppose that the statement holds for $1 \leq n \leq k-1$. Then for $n=k$, we can apply the Jacobi Identity \eqref{Jacobi Identity} to obtain
\begin{eqnarray}
\label{Jacobi identity_special form}
\left| X^{(k)} \right|_{ij}=
\left|
X_{\widehat{l},~\!\widehat{m}}^{(k)}
\right|_{i,j}
\!\! -
\left|
X_{\widehat{l},~\!\widehat{j}}^{(k)}
\right|_{i,m}
\left|
X_{\widehat{i},~\!\widehat{j}}^{(k)}
\right|_{l,m}^{-1}
\left|
X_{\widehat{i},~\!\widehat{m}}^{(k)}
\right|_{l,j}.    
\end{eqnarray}
Since $X_{\widehat{l},~\!\widehat{m}}^{(k)}$, $X_{\widehat{l},~\!\widehat{j}}^{(k)}$, 
$X_{\widehat{i},~\!\widehat{j}}^{(k)}$, and $X_{\widehat{i},~\!\widehat{m}}^{(k)}$
are the $(k-1) \times (k-1)$ submatrices of $X^{(k)}$,
we can conclude that 
 $\left|X_{\widehat{l},~\!\widehat{m}}^{(k)}\right|_{i,j}$, $\left|X_{\widehat{l},~\!\widehat{j}}^{(k)}\right|_{i,m}$, $\left|X_{\widehat{i},~\!\widehat{j}}^{(k)}\right|_{l,m}$, and $\left|X_{\widehat{i},~\!\widehat{m}}^{(k)}\right|_{l,j}$ are all in the same form as $\left|X^{(k-1)}\right|_{ij}$ (Cf: \eqref{Quasideterminant_special form} for $n=k-1$).
 This fact implies that $\left| X^{(k)} \right|_{ij}$ is also in the form of \eqref{Quasideterminant_special form} due to simple arithmetic operations of each terms in the RHS of \eqref{Jacobi identity_special form}.
On the other hand, we can find in the Ultrahyperbolic space case that 
\begin{eqnarray}
J_{n+1}&\!\!\!\!=&\!\!\!\!-\Psi_{n}\Lambda_{n}\Psi_{n}^{-1}J_{n},  \\
\Psi_{n}
&\!\!\!\!=&\!\!\!\! \left| 
\begin{array}{cccc}
\!\psi_{1} & \!\!\psi_{2} & \!\!\!\cdots & \!\!\!\psi_{n}  \\
\!\psi_{1}\Lambda_{1} & \!\!\psi_{2}\Lambda_{2} & \!\!\!\cdots & \!\!\!\psi_{n}\Lambda_{n}  \\
\!\vdots & \!\!\vdots & \!\!\! \ddots & \!\!\!\vdots \\
\!\psi_{1}\Lambda_{1}^{n-1} & \!\!\psi_{2}\Lambda_{2}^{n-1} & \!\!\!\cdots & \!\!\!\fbox{$\psi_{n}\Lambda_{n}^{n-1}$} 
\end{array}
\!\right|,~
\psi_i
:=\left(
\begin{array}{cc}
\!A_i & \!\overline{B}_i \\
\!-B_i & \!\overline{A}_i
\end{array}
\right),~
\Lambda_i
:=\left(
\begin{array}{cc}
\lambda_i & 0 \\
0 & \overline{\lambda}_i
\end{array}
\right). \label{Psi_n_U}~~~~~~~~~
\end{eqnarray}
This fact implies that $\Psi_n$ is exactly in the form of \eqref{Quasideterminant_special form} because the components $\psi_i\Lambda_i$ are in the same form as \eqref{x_pq_special form}.
\newtheorem{prop_6.2}[prop_6.1]{Proposition}
\begin{prop_6.2}
\begin{eqnarray}
\label{J_n+1 J_n+1^{dagger}}
J_{n+1}J_{n+1}^{\dagger}=J_{n+1}^{\dagger}J_{n+1}=\prod_{i=1}^{n}|\lambda_i|^2I,~~ I : 2\times 2~ \mbox{identity matrix}.
\end{eqnarray}
\end{prop_6.2}
{\bf (\emph{Proof})}
\\
For $n=1$,
\begin{eqnarray*}
&& \Lambda_{1}\Lambda_{1}^{\dagger}=\left| \lambda_{1} \right|^{2}I, \nonumber \\
&&\psi_{1}\psi_{1}^{\dagger}
=
\left(
\begin{array}{cc}
A_1 & \overline{B}_1 \\
-B_{1} & \overline{A}_1
\end{array}
\right)
\left(
\begin{array}{cc}
\overline{A}_1 & -\overline{B}_1 \\
B_{1} & A_1
\end{array}
\right)
=(|A_1|^2+|B_1|^2)I
=\det{(\psi_{1})}I
=\psi_{1}^{\dagger}\psi_{1},   \nonumber \\
&\Longrightarrow&
J_{2}J_{2}^{\dagger}
=\psi_{1}\Lambda_{1}\underbrace{\psi_{1}^{-1}\psi_{1}^{-\dagger}}\Lambda_{1}^{\dagger}\psi_{1}^{\dagger} 
=
\psi_{1}\underbrace{(\Lambda_{1}\Lambda_{1}^{\dagger})}\psi_{1}^{\dagger}(\psi_{1}^{\dagger}\psi_{1})^{-1}
=\Lambda_{1}\Lambda_{1}^{\dagger}(\psi_{1}\psi_{1}^{\dagger})(\psi_{1}^{\dagger}\psi_{1})^{-1}
=|\lambda_{1}|^2I    \nonumber  \\
&& ~~~~~~~~~~~~~ (\det{(\psi_{1})})^{-1}I ~~~~~~~~~~~ \left| \lambda_{1} \right|^2I  \nonumber \\
&& ~~\mbox{which satisfies}~ \eqref{J_n+1 J_n+1^{dagger}}.  \nonumber 
\end{eqnarray*}
Suppose that the statement holds for $1\leq n \leq k-1$,~that is, 
$J_{k}J_{k}^{\dagger}=J_{k}^{\dagger}J_{k}=\prod\limits_{i=1}^{k-1}|\lambda_i|^2I$.
Then for $n=k$, we have
\begin{eqnarray*}
J_{k+1}J_{k+1}^{\dagger}
&\!\!\!\!=&\!\!\!\!
\Psi_{k}\Lambda_{k}\Psi_{k}^{-1}\underbrace{J_{k}J_{k}^{\dagger}}\Psi_{k}^{-\dagger}\Lambda_{k}^{\dagger}\Psi_{k}^{\dagger}
=
	\Psi_{k}\Lambda_{k}\underbrace{(\Psi_{k}^{-1}\Psi_{k}^{-\dagger})}\Lambda_{k}^{\dagger}\Psi_{k}^{\dagger}
	(J_{k}J_{k}^{\dagger})  \nonumber \\ 
&& ~~~~~~~~~~\prod_{i=1}^{k-1}|\lambda_i|^2I ~~~~~~~~~~~~~~~~ (\det(\Psi_{k}))^{-1}I ~~\mbox{by \eqref{Psi_n_U} and \eqref{Quasideterminant_special form}}
\nonumber \\
&\!\!\!\!=\!\!\!\!&
	\Psi_{k}\underbrace{(\Lambda_{k}\Lambda_{k}^{\dagger})}(\Psi_{k}^{-1}\Psi_{k}^{-\dagger})\Psi_{k}^{\dagger}
	(J_{k}J_{k}^{\dagger})
=(\Lambda_{k}\Lambda_{k}^{\dagger})(\Psi_{k}\Psi_{k}^{-1}\Psi_{k}^{-\dagger}\Psi_{k}^{\dagger})(J_{k}J_{k}^{\dagger})  \nonumber \\
&& ~~~~~~ |\lambda_{k}|^2I ~~\mbox{by \eqref{Psi_n_U}} \\
&\!\!\!\!=&\!\!\!\!
(\Lambda_k\Lambda_k^{\dagger})(J_kJ_k^{\dagger})
=(|\lambda_{k}|^2I)(\prod_{i=1}^{k-1}|\lambda_i|^2I) 
=\prod_{i=1}^{k}|\lambda_i|^2I.
\nonumber
\end{eqnarray*}
\hfill$\Box$ 
\\
Now we find that \eqref{J_n+1 J_n+1^{dagger}} implies
\begin{eqnarray*}
\left[(\partial_{\mu}J_{n+1})J_{n+1}^{-1}\right]^{\dagger}
\!= \!
J_{n+1}^{-\dagger}(\partial_{\mu}J_{n+1}^{\dagger})
\!= \!
\left[\prod_{i=1}^{n}|1/\lambda_i|^2J_{n+1}\right]
\!\!\left[\prod_{i=1}^{n}|\lambda_i|^2\left(\partial_{\mu}J_{n+1}^{-1}\right)\right] 
\!= \!
-\!\left(\partial_{\mu}J_{n+1}^{-1}\right)\!J_{n+1}^{-1}.   \nonumber 
\end{eqnarray*}
Combining with \eqref{Tr(A_mu)=0}, we can conclude the following important result.
\newtheorem{thm_6.3}[prop_6.1]{Theorem}  
\begin{thm_6.3}
\label{Thm_6.3}
{\color{white}==}\\
The gauge fields 
\begin{eqnarray}
\label{Real representation of gauge fields_U_muti-soliton} 
&&A_{1}^{(n+1)} = A_{3}^{(n+1)}
=\frac{-1}{2}
\left[(\partial_{1}J_{n+1})J_{n+1}^{-1}+(\partial_{3}J_{n+1})J_{n+1}^{-1} \right],   \\
&&A_{2}^{(n+1)}=A_{4}^{(n+1)}
=\frac{-1}{2}
\left[(\partial_{2}J_{n+1})J_{n+1}^{-1} + (\partial_{4}J_{n+1})J_{n+1}^{-1} \right]  
\end{eqnarray}
on the Ultrahyperbolic space $\mathbb{U}$
are all anti-hermitian and traceless. Therefore, the gauge group can be $G=\mathrm{SU}(2)$.
\end{thm_6.3}

\subsection{Asymptotic behavior of the ASDYM Multi-Solitons, $G=\mathrm{SU}(2)$ or $\mathrm{SL(2, \mathbb{C})}$}
\label{Section 6.2}

{\bf $n$-Soliton Solution}
\medskip \\
Let us consider a simplified version of \eqref{Multi-Soliton Solutions_J_n+1_Candidate} as the following form: 
\begin{eqnarray}
\label{n-Soliton Solutions_J_n+1}
J_{n+1}=
\left|
\begin{array}{cccc}
\psi_1&\cdots&\psi_n& 1\\
\psi_1\Lambda_1&\cdots &\psi_n\Lambda_n& 0\\
\vdots   && \vdots& \vdots\\
\psi_1\Lambda_1^{n}&\cdots& \psi_n\Lambda_n^{n}& \fbox{$0$}
\end{array}\right|, ~ 
\begin{array}{l}
\psi_i=\left(
\begin{array}{cc}
\!a_i e^{L_i} & \overline{b}_i~\!e^{-\overline{L}_i}  \\ 
\!\!\!-b_i e^{-L_i} & \!\!\overline{a}_i~\! e^{\overline{L}_i}
\end{array}\!\!\right),
\Lambda_i=
\left(
\begin{array}{cc}
\!\!\lambda_i^{(+)} & \!\!\!0 \\
\!\!0 & \!\!\!\lambda_i^{(-)}
\end{array}
\!\!\right), ~~ \\
L_i
=\lambda_i^{(+)} \alpha_i z
+\beta_i\widetilde{z}
+\lambda_i^{(+)} \beta_i w
+\alpha_i\widetilde{w}, \\
a_i, b_i, \alpha_i,\beta_i, \lambda_i^{(+)},\lambda_i^{(-)}\in \mathbb{C},~
i=1,2,\cdots,n.
\end{array}
\end{eqnarray}
Here we redefine $(\lambda_i, \mu_i):=(\lambda_{i}^{(+)}, \lambda_{i}^{(-)})$ for convenience in the discussion of multi-soliton scattering.
For $n=1$, the $J$-matrix can be referred to \eqref{Reduced 1-soliton_3 real spaces} and it gives a single soliton wall. Our aim is to show that for any $n \geq 2$, the $J$-matrix \eqref{n-Soliton Solutions_J_n+1} actually gives $n$ intersecting soliton walls. In general, we ought to show that the distribution of Lagrangian density possesses $n$ overlapping principal peaks within the soliton scattering region, however, it does not seem to be a easy job to verify this. Equivalently, we can consider the asymptotic behavior of the Lagrangian density and show that it indeed possesses $n$ isolated principal peaks in a region far away enough from the scattering region, called the asymptotic region. 
To simplify the problem, we assume that $L_i$ $(i=1,2,\cdots,n)$ are mutually independent. In other words, we just consider the pure soliton scattering and exclude any cases of resonance process.
\medskip \\
{\bf Asymptotic behavior of $n$-soliton}
\medskip \\ 
Inspired by the technique \cite{MaSa} (Cf: Appendix \ref{Appendix A}) for dealing with 
the asymptotic behavior of the KP multi-solitons,
we follow a similar procedure to show that
the quasi-Wronskian type solution \eqref{n-Soliton Solutions_J_n+1} 
exactly satisfy the requirement of the ASDYM multi-solitons. (Cf: Definition of ASDYM Soliton \eqref{Def_ASDYM Multi-Soliton}).
It is actually a new attempt 
because the elements $\psi_{i}\Lambda_{i}$ in the quasi-Wronskian  
$J_{n+1}$ are $2\times 2$ matrices rather than a scalar function. 
If the reader prefer using the Wronskian determinants, one can convert the quasi-Wronskian to the ratios of Wronskians in the following way :
\begin{eqnarray}
\label{1 quasideterminant to 4 quasideterminants}
\begin{vmatrix}
A\!\!\!\!\!& \begin{array}{cc} B & C \end{array}\\
\begin{array}{c}
D\!\!\!\!\!\\ E\!\!\!\!\!
\end{array}
&\fbox{$
	\begin{array}{cc}
	f & g \\
	h & i
	\end{array}$}
\end{vmatrix} 
&\!\! = & \!\!
\left(
\!\begin{array}{cc}
\begin{vmatrix}
A  & B \\
D  & \fbox{$f$}
\end{vmatrix}&
\begin{vmatrix}
A & C \\
D & \fbox{$g$}
\end{vmatrix}
\\
\begin{vmatrix}
A & B \\
E & \fbox{$h$}\\
\end{vmatrix}&
\begin{vmatrix}
A & C \\
E & \fbox{$i$}\\
\end{vmatrix}
\end{array}
\!\right)  \\
&\!\! =& \!\!
\!\frac{1}{\det(A)}
\left(
\!\begin{array}{cc}
\det\begin{pmatrix}
A  & B \\
D  & f
\end{pmatrix} &
\det\begin{pmatrix}
A & C \\
D & g
\end{pmatrix}
\\
\det\begin{pmatrix}
A & B \\
E & h \\
\end{pmatrix}&
\det\begin{pmatrix}
A & C \\
E & i \\
\end{pmatrix}
\end{array}
\!\!\right), ~~~
\end{eqnarray} 
where the size of matrices are 
$A: k \times k, ~ B, C : k \times 1, ~ D, E : 1 \times k,$ 
and $f, g, h, i$ are $1 \times 1 ~$ scalar elements.

Let us begin with analyzing the asymptotic behavior of \eqref{n-Soliton Solutions_J_n+1}.
Firstly, we pick an $I\in\left\{1, 2, ..., n\right\}$ 
and keep $L_I$ (and $\overline{L}_I$) to be finite. 
In fact, this practice is equivalent to consider a comoving frame with respect to
the $I$-th 1-soliton solution.
If we take the asymptotic limit $r:=[~\!\sum_{\mu}(x^{\mu})^{2}~\!]^{1/2}
\rightarrow \infty$.  
Then for $i\neq I$, we have
\begin{eqnarray}
\label{Asymptotic of ReL_i}
\mbox{Re$ L_i$ ($=$ Re$\overline{L}_i$)}  \longrightarrow
\left\{
\begin{array}{l}
\mbox{(i)}~~ +\infty ~~~ (\vert e^{L_i} \vert =\vert e^{\overline{L}_i}\vert \rightarrow \infty)   \\
\mbox{(ii)}~-\infty  ~~~ (\vert e^{L_i} \vert = \vert e^{\overline{L}_i}\vert \rightarrow ~0~)
\end{array}.
\right.
\end{eqnarray}
On the other hand, we can expand $J_{n+1}$ by \eqref{1 quasideterminant to 4 quasideterminants} and use the multiplication rule \eqref{right multiplication law} to eliminate common factors in the first column and second column of $(\psi_i, \psi_i\Lambda_i, ...,\psi_i\Lambda_i^{n})^T$ $(i \neq I)$, respectively. Now we can obtain two equivalent expressions of $J_{n+1}$ :
\begin{eqnarray*}
J_{n+1} 
=
\left|
\begin{array}{cccccc}
\!\widetilde{\psi}_1 & \!\!\cdots & \!\!\psi_I & \!\!\cdots & \!\!\widetilde{\psi}_n & \!1 \\
\!\widetilde{\psi}_1 \Lambda_1 & \!\!\cdots & \!\!\psi_I\Lambda_I & \!\!\cdots & \!\!\widetilde{\psi}_n\Lambda_n & \!0 \\
\!\vdots & \!\!\ddots & \!\!\vdots & \!\!\ddots & \!\!\vdots & \!\vdots \\
\!\widetilde{\psi}_1\Lambda_1^n & \!\!\cdots & \!\!\psi_I\Lambda_I^n & \!\!\cdots & \!\!\widetilde{\psi}_n\Lambda_n^n & \!\fbox{0}
\end{array}
\!\right|,
~
~\widetilde{\psi}_i=
\left\{
\begin{array}{l}
\mbox{(i)}~
\left(
\begin{array}{cc}
1 & (\overline{b_i} / \overline{a_i})e^{-2\overline{L_i}} \\
-(b_i/a_i)e^{-2L_i} & 1 
\end{array}
\!\!\right)  \\
\mbox{(ii)}~\!
\left(
\begin{array}{cc}
(a_i/b_i)e^{2L_i} & 1 \\
-1 & (\overline{a_i}/\overline{b_i})e^{2\overline{L_i}} 
\end{array}
\right)
\end{array}.
\right.  \nonumber 
\end{eqnarray*}
By \eqref{Asymptotic of ReL_i}, we have
\begin{eqnarray}
\label{Asymptotic form of J_n+1_C_i}
	J_{n+1} ~\stackrel{r \rightarrow \infty}{\longrightarrow}~
	\left|
	\begin{array}{cccccc}
		C_1 & \!\!\cdots & \!\!\psi_I & \!\!\cdots & \!\!C_n & 1 \\
		C_1 \Lambda_1 & \!\!\cdots & \!\!\psi_I\Lambda_I & \!\!\cdots & \!\!C_n\Lambda_n & 0 \\
		\vdots & \!\!\ddots & \!\!\vdots & \!\!\ddots & \!\!\vdots & \vdots \\
		C_1\Lambda_1^n & \!\!\cdots & \!\!\psi_I\Lambda_I^n & \!\!\cdots & \!\!C_n\Lambda_n^n & \fbox{0}
	\end{array}
	\right|,
	~
	~C_i=
	\left\{
	\begin{array}{l}
	\mbox{(i)}~~
	\left(
	\begin{array}{cc}
		1 & 0 \\
		0 & 1 
	\end{array}
	\right)  \\
	\mbox{(ii)}~\!
	\left(
	\begin{array}{cc}
		0 & 1 \\
	   -1 & 0 
	\end{array}
	\right)
	\end{array}.
	\right. ~~~~
\end{eqnarray}
Next, our goal is to remove all the $C_{i}$
from \eqref{Asymptotic form of J_n+1_C_i} such that 
the remaining elements  
are just the $I$-th column and the constant matrices $\Lambda_{i} ~ (i \neq I)$. 
The case (i) is trivial, while the case (ii) can be done by using the following relation
\begin{eqnarray*}
C_i \Lambda_i^k = 
\left(
\begin{array}{cc}
0 & 1 \\
-1 & 0 
\end{array}
\right)
\left(
\begin{array}{cc}
(\lambda_i^{(+)})^k & 0 \\
0 & (\lambda_i^{(-)})^k
\end{array}
\right)
=
\left(
\begin{array}{cc}
(\lambda_i^{(-)})^k & 0 \\
0 & (\lambda_i^{(+)})^k
\end{array}
\right)
\left(
\begin{array}{cc}
0&1 \\
-1&0 
\end{array}
\right),   \nonumber 
\end{eqnarray*}
and \eqref{right multiplication law} to remove all the right common factors $C_{i,i \neq I}$. 
The explicit result is
\begin{eqnarray}
	&&J_{n+1} ~\stackrel{r \rightarrow \infty}{\longrightarrow}~ 
	\widetilde{J}_{n+1}^{~\!(I)}:=
	\left|
	\begin{array}{cccccc}
		1 & \cdots & \psi_I & \cdots & 1 & 1 \\
		\Lambda_1^{(\pm)} & \cdots & \psi_I\Lambda_I & \cdots & \Lambda_n^{(\pm)} & 0 \\
		\vdots & \ddots & \vdots & \ddots & \vdots & \vdots \\
	    \Lambda_1^{(\pm)n} & \cdots & \psi_I\Lambda_I^n & \cdots & \Lambda_n^{(\pm)n} & \fbox{0}
	\end{array}
	\right|,    \label{Asymptotic J_n+1}  \\
	&& \Lambda_{i,~ i \neq I}^{(\pm)}:=
	\left(
	\begin{array}{cc}
		\lambda_i^{(\pm)}&0 \\
		0&\lambda_i^{(\mp)}
	\end{array}
	\right)
	=
	\left\{
	\begin{array}{ll}
	\mbox{(i)}~\!	\left(
		\begin{array}{cc}
			\lambda_i^{(+)}&0 \\
			0&\lambda_i^{(-)}
		\end{array}
		\right),
		& \mbox{Re}L_i\rightarrow +\infty  \\
		\mbox{(ii)} \left(
		\begin{array}{cc}
			\lambda_i^{(-)}&0 \\
			0&\lambda_i^{(+)}
		\end{array}
		\right),
		& \mbox{Re}L_i\rightarrow -\infty
	\end{array}.
	\right.  ~~~~
\end{eqnarray}
Now we need to to show that the asymptotic form $\widetilde{J}_{n+1}^{~\!(I)}$ \eqref{Asymptotic J_n+1} is essentially the same as the 1-soliton solution \eqref{Reduced 1-soliton_3 real spaces} up to a constant multiple and a replacement of $\psi$ by $\psi_I$.
Without loss of generality, we can 
consider $I=1$ for convenience. 
The discussions for $I \neq 1$ cases are all the same as $I=1$ 
due to the permutation property of 
the quasideterminants (Proposition \ref{Prop_2.3}).
By applying the properties of quasideterminants (Cf: Section \ref{Section 2})
to $\widetilde{J}_{n+1}^{~\!(I=1)}$, 
it can be represented as the following beautiful form.
\newtheorem{prop_6.4}[prop_6.1]{Proposition}   
\begin{prop_6.4}
{\bf (Asymptotic form of the $n$-soliton solutions)}   \label{Prop_6.4}
\begin{eqnarray}
\label{Asymptotic form of J_{n+1}_psi_1}
\widetilde{J}_{n+1}^{~\!(I=1)}
=
\left|
\begin{array}{ccccc}
\!\!\psi_1 & \!\!1 & \!\!\!\!\cdots & \!\!\!\!1 & \!\!\!\!1 \\
\!\!\psi_1\Lambda_1 & \!\!\Lambda_2^{(\pm)} & \!\!\!\!\cdots & \!\!\!\!\Lambda_n^{(\pm)} & \!\!\!\!0 \\
\!\!\vdots & \!\!\vdots & \!\!\!\!\ddots & \!\!\!\!\vdots & \!\!\!\!\vdots \\
\!\!\psi_1\Lambda_1^{n-1} & \!\!\Lambda_2^{(\pm)n-1} & \!\!\!\!\cdots & \!\!\!\!\Lambda_{n}^{(\pm)n-1} & \!\!\!\!0 \\
\!\!\psi_1\Lambda_1^{n} & \!\!\Lambda_2^{(\pm)n} & \!\!\!\!\cdots & \!\!\!\!\Lambda_{n}^{(\pm)n} & \!\!\!\!\fbox{$0$}
\end{array}
\!\right|
=-\widetilde{\Psi}_{n}^{~\!(I=1)} \Lambda_1 (\widetilde{\Psi}_{n}^{(~\!I=1)})^{-1} D_{n}^{~\!(I=1)}, ~n \geq 2, ~~~
\end{eqnarray}
where
\begin{eqnarray}
\label{Qn}
\widetilde{\Psi}_{n}^{~\!(I=1)}\!\!\!&=&\!\!\!
\left(
\begin{array}{cc}
\prod\limits_{i=2}^{n}(\lambda_1^{(+)}-\lambda_i^{(\pm)})\!~a_1e^{L_1}
&
\prod\limits_{i=2}^{n}(\lambda_1^{(-)}-\lambda_i^{(\pm)})~\!\overline{b}_1~\!e^{-\overline{L}_1}\!
\\
\!\!\!-\prod\limits_{i=2}^{n}(\lambda_1^{(+)}-\lambda_i^{(\mp)})~\!b_1e^{-L_1}
&
\!\!\prod\limits_{i=2}^{n}(\lambda_1^{(-)}-\lambda_i^{(\mp)})~\!\overline{a}_1~\!e^{\overline{L}_1}\!
\end{array}
\right),\!\! ~~~~~~~~~\\
D_n^{~\!(I=1)}\!\!\!&=&\!\!\! (-1)^{n-1} \prod_{i=2}^{n}\Lambda_i^{(\pm)}.
\label{Dn}
\end{eqnarray}
\end{prop_6.4}
{\bf (\emph{Proof})}
\medskip \\
By the Jacobi identity \eqref{Jacobi Identity} and the right multiplication rule \eqref{right multiplication law}, we have
\begin{eqnarray}
\widetilde{J}_{n+1}^{~\!(I=1)}
&\!\!\!=& \!\!\!
\left|
\begin{array}{c:ccc:c}
\psi_1 & 1 & \cdots & 1 & 1 \\
\hdashline
\psi_1\Lambda_1 & \Lambda_2^{(\pm)} & \cdots & \Lambda_n^{(\pm)} & 0 \\
\vdots & \vdots & \ddots & \vdots & \vdots \\
\psi_1\Lambda_1^{n-1} & \Lambda_2^{(\pm)n-1} & \cdots & \Lambda_{n}^{(\pm)n-1} & 0 \\
\hdashline
\psi_1\Lambda_1^{n} & \Lambda_2^{(\pm)n} & \cdots & \Lambda_{n}^{(\pm)n} & \fbox{0}
\end{array}
\right|  \nonumber  \\
&\!\!\!=& \!\!\!
-\left|
\begin{array}{cccc}
\psi_1 & 1 & \cdots & 1 \\
\psi_1\Lambda_1 & \Lambda_2^{(\pm)} & \cdots & \Lambda_n^{(\pm)} \\
\vdots & \vdots & \ddots & \vdots \\
\psi_1\Lambda_1^{n-2} & \Lambda_2^{(\pm)n-2} & \cdots & \Lambda_{n}^{(\pm)n-2} \\
\fbox{$\psi_1\Lambda_1^{n-1}$} & \Lambda_2^{(\pm)n-1} & \cdots & \Lambda_{n}^{(\pm)n-1}
\end{array}
\right|\Lambda_{1}
\underbrace{\left|
\begin{array}{cccc}
\fbox{$\psi_1$} & 1 & \cdots & 1 \\
\psi_1\Lambda_1 & \Lambda_2^{(\pm)} & \cdots & \Lambda_n^{(\pm)} \\
\vdots & \vdots & \ddots & \vdots \\
\psi_1\Lambda_1^{n-2} & \Lambda_2^{(\pm)n-2} & \cdots & \Lambda_{n}^{(\pm)n-2} \\
\psi_1\Lambda_1^{n-1} & \Lambda_2^{(\pm)n-1} & \cdots & \Lambda_{n}^{(\pm)n-1}
\end{array}
\right|^{-1}}. \nonumber  \\
&& ~~~~~~~~~~~~~~~~~~~~~~~~~~~~~~~~~~~~~~~~~~~~~~~~~~~~~~~~~~~~~~~~~\cdots\cdots (*)   \nonumber 
\end{eqnarray}
\begin{eqnarray}
&& \!\!\!\!\!\!\!\!\!\!\!\!\!\!\!\!\!\! \mbox{By the homological relation \eqref{Homological relation}},  \nonumber \\
(*)
&\!\!\!=& \!\!\!
\left|
\begin{array}{cccc}
\psi_1 & 1 & \cdots & 1 \\
\psi_1\Lambda_1 & \Lambda_2^{(\pm)} & \cdots & \Lambda_n^{(\pm)} \\
\vdots & \vdots & \ddots & \vdots \\
\psi_1\Lambda_1^{n-2} & \Lambda_2^{(\pm)n-2} & \cdots & \Lambda_{n}^{(\pm)n-2} \\
\fbox{$\psi_1\Lambda_1^{n-1}$} & \Lambda_2^{(\pm)n-1} & \cdots & \Lambda_{n}^{(\pm)n-1}
\end{array}
\right|^{-1}
\underbrace{\left|
\begin{array}{cccc}
\fbox{0} & 1 & \cdots & 1 \\
0 & \Lambda_2^{(\pm)} & \cdots & \Lambda_n^{(\pm)} \\
\vdots & \vdots & \ddots & \vdots  \\
0 & \Lambda_2^{(\pm)n-2} & \cdots & \Lambda_{n}^{(\pm)n-2}  \\
1 & \Lambda_2^{(\pm)n-1} & \cdots & \Lambda_{n}^{(\pm)n-1}
\end{array}
\right|^{-1}}   \nonumber \\
&&\mbox{(By the permutation property (Proposition \ref{Prop_2.3}) and the inverse relation \eqref{Inverse relation})} \nonumber \\
&\!\!\!=& \!\!\!
\left|
	\begin{array}{cccc}
	\psi_1 & 1 & \cdots & 1 \\
	\psi_1\Lambda_1 & \Lambda_2^{(\pm)} & \cdots & \Lambda_n^{(\pm)} \\
	\vdots & \vdots & \ddots & \vdots \\
	\psi_1\Lambda_1^{n-2} & \Lambda_2^{(\pm)n-2} & \cdots & \Lambda_{n}^{(\pm)n-2} \\
	\fbox{$\psi_1\Lambda_1^{n-1}$} & \Lambda_2^{(\pm)n-1} & \cdots & \Lambda_{n}^{(\pm)n-1}
	\end{array}
	\right|^{-1}
\left|
\begin{array}{cccc}
1 & \cdots & 1 & 1 \\
\Lambda_2^{(\pm)} & \cdots & \Lambda_n^{(\pm)} & 0 \\
\vdots & \ddots & \vdots & \vdots \\
\Lambda_2^{(\pm)n-2} & \cdots & \Lambda_{n}^{(\pm)n-2} & 0 \\
\Lambda_2^{(\pm)n-1} & \cdots & \Lambda_{n}^{(\pm)n-1} & \fbox{0}
\end{array}
\right|. ~~~~ \nonumber 
\end{eqnarray}
We can conclude that 
\begin{eqnarray}
\widetilde{J}_{n+1}^{~\!(I=1)}
&\!\!\!\!=\!\!\!\!&-\widetilde{\Psi}_{n}^{~\!(I=1)} \Lambda_1 (\widetilde{\Psi}_{n}^{~\!(I=1)})^{-1} D_{n}^{~\!(I=1)},   \nonumber  \\
\widetilde{\Psi}_{n}^{~\!(I=1)}
&\!\!\!\!:=\!\!\!\!&
\left|
\begin{array}{cccc}
\psi_1 & 1 & \cdots & 1 \\
\psi_1\Lambda_1 & \Lambda_2^{(\pm)} & \cdots & \Lambda_n^{(\pm)} \\
\vdots & \vdots & \ddots & \vdots \\
\psi_1\Lambda_1^{n-2} & \Lambda_2^{(\pm)n-2} & \cdots & \Lambda_{n}^{(\pm)n-2} \\
\fbox{$\psi_1\Lambda_1^{n-1}$} & \Lambda_2^{(\pm)n-1} & \cdots & \Lambda_{n}^{(\pm)n-1}
\end{array}
\right|,~
D_{n}^{~\!(I=1)}
:=
\left|
\begin{array}{cccc}
1 & \!\! \cdots & \!\!1 & 1 \\
\Lambda_2^{(\pm)} & \!\!\cdots & \!\!\Lambda_n^{(\pm)} & 0 \\
\vdots & \!\! \ddots & \!\! \vdots & \vdots \\
\Lambda_2^{(\pm)n-2} & \!\! \cdots & \!\! \Lambda_{n}^{(\pm)n-2} & 0 \\
\Lambda_2^{(\pm)n-1} & \!\! \cdots & \!\! \Lambda_{n}^{(\pm)n-1} & \fbox{0}
\end{array}
\right|.  \nonumber ~~~~~
\end{eqnarray}
Now let us show that $\widetilde{\Psi}_{n}^{~\!(I=1)},D_{n}^{~\!(I=1)}$ can be expanded as \eqref{Qn} and \eqref{Dn}, respectively.
This can be done by mathematical induction.
For $n=2$, the statement is clearly true because 
we can easily check that 
\begin{eqnarray}
	&&\!\!\!\!\!\!\!\!\!\!\!\!\!\!\!\!\!\! \widetilde{\Psi}_{2}^{~\!(I=1)}
	=
	\left|
	\begin{array}{cc}
		\psi_{1} & 1  \\
		\fbox{ $\psi_{1} \Lambda_{1}$} & \Lambda_{2}^{(\pm)}
	\end{array}
	\right|   \nonumber  \\
	&& =\psi_{1}\Lambda_{1}-\Lambda_2^{(\pm)}\psi_1 
	=
	\left(
	\begin{array}{cc}
	(\lambda_1^{(+)}-\lambda_2^{(\pm)})\!~a_1e^{L_1}
	&
	(\lambda_1^{(-)}-\lambda_2^{(\pm)})~\!\overline{b}_1~\!e^{-\overline{L}_1}\!
	\\
	\!\!-(\lambda_1^{(+)}-\lambda_2^{(\mp)})~\!b_1e^{-L_1}
	&
	\!\!(\lambda_1^{(-)}-\lambda_2^{(\mp)})~\!\overline{a}_1~\!e^{\overline{L}_1}\!
	\end{array}
	\right),  \nonumber \\
	&&\!\!\!\!\!\!\!\!\!\!\!\!\!\!\!\!\!\! D_{2}^{~\!(I=1)} =
	\left|
	\begin{array}{cc}
	    1 & 1  \\
	   \Lambda_{2}^{(\pm)} & \fbox{0}
	\end{array}
	\right|
	=	-\Lambda_2^{(\pm)}.  \nonumber 
\end{eqnarray}
Suppose that the statement holds for $2 \leq n \leq k$. Then for $n=k+1$, we can apply the Jacobi identity \eqref{Jacobi Identity} and the right multiplication rule \eqref{right multiplication law} to $\widetilde{\Psi}_{k+1}^{~\!(I=1)}$ and get 
\begin{eqnarray}
	\widetilde{\Psi}_{k+1}^{~\!(I=1)}
	&\!\!\!= &\!\!\!
	\left|
	\begin{array}{c:ccc:c}
		\psi_1 & 1 & \cdots & 1 & 1 \\
		\hdashline
		\psi_1\Lambda_1 & \Lambda_2^{(\pm)} & \cdots & \Lambda_{k}^{(\pm)} & \Lambda_{k+1}^{(\pm)} \\
		\vdots & \vdots & \ddots & \vdots & \vdots \\
		\psi_1\Lambda_1^{k-1} & \Lambda_2^{(\pm)k-1} & \cdots & \Lambda_{k}^{(\pm)k-1} & \Lambda_{k+1}^{(\pm)k+1}  \\
		\hdashline
		\fbox{$\psi_1\Lambda_1^{k}$} & \Lambda_2^{(\pm)k} & \cdots & \Lambda_{k}^{(\pm)k} & \Lambda_{k+1}^{(\pm)k}
	\end{array}    
	\right|  \nonumber \\
	&\!\!\!=& \!\!\!
	\left|
	\begin{array}{ccccc}
		\psi_1 & 1 & \cdots & 1 \\
		\psi_1\Lambda_1 & \Lambda_2^{(\pm)} & \cdots & 
		\Lambda_{k}^{(\pm)} \\
		\vdots & \vdots & \ddots & \vdots\\
		\!\!\fbox{$\psi_1 \Lambda_1^{k-1}$}\!\ & \!\!\Lambda_2^{(\pm)k-1}\!\! & \!\!\cdots\!\! & 
		\!\!\Lambda_{k}^{(\pm)k-1}\!\!
	\end{array}
	\right| \Lambda_1   \nonumber \\
	&\!\!\!-& \!\!\!
	\left|
	\begin{array}{ccc}
		\!\!1 & \!\! \cdots & \!\!\! 1 \\
		\!\!\Lambda_2^{(\pm)} & \!\!\cdots & \!\!\!\Lambda_{k+1}^{(\pm)} \\
		\!\!\vdots & \!\!\ddots & \!\!\vdots \\
		\!\!\Lambda_2^{(\pm)k-1}\!\! & \!\!\cdots\!\! & \!\!\!\fbox{$\Lambda_{k+1}^{(\pm)k-1}$}\!\!
	\end{array}
	\right|
	\Lambda_{k+1}^{(\pm)}
	\underbrace{\left|
	\begin{array}{ccc}
		\!\!1 & \!\!\cdots & \!\!\!\fbox{$1$} \\
		\!\!\Lambda_2^{(\pm)} & \!\!\cdots & \!\!\!\Lambda_{k+1}^{(\pm)} \\
		\!\!\vdots & \!\!\ddots & \!\!\!\vdots \\
		\!\!\Lambda_2^{(\pm)k-1}\!\! & \!\!\cdots\!\! & \!\!\!\Lambda_{k+1}^{(\pm)k-1}\!\!
	\end{array}
	\right|^{-1}
	\!
	\left|
	\begin{array}{ccccc}
	    \!\!\fbox{$\psi_1$} & \!\!1 & \!\! \cdots & \!\!\!1 \\
		\!\!\psi_1\Lambda_1 & \!\!\Lambda_2^{(\pm)} & \!\!\cdots & \!\!\Lambda_{l}^{(\pm)} \\
		\!\!\vdots & \!\!\vdots & \!\!\ddots & \!\!\!\vdots \\
		\!\!\psi_1 \Lambda_1^{k-1}\!\! & \!\!\Lambda_2^{(\pm)k-1}\!\! & \!\!\!\cdots\!\! & \!\!\!\Lambda_{k}^{(\pm)k-1}\!\!
	\end{array}
	\right|}.  \nonumber   \\
&& ~~~~~~~~~~~~~~~~~~~~~~~~~~~~~~~~~~~~~~~~~~~~~~~~~~~~~~~~ \cdots\cdots (**)  \nonumber 
\end{eqnarray}
By applying the homological relation \eqref{Homological relation} to $(**)$, we have
\begin{eqnarray}
	&&\!\!\!\!\!\!\!\!\!\! \left|
	\begin{array}{ccc}
		1 & \!\!\cdots & \!\!\fbox{$1$} \\
		\Lambda_2^{(\pm)} & \!\!\cdots & \!\!\Lambda_{k+1}^{(\pm)} \\
		\vdots & \!\!\ddots & \!\!\vdots \\
		\Lambda_2^{(\pm)k-1} & \!\!\cdots & \!\!\Lambda_{k+1}^{(\pm)k-1}
	\end{array}
	\right|
	=
	\underbrace{\left|
	\begin{array}{cccc}
		1 & \!\!\cdots & \!\!1 & \fbox{$0$} \\
		\Lambda_2^{(\pm)} & \!\!\cdots & \!\!\Lambda_k^{(\pm)} & 0 \\
		\vdots & \!\!\ddots & \!\!\vdots & 0 \\
		\Lambda_2^{(\pm)k-1} & \!\!\cdots & \!\!\Lambda_k^{(\pm)k-1} & 1
	\end{array}
	\right|}
	\left|
	\begin{array}{ccc}
		1 & \cdots & 1 \\
		\Lambda_2^{(\pm)} & \!\!\cdots & \!\!\Lambda_{k+1}^{(\pm)} \\
		\vdots & \!\!\ddots & \!\!\vdots\\
		\Lambda_2^{(\pm)k-1} & \!\!\cdots & \!\!\fbox{$\Lambda_{k+1}^{(\pm)k-1}$}
	\end{array}
	\right|,    \nonumber \\
	&&~~~~~~~~~~~~~~~~~~~~~~~~~~~~~~~~~~~\cdots\cdots(***)  \nonumber
	\end{eqnarray}
	\begin{eqnarray}
	&&\!\!\!\!\!\!\!\!\!\!\left|
	\begin{array}{ccccc}
		\!\!\fbox{$\psi_1$} & \!\!1 & \!\!\cdots & \!\!1 \\
		\!\!\psi_1\Lambda_1 & \!\!\Lambda_2^{(\pm)} & \!\!\cdots & \!\!\Lambda_{k}^{(\pm)} \\
		\!\!\vdots & \!\!\vdots & \!\!\ddots & \!\!\vdots \\
		\!\!\psi_1 \Lambda_1^{k-1}\!\! & \!\!\Lambda_2^{(\pm)k-1}\!\! & 
		\!\!\cdots\!\! & \!\!\Lambda_{k}^{(\pm)k-1}\!\!
	\end{array}
	\right|
	=
	\underbrace{\left|
	\begin{array}{cccc}
		\!\!\fbox{$0$} & \!\!1 & \!\!\cdots & 1 \\
		\!\!0 & \!\!\Lambda_2^{(\pm)} & \!\!\cdots & \!\!\Lambda_{k}^{(\pm)}\!\! \\
		\!\!0 & \!\!\vdots & \!\!\ddots & \!\!\vdots \\
		\!\!1 & \!\!\Lambda_2^{(\pm)k-1}\!\! & \!\!\cdots\!\! & \!\!\Lambda_{k}^{(\pm)k-1}\!\!
	\end{array}
	\right|}
	\left|
	\begin{array}{ccccc}
		\!\!\psi_1 & \!\!1 & \!\!\cdots\!\! & \!\!1 \\
		\!\!\psi_1\Lambda_1\!\! & \!\!\Lambda_2^{(\pm)} &
		\!\!\cdots\!\! & \!\!\Lambda_{k}^{(\pm)}\!\! \\
		\!\!\vdots & \!\!\vdots & \!\!\ddots & \!\!\vdots \\
		\!\!\fbox{$\psi_1 \Lambda_1^{k-1}$}\!\! & \!\!\Lambda_2^{(\pm)k-1}\!\! & \!\!\cdots\!\! & 
		\!\!\Lambda_{k}^{(\pm)k-1}\!\!
	\end{array}
	\right|.  \nonumber  \\
	&&~~~~~~~~~~~~~~~~~~~~~~~~~~~~~~~~~~~~~~~~~\cdots\cdots(****)  \nonumber
\end{eqnarray}
Substituting the above two equations into $(**)$, using the fact that $(***) = (****)$ and
$[\Lambda_i,\Lambda_j]=0$ for all $i,j$, we can conclude the following recurrence relation :
\begin{eqnarray}
\widetilde{\Psi}_{k+1}^{~\!(I=1)}
&\!\!\!\!=& \!\!\!\!
\left|
\begin{array}{ccccc}
\psi_1 & 1 & \cdots & 1 \\
\psi_1\Lambda_1 & \Lambda_2^{(\pm)} & \cdots & 
\Lambda_{k}^{(\pm)} \\
\vdots & \vdots & \ddots & \vdots\\
\!\!\fbox{$\psi_1 \Lambda_1^{k-1}$}\!\ & \!\!\Lambda_2^{(\pm)k-1}\!\! & \!\!\cdots\!\! & 
\!\!\Lambda_{k}^{(\pm)k-1}\!\!
\end{array}
\right| \Lambda_1 
-
\Lambda_{k+1}^{(\pm)}
\left|
\begin{array}{ccccc}
\psi_1 & 1 & \cdots & 1 \\
\psi_1\Lambda_1 & \Lambda_2^{(\pm)} & \cdots & 
\Lambda_{k}^{(\pm)} \\
\vdots & \vdots & \ddots & \vdots\\
\!\!\fbox{$\psi_1 \Lambda_1^{k-1}$}\!\ & \!\!\Lambda_2^{(\pm)k-1}\!\! & \!\!\cdots\!\! & 
\!\!\Lambda_{k}^{(\pm)k-1}\!\!
\end{array}
\right|   \nonumber  \\
&=&\widetilde{\Psi}_{k}^{~\!(I=1)}\Lambda_{1} - \Lambda_{k+1}^{(\pm)}~\!  \widetilde{\Psi}_{k}^{~\!(I=1)} \nonumber \\
&& \mbox{By \eqref{Qn} for $n=k$}    \nonumber \\
&\!\!\!\!=& \!\!\!\!
\left(
\begin{array}{cc}
\prod\limits_{i=2}^{k}(\lambda_1^{(+)}-\lambda_i^{(\pm)})\!~a_1e^{L_1}
&
\prod\limits_{i=2}^{k}(\lambda_1^{(-)}-\lambda_i^{(\pm)})~\!\overline{b}_1~\!e^{-\overline{L}_1}\!
\\
\!\!\!-\prod\limits_{i=2}^{k}(\lambda_1^{(+)}-\lambda_i^{(\mp)})~\!b_1e^{-L_1}
&
\!\!\prod\limits_{i=2}^{k}(\lambda_1^{(-)}-\lambda_i^{(\mp)})~\!\overline{a}_1~\!e^{\overline{L}_1}\!
\end{array}
\right)
\left(
\begin{array}{cc}
\lambda_{1}^{(+)} & 0 \\
0 & \lambda_{1}^{(-)}
\end{array}
\right)   \nonumber \\
&&\!\!-
\left(
\begin{array}{cc}
\lambda_{k+1}^{(\pm)} & 0 \\
0 & \lambda_{k+1}^{(\pm)}
\end{array}
\right)
\left(
\begin{array}{cc}
\prod\limits_{i=2}^{k}(\lambda_1^{(+)}-\lambda_i^{(\pm)})\!~a_1e^{L_1}
&
\prod\limits_{i=2}^{k}(\lambda_1^{(-)}-\lambda_i^{(\pm)})~\!\overline{b}_1~\!e^{-\overline{L}_1}\!
\\
\!\!\!-\prod\limits_{i=2}^{k}(\lambda_1^{(+)}-\lambda_i^{(\mp)})~\!b_1e^{-L_1}
&
\!\!\prod\limits_{i=2}^{k}(\lambda_1^{(-)}-\lambda_i^{(\mp)})~\!\overline{a}_1~\!e^{\overline{L}_1}\!
\end{array}
\right)   \nonumber \\
	&\!\!\!\!=& \!\!\!\!
	\left(
	\begin{array}{cc}
	\prod\limits_{i=2}^{k+1}(\lambda_1^{(+)}-\lambda_i^{(\pm)})\!~a_1e^{L_1}
	&
	\prod\limits_{i=2}^{k+1}(\lambda_1^{(-)}-\lambda_i^{(\pm)})~\!\overline{b}_1~\!e^{-\overline{L}_1}\!
	\\
	\!\!\!-\prod\limits_{i=2}^{k+1}(\lambda_1^{(+)}-\lambda_i^{(\mp)})~\!b_1e^{-L_1}
	&
	\!\!\prod\limits_{i=2}^{k+1}(\lambda_1^{(-)}-\lambda_i^{(\mp)})~\!\overline{a}_1~\!e^{\overline{L}_1}\!
	\end{array}
	\right). \nonumber 
\end{eqnarray}
As for $D_n^{~\!(I=1)}$, the proof is very similar to the above. We can use the Jacobi identity \eqref{Jacobi Identity}, the right multiplication rule \eqref{right multiplication law}, the homological relation \eqref{Homological relation}, and the fact that $\left[ \Lambda_{i}, \Lambda_{j}\right]=0$ for all $i, j$ to get
\begin{eqnarray}
	D_n^{~\!(I=1)}
	&\!\!\!\!=& \!\!\!\!
	\left|
	\begin{array}{ccc:c:c}
	1 & \cdots & 1 & 1 & 1 \\
	\hdashline
	\Lambda_2^{(\pm)} & \cdots  & \Lambda_{n-1}^{(\pm)} & \Lambda_n^{(\pm)} & 0 \\
	\vdots & \ddots & \vdots & \vdots & \vdots \\
	\Lambda_2^{(\pm)n-2} & \cdots & \Lambda_{n-1}^{(\pm)n-2} & \Lambda_{n}^{(\pm)n-2} & 0 \\
	\hdashline
	\Lambda_2^{(\pm)n-1} & \cdots & \Lambda_{n-1}^{(\pm)n-1} & \Lambda_{n}^{(\pm)n-1} & \fbox{0}
	\end{array}
	\right|   \nonumber \\
	&\!\!\!\!=& \!\!\!\!
	-\left|
	\begin{array}{ccc}
		\!\!1 & \!\!\!\cdots & \!\!\!1 \\
		\!\!\Lambda_2^{(\pm)} & \!\!\!\cdots & \!\!\!\Lambda_{n}^{(\pm)} \\
		\!\!\vdots & \!\!\!\ddots & \!\!\!\vdots \\
        \!\!\Lambda_2^{(\pm)n-2}\!\! & \!\!\!\cdots\!\! & \!\!\!\fbox{$\Lambda_{n}^{(\pm)n-2}$}\!\!
	\end{array}
	\right|
	\!
	\Lambda_{n}^{(\pm)}
	\!
	\left|
	\begin{array}{ccc}
		\!\!1 & \!\!\!\cdots & \!\!\!\fbox{$1$} \\
		\!\!\Lambda_2^{(\pm)} & \!\!\!\cdots & \!\!\!\Lambda_{n}^{(\pm)} \\
		\!\!\vdots & \!\!\ddots & \vdots \\
		\!\!\Lambda_2^{(\pm)n-2}\!\! & \!\!\!\cdots\!\! & \!\!\!\Lambda_{n}^{(\pm)n-2}\!\!
	\end{array}
	\right|^{-1}  \nonumber \\
	&\!\!\!\! = \!\!\!\!&
	-\!
	\Lambda_{n}^{(\pm)}\!
	\left|
	\begin{array}{cccc}
		\!\!1 & \!\!\cdots & \!\!\!1\!\! & \!\!\!\fbox{$0$} \\
		\!\!\Lambda_2^{(\pm)} & \!\!\cdots & \!\!\!\Lambda_{n-1}^{(\pm)} & \!\!\!0 \\
		\!\!\vdots & \!\!\ddots & \!\!\!\vdots & \!\!\!0 \\
		\!\!\Lambda_2^{(\pm)n-2}\!\! & \!\!\cdots\!\! & \!\!\!\Lambda_{n-1}^{(\pm)n-2}\!\! & \!\!\!1
	\end{array}
	\!\right|^{-1}.   \nonumber 
\end{eqnarray}
By continuing the same process $n-1$ times, 
we can prove the statement \eqref{Dn}. 
\hfill $\Box$
\medskip \\
In general case, we can conclude the following result for the asymptotic form :
\begin{eqnarray}
\label{Asymptotic form of J_{n+1}_psi_I}
\widetilde{J}_{n+1}^{~\!(I)}
=
\left|
\begin{array}{cccccc}
1 & \cdots & \psi_I & \cdots & 1 & 1 \\
\Lambda_1^{(\pm)} & \cdots & \psi_I\Lambda_I & \cdots & \Lambda_n^{(\pm)} & 0 \\
\vdots & \ddots & \vdots & \ddots & \vdots & \vdots \\
\Lambda_1^{(\pm)n} & \cdots & \psi_I\Lambda_I^n & \cdots & \Lambda_n^{(\pm)n} & \fbox{0}
\end{array}
\right|
=- \widetilde{\Psi}_{n}^{~\!(I)}\Lambda_{I} (\widetilde{\Psi}_{n}^{~\!(I)})^{-1} D_{n}^{~\!(I)}, ~n \geq 2 ~~~~~ 
\end{eqnarray}
where
\begin{eqnarray}
\label{Qn_I}
\widetilde{\Psi}_{n}^{~\!(I)}\!\!\!&=&\!\!\!
\left(
\begin{array}{cc}
\!\prod\limits_{i=1,i \neq I}^{n}(\lambda_{I}^{(+)}-\lambda_{i}^{(\pm)})~\! a_Ie^{L_I}
&
\!\prod\limits_{i=1,i \neq I}^{n}(\lambda_{I}^{(-)}-\lambda_{i}^{(\pm)})~\!\overline{b}_I~\! e^{-\overline{L}_I}\!
\\
\!\!\!-\!\!\!\prod\limits_{i=1,i \neq I}^{n}(\lambda_{I}^{(+)}-\lambda_{i}^{(\mp)})~\!b_{I}e^{-L_I}
&
\!\!\!\prod\limits_{i=1,i \neq I}^{n}(\lambda_{I}^{(-)}-\lambda_{i}^{(\mp)})~\!\overline{a}_I~\! e^{\overline{L}_I}\!
\end{array}
\right),~~~~~~~~~\\
D_n^{~\!(I)}\!\!\!&=&\!\!\! (-1)^{n-1} \prod_{i=1,i \neq I}^{n}\Lambda_{i}^{(\pm)}.
\label{Dn_I}
\end{eqnarray}
\bigskip \\
{\bf Asymptotic form of $n$-soliton type distribution of Lagrangian density}
\medskip \\
Now we can find that the asymptotic form 
$\widetilde{J}_{n+1}^{~\!(I)} \sim \widetilde{\Psi}_{n}^{~\!(I)}\Lambda_I(\widetilde{\Psi}_{n}^{~\!(I)})^{-1}$ 
is in a very similar form as the 1-soliton solution \eqref{Reduced 1-soliton_3 real spaces} 
with a replacement of $(\psi, \Lambda)$ by $(\widetilde{\Psi}_{n}^{~\!(I)}, \Lambda_{I})$. 
If we impose additional conditions on $\widetilde{\Psi}_{n}^{~\!(I)}$, for instance 
$(\lambda_{i}^{(+)},\lambda_{i}^{(-)})=(\lambda_i,\overline{\lambda}_i)$ for all $i$ (not the unique choice), 
$\widetilde{\Psi}_{n}^{~\!(I)}$ can be simplified in a more concise form :
\begin{eqnarray}
\label{Qn_I_reduced form}
\widetilde{\Psi}_{n}^{~\!(I)}\!\!\!&=&\!\!\!\left(
\begin{array}{cc}
a_I^{~\!\prime} e^{L_I} & \overline{b}_I^{~\!\prime} ~\!e^{-\overline{L}_I} \\
-b_I^{~\!\prime} e^{-L_I} & \overline{a}_I^{~\!\prime}~\! e^{\overline{L}_I}
\end{array}
\right), \\
&&\left\{
\begin{array}{l}
\mbox{(i)}~\!  
\left\{
\begin{array}{l}
a_I^{~\!\prime}:=\prod\limits_{i=1,i \neq I}^{n}(\lambda_{I} - \lambda_{i})\!~a_{I} \\
b_I^{~\!\prime}:=\prod\limits_{i=1,i \neq I}^{n}(\lambda_{I} - \overline{\lambda}_{i})\!~b_{I}
\end{array}
\right.,~
\begin{array}{l}
\mbox{Re}L_i \rightarrow +\infty
\end{array}
\\
\mbox{(ii)}
\left\{
\begin{array}{l}
a_I^{~\!\prime}:=\prod\limits_{i=1,i \neq I}^{n}(\lambda_{I} - \overline{\lambda}_{i})\!~a_{I} \\
b_I^{~\!\prime}:=\prod\limits_{i=1,i \neq I}^{n}(\lambda_{I} - \lambda_{i})\!~b_{I}
\end{array}
\right.,~
\begin{array}{l}
\mbox{Re}L_i \rightarrow -\infty
\end{array}
\end{array}.
\right.  \label{a_I_prime, b_I_prime}
\end{eqnarray}
Comparing $\widetilde{\Psi}_{n}^{~\!(I)}$ with $\psi_{I}$ in \eqref{n-Soliton Solutions_J_n+1},
we find that the only difference between $\widetilde{\Psi}_{n}^{~\!(I)}$ and $\psi_I$ is  the constants $a_I^{~\!\prime}, b_I^{~\!\prime}$ and $a_I, b_I$. This difference leads to a position shift of the principal peak of Lagrangian density (Cf: \eqref{Reduced Lagrangian density_special_3 real spaces}), and we call it phase shift.  
On the other hand, we also find that the gauge fields (given by $\widetilde{J}_{n+1}^{~\!(I)}
=-\widetilde{\Psi}_{n}^{~\!(I)} \Lambda_{I} (\widetilde{\Psi}_{n}^{~\!(I)})^{-1} D_{n}^{~\!(I)}$) are independent of the constant matrix $D_{n}^{~\!(I)}$ because
\begin{eqnarray*}
\left(\partial_{\mu}\widetilde{J}_{n+1}^{~\!(I)}\right)(\widetilde{J}_{n+1}^{~\!(I)})^{-1}
=
\partial_{\mu}\left( \widetilde{\Psi}_{n}^{~\!(I)}\Lambda_{I}(\widetilde{\Psi}_{n}^{~\!(I)})^{-1} \right)
\widetilde{\Psi}_{n}^{~\!(I)}\Lambda_{I}^{-1}(\widetilde{\Psi}_{n}^{~\!(I)})^{-1}.   \nonumber 
\end{eqnarray*} 
Now we can come to a conclusion that the Lagrangian density Tr$F_{\mu\nu}F^{\mu\nu}$ (given by $\widetilde{J}_{n+1}^{~\!(I)}$) is almost the same as \eqref{Reduced Lagrangian density_special_3 real spaces} except for a replacement of $(\psi, \Lambda)$ by $(\psi_I, \Lambda_I)$ and a phase shift factor.  
In other words, if we consider the comoving frame with respect to the $I$-th 1-soliton solution, the asymptotic form of the $n$-soliton solution inherits almost the same features from the $I$-th 1-soliton solution except for a phase shift factor.
The remaining mission is to calculate the phase shift factor explicitly and determine whether it is real-valued or not. 
\medskip \\
{\bf Phase shift}
\medskip \\
Firstly, we take the split signature for example. It is obviously that \eqref{Qn_I_reduced form}  satisfies the requirement of the Ultrahyperbolic space $\mathbb{U}$ because
$(\lambda_{i}^{(+)},\lambda_{i}^{(-)})=(\lambda_i,\overline{\lambda}_i)$ for all $i$ (Cf: Table \ref{table_2}). 
By \eqref{Reduced Lagrangian density_special_3 real spaces}, we have
\begin{eqnarray*}
{\mbox{Tr}} F_{\mu \nu}F^{\mu \nu}=
8\left[(\alpha_I\overline{\beta}_I-\overline{\alpha}_I\beta_I)
(\lambda_I -\overline{\lambda}_I)  \right]^2
\left(2{\mbox{sech}}^2 X_I^{\prime}-3{\mbox{sech}}^4 X_I^{\prime}\right),   \nonumber 
\end{eqnarray*}
where 
$X_I^{\prime}=L_I + \overline{L}_I +\log\left|a_I^\prime/b_I^\prime\right|
=L_I + \overline{L}_I +\log\left|a_I/b_I\right|+\Delta_{I}
=X_I+\Delta_{I}$.  
The phase shift can be calculated explicitly by \eqref{Qn_I} and it is real-valued :
\begin{eqnarray}
\label{Phase shift_G=SU(2)_U}
\Delta_{I}
=
\sum_{i=1,i\neq I}^{n}\varepsilon_i^{(\pm)}
\log 
\left|
\frac{\lambda_I-\lambda_i}{\lambda_I-\overline{\lambda}_i} 
\right|,~~
\left\{
\begin{array}{l}
\varepsilon_i^{(+)}:=+1, ~~ X_{i,i \neq I} \rightarrow +\infty \\
\varepsilon_i^{(-)}:=-1, ~~ X_{i,i \neq I} \rightarrow -\infty
\end{array}.
\right.
\end{eqnarray}
 
For the Euclidean signature, we can take the condition $(\lambda_{i}^{(+)},\lambda_{i}^{(-)})=(\lambda_i,-1/\overline{\lambda}_i)$ for all $i$ (Cf: Table \ref{table_2}) to satisfy the requirement of the Euclidean space $\mathbb{E}$. 
By \eqref{Qn_I}, the resulting phase shift factor is also real-valued :
\begin{eqnarray}
\label{Phase shift_G=SL(2,C)_E}
\Delta_{I}
=
\sum_{i=1, i\neq I}^{n}\varepsilon_i^{(\pm)}
\log 
\left|
\frac{\lambda_I-\lambda_i}{1+\lambda_I\overline{\lambda}_i} 
\right|,~~
\left\{
\begin{array}{l}
\varepsilon_i^{(+)}:=+1, ~~ X_{i,i \neq I} \rightarrow +\infty  \\
\varepsilon_i^{(-)}:=-1, ~~ X_{i,i \neq I} \rightarrow -\infty
\end{array}.
\right.
\end{eqnarray}
For the Minkowski signature, if we don't impose any condition 
on $(\lambda_{i}^{(+)},\lambda_{i}^{(-)})$ as mentioned in Table \ref{table_2} ($(\lambda_{i}^{(+)},\lambda_{i}^{(-)})=(\lambda_i, \mu_i)$), the phase shift factor is complex-valued in general. 
This shortcoming can be solved immediately, 
for example we can take $(\lambda_{i}^{(+)},\lambda_{i}^{(-)})=(\lambda_i,\overline{\lambda}_i)$ or $(\lambda_i,-1/\overline{\lambda}_i)$ like that 
in $\mathbb{U}$ and $\mathbb{E}$, respectively. 
\medskip \\
{
{\bf Reduction to (1+1)-dimensional real slice of $\mathbb{U}$}
\medskip \\
Taking the Ultrahyperbolic space $\mathbb{U}$ for example, we can impose the condition $x^{1}=t$, $x^{2}=x^{4}=0$, $x^{3}=x$ on the space-time coordinates of \eqref{l_imu_U} such that
\begin{eqnarray}
L_i\!\!\!&=& \!\!\!
\frac{1}{\sqrt{2}}
\left(
(\lambda_i\alpha_i+\beta_i)t
+(\lambda_i\alpha_i-\beta_i)x
\right),~i=1,2,...,n.
\end{eqnarray} 
Then we can fix an $I \in \left\{1,2,...,n \right\}$ and choose a complex number $l_{I}$ such that $L_{I}=l_{I}$ and $X_{I}$ is finite. More precisely,  
\begin{eqnarray}
L_{I}&\!\!\!=& \!\!\!\!
\frac{1}{\sqrt{2}}
\left(
(\lambda_I\alpha_I+\beta_I)t
+(\lambda_I\alpha_I-\beta_I)x
\right)
=l_{I} \in \mathbb{C},  \label{L_I=l_I_2D} \\ 
X_{I}&\!\!\!\!=& \!\!\!\! 
l_I + \overline{l}_I +\log\left|a_I/b_I\right|
\in \mathbb{R}.  \label{X_I=0_2D}
\end{eqnarray}
Substituting \eqref{L_I=l_I_2D} into $L_{i}$ and $X_i$ for all $i \neq I$,
we have 
\begin{eqnarray}
X_{i}=
L_i + \overline{L}_i +\log\left|a_i/b_i\right|,~\mbox{where}~
L_i=
\left(\frac{\lambda_i\alpha_i-\beta_i}{\lambda_I\alpha_I-\beta_I}\right)l_I
+\sqrt{2}\left(
\frac{\lambda_i\alpha_i\beta_I-\lambda_I\alpha_I\beta_i}{\lambda_I\alpha_I-\beta_I}
\right)\!t. ~~~
\end{eqnarray}
Now we find that 
\begin{eqnarray}
\left\{
\begin{array}{l}
X_{I} = \mbox{finite} \\
X_{i} \rightarrow \pm \infty ~ \mbox{or} ~\mp \infty 
\end{array}
\right.
~~\mbox{when}~~  t \rightarrow \pm \infty.
\end{eqnarray}
In other words, we choose a comoving frame related to the $I$-th soliton. In this comoving frame, the asymptotic behavior of the $i$-th soliton is dominated by the real number
\begin{eqnarray} 
\frac{\lambda_i\alpha_i\beta_I-\lambda_I\alpha_I\beta_i}{\lambda_I\alpha_I-\beta_I}
+
\frac{\overline{\lambda}_i\overline{\alpha}_i\overline{\beta}_I-\overline{\lambda}_I\overline{\alpha}_I\overline{\beta}_i}{\overline{\lambda}_I\overline{\alpha}_I-\overline{\beta}_I},
\end{eqnarray}
and Tr$F_{\mu\nu}F^{\mu\nu ~\!(i)} \sim 2\mbox{sech}^2X_{i}-3\mbox{sech}^4X_{i} \rightarrow 0$ as $t \rightarrow \pm \infty$.
On the other hand, the whole asymptotic distribution of Lagrangian density behaves as 
\begin{eqnarray}
\label{X_I=Delta_2D}
\mbox{Tr}F_{\mu\nu}F^{\mu\nu ~\!} &\!\!\!\! \sim & \!\!\!\!  2\mbox{sech}^2X^{\prime}_{I}-3\mbox{sech}^4X^{\prime}_{I}, ~~ \nonumber \\
X^{\prime}_{I}    
&\!\!\!\!=& \!\!\!\!
\left\{
\begin{array}{l}
X_{I} + \Delta_{I}^{(+)}  ~ \mbox{ when } ~ t \rightarrow +\infty \\ 
X_{I} + \Delta_{I}^{(-)}  ~ \mbox{ when } ~ t \rightarrow -\infty
\end{array},
\right. ~~ \mbox{In fact}, ~\Delta_{I}^{(-)}=-\Delta_{I}^{(+)}.  
\end{eqnarray}
We find that the position of the principal peak is shifted by $\left|\Delta_{I}^{(\pm)}\right|$ and hence we call $\Delta_I^{(\pm)}$ the phase shift. 

To make the discussions of the multi-soliton scattering more clearly, let us consider the situations of 3-soliton scattering for simplicity.    
If we choose the parameters $\lambda_i$, $\alpha_i$, $\beta_i$, $i=1, 2, 3$ such that $X_1$, $X_2$, $X_3$ are presented as the dash line in the Figure \ref{3-soliton} and the resulting phase shifts are
\begin{eqnarray}
\Delta_{1}^{(+)}
&\!\!\!\!=\!\!\!\!&
-\Delta_{1}^{(-)}
=
-\log\left| \frac{\lambda_1-\lambda_2}{\lambda_1-\overline{\lambda}_2} \right|
-\log\left| \frac{\lambda_1-\lambda_3}{\lambda_1-\overline{\lambda}_3} \right|,~~~~   \\
\Delta_{2}^{(+)}
&\!\!\!\!=\!\!\!\!&
-\Delta_{2}^{(-)}
=
+\log\left| \frac{\lambda_2-\lambda_1}{\lambda_2-\overline{\lambda}_1} \right|
-\log\left| \frac{\lambda_2-\lambda_3}{\lambda_3-\overline{\lambda}_3} \right|,~~~~ \\
\Delta_{3}^{(+)}
&\!\!\!\!=\!\!\!\!&
-\Delta_{3}^{(-)}
=
+\log\left| \frac{\lambda_3-\lambda_1}{\lambda_3-\overline{\lambda}_1} \right|
+\log\left| \frac{\lambda_3-\lambda_2}{\lambda_3-\overline{\lambda}_2} \right|,~~~~ 
\end{eqnarray}
then all the situations of the 3-soliton scattering can be presented as the following Figure \ref{3-soliton}. Without loss of generality, here we just show one of the situations : $\Delta_{1}^{(+)}<0$, $\Delta_{2}^{(+)}<0$, and $\Delta_{3}^{(+)}>0$. 
\bigskip \\
\begin{figure}[h]
	\begin{center}
		\includegraphics[width=100mm]{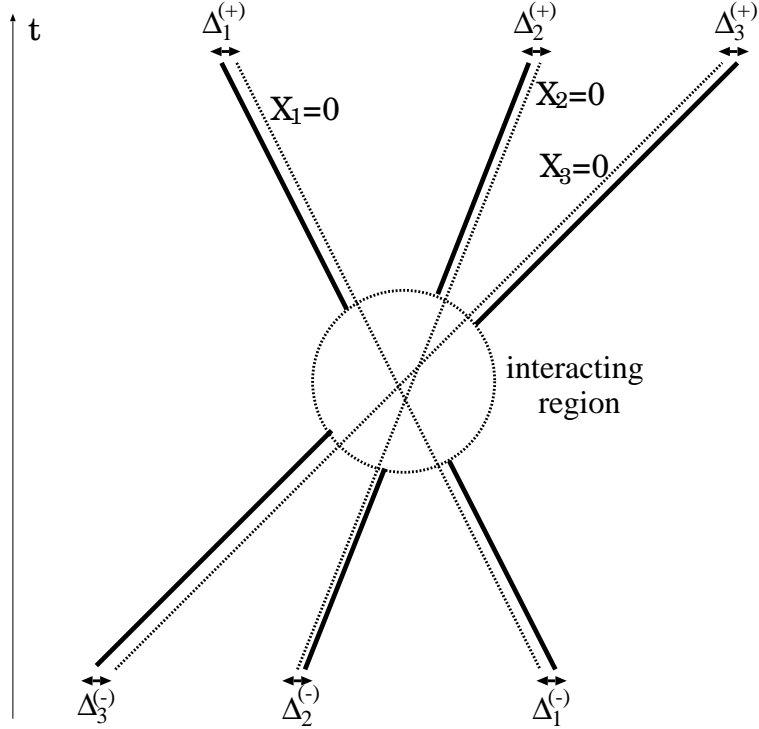}
	\end{center}
	\caption{3-soliton scattering}
	\label{3-soliton}
\end{figure}
\\
{\bf Summary}
\smallskip \\
In this section, we analyze the asymptotic behavior of Lagrangian density for $n$-soliton solutions \eqref{n-Soliton Solutions_J_n+1} and find that it can be interpreted as 
$n$ isolated soliton walls in the asymptotic region. 
Therefore, we can come to a conclusion that the Lagrangian density of $n$-soliton solutions
would possess $n$ overlapping principal peaks within the soliton-scattering region and the corresponding interpretation is $n$ intersecting soliton walls. 
This fact is a well-known feature for the KP multi-solitons, but a new insight for 
the ASDYM multi-solitons.  
Combining with Theorem \ref{Thm_6.3}, the $n$ intersecting soliton walls can be embedded into $G=\mathrm{SU}(2)$ gauge theory on the Ultrahyperbolic 
space $\mathbb{U}$ 
and hence they can be interpreted as some physical objects (e.g. $n$ intersecting branes) in open $\mathrm{N=2}$ string theories.

\newpage

\section{ASDYM 1-Solitons ($G=\mathrm{SU}(3)$) and Multi-Solitons ($G=\mathrm{SL}(3,\mathbb{C})$)}
\label{Section 7}

Inspired directly by the results of previous section, a nature question to ask is whether the soliton wall can be found in $G=\mathrm{SL}(3,\mathbb{C})$ gauge theory or not.
Firstly, we propose an ansatz to construct $G=\mathrm{SU}(3)$ 1-soliton solution on the Ultrahyperbolic space $\mathbb{U}$ which gives one single soliton wall successfully.
For such kind of $G=\mathrm{SU}(3)$ 1-soliton solutions, we apply $n$ iterations of the Darboux transformation \eqref{Darboux transf} to construct $n$-soliton solution and show that the resulting $n$-soliton solution gives $n$ intersecting soliton walls as well. Moreover, we also show that the gauge group can be $G=\mathrm{SU}(3)$ for each soliton wall in the asymptotic region. 

\subsection{ASDYM 1-Solitons on the Ultrahyperbolic space $\mathbb{U}$, $G=\mathrm{SU}(3)$}
\label{Section 7.1}

{\bf 1-Soliton Solution}
\medskip \\
Let us begin with the following ansatz of $\psi(\Lambda)$ that satisfies the initial linear system \eqref{Initial linear system_J_1=I} and consider the corresponding 1-soliton solution :
\begin{eqnarray}
\label{Soliton solution_U_G=SU(3)}
J\!\!\!\!&=\!\!\!\!&
\left|
\begin{array}{ccc}
\!\psi & \!\!1 \\
\!\psi\Lambda & \!\!\fbox{0}
\end{array}
\right|
=
-\psi\Lambda \psi^{-1} ~~
\left\{
\begin{array}{l}
\psi
:=\left(
\begin{array}{ccc}
a_1e^{L}+a_2e^{-L}
& 
\overline{b}_1e^{\overline{L}} + \overline{b}_2e^{-\overline{L}}
&
0
\\ 
\!\!\!\!-b_1e^{L} - b_2e^{-L}
& 
\overline{a}_1e^{\overline{L}}
+\overline{a}_2e^{-\overline{L}}  
&
c_1e^{L} + c_2e^{-L}
\\
0
&
\!\!\!\!-\overline{c}_1e^{\overline{L}} - \overline{c}_2e^{-\overline{L}}
&
a_1e^{L}+a_2e^{-L}
\end{array}
\!\!\right),~~ 
\\ 
\Lambda:=
\left(
\begin{array}{ccc}
\lambda & 0 & 0\\
0 & \overline{\lambda} & 0 \\
0 & 0 & \lambda
\end{array}
\right)
\end{array}
\right.  \\
&&\mbox{where}~
L=(\lambda \alpha) z
+\beta\widetilde{z}
+(\lambda \beta) w
+\alpha \widetilde{w},~~ z,\widetilde{z}, w, \widetilde{w} \in \mathbb{R}   \nonumber \\
&&~~~~~~~~~~~=\frac{1}{\sqrt{2}}
\left[
(\lambda\alpha+\beta)x^1
+(\lambda\beta-\alpha)x^2
+(\lambda\alpha-\beta)x^3
+(\lambda\beta+\alpha)x^4 
\right],    ~~~~~~~~~~~~~ \nonumber  \\
&&~~~~~~~~~~~~~~~~ a_1, a_2, b_1, b_2, \alpha, \beta, \lambda \in \mathbb{C}, \nonumber 
\end{eqnarray} 
or equivalently,
\begin{eqnarray*}
J&\!\!\!=&\!\!\! \frac{-1}{\Delta}
\!\left(\!
\begin{array}{ccc}
\lambda(|A|^2+|C|^2)+\overline{\lambda}|B|^2 & -(\lambda-\overline{\lambda})A\overline{B} & (\lambda-\overline{\lambda})\overline{B}C \\
 -(\lambda-\overline{\lambda})\overline{A}B & \lambda(|B|^2+|C|^2)+\overline{\lambda}|A|^2 & (\lambda-\overline{\lambda})\overline{A}C \\
(\lambda-\overline{\lambda})B\overline{C} & (\lambda-\overline{\lambda})A\overline{C} & \lambda(|A|^2+|B|^2)+\overline{\lambda}|C|^2
\end{array}
\!\!\!\right), ~~~~~~~ \nonumber \\
&& \!\Delta:=|A|^2\!+\!|B|^2\!+\!|C|^2,~ A:=a_1e^{L}\!+\!a_2e^{-L},~ B:= b_1e^{L}\!+\!b_2e^{-L},~ C:=c_1e^{L}\!+\!c_2e^{-L}.   \nonumber 
\end{eqnarray*}
We find that
\begin{eqnarray*}
\det(J)=-\left|\lambda\right|^2\lambda, ~~J^{\dagger}J=JJ^{\dagger}=\left|\lambda\right|^2I, ~I : 3 \times 3 ~\mbox{identity matrix}.   \nonumber 
\end{eqnarray*}
By Jacobi's formula \eqref{Jacobi's formula} and Proposition \ref{Prop_3.9} ($h=J$, $\widetilde{h}=I$), we can show that the gauge fields are traceless and anti-hermitian, therefore, the gauge group can be $G=\mathrm{SU}(3)$ on the Ultrahyperbolic space $\mathbb{U}$.  
In fact, we also can calculate the gauge fields explicitly by substituting the $J$-matrix \eqref{Soliton solution_U_G=SU(3)} into \eqref{Gauge Fields_U_A_1=A_3, A_2=A_4} to get the results as follows :
\medskip \\
{\bf Gauge Fields}
\begin{eqnarray}
\label{Gauge fields_U_soliton_G=SU(3)}
A_{\mu}
&\!\!\!\!=& \!\!\!\!
\frac{\sqrt{2}(\lambda-\overline{\lambda})}{\Delta^2}
\!\left(
\!\!\!\begin{array}{c:c:c} 
\begin{array}{l}
~(\overline{p}_{\mu}^{(1)}A-\overline{p}_{\mu}^{(2)}C)B \\
\!\!\!+(p_{\mu}^{(1)}\overline{A}-p_{\mu}^{(2)}\overline{C})\overline{B}
\end{array} 
&
\begin{array}{l}
~(\overline{p}_{\mu}^{(1)}A-\overline{p}_{\mu}^{(2)}C)A \\
\!\!\!-(p_{\mu}^{(1)}\overline{B}-p_{\mu}^{(3)}\overline{C})\overline{B}
\end{array} 
&
\begin{array}{l}
\!\!\!-(\overline{p}_{\mu}^{(1)}A-\overline{p}_{\mu}^{(2)}C)C \\
\!\!\!-(p_{\mu}^{(2)}\overline{B}-p_{\mu}^{(3)}\overline{A})\overline{B}
\end{array} 
\smallskip
\\
\hdashline
\smallskip 
\begin{array}{l}
\!\!\!-(\overline{p}_{\mu}^{(1)}B-\overline{p}_{\mu}^{(3)}C)B \\
\!\!\!+(p_{\mu}^{(1)}\overline{A}-p_{\mu}^{(2)}\overline{C})\overline{A}
\end{array}
&
\begin{array}{l}
\!\!\!-(\overline{p}_{\mu}^{(1)}B-\overline{p}_{\mu}^{(3)}C)A \\
\!\!\!-(p_{\mu}^{(1)}\overline{B}-p_{\mu}^{(3)}\overline{C})\overline{A}
\end{array}
&
\begin{array}{l}
~(\overline{p}_{\mu}^{(1)}B-\overline{p}_{\mu}^{(3)}C)C \\
\!\!\!-(p_{\mu}^{(2)}\overline{B}-p_{\mu}^{(3)}\overline{A})\overline{A}
\end{array}
\\
\hdashline 
\begin{array}{l}
\!\!\!-(\overline{p}_{\mu}^{(2)}B-\overline{p}_{\mu}^{(3)}A)B \\
\!\!\!-(p_{\mu}^{(1)}\overline{A}-p_{\mu}^{(2)}\overline{C})\overline{C}
\end{array}
&
\begin{array}{l}
\!\!\! -(\overline{p}_{\mu}^{(2)}B-\overline{p}_{\mu}^{(3)}A)A \\
\!\!\! +(p_{\mu}^{(1)}\overline{B}-p_{\mu}^{(3)}\overline{C})\overline{C}
\end{array}
& 
\begin{array}{l}
~(\overline{p}_{\mu}^{(2)}B-\overline{p}_{\mu}^{(3)}A)C \\
\!\!\!+(p_{\mu}^{(2)}\overline{B}-p_{\mu}^{(3)}\overline{A})\overline{C}
\end{array}
\end{array}  
\!\!\right)~~~~~~~ \\
&&
p_{\mu}^{(i)}~(i=1,2,3):=
\left\{
\begin{array}{l}
\alpha\varepsilon_0^{(i)} ~~~~~
{\mbox{if}} ~~ \mu=1,3 \\
\beta\varepsilon_0^{(i)} ~~~~~
{\mbox{if}} ~~ \mu=2,4 \nonumber \\
\varepsilon_0^{(1)}:=a_1b_2 - a_2b_1,~\varepsilon_0^{(2)}:=b_1c_2 - b_2c_1,~ \varepsilon_0^{(3)}:=c_1a_2 - c_2a_1
\end{array}
\right. \\
&&(~\!\mbox{In fact},~\varepsilon_{0}^{(i)}~ \mbox{satisfy a cyclic relation}:~\varepsilon_{0}^{(1)}C+\varepsilon_{0}^{(2)}A+\varepsilon_{0}^{(3)}B=0. ~\!) \nonumber 
\end{eqnarray}
Note that the gauge fields $A_{\mu}$ are nonzero if $\lambda \notin \mathbb{R}$. Furthermore, all of them are anti-hermitian and traceless, therefore, the group group is actually $G=\mathrm{SU}(3)$. 
After a bit lengthy calculation, we can obtain the Lagrangian density :
\medskip \\
{\bf{Periodic 1-Soliton type distribution of Lagrangian Density}}
\begin{eqnarray}
\label{Lagrangian density_U_G=SU(3)}
{\mbox{Tr}} F_{\mu \nu}F^{\mu \nu}\!\!\!&=&\!\!\!
8\left[(\alpha\overline{\beta} \!-\! \overline{\alpha}\beta)
(\lambda \!-\!\overline{\lambda})\right]^2   
\left[\left|\varepsilon_0^{(1)} \right|^2+ 
\left|\varepsilon_0^{(2)} \right|^2+
\left|\varepsilon_0^{(3)} \right|^2 \right]
\nonumber \\
&&\times\left\{\frac{2\varepsilon_1\widetilde{\varepsilon}_1
	\sinh^2 X_1
	-2\left|\varepsilon_2 \right|^2
	\sinh^2 X_2 
	-\left[\left|\varepsilon_0^{(1)} \right|^2+ 
	\left|\varepsilon_0^{(2)} \right|^2+
	\left|\varepsilon_0^{(3)} \right|^2 \right]}
{\left[(\varepsilon_1\widetilde{\varepsilon}_1)^{\frac12}
	\cosh X_1 
	+\left|\varepsilon_2 \right|
	\cosh X_2\right]^4} 
\right\},~~~~~~~~	\\
&&X_1=\overline{L} + L+\dfrac12 \log(\varepsilon_1/\widetilde{\varepsilon}_1),~
X_2=\overline{L} - L+\dfrac12 \log(\varepsilon_2/\overline{\varepsilon}_2),  \label{X_1, X_2_SU(3)}\\
&&
\left\{
\begin{array}{l}
\varepsilon_0^{(1)}:=a_1 b_2 - a_2b_1,~ \varepsilon_0^{(2)}:=b_1 c_2 - b_2c_1,~ \varepsilon_0^{(3)}:=c_1 a_2 - c_2a_1,~  \\
\varepsilon_1=\left|a_1\right|^2 + \left|b_1 \right|^2 + \left|c_1 \right|^2,~
\widetilde{\varepsilon}_1=\left|{a}_2 \right|^2 + \left|{b_2} \right|^2 + \left|{c_2} \right|^2 \in \mathbb{R},  \\
\varepsilon_2=a_1\overline{a}_2 + b_1\overline{b}_2 + c_1\overline{c}_2.
\end{array}  \label{epsilons_U_G=SU(3)}
\right.\\
&&\!\!\!\!\!\!\!\!\!\!\!\mbox{(Note that the Lagrangian density is nonzero if $\alpha\overline{\beta} \notin \mathbb{R}$, $\lambda \notin \mathbb{R}$, and $\varepsilon_{0}^{(i)} \neq 0$.)} \nonumber 
\end{eqnarray}
{\bf Pure 1-Soliton type distribution of Lagrangian density}
\medskip \\
In fact, $\varepsilon_{0}^{(i)},~ \varepsilon_{1},~ \varepsilon_{2}$ satisfy the relation : 
\begin{eqnarray}
\label{Relation of epsilons_U_G=SU(3)}
\varepsilon_1\widetilde{\varepsilon}_1 -\vert \varepsilon_2\vert^2 
=\left|\varepsilon_0^{(1)} \right|^2+ 
\left|\varepsilon_0^{(2)} \right|^2+
\left|\varepsilon_0^{(3)} \right|^2. 
\end{eqnarray}
If we choose suitable parameters $a_1, a_2, b_1, b_2, c_1, c_2$ in \eqref{Soliton solution_U_G=SU(3)} such that $\varepsilon_{2}=0$, then the periodic part of the Lagrangian density \eqref{Lagrangian density_U_G=SU(3)} can be removed completely. On the other hand, all the epsilon factors in the numerator and denominator of \eqref{Lagrangian density_U_G=SU(3)} can be cancel out and then we get a more concise form :
\begin{eqnarray}
\label{Reduced Lagrangian density_U_G=SU(3)}
{\mbox{Tr}} F_{\mu \nu}F^{\mu \nu}&\!\!\!\!=& \!\!\!\!
8\left[(\alpha\overline{\beta}-\overline{\alpha}\beta)
(\lambda -\overline{\lambda})  \right]^2
\left(2{\mbox{sech}}^2 X-3{\mbox{sech}}^4 X\right),  \\
X&\!\!\!\!=& \!\!\!\! L + \overline{L} + \frac{1}{2}\log\left[(\left|a_1 \right|^2 + \left|b_1 \right|^2 + \left|c_1 \right|^2) / (\left|a_2 \right|^2+\left|b_2 \right|^2 +\left|c_2 \right|^2)\right].  
\end{eqnarray}
Note that the slight difference between the Lagrangian density \eqref{Reduced Lagrangian density_U_G=SU(3)} and \eqref{Reduced Lagrangian density_U} is a phase shift factor. Therefore, we can interpret it as soliton wall of $G=\mathrm{SU}(3)$ version. 
In particular, we can choose $a_1=a, b_2=b, c_2=c,a_2=b_1=c_1=0$ to satisfy $\varepsilon_{2}=0$ (Cf: \eqref{epsilons_U_G=SU(3)}). In this case, the 1-soliton solution is
\begin{eqnarray}
\label{Reduced Soliton solution_U_G=SU(3)}
J\!\!\!\!&=\!\!\!\!&
\left|
\begin{array}{ccc}
\!\psi & \!\!1 \\
\!\psi\Lambda & \!\!\fbox{0}
\end{array}
\right|
=
-\psi\Lambda \psi^{-1},~
\psi
=\left(
\begin{array}{ccc}
\!\!ae^{L} & \overline{b}e^{-\overline{L}} & 0 \\ 
\!\!- be^{-L} & \overline{a}e^{\overline{L}}  & ce^{-L} \\
\!\!0 & -\overline{c}e^{-\overline{L}} & ae^{L} 
\end{array}
\!\!\right),~ 
\Lambda=
\left(
\begin{array}{ccc}
\lambda & 0 & 0\\
0 & \overline{\lambda} & 0 \\
0 & 0 & \lambda
\end{array}
\right). ~~~~~~~~
\end{eqnarray}
\subsection{Asymptotic behavior of the ASDYM Multi-Solitons, $G=\mathrm{SL(3, \mathbb{C})}$}
\label{Section 7.2}

Let us prepare $n$ different $\psi_i(\Lambda_i)$  $(i=1,2,...n)$ that are all in the same form of \eqref{Reduced Soliton solution_U_G=SU(3)}. After applying $n$ iterations of the Darboux transformation \eqref{Darboux transf}, we get the following $n$-soliton solution.
\medskip \\ 
{\bf $n$-Soliton Solution} 
\begin{eqnarray}
\label{n-Soliton Solutions_J_n+1_G=SU(3)}
J_{n+1}=
\left|
\begin{array}{cccc}
\psi_1&\cdots&\psi_n& 1\\
\psi_1\Lambda_1&\cdots &\psi_n\Lambda_n& 0\\
\vdots   && \vdots& \vdots\\
\psi_1\Lambda_1^{n}&\cdots& \psi_n\Lambda_n^{n}& \fbox{$0$}
\end{array}\right|, ~~
\left\{
\begin{array}{l}
\psi_i
=\left(
\begin{array}{ccc}
\!\!a_ie^{L_i} & \overline{b_i}e^{-\overline{L_i}} & 0 \\ 
\!\!- b_ie^{-L_i} & \overline{a_i}e^{\overline{L_i}}  & c_ie^{-L_i} \\
\!\!0 & -\overline{c_i}e^{-\overline{L_i}} & a_ie^{L_i} 
\end{array}
\!\!\right),~ \\
\Lambda_i=
\left(
\begin{array}{ccc}
\lambda_i & 0 & 0\\
0 & \overline{\lambda_i} & 0 \\
0 & 0 & \lambda_i
\end{array}
\right), ~~~ \\
L_i
=\lambda_i \alpha_i z
+\beta_i\widetilde{z}
+\lambda_i \beta_i w
+\alpha_i\widetilde{w}, \\
a_i, b_i, \alpha_i,\beta_i, \lambda_i \in \mathbb{C},~
i=1,2,\cdots,n.
\end{array}
\right. ~~~~
\end{eqnarray}
Our aim is to discuss the asymptotic behavior of the $n$-soliton solution \eqref{n-Soliton Solutions_J_n+1_G=SU(3)} and show that it can be interpreted as $n$ intersecting soliton walls.

Firstly, we pick an $I\in\left\{1, 2, ..., n\right\}$
and keep $L_I$ (and $\overline{L}_I$) to be finite. 
In fact, this practice is equivalent to consider a comoving frame with respect to
the $I$-th 1-soliton solution. 
On the other hand, we can use the multiplication rule \eqref{right multiplication law} to eliminate common factors in each column of $(\psi_i, \psi_i\Lambda_i, ... \psi_i\Lambda_i^{n})^{T}$ ($i \neq I$) to get the following two equivalent expressions of the $n$-soliton solution \eqref{n-Soliton Solutions_J_n+1_G=SU(3)} :
\begin{eqnarray}
J_{n+1} 
&\!\!\!=& \!\!\!
\left|
\begin{array}{cccccc}
\widetilde{\psi}_1 & \!\!\cdots & \!\!\psi_I & \!\!\cdots & \!\!\widetilde{\psi}_n & 1 \\
\widetilde{\psi}_1 \Lambda_1 & \!\!\cdots & \!\!\psi_I\Lambda_I & \!\!\cdots & \!\!\widetilde{\psi}_n\Lambda_n & 0 \\
\vdots & \!\!\ddots & \!\!\vdots & \!\!\ddots & \!\!\vdots & \vdots \\
\widetilde{\psi}_1\Lambda_1^n & \!\!\cdots & \!\!\psi_I\Lambda_I^n & \!\!\cdots & \!\!\widetilde{\psi}_n\Lambda_n^n & \fbox{0}
\end{array}
\right|,   \nonumber 
~\\
~\widetilde{\psi}_{i,~\! i \neq I}
&\!\!\! :=& \!\!\!
\left\{
\begin{array}{l}
\mbox{(i)}~
\left(
\begin{array}{ccc}
1 & (\overline{b}_i / \overline{a}_i)e^{-2\overline{L}_i} & 0 \\
-(b_i/a_i)e^{-2L_i} & 1  & (c_i/a_i)e^{-2L_i}  \\
0 & -(\overline{c}_i / \overline{a}_i)e^{-2\overline{L}_i} & 1
\end{array}
\right)  \\
\mbox{(ii)}~\!
\left(
\begin{array}{ccc}
(a_ie^{2L_i} & \overline{b}_i & 0 \\
-1 & \overline{a}_ie^{2\overline{L}_i} & 1  \\
0 & -\overline{c}_i & (a_i/c_i)e^{2L_i}
\end{array}
\right)
\end{array}.
\right.  \nonumber 
\end{eqnarray}
By using the same technique as \eqref{Asymptotic of ReL_i} to analyze the asymptotic behavior, we have
\begin{eqnarray}
\label{Asymptotic form of J_n+1_C_i_G=SU(3)}
J_{n+1} ~\stackrel{r \rightarrow \infty}{\longrightarrow}~
\left|
\begin{array}{cccccc}
C_1 & \!\!\cdots & \!\!\psi_I & \!\!\cdots & \!\!C_n & 1 \\
C_1 \Lambda_1 & \!\!\cdots & \!\!\psi_I\Lambda_I & \!\!\cdots & \!\!C_n\Lambda_n & 0 \\
\vdots & \!\!\ddots & \!\!\vdots & \!\!\ddots & \!\!\vdots & \vdots \\
C_1\Lambda_1^n & \!\!\cdots & \!\!\psi_I\Lambda_I^n & \!\!\cdots & \!\!C_n\Lambda_n^n & \fbox{0}
\end{array}
\right|,  \nonumber 
~
~C_{i,~\!i \neq I}=
\left\{
\begin{array}{l}
\mbox{(i)}~~
\left(
\begin{array}{ccc}
1 & 0 & 0 \\
0 & 1 & 0 \\
0 & 0 & 1
\end{array}
\right)  \\
\mbox{(ii)}~\!
\left(
\begin{array}{ccc}
0 & \overline{b}_i & 0 \\
-1 & 0 & 1 \\
0 & -\overline{c}_i & 0
\end{array}
\right)
\end{array}.
\right. ~
\end{eqnarray}
Since
\begin{eqnarray}
C_i \Lambda_i^k = 
\left(
\begin{array}{ccc}
0 & \overline{b}_i & 0  \\
-1 & 0 & 1 \\
0 & -\overline{c}_i & 0 
\end{array}
\right)
\left(
\begin{array}{ccc}
\lambda_i^k & 0 & 0 \\
0 & \overline{\lambda}_i^k & 0 \\
0 & 0 & \lambda_i^k
\end{array}
\right)
=
\left(
\begin{array}{ccc}
\overline{\lambda}_i^k & 0 & 0 \\
0 &\lambda_i^k & 0 \\
0 & 0 &  \overline{\lambda}_i^k 
\end{array}
\right)
\left(
\begin{array}{ccc}
0 & \overline{b}_i & 0  \\
-1 & 0 & 1 \\
0 & -\overline{c}_i & 0 
\end{array}
\right),   \nonumber 
\end{eqnarray}
we can use the right multiplication rule \eqref{right multiplication law} to remove all the right common factors $C_{i, i \neq I}$. 
That is,
\begin{eqnarray}
&&J_{n+1} ~\stackrel{ r \rightarrow \infty}{\longrightarrow}~ 
\widetilde{J}_{n+1}^{~\!(I)}:=
\left|
\begin{array}{cccccc}
1 & \cdots & \psi_I & \cdots & 1 & 1 \\
\Lambda_1^{(\pm)} & \cdots & \psi_I\Lambda_I & \cdots & \Lambda_n^{(\pm)} & 0 \\
\vdots & \ddots & \vdots & \ddots & \vdots & \vdots \\
\Lambda_1^{(\pm)n} & \cdots & \psi_I\Lambda_I^n & \cdots & \Lambda_n^{(\pm)n} & \fbox{0}
\end{array}
\right|,  \nonumber  \\
&& \Lambda_{i,~ i \neq I}^{(\pm)}:=
\left(
\begin{array}{ccc}
\lambda_i^{(\pm)} & 0 & 0 \\
0 & \lambda_i^{(\mp)} & 0 \\
0 & 0 & \lambda_i^{(\pm)}
\end{array}
\right)
=
\left\{
\begin{array}{ll}
\mbox{(i)}~\!	\left(
\begin{array}{ccc}
\lambda_i^{(+)} & 0 & 0 \\
0 & \lambda_i^{(-)} & 0 \\
0 & 0 & \lambda_i^{(+)}
\end{array}
\right),
& \mbox{Re}L_i\rightarrow +\infty  \\
\mbox{(ii)} \left(
\begin{array}{ccc}
\lambda_i^{(-)} & 0 & 0 \\
0 & \lambda_i^{(+)} & 0 \\
0 & 0 & \lambda_i^{(-)}
\end{array}
\right),
& \mbox{Re}L_i\rightarrow -\infty
\end{array},
\right. \nonumber ~~~~~~ \\
&&~~~~~~~~~~~~~~~\mbox{where}~~ (\lambda_i^{(+)}, \lambda_i^{(-)}):=(\lambda_i, \overline{\lambda_i}).  \nonumber 
\end{eqnarray}
By a similar proof of Proposition \ref{Prop_6.4}, we can get the following asymptotic form $\widetilde{J}_{n+1}^{~\!(I)}$ : 
\begin{eqnarray}
\label{Asymptotic form of J_{n+1}_psi_I_G=SU(3)}
\widetilde{J}_{n+1}^{~\!(I)}
=
\left|
\begin{array}{cccccc}
1 & \cdots & \psi_I & \cdots & 1 & 1 \\
\Lambda_1^{(\pm)} & \cdots & \psi_I\Lambda_I & \cdots & \Lambda_n^{(\pm)} & 0 \\
\vdots & \ddots & \vdots & \ddots & \vdots & \vdots \\
\Lambda_1^{(\pm)n} & \cdots & \psi_I\Lambda_I^n & \cdots & \Lambda_n^{(\pm)n} & \fbox{0}
\end{array}
\right|
=-\widetilde{\Psi}_{n}^{~\!(I)} \Lambda_{I} (\widetilde{\Psi}_{n}^{~\!(I)})^{-1} D_{n}^{~\!(I)}, ~n \geq 2 ~~~~~ 
\end{eqnarray}
where
\begin{eqnarray*}
\label{Qn_I_G=SU(3)}
&&\!\!\!\!\!\!\!\!\!\! \widetilde{\Psi}_{n}^{~\!(I)}
\!=\!\left(
\begin{array}{ccc}
\!\!\!\prod\limits_{i=1,i \neq I}^{n}(\lambda_{I}^{(+)}\!\!-\!\lambda_{i}^{(\pm)})~\!a_Ie^{L_I}
&
\!\prod\limits_{i=1,i \neq I}^{n}(\lambda_{I}^{(-)}\!\!-\!\lambda_{i}^{(\pm)})~\!\overline{b}_I~\! e^{-\overline{L}_I}\!
&
\!\!\! 0
\\
\!\! -\!\!\!\prod\limits_{i=1,i \neq I}^{n}(\lambda_{I}^{(+)}\!\!-\!\lambda_{i}^{(\mp)})~\!b_{I}e^{-L_I}
&
\!\!\! \prod\limits_{i=1,i \neq I}^{n}(\lambda_I^{(-)}\!\!-\!\lambda_{i}^{(\mp)})~\!\overline{a}_I~\! e^{\overline{L}_I}\!
&
\! \prod\limits_{i=1,i \neq I}^{n}(\lambda_{I}^{(+)}\!\!-\!\lambda_{i}^{(\mp)})~\!c_{I}e^{-L_I}
\\
\!\!\! 0
&
\!\! -\!\!\!\prod\limits_{i=1,i \neq I}^{n}(\lambda_{I}^{(-)}\!\!-\!\lambda_{i}^{(\pm)})~\!\overline{c}_I~\! e^{-\overline{L}_I}\!
&
\!\!\! \prod\limits_{i=1,i \neq I}^{n}(\lambda_{I}^{(+)}\!\!-\!\lambda_{i}^{(\pm)})~\!a_{I}e^{L_I}
\end{array}
\!\!\!\!\right), ~~~~~~~~~ \nonumber \\
&&\!\!\!\!\!\!\!\!\!\!\! D_n^{~\!(I)}
\!=\! (-1)^{n-1} \prod\limits_{i=1,i \neq I}^{n}\Lambda_{i}^{(\pm)}.  \nonumber 
\end{eqnarray*}
Note that we can rewrite $\widetilde{\Psi}_{n}^{~\!(I)}$ as
\begin{eqnarray}
\label{Qn_I_reduced form_G=SU(3)}
\widetilde{\Psi}_{n}^{~\!(I)}
\!\!\!&=&\!\!\!\left(
\begin{array}{ccc}
a_I^{~\!\prime} e^{L_I} & \overline{b}_I^{~\!\prime} ~\!e^{-\overline{L}_I} & 0 \\
-b_I^{~\!\prime} e^{-L_I} & \overline{a}_I^{~\!\prime}~\! e^{\overline{L}_I} & c_I^{~\!\prime} e^{-L_I} \\
0 & -\overline{c}_I^{~\!\prime}~\! e^{-\overline{L}_I} & a_I^{~\!\prime} e^{L_I}
\end{array}
\right)
\end{eqnarray}
which is in the same form as \eqref{Reduced Soliton solution_U_G=SU(3)}. In addition,
\begin{eqnarray}
\label{Det(widetilde(J_n+1)^(I)}
\det(\widetilde{J}_{n+1}^{~\!(I)})
&\!\!\!=& \!\!\!
\det\left(-\widetilde{\Psi}_{n}^{~\!(I)}\Lambda_I (\widetilde{\Psi}_{n}^{~\!(I)})^{-1}D_{n}^{~\!(I)}\right)
=\det\left(-\Lambda_ID_{n}^{~\!(I)}\right)
=
\det\left((-1)^{n}\Lambda_{I}\!\!\prod_{i=1,i \neq I}^{n}\Lambda_{i}^{(\pm)}\right)  \nonumber \\
&\!\!\!=&\!\!\!
(-1)^{n}\prod_{i=1}^{n}\left| \lambda_i \right|^2 \lambda_{i}^{(\pm)}, 
~~\left(\lambda_{i}^{(+)}, \lambda_{i}^{(-)}\right):=
\left\{
\begin{array}{l}
\left(\lambda_i, \overline{\lambda}_i\right), ~~ \mbox{if}~~ i \neq I \\
\left(\lambda_I, \lambda_I \right), ~~\mbox{otherwise}
\end{array}.
\right.
\end{eqnarray}
We also find that
\begin{eqnarray}
\label{Psi^{dagger}Psi_SU(3)}
\left(\widetilde{\Psi}_{n}^{~\!(I)}\right)^{\dagger}\widetilde{\Psi}_{n}^{~\!(I)}
=
\left(\!\!
\begin{array}{ccc}
\begin{array}{l}
~~\!\left| a_{I}^{~\!\prime}\right|^2e^{L_{I}+\overline{L}_{I}}    \\
\!\!+\left| b_{I}^{~\!\prime}\right|^2e^{-(L_{I}+\overline{L}_{I})}
\end{array}
& \!\!\!\!0 & \!\!\!\!\!\!\! -\overline{b}_{I}^{~\!\prime}c_{I}^{~\!\prime}e^{-(L_{I}+\overline{L}_{I})} \\
0 & \!\!\!\!\begin{array}{l}
~\!\left| a_{I}^{~\!\prime}\right|^2 e^{L_{I}+\overline{L}_{I}} \\
\!\!\!\!+ \left(\left| b_{I}^{~\!\prime}\right|^2\!+\!\left| c_{I}^{~\!\prime} \right|^2\right)e^{-(L_{I}+\overline{L}_{I})}
\end{array} & \!\!\!\!0 \\
\!\!-b_{I}\overline{c}_{I}^{~\!\prime}e^{-(L_{I}+\overline{L}_{I})} & \!\!\!\!0 & 
\begin{array}{l}
~\!\left| a_{I}^{~\!\prime}\right|^2e^{L_{I}+\overline{L}_{I}}  \\
\!\!\!\!+ \left| c_{I}^{~\!\prime}\right|^2e^{-(L_{I}+\overline{L}_{I})}
\end{array}
\end{array}
\!\!\!\right)~~~
\end{eqnarray}
is commutative with $\Lambda_{I}$ and $\Lambda_{I}^{\dagger}$ by direct calculation. \\
Now we have
\begin{eqnarray}
\widetilde{J}_{n+1}^{~\!(I)~\!\dagger}\widetilde{J}_{n+1}^{~\!(I)}
&\!\!\!=& \!\!\! 
D_{n}^{~\!(I)~\!\dagger}(\widetilde{\Psi}_{n}^{~\!(I)})^{-\dagger}\Lambda_{I}^{\dagger}
\underbrace{(\widetilde{\Psi}_{n}^{~\!(I)})^{\dagger}
\widetilde{\Psi}_{n}^{~\!(I)}}\Lambda_{I}(\widetilde{\Psi}_{n}^{~\!(I)})^{-1}D_{n}^{~\!(I)} \nonumber  \\
&& ~~~~~~~~~~~~(\mbox{$(\widetilde{\Psi}_{n}^{~\!(I)})^{\dagger}
	\widetilde{\Psi}_{n}^{~\!(I)}$ is commutative with $\Lambda_I$ by \eqref{Psi^{dagger}Psi_SU(3)}}.) \nonumber \\
&\!\!\!=& \!\!\!  
(D_{n}^{~\!(I)})^{\dagger}(\widetilde{\Psi}_{n}^{~\!(I)})^{-\dagger}
\underbrace{\left(\Lambda_{I}^{\dagger}\Lambda_{I}\right)}
\left(
(\widetilde{\Psi}_{n}^{~\!(I)})^{\dagger}
\widetilde{\Psi}_{n}^{~\!(I)}\right)
(\widetilde{\Psi}_{n}^{~\!(I)})^{-1}D_{n}^{~\!(I)} \nonumber \\
&& ~~~~~~~~~~~~~~~~~~~~~~ \left| \lambda_{I} \right|^2 I_{3 \times 3}  \nonumber \\
&\!\!\!=& \!\!\!  
(D_{n}^{~\!(I)})^{\dagger}
\left((\widetilde{\Psi}_{n}^{~\!(I)})^{-\dagger}(\widetilde{\Psi}_{n}^{~\!(I)})^{\dagger}\right)
\left(\widetilde{\Psi}_{n}^{~\!(I)}(\widetilde{\Psi}_{n}^{~\!(I)})^{-1}\right)
D_{n}^{~\!(I)} 
\left(\Lambda_{I}^{\dagger}\Lambda_{I}\right) \nonumber \\
&\!\!\!=& \!\!\! 
(D_{n}^{~\!(I)})^{\dagger}D_{n}^{~\!(I)}\Lambda_{I}^{\dagger}\Lambda_{I}
=
\left(\prod_{i=1, i \neq I}^{n}(\Lambda_{i}^{(\pm)})^{\dagger}\Lambda_{i}^{(\pm)}
\right)\Lambda_{I}^{\dagger}\Lambda_{I}  \nonumber  \\
&\!\!\!=& \!\!\!\prod_{i=1}^{n}\left| \lambda_i \right|^2 I_{3 \times 3}
=\widetilde{J}_{n+1}^{~\!(I)}\widetilde{J}_{n+1}^{~\!(I)~\!\dagger}
\label{widetilde(J_n+1)^(I dagger)widetilde(J_n+1)^(I)}
\end{eqnarray}
By applying Jacobi's formula \eqref{Jacobi's formula} and Proposition \ref{Prop_3.9} to
\eqref{Det(widetilde(J_n+1)^(I)} and \eqref{widetilde(J_n+1)^(I dagger)widetilde(J_n+1)^(I)} respectively, we can find that the gauge fields (given by $\widetilde{J}_{n+1}^{~\!(I)}$) are all traceless and anti-hermitian. Therefore, the gauge group can be $G=\mathrm{SU}(3)$ in the asymptotic region.
On the other hand, these gauge fields (given by $\widetilde{J}_{n+1}^{~\!(I)}$) are also in the same form as those given by \eqref{Reduced Soliton solution_U_G=SU(3)} because $\widetilde{\Psi}_{n}^{~\!(I)}$ is in the same form as $\psi$ and
\begin{eqnarray}
\left(\partial_{\mu}\widetilde{J}_{n+1}^{~\!(I)}\right)(\widetilde{J}_{n+1}^{~\!(I)})^{-1}
=
\partial_{\mu}\left( \widetilde{\Psi}_{n}^{~\!(I)}\Lambda_{I}(\widetilde{\Psi}_{n}^{~\!(I)})^{-1} \right)
\widetilde{\Psi}_{n}^{~\!(I)}\Lambda_{I}^{-1}(\widetilde{\Psi}_{n}^{~\!(I)})^{-1}   \nonumber 
\end{eqnarray}
is independent of constant matrix $D_{n}^{~\!(I)}$. Therefore, the Lagrangian density (given by $\widetilde{J}_{n+1}^{~\!(I)}$) is in the same form as that given by the $I$-th 1-soliton solution except for a phase shift factor. More precisely, the Lagrangian density is
\begin{eqnarray}
{\mbox{Tr}} F_{\mu \nu}F^{\mu \nu}
&\!\!\!=& \!\!\!
8\left[(\alpha_I\overline{\beta}_I-\overline{\alpha}_I\beta_I)
(\lambda_I -\overline{\lambda}_I)  \right]^2
\left(2{\mbox{sech}}^2 X_I^{\prime}-3{\mbox{sech}}^4 X_I^{\prime}\right),  \\
&&\!\!\! 
X_I^{\prime}=L_I + \overline{L}_I + \frac{1}{2}\log\left(\left|a_I^{~\!\prime}\right|^2/ \left(\left|b_I^{~\!\prime}\right|^2 + \left|c_I^{~\!\prime}\right|^2 \right) \right)  \nonumber  \\
&&~~=L_I + \overline{L}_I +
\frac{1}{2}\log\left(\left|a_I\right|^2/ \left(\left|b_I\right|^2 + \left|c_I\right|^2 \right) \right)
+\Delta_{I}
=X_I+\Delta_{I}. ~~~~~~~~
\end{eqnarray} 
The phase shift $\Delta_I$ can be calculated explicitly by \eqref{Qn_I_reduced form_G=SU(3)} as
\begin{eqnarray}
\label{Phase shift_G=SU(3)_U}
\Delta_{I}
=
\frac{1}{2}
\log 
\left[
\frac{\prod\limits_{i=1, i \neq I}^{n}\left|\lambda_I-\lambda_i^{(\pm)}\right|^2(\left| b_I\right|^2 + \left| c_I\right|^2)}
{\prod\limits_{i=1, i \neq I}^{n}\left|(\lambda_I-{\lambda}_i^{(\mp)})\right|^2\left|b_I\right|^2 +\prod\limits_{i=1, i \neq I}^{n}\left|(\lambda_I-\lambda_i^{(\mp)})\right|^2\left|c_I\right|^2} 
\right].
\end{eqnarray}
It is real-valued and depends on $2^{n-1}$ choices of $(\lambda_i^{(+)}, \lambda_i^{(-)}):=(\lambda_i, \overline{\lambda}_i)$.
Now we can conclude that the $n$-soliton solution \eqref{n-Soliton Solutions_J_n+1_G=SU(3)} gives $n$ intersecting soliton walls which can be embedded into $G=\mathrm{SL}(3,\mathbb{C})$ gauge theory.
Moreover, the gauge group can be $G=\mathrm{SU}(3)$ for each soliton wall in the asymptotic region.

\newpage

\section{Conclusion and Future work}
\label{Section 8}

In this thesis, we constructed the ASDYM 1-solitons and multi-solitons on three kinds of 4-dimensional real spaces by applying the Darboux transformation \eqref{Darboux transf_(J_k+1, Phi_k+1)}.
We proposed ansatzes for $G=\mathrm{SL}(2,\mathbb{C})$ 1-soliton solutions and an ansatz for $G=\mathrm{SL}(3,\mathbb{C})$ 1-soliton solution, respectively (Cf: \eqref{Soliton solution_U}, \eqref{Soliton solution_E}, \eqref{Soliton solution_M}, \eqref{Soliton solution_U_G=SU(3)}). The resulting Lagrangian density is real-valued on each space even if the gauge group is non-unitary.
Especially for the split signature $(+,+,-,-)$, we found that the gauge fields are all anti-hermitian and traceless (Cf: \eqref{Gauge fields_U_soliton}, \eqref{Gauge fields_U_soliton_G=SU(3)}), therefore, the gauge group can be $G=\mathrm{SU}(2)$ and $G=\mathrm{SU}(3)$ on the Ultrahyperbolic space $\mathbb{U}$. 

\begin{table}[h]
	\!\!\!\!\caption{}
	\label{Table_3}
	\begin{tabular}{|c|c|c|c|}
		\hline
		4D Real Spaces & $\mathbb{U}$  &  $\mathbb{E}$ & $\mathbb{M}$  \\ 
		(Metric) & $(+,+,-,-)$ & $(+,+,+,+)$ & $(+,-,-,-)$  \\
		\hline
		\hline
		Gauge Group of   & $G=\mathrm{SU(2)}$ &  $G=\mathrm{SL(2,\mathbb{C})}$ & $G=\mathrm{SL(2, \mathbb{C})}$ \\ 
		the 1-Soliton (Full Region) & $G=\mathrm{SU(3)}$ &  $G=\mathrm{SL(3,\mathbb{C})}$ & $G=\mathrm{SL(3, \mathbb{C})}$ \\
		\hline
		Lagrangian Density & Real-valued & Real-valued & Real-valued \\ 
		\hline \hline
		Gauge Group of  & $G=\mathrm{SU(2)}$ &  $G=\mathrm{SL(2,\mathbb{C})}$ & $G=\mathrm{SL(2, \mathbb{C})}$ \\ 
		the Multi-Soliton (Asymptotic Region)   &  $G=\mathrm{SU(3)}$ &  $G=\mathrm{SL(3,\mathbb{C})}$ & $G=\mathrm{SL(3, \mathbb{C})}$ \\
		\hline
		Lagrangian Density & Real-valued & Real-valued & Real-valued  \\
		\hline \hline
		Gauge Group of  & $G=\mathrm{SU(2)}$ &  $G=\mathrm{SL(2,\mathbb{C})}$ & $G=\mathrm{SL(2, \mathbb{C})}$ \\ 
		the Multi-Soliton (Full Region)   & $G=\mathrm{SU(3)}$~? &  $G=\mathrm{SL(3,\mathbb{C})}$ & $G=\mathrm{SL(3, \mathbb{C})}$ \\
		\hline
		Lagrangian Density & Real-valued & ? & ? \\
		\hline 
	\end{tabular}
\end{table}

By imposing suitable conditions on the 1-soliton solutions, we obtained reduced versions (Cf: \eqref{Reduced 1-soliton_3 real spaces}, \eqref{Reduced Soliton solution_U_G=SU(3)}) of them, and the resulting Lagrangian densities become more beautiful and concise as the following forms : 
\begin{eqnarray}
{\mbox{Tr}} F_{\mu\nu}F^{\mu\nu}
=C\left(
2{\mbox{sech}}^2 X-3{\mbox{sech}}^4 X
\right),  \nonumber 
\end{eqnarray} 
where $C$ is a real constant depending on the signature of different real spaces.
Because the principal peak of the resulting Lagrangian density is localized on a 3-dimensional hyperplane in 4-dimensional real space,
we call them soliton walls to distinguish from the domain walls.
By applying $n$ iterations of the Darboux transformation \eqref{Darboux transf}, we obtained the $n$-soliton solution \eqref{n-Soliton Solutions_J_n+1} and \eqref{n-Soliton Solutions_J_n+1_G=SU(3)}, respectively.
We showed that the  $n$-soliton solutions can be interpreted as $n$ intersecting soliton walls. For split the signature, we even showed that the gauge group can be $G=\mathrm{SU}(2)$ for the whole $n$ intersecting soliton walls, and $G=\mathrm{SU}(3)$ for each soliton wall in the asymptotic region, respectively.  

Finally, we propose several interesting future works as the ending of this thesis.
For the split signature, 
the ASDYM equations are equations of motion of effective action for open $\mathrm{N=2}$ string theories \cite{OoVa1, OoVa2}. Therefore, it would be an important topic to understand the role and the applications of our soliton walls in open $\mathrm{N=2}$ string theories.
For the Euclidean signature,
we find that the unitary symmetry is lost under the $S$-gauge. 
Therefore, we have to generalize the Darboux transformation to a version without the constraint of the $S$-gauge so that the unitary symmetry can be preserved on the Euclidean space $\mathbb{E}$.
For the Minkowski space, 
although the unitary symmetry is lost for the pure ASD Yang-Mills theory,
we can modify the Darboux transformation and apply it to the Bogomol'nyi equation \cite{Bogomolny} to construct new solitons in the Yang-Mills-Higgs theory.
It would be an interesting but challenging work. 
On the other hand, the Ernst equation \cite{Ernst1, Ernst2, L.Witten} (axisymmetric Einstein equation) on the $n$-dimensional Minkowski space is equivalent to the reduced Yang equation \cite{MaWo} for $G=\mathrm{GL}(n-2, \mathbb{R})$. The application of $J$-matrix formulation to the axisymmetric gravitational field is another topic worthy studying.

\newpage

\appendix

\section{The asymptotic behavior of the KP Multi-Soliton}
\label{Appendix A}
In this appendix, we verify that the KP multi-soliton exactly satisfies the Property \ref{Property 3} of solitons mentioned in Section \ref{Section 1}. Here we follow the technique developed by the authors of \cite{MaSu}. The KP $n$-soliton is defined by
\begin{eqnarray}
u(x,y,t)&\!\!\!\!=& \!\!\!\!
U(X_1, X_2, ..., X_n)
:= 
2\frac{\partial^{2}}{\partial x^{2}}(\log\tau_{n}), ~~ \label{KP n-Soliton} \\
\tau_{n}
&\!\!\!\!:=& \!\!\!\!
\mbox{Wr}(f_1, f_2, ..., f_n)  
:=
\left|
\begin{array}{ccccc}
f_1^{(0)} & f_2^{(0)} & \cdots & f_n^{(0)} \\
f_1^{(1)} & f_2^{(1)} & \cdots & f_n^{(1)} \\
\vdots & \vdots & \ddots & \vdots \\
f_1^{(n-1)} & f_2^{(n-1)} & \cdots & f_n^{(n-1)}
\end{array}
\right|, ~ f_i^{(m)}:=\frac{\partial^{m}f_i}{\partial x^{m}}, ~~~~~~~~  \label{Wr(f_i)} \\
f_i&\!\!\!\!:=& \!\!\!
\mbox{cosh}X_i,~~
X_i:=\frac{\kappa_{i1}-\kappa_{i2}}{2} x + \frac{\kappa_{i1}^{2}-\kappa_{i2}^{2}}{2}y + \frac{\kappa_{i1}^{3}-\kappa_{i2}^{3}}{2}t + \frac{\delta_{i1}-\delta_{i2}}{2}.
\end{eqnarray}
For convenience, we introduce a new notation $\alpha_i^{(\pm)}$ defined by
\begin{eqnarray}
(\alpha_i^{(+)}, \alpha_i^{(-)}):=(\alpha_i, -\alpha_i),~~
\alpha_i:=\frac{\kappa_{i1}-\kappa_{i2}}{2},~ i=1,2,...,n
\end{eqnarray}
and rewrite $f_i^{(m)}$ by
\begin{eqnarray}
f_i^{(m)}
&\!\!\!\!=& \!\!\!\! \frac{1}{2}\left(\alpha_i^{(+)m}e^{X_i} + \alpha_i^{(-)m}e^{-X_i} \right)  \\
&\!\!\!\!=& \!\!\!\! \frac{1}{2}e^{X_i}\left(\alpha_i^{(+)m} + \alpha_i^{(-)m}e^{-2X_i} \right)
=\frac{1}{2}e^{-X_i}\left(\alpha_i^{(+)m}e^{2X_i} + \alpha_i^{(-)m} \right).  
\end{eqnarray}
Since the factors $e^{X}$, $e^{-X}$ have no contribution to \eqref{KP n-Soliton} and hence  we can obtain an equivalent expression of \eqref{KP n-Soliton} as
\begin{eqnarray}
U(X_1, X_2, ..., X_n)
&\!\!\!\!=& \!\!\!\!
2\frac{\partial^{2}}{\partial x^{2}}(\log \widetilde{\tau}_{n}), ~~ 
\widetilde{\tau}_{n}
:=
\left|
\begin{array}{ccccc}
g_1^{(0)} & g_2^{(0)} & \cdots & g_n^{(0)} \\
g_1^{(1)} & g_2^{(1)} & \cdots & g_n^{(1)} \\
\vdots & \vdots & \ddots & \vdots \\
g_1^{(n-1)} & g_2^{(n-1)} & \cdots & g_n^{(n-1)}
\end{array}
\right|,   \label{Wr(g_i)}  \\
g_i^{(m)}&\!\!\!\! :=& \!\!\!\!
\left\{
\begin{array}{l}
\alpha_i^{(+)m}e^{X_i} + \alpha_i^{(-)m}e^{-X_i} ~~~\! \mbox{if} \!~~~ X_i ~\mbox{is finite}  \\
\alpha_i^{(+)m} + \alpha_i^{(-)m}e^{-2X_i}  ~~ \longrightarrow ~ \alpha_i^{(+)m} ~~ \mbox{as} ~~ X_i \rightarrow  +\infty  \\
\alpha_i^{(+)m}e^{2X_i} + \alpha_i^{(-)m}  \!~~~~ \longrightarrow ~ \alpha_i^{(-)m} ~~ \mbox{as} ~~ X_i \rightarrow  -\infty
\end{array}.
\right. ~~~~~~
\end{eqnarray}
Now let us pick $I \in \left\{ 1,2,...,n\right\}$ such that 
\begin{eqnarray}
\left\{
\begin{array}{l}
X_{I} = \mbox{finite} \\
X_{i} \rightarrow \pm \infty ~ \mbox{or} ~\mp \infty 
\end{array}
\right.
~~\mbox{when}~~  t \rightarrow \pm \infty.
\end{eqnarray}
In other words, we choose a comoving frame related to the $I$-th KP soliton, and then 
\begin{eqnarray}
\label{KP multi-soliton_asymptotic}
U(X_1, X_2, ..., X_n) ~ \stackrel{t \rightarrow \pm \infty}{\longrightarrow}~
2\frac{\partial^{2}}{\partial x^{2}}\log\widetilde{\tau}_n^{(I)}, 
\end{eqnarray}
\begin{eqnarray}
\widetilde{\tau}_n^{(I)}
&\!\!\!\!:=& \!\!\!\!
\left|
\begin{array}{ccccccc}
\!1 & \!\!\cdots & \!\!1 & e^{X_I}+e^{-X_I} & 1 & \!\!\cdots & \!\!1 \\
\!\alpha_1^{(\pm)} & \!\!\cdots & \!\!\alpha_{I-1}^{(\pm)} & \alpha_I^{(+)}e^{X_I}+\alpha_I^{(-)}e^{-X_I} & \alpha_{I+1}^{(\pm)} & \!\!\cdots & \!\!\alpha_n^{(\pm)}  \\
\!\alpha_1^{(\pm)2} & \!\!\cdots & \!\!\alpha_{I-1}^{(\pm)2} & \alpha_I^{(+)2}e^{X_I}+\alpha_I^{(-)2}e^{-X_I} & \alpha_{I+1}^{(\pm)2} & \!\!\cdots & \!\!\alpha_n^{(\pm)2}  \\
\!\vdots & \!\!\ddots & \!\!\vdots & \vdots & \vdots & \!\!\ddots & \!\!\vdots \\
\!\alpha_1^{(\pm)n-1} & \!\!\cdots & \!\!\alpha_{I-1}^{(\pm)n-1} & \alpha_I^{(+)n-1}e^{X_I}+\alpha_I^{(-)n-1}e^{-X_I} & \alpha_{I+1}^{(\pm)n-1} & \!\!\cdots & \!\!\alpha_n^{(\pm)n-1}  \\
\end{array}
\!\!\right| ~~~~~~~~~~ \\
&\!\!\!\!=&\!\!\!\!
A_I^{(+)}e^{X_I}+A_I^{(-)}e^{-X_I}   \\
&\!\!\!\!=&\!\!\!\!
\left( A_I^{(+)}A_I^{(-)} \right)^{\frac{1}{2}}
\mbox{cosh}\left(X_I+\frac{1}{2}\log\left(A_I^{(+)}/A_I^{(-)} \right) \right),
\end{eqnarray}
$A_I^{(\pm)}$ are the Vandermonde determinants defined by 
\begin{eqnarray}
A_I^{(+)}&\!\!\!\!:=& \!\!\!\!  
\left|
\begin{array}{ccccc}
\!\!1 & \!\!\cdots & \!\!1 & \!\!\cdots & \!\!1 \\
\!\!\alpha_1^{(\pm)} & \!\!\cdots & \!\!\alpha_{I}^{(+)} & \!\!\cdots & \!\!\alpha_n^{(\pm)} \\
\!\!\alpha_1^{(\pm)2} & \!\!\cdots & \!\!\alpha_{I}^{(+)2} & \!\!\cdots & \!\!\alpha_n^{(\pm)2} \\
\!\!\vdots & \!\!\ddots & \!\!\vdots & \!\!\ddots & \!\!\vdots \\
\!\!\alpha_1^{(\pm)n-1} & \!\!\cdots & \!\!\alpha_{I}^{(+)n-1} & \!\!\cdots & \!\!\alpha_n^{(\pm)n-1}
\end{array}
\!\!\right|, ~ A_I^{(-)}:=
\left|
\begin{array}{ccccc}
\!\!1 & \!\!\cdots & \!\!1 & \!\!\cdots & \!\!1 \\
\!\!\alpha_1^{(\pm)} & \!\!\cdots & \!\!\alpha_{I}^{(-)} & \!\!\cdots & \!\!\alpha_n^{(\pm)} \\
\!\!\alpha_1^{(\pm)2} & \!\!\cdots & \!\!\alpha_{I}^{(-)2} & \!\!\cdots & \!\!\alpha_n^{(\pm)2} \\
\!\!\vdots & \!\!\ddots & \!\!\vdots & \!\!\ddots & \!\!\vdots \\
\!\!\alpha_1^{(\pm)n-1} & \!\!\cdots & \!\!\alpha_{I}^{(-)n-1} & \!\!\cdots & \!\!\alpha_n^{(\pm)n-1}
\end{array}
\!\!\right|,
\nonumber 
\end{eqnarray}
and hence
\begin{eqnarray}
\frac{A_I^{(+)}}{A_I^{(-)}}
=\prod\limits_{i=1, i \neq I}^{n} \left(\frac{\alpha_I^{(+)}-\alpha_i^{(\pm)}}{\alpha_I^{(-)}-\alpha_i^{(\pm)}} \right).
\end{eqnarray}
Since the factor $\left( A_I^{(+)}A_I^{(-)} \right)^{\frac{1}{2}}$
has no contribution to \eqref{KP multi-soliton_asymptotic}, we can conclude that
\begin{eqnarray}
U(X_1, X_2, ..., X_n) &\!\!\!\! ~ \stackrel{t \rightarrow \pm \infty}{\longrightarrow}~ & \!\!\!\! 
2\frac{\partial^{2}}{\partial x^{2}}\log\mbox{cosh}\left( X_I + \Delta_I^{(\pm)} \right) \\
&& \!\!\!\!\!\!\!\!\!\!\!\! =
\frac{(\kappa_{I1}-\kappa_{I2})^{2}}{2}\mbox{sech}^{2}\left( X_I + \Delta_I^{(\pm)} \right)
=U_{I}(X_I+\Delta_I^{(\pm)}), ~~~~~~~~~~~~
\end{eqnarray}
where the phase shift 
\begin{eqnarray}
\Delta_I^{(\pm)}
:=\frac{1}{2}\sum\limits_{i=1, i \neq I}^{n}\log \left(\frac{\alpha_I^{(+)}-\alpha_i^{(\pm)}}{\alpha_I^{(-)}-\alpha_i^{(\pm)}} \right)  
\end{eqnarray}
depends on $2^{n-1}$ choices of 
\begin{eqnarray}
\left(\alpha_1^{(\pm)}, \alpha_2^{(\pm)}, ..., \alpha_{I-1}^{(\pm)}, \alpha_{I+1}^{(\pm)}, ..., \alpha_n^{(\pm)}\right)
:=
\left( \pm\alpha_1, \pm\alpha_2, ..., \pm\alpha_{I-1}, \pm\alpha_{I+1}, ..., \pm\alpha_n \right).
\end{eqnarray}

\newpage

\section{Quasi-Wronskian form of $L_{n+1}(\Phi_{n+1})$ and $M_{n+1}(\Phi_{n+1})$} 
\label{Appendix B}
In this appendix, we introduce some new results that are independent of the collaborative paper \cite{GiHaHuNi}.
Firstly, we show that the solution $\Phi_{n+1}$, generated by applying $n$ iterations of the Darboux transformation (Cf: \eqref{Darboux transf_(J_k+1, Phi_k+1)}), has a sum representation (Cf: Proposition \ref{Prop_B1}). We use this property to show that 
\begin{eqnarray}
\label{L_n+1(Phi_n+1)}
L_{n+1}(\Phi_{n+1})
&:=&
[\partial_{w}-(\partial_{w}J_{n+1})J_{n+1}^{-1}]\Phi_{n+1}-(\partial_{\widetilde{z}}\Phi_{n+1})\zeta   \\
M_{n+1}(\Phi_{n+1})
&:=&
[\partial_{z}-(\partial_{z}J_{n+1})J_{n+1}^{-1}]\Phi_{n+1}-(\partial_{\widetilde{w}}\Phi_{n+1})\zeta   \label{M_n+1(Phi_n+1)}  
\end{eqnarray}
are in the form of the quasi-Wronskian (Cf: Theorem \ref{Thm_B2}) in which the components are $(\psi_{i}, \Lambda_{i})$ and $L_{1}(\phi)\zeta^{i}$.   
\newtheorem{prop_B1}{Proposition}[section]
\begin{prop_B1}
	{\bf (Sum representation of $\Phi_{n+1}$)} \label{Prop_B1}
\begin{eqnarray}
\Phi_{n+1}
:=
\left|
\begin{array}{ccccc}
\psi_{1} & \psi_{2} & \!\!\cdots & \!\! \psi_{n} & \phi \\
\vdots & \vdots & \!\! \ddots & \!\! \vdots & \!\! \vdots  \\
\psi_{1}\Lambda_{1}^{k} & \psi_{2}\Lambda_{2}^{k} & \!\! \cdots & \!\! \psi_{n}\Lambda_{n}^{k} & \phi \zeta^{k} \\
\vdots & \vdots & \!\! \ddots & \!\! \vdots & \!\! \vdots  \\
\psi_{1}\Lambda_{1}^{n} & \psi_{2}\Lambda_{2}^{n} & \!\! \cdots & \!\! \psi_{n}\Lambda_{n}^{n} & \fbox{$\phi\zeta^{n}$}
\end{array}\right|
=
\phi\zeta^{n}
+\sum_{k=0}^{n-1}
\left|
\begin{array}{ccccc}
\psi_{1} & \!\!\cdots & \!\! \psi_{n} & 0 \\
\vdots & \!\! \ddots & \!\! \vdots & \!\! \vdots  \\
\psi_{1}\Lambda_{1}^{k} & \!\! \cdots & \!\! \psi_{n}\Lambda_{n}^{k} & 1 \\
\vdots & \!\! \ddots & \!\! \vdots & \!\! \vdots  \\
\psi_{1}\Lambda_{1}^{n} & \!\! \cdots & \!\! \psi_{n}\Lambda_{n}^{n} & \fbox{$0$}
\end{array}
\right|
\phi\zeta^{k} \nonumber 
\end{eqnarray}
\end{prop_B1}
{\bf(\emph{Proof})}
\smallskip \\
For convenience, we use the notation of \eqref{Nimmo notation} in this proof.
\begin{eqnarray}
\Phi_{n+1} \!\!\!\! 
&=& \!\!\!\!
\phi\zeta^{n}
+
\left|
\begin{array}{cc}
{\bm{\psi}} & \phi \\
\vdots & \vdots \\ 
{\bm{\psi}}{\bm{\Lambda}}^{n-2} & \phi\zeta^{n-2} \\
{\bm{\psi}}{\bm{\Lambda}}^{n-1} & \phi\zeta^{n-1} \\
{\bm{\psi}}{\bm{\Lambda}}^{n} & \fbox{0}
\end{array}
\right|   
=
\phi\zeta^{n}
+
\left|
\begin{array}{cc}
{\bm{\psi}} & 0 \\
\vdots & \vdots \\ 
{\bm{\psi}}{\bm{\Lambda}}^{n-2} & 0 \\
{\bm{\psi}}{\bm{\Lambda}}^{n-1} & 1 \\
{\bm{\psi}}{\bm{\Lambda}}^{n} & \fbox{0}
\end{array}
\right|
\left|
\begin{array}{cc}
{\bm{\psi}} & \phi \\
\vdots & \vdots \\ 
{\bm{\psi}}{\bm{\Lambda}}^{n-2} & \phi\zeta^{n-2} \\
{\bm{\psi}}{\bm{\Lambda}}^{n-1} & \fbox{$\phi\zeta^{n-1}$} \\
{\bm{\psi}}{\bm{\Lambda}}^{n} & 0
\end{array}
\right|   \nonumber  \\
&&{\mbox{(Here we use the homological relation \eqref{Homological relation}.})} \nonumber \\
&=& \!\!\!\!
\phi\zeta^{n}
+
\left|
\begin{array}{cc}
{\bm{\psi}} & 0 \\
\vdots & \vdots \\ 
{\bm{\psi}}{\bm{\Lambda}}^{n-2} & 0 \\
{\bm{\psi}}{\bm{\Lambda}}^{n-1} & 1 \\
{\bm{\psi}}{\bm{\Lambda}}^{n} & \fbox{0}
\end{array}
\right|\phi\zeta^{n-1}
+
\left|
\begin{array}{cc}
{\bm{\psi}} & 0 \\
\vdots & \vdots \\ 
{\bm{\psi}}{\bm{\Lambda}}^{n-2} & 0 \\
{\bm{\psi}}{\bm{\Lambda}}^{n-1} & 1 \\
{\bm{\psi}}{\bm{\Lambda}}^{n} & \fbox{0}
\end{array}
\right|
\left|
\begin{array}{cc}
{\bm{\psi}} & \phi \\
\vdots & \vdots \\ 
{\bm{\psi}}{\bm{\Lambda}}^{n-2} & \phi\zeta^{n-2} \\
{\bm{\psi}}{\bm{\Lambda}}^{n-1} & \fbox{0} \\
{\bm{\psi}}{\bm{\Lambda}}^{n} & 0
\end{array}
\right|  \nonumber \\
&=& \!\!\!\!
\phi\zeta^{n}
+
\left|
\begin{array}{cc}
{\bm{\psi}} & 0 \\
\vdots & \vdots \\ 
{\bm{\psi}}{\bm{\Lambda}}^{n-2} & 0 \\
{\bm{\psi}}{\bm{\Lambda}}^{n-1} & 1 \\
{\bm{\psi}}{\bm{\Lambda}}^{n} & \fbox{0}
\end{array}
\right|\phi\zeta^{n-1}
+
\left|
\begin{array}{cc}
{\bm{\psi}} & 0 \\
\vdots & \vdots \\ 
{\bm{\psi}}{\bm{\Lambda}}^{n-2} & 1 \\
{\bm{\psi}}{\bm{\Lambda}}^{n-1} & 0 \\
{\bm{\psi}}{\bm{\Lambda}}^{n} & \fbox{0}
\end{array}
\right|
\left|
\begin{array}{cc}
{\bm{\psi}} & \phi \\
\vdots & \vdots \\ 
{\bm{\psi}}{\bm{\Lambda}}^{n-2} & \fbox{$\phi\zeta^{n-2}$} \\
{\bm{\psi}}{\bm{\Lambda}}^{n-1} & 0 \\
{\bm{\psi}}{\bm{\Lambda}}^{n} & 0
\end{array}
\right|    \nonumber \\  \nonumber  \\
&&
{\mbox{(Here we use the equivalent expressions of the homological relation \eqref{Homological relation}.)}} \nonumber 
\end{eqnarray}
\begin{eqnarray}
&=& \!\!\!\!
\phi\zeta^{n}
\!+
\! \left|
\begin{array}{cc}
{\bm{\psi}} & 0 \\
\vdots & \vdots \\ 
{\bm{\psi}}{\bm{\Lambda}}^{n-2} & 0 \\
{\bm{\psi}}{\bm{\Lambda}}^{n-1} & 1 \\
{\bm{\psi}}{\bm{\Lambda}}^{n} & \fbox{0}
\end{array}
\right|\phi\zeta^{n-1}
\!+
\!\left|
\begin{array}{cc}
{\bm{\psi}} & 0 \\
\vdots & \vdots \\ 
{\bm{\psi}}{\bm{\Lambda}}^{n-2} & 1 \\
{\bm{\psi}}{\bm{\Lambda}}^{n-1} & 0 \\
{\bm{\psi}}{\bm{\Lambda}}^{n} & \fbox{0}
\end{array}
\right|\phi\zeta^{n-2}
\!+
\!\left|
\begin{array}{cc}
{\bm{\psi}} & 0 \\
\vdots & \vdots \\ 
{\bm{\psi}}{\bm{\Lambda}}^{n-3} & \! 0 \\
{\bm{\psi}}{\bm{\Lambda}}^{n-2} & \! 1 \\
{\bm{\psi}}{\bm{\Lambda}}^{n-1} & \! 0 \\
{\bm{\psi}}{\bm{\Lambda}}^{n} & \! \fbox{0}
\end{array}
\right|
\left|
\begin{array}{cc}
{\bm{\psi}} & \phi \\
\vdots & \vdots \\ 
{\bm{\psi}}{\bm{\Lambda}}^{n-3} & \!\phi\zeta^{n-3} \\
{\bm{\psi}}{\bm{\Lambda}}^{n-2} & \!\fbox{0} \\
{\bm{\psi}}{\bm{\Lambda}}^{n-1} & \!0 \\
{\bm{\psi}}{\bm{\Lambda}}^{n} & \!0
\end{array}
\right|   \nonumber  \\
&=& \!\!\!\!
\phi\zeta^{n}
\!+
\!\left|
\begin{array}{cc}
{\bm{\psi}} & 0 \\
\vdots & \vdots \\ 
{\bm{\psi}}{\bm{\Lambda}}^{n-2} & 0 \\
{\bm{\psi}}{\bm{\Lambda}}^{n-1} & 1 \\
{\bm{\psi}}{\bm{\Lambda}}^{n} & \fbox{0}
\end{array}
\right|\phi\zeta^{n-1}
\!+
\!\left|
\begin{array}{cc}
{\bm{\psi}} & 0 \\
\vdots & \vdots \\ 
{\bm{\psi}}{\bm{\Lambda}}^{n-2} & 1 \\
{\bm{\psi}}{\bm{\Lambda}}^{n-1} & 0 \\
{\bm{\psi}}{\bm{\Lambda}}^{n} & \fbox{0}
\end{array}
\right|\phi\zeta^{n-2}
\!+
\!\left|
\begin{array}{cc}
{\bm{\psi}} & 0 \\
\vdots & \vdots \\ 
{\bm{\psi}}{\bm{\Lambda}}^{n-3} & \! 1 \\
{\bm{\psi}}{\bm{\Lambda}}^{n-2} & \! 0 \\
{\bm{\psi}}{\bm{\Lambda}}^{n-1} & \!0 \\
{\bm{\psi}}{\bm{\Lambda}}^{n} & \! \fbox{0}
\end{array}
\right|
\left|
\begin{array}{cc}
{\bm{\psi}} & \phi \\
\vdots & \vdots \\ 
{\bm{\psi}}{\bm{\Lambda}}^{n-3} & \!\!\! \fbox{$\phi\zeta^{n-3}$} \\
{\bm{\psi}}{\bm{\Lambda}}^{n-2} & \!\!\! 0 \\
{\bm{\psi}}{\bm{\Lambda}}^{n-1} & \!\!\! 0 \\
{\bm{\psi}}{\bm{\Lambda}}^{n} & \!\!\! 0
\end{array}
\right|   \nonumber  \\
&&\mbox{(Continuing the same process, we can complete the proof.)}  \nonumber  \\
&=& \!\!\!\! \phi\zeta^{n}
+
\left|
\begin{array}{cc}
{\bm{\psi}} & 0 \\
\vdots & \vdots \\ 
{\bm{\psi}}{\bm{\Lambda}}^{n-2} & 0 \\
{\bm{\psi}}{\bm{\Lambda}}^{n-1} & 1 \\
{\bm{\psi}}{\bm{\Lambda}}^{n} & \fbox{0}
\end{array}
\right|\phi\zeta^{n-1}
+
\left|
\begin{array}{cc}
{\bm{\psi}} & 0 \\
\vdots & \vdots \\ 
{\bm{\psi}}{\bm{\Lambda}}^{n-2} & 1 \\
{\bm{\psi}}{\bm{\Lambda}}^{n-1} & 0 \\
{\bm{\psi}}{\bm{\Lambda}}^{n} & \fbox{0}
\end{array}
\right|\phi\zeta^{n-2}
+
...
+
\left|
\begin{array}{cc}
{\bm{\psi}} & 1 \\
\vdots & \vdots \\ 
{\bm{\psi}}{\bm{\Lambda}}^{n-2} & 0 \\
{\bm{\psi}}{\bm{\Lambda}}^{n-1} & 0 \\
{\bm{\psi}}{\bm{\Lambda}}^{n} & \fbox{0}
\end{array}
\right|\phi  \nonumber 
\end{eqnarray}
$\hfill\Box$
\newtheorem{thm_B2}[prop_B1]{Theorem}
\begin{thm_B2} 
	{\color{white}Theorem B.2}  \label{Thm_B2}  \\
Let $L_{n+1}(\Phi_{n+1})$,~ $M_{n+1}(\Phi_{n+1})$ be defined as \eqref{L_n+1(Phi_n+1)}, \eqref{M_n+1(Phi_n+1)}, then 
\begin{eqnarray}
\label{L_n+1(Phi_n+1)_Q}
L_{n+1}(\Phi_{n+1})
&\!\!\!\!=& \!\!\!\!
\left|
\begin{array}{ccccc}
\psi_{1} & \psi_{2} & \cdots & \psi_{n} & L_{1}(\phi) \\
\psi_{1}\Lambda_{1} & \psi_{2}\Lambda_{2} & \cdots & \psi_{n}\Lambda_{n} & L_{1}(\phi)\zeta \\
\vdots & \vdots & \ddots & \vdots & \vdots \\
\psi_{1}\Lambda_{1}^{n} & \psi_{2}\Lambda_{2}^{n} & \cdots & \psi_{n}\Lambda_{n}^{n} & \fbox{$L_{1}(\phi)\zeta^{n}$}
\end{array}
\right|    \\
M_{n+1}(\Phi_{n+1})
&\!\!\!\!=& \!\!\!\!
\left|
\begin{array}{ccccc}
\psi_{1} & \psi_{2} & \cdots & \psi_{n} & M_{1}(\phi) \\
\psi_{1}\Lambda_{1} & \psi_{2}\Lambda_{2} & \cdots & \psi_{n}\Lambda_{n} & M_{1}(\phi)\zeta \\
\vdots & \vdots & \ddots & \vdots & \vdots \\
\psi_{1}\Lambda_{1}^{n} & \psi_{2}\Lambda_{2}^{n} & \cdots & \psi_{n}\Lambda_{n}^{n} & \fbox{$M_{1}(\phi)\zeta^{n}$}
\end{array}
\right| \label{M_n+1(Phi_n+1)_Q}
\end{eqnarray}
\end{thm_B2}
{\bf (\emph{Proof})}
\\
For convenience, we use the notation of \eqref{Nimmo notation} in this proof. By substituting Theorem \ref{Thm_4.6} and Proposition \ref{Prop_B1} into \eqref{L_n+1(Phi_n+1)}, we have \\
\begin{eqnarray}
&&L_{n+1}(\Phi_{n+1}) \nonumber \\
&=&[\partial_{w}-(\partial_{w}J_{n+1})J_{n+1}^{-1}]\Phi_{n+1}-(\partial_{\widetilde{z}}\Phi_{n+1})\zeta  \nonumber  \\
&=&
\left[
\partial_{w}
-
(\partial_{w}J_{1})J_{1}^{-1} 
+ 
\partial_{\widetilde{z}}
\left|
\begin{array}{cc}
{\bm{\psi}} & 0 \\
{\bm{\psi}}{\bm{\Lambda}} & 0 \\
\vdots & \vdots \\
{\bm{\psi}}{\bm{\Lambda}}^{n-1} & 1 \\ 
{\bm{\psi}}{\bm{\Lambda}}^{n} & \fbox{0} 
\end{array}
\right|~
\right]
\left(
\phi\zeta^{n} 
+
\sum_{k=0}^{n-1}
\left|
\begin{array}{cc}
{\bm{\psi}} & 0 \\
\vdots & \vdots \\
{\bm{\psi}}{\bm{\Lambda}}^{k} & 1 \\
\vdots & \vdots \\
{\bm{\psi}}{\bm{\Lambda}}^{n} & \fbox{0} 
\end{array}
\right|\phi\zeta^{k}
\right) \nonumber \\
&&
-\partial_{\widetilde z}
\left(
\phi\zeta^{n} 
+
\sum_{k=0}^{n-1}
\left|
\begin{array}{cc}
{\bm{\psi}} & 0 \\
\vdots & \vdots \\
{\bm{\psi}}{\bm{\Lambda}}^{k} & 1 \\
\vdots & \vdots \\
{\bm{\psi}}{\bm{\Lambda}}^{n} & \fbox{0} 
\end{array}
\right|\phi\zeta^{k}
\right) \zeta  \nonumber \\
&=&
(\partial_{w}\phi)\zeta^{n}
+
\sum_{k=0}^{n-1}
\left(
\partial_{w}
\left|
\begin{array}{cc}
{\bm{\psi}} & 0 \\
\vdots & \vdots \\
{\bm{\psi}}{\bm{\Lambda}}^{k} & 1 \\
\vdots & \vdots \\
{\bm{\psi}}{\bm{\Lambda}}^{n} & \fbox{0} 
\end{array}
\right|
\right)\phi\zeta^{k}
+
\sum_{k=0}^{n-1}
\left|
\begin{array}{cc}
{\bm{\psi}} & 0 \\
\vdots & \vdots \\
{\bm{\psi}}{\bm{\Lambda}}^{k} & 1 \\
\vdots & \vdots \\
{\bm{\psi}}{\bm{\Lambda}}^{n} & \fbox{0} 
\end{array}
\right|
(\partial_{w}\phi)\zeta^{k}   \nonumber  \\
&&
-(\partial_{w}J_{1})J_{1}^{-1}\phi\zeta^{n}
-(\partial_{w}J_{1})J_{1}^{-1}
\sum_{k=0}^{n-1}
\left|
\begin{array}{cc}
{\bm{\psi}} & 0 \\
\vdots & \vdots \\
{\bm{\psi}}{\bm{\Lambda}}^{k} & 1 \\
\vdots & \vdots \\
{\bm{\psi}}{\bm{\Lambda}}^{n} & \fbox{0} 
\end{array}
\right|\phi\zeta^{k}    \nonumber  \\
&&
+
\left(
\partial_{\widetilde{z}}
\left|
\begin{array}{cc}
{\bm{\psi}} & 0 \\
{\bm{\psi}}{\bm{\Lambda}} & 0 \\
\vdots & \vdots \\
{\bm{\psi}}{\bm{\Lambda}}^{n-1} & 1 \\ 
{\bm{\psi}}{\bm{\Lambda}}^{n} & \fbox{0} 
\end{array}
\right|~
\right)\phi\zeta^{n} 
+
\left(
\partial_{\widetilde{z}}
\left|
\begin{array}{cc}
{\bm{\psi}} & 0 \\
{\bm{\psi}}{\bm{\Lambda}} & 0 \\
\vdots & \vdots \\
{\bm{\psi}}{\bm{\Lambda}}^{n-1} & 1 \\ 
{\bm{\psi}}{\bm{\Lambda}}^{n} & \fbox{0} 
\end{array}
\right|
\right)
\sum_{k=0}^{n-1}
\left|
\begin{array}{cc}
{\bm{\psi}} & 0 \\
\vdots & \vdots \\
{\bm{\psi}}{\bm{\Lambda}}^{k} & 1 \\
\vdots & \vdots \\
{\bm{\psi}}{\bm{\Lambda}}^{n} & \fbox{0} 
\end{array}
\right|
\phi\zeta^{k} \nonumber \\
&&
-(\partial_{\widetilde{z}}\phi)\zeta^{n+1}
-
\sum_{k=0}^{n-1}
\left(
\partial_{\widetilde{z}}
\left|
\begin{array}{cc}
{\bm{\psi}} & 0 \\
\vdots & \vdots \\
{\bm{\psi}}{\bm{\Lambda}}^{k} & 1 \\
\vdots & \vdots \\
{\bm{\psi}}{\bm{\Lambda}}^{n} & \fbox{0} 
\end{array}
\right|
\right)\phi\zeta^{k+1}
-
\sum_{k=0}^{n-1}
\left|
\begin{array}{cc}
{\bm{\psi}} & 0 \\
\vdots & \vdots \\
{\bm{\psi}}{\bm{\Lambda}}^{k} & 1 \\
\vdots & \vdots \\
{\bm{\psi}}{\bm{\Lambda}}^{n} & \fbox{0} 
\end{array}
\right|
(\partial_{\widetilde{z}}\phi)\zeta^{k+1}   \nonumber  
\end{eqnarray}
\begin{eqnarray}
&=&
\left[
\partial_{w}\phi - (\partial_{w}J_{1})J_{1}^{-1}\phi - (\partial_{\widetilde{z}}\phi)\zeta
\right]\zeta^{n} \nonumber  \\
&+&
\!\!\! \sum_{k=0}^{n-1}
\left|
\begin{array}{cc}
{\bm{\psi}} & 0 \\
\vdots & \vdots \\
{\bm{\psi}}{\bm{\Lambda}}^{k} & 1 \\
\vdots & \vdots \\
{\bm{\psi}}{\bm{\Lambda}}^{n} & \fbox{0} 
\end{array}
\right|
\left[
\partial_{w}\phi - (\partial_{w}J_{1})J_{1}^{-1}\phi - (\partial_{\widetilde{z}}\phi)\zeta
\right]\zeta^{k}  \nonumber \\
&+&
\!\!\! \sum_{k=0}^{n-1}
\left\{
\begin{array}{ccc}
\partial_{w}
\left|
\begin{array}{cc}
{\bm{\psi}} & 0 \\
\vdots & \vdots \\
{\bm{\psi}}{\bm{\Lambda}}^{k} & 1 \\
\vdots & \vdots \\
{\bm{\psi}}{\bm{\Lambda}}^{n} & \fbox{0} 
\end{array}
\right|
+
\partial_{\widetilde{z}}
\left|
\begin{array}{cc}
{\bm{\psi}} & 0 \\
{\bm{\psi}}{\bm{\Lambda}} & 0 \\
\vdots & \vdots \\
{\bm{\psi}}{\bm{\Lambda}}^{n-1} & 1 \\ 
{\bm{\psi}}{\bm{\Lambda}}^{n} & \fbox{0} 
\end{array}
\right|
\left|
\begin{array}{cc}
{\bm{\psi}} & 0 \\
\vdots & \vdots \\
{\bm{\psi}}{\bm{\Lambda}}^{k} & 1 \\
\vdots & \vdots \\
{\bm{\psi}}{\bm{\Lambda}}^{n} & \fbox{0} 
\end{array}
\right|
-\partial_{\widetilde z}
\left|
\begin{array}{cc}
{\bm{\psi}} & 0 \\
\vdots & \vdots \\
{\bm{\psi}}{\bm{\Lambda}}^{k-1} & 1 \\
\vdots & \vdots \\
{\bm{\psi}}{\bm{\Lambda}}^{n} & \fbox{0} 
\end{array}
\right|
\\
-
(\partial_{w}J_{1})J_{1}^{-1}
\left|
\begin{array}{cc}
{\bm{\psi}} & 0 \\
\vdots & \vdots \\
{\bm{\psi}}{\bm{\Lambda}}^{k} & 1 \\
\vdots & \vdots \\
{\bm{\psi}}{\bm{\Lambda}}^{n} & \fbox{0} 
\end{array}
\right|
+
\left|
\begin{array}{cc}
{\bm{\psi}} & 0 \\
\vdots & \vdots \\
{\bm{\psi}}{\bm{\Lambda}}^{k} & 1 \\
\vdots & \vdots \\
{\bm{\psi}}{\bm{\Lambda}}^{n} & \fbox{0} 
\end{array}
\right|(\partial_{w}J_{1})J_{1}^{-1}
\end{array}
\right\}\phi\zeta^{k} \nonumber \\
&&~~~~~~~~~~~{\mbox{(Sum of these terms in the curly bracket is zero by Lemma \ref{Lem_4.4}.)}}  \nonumber \\
&=&
L_{1}(\phi)\zeta^{n}
+
\sum_{k=0}^{n-1}
\left|
\begin{array}{cc}
{\bm{\psi}} & 0 \\
\vdots & \vdots \\
{\bm{\psi}}{\bm{\Lambda}}^{k} & 1 \\
\vdots & \vdots \\
{\bm{\psi}}{\bm{\Lambda}}^{n} & \fbox{0} 
\end{array}
\right|
L_{1}(\phi)\zeta^{k}  \nonumber \\
&=&
\left|
\begin{array}{ccccc}
\psi_{1} & \psi_{2} & \cdots & \psi_{n} & L_{1}(\phi) \\
\psi_{1}\Lambda_{1} & \psi_{2}\Lambda_{2} & \cdots & \psi_{n}\Lambda_{n} & L_{1}(\phi)\zeta \\
\vdots & \vdots & \ddots & \vdots & \vdots \\
\psi_{1}\Lambda_{1}^{n} & \psi_{2}\Lambda_{2}^{n} & \cdots & \psi_{n}\Lambda_{n}^{n} & \fbox{$L_{1}(\phi)\zeta^{n}$}   
\end{array}
\right| ~~~ {\mbox{by Proposition \ref{Prop_B1}.}}  \nonumber
\end{eqnarray}
By the same process, we can obtain the result of $M_{n+1}(\Phi_{n+1})$.
$\hfill\Box$ 
\medskip \\
Note that if $L_1(\phi)=M_1(\phi)=0$, it implies $L_{n+1}(\Phi_{n+1})=M_{n+1}(\Phi_{n+1})=0$ directly. Therefore, we showed the form-invariant property of the linear system ( given by $n$ iterations of the Darboux transformation) in another way. The conclusion is the following.
\newtheorem{rem_B3}[prop_B1]{Remark}
\begin{rem_B3}
	{\color{white}Remark B3}  \label{Rem_B3} \\
Assume that $\psi_{i}(\Lambda_{i}), ~i=1,2,...,n$ satisfy the initial linear system $L_1(\phi)=M_1(\phi)=0$ \eqref{Initial linear system}. 
Then $(J_{n+1}, \Phi_{n+1})$ pair, generated by applying $n$ iterations of the Darboux transformation \eqref{Darboux transf}, satisfies the following linear system
\begin{eqnarray}
\left\{
\begin{array}{c}
L_{n+1}(\Phi_{n+1})= [\partial_{w}-(\partial_{w}J_{n+1})J_{n+1}^{-1}]\Phi_{n+1}-(\partial_{\widetilde{z}}\Phi_{n+1})\zeta = 0 \\
M_{n+1}(\Phi_{n+1})=	
[\partial_{z}-(\partial_{z}J_{n+1})J_{n+1}^{-1}]\Phi_{n+1}-(\partial_{\widetilde{w}}\Phi_{n+1})\zeta = 0   
\end{array}.
\right.   \nonumber
\end{eqnarray}
\end{rem_B3}


\section{Calculation of Lagrangian density}
\label{Appendix C}
\subsection*{Yang's $J$-matrix}
\begin{eqnarray}
\label{J_2_Appendix}
J
&\!\!\!=& \!\!\!
-\psi\Lambda\psi^{-1}~~ \nonumber \\
&&
~(\psi
=
\left(
\begin{array}{cc}
A & C
\\ 
B & D
\end{array}\right),~~
\Lambda
=
\left(
\begin{array}{cc}
\lambda & 0 \\
0 & \mu
\end{array}\right)) \nonumber   \\
&\!\!\!=&  \!\!\!
\frac{-1}{AD-BC}\left(
\begin{array}{cc}
\lambda AD - \mu BC &  (\mu-\lambda)AC
\\ 
(\lambda - \mu)BD
& 
\mu AD - \lambda BC
\end{array}\right).
\end{eqnarray}

\subsection*{Derivative of $J$-matrix}

\begin{eqnarray}
\partial_{m}J
&\!\!\!\!=& \!\!\!\! 
\frac{\lambda-\mu}{(AD-BD)^2}
\left(
\begin{array}{cc}
\widetilde{P}_mAB - P_mCD & -\widetilde{P}_mA^2 + P_mC^2   \\ 
\widetilde{P}_mB^2 - P_mD^2 & -\widetilde{P}_mAB + P_mCD 
\end{array}\right), \\
&& \!\!
P_m:=A(\partial_{m}B) - (\partial_{m}A)B,~~ 
\widetilde{P}_m:=C(\partial_{m}D) - (\partial_{m}C)D,~~ m=z, w.  \nonumber 
\end{eqnarray}

\subsection*{Gauge Field~~( $m:=z,w$)}  

\begin{eqnarray}
\label{A_m_Appendix_ABCD}
{A_{m}}
&\!\!\!\!:=& \!\!\!\!
-(\partial_{m}J)J^{-1}    \nonumber \\
&\!\!\!=& \!\!\!
\frac{\lambda - \mu}{(AD-BC)^2}
\!\left(
\begin{array}{cc}
\!\!\!(1/\mu)\widetilde{P}_mAB - (1/\lambda) P_mCD  &  - (1/\mu) \widetilde{P}_mA^2 + (1/\lambda) P_mC^2  \\ 
\!\!\!(1/\mu) \widetilde{P}_mB^2 -(1/\lambda) P_mD^2 & -(1/\mu)\widetilde{P}_mAB + (1/\lambda) P_mCD
\end{array}
\!\!\!\right). ~~~~~~~
\end{eqnarray}
If we take $A, B, C, D$ as 
\begin{eqnarray}
\label{ABCD}
\begin{array}{ll}
\left(
\begin{array}{cc}
	\!\!A & \!\!C \\
	\!\!B & \!\!D
\end{array}
\!\right)
=\left(
\begin{array}{cc}
	\!a_1e^{L}+a_2e^{-L} & \!\!c_1e^{M}+c_2e^{-M} \\
	\!b_1e^{L}+b_2e^{-L} & \!\!d_1e^{M}+d_2e^{-M}
\end{array}\!\!\right), &
\begin{array}{l}
\!\!\!\!L:=(\lambda \alpha) z
+\beta \widetilde{z}
+(\lambda \beta) w
+\alpha\widetilde{w} \\
\!\!\!\!M:=(\mu\gamma) z
+\delta \widetilde{z}
+(\mu\delta) w
+\gamma \widetilde{w} 
\end{array}
\end{array}.~~~
\end{eqnarray}
Then 
\begin{eqnarray}
P_m&\!\!\!\!=& \!\!\!\!
(a_1e^{L}+a_2e^{-L})\left[\partial_{m}(b_1e^{L}+b_2e^{-L})\right]
-\left[\partial_{m}(a_1e^{L}+a_2e^{-L})\right](b_1e^{L}+b_2e^{-L})  \nonumber   \\
&\!\!\!\!=& \!\!\!\!
\left\{
\begin{array}{l}
2\lambda\alpha(a_2b_1 - a_1b_2),~~~~~~\mbox{if} ~~m=z \\
2\lambda\beta(a_2b_1 - a_1b_2),~~~~~~ \mbox{if} ~~m=w
\end{array}
\right.,
\end{eqnarray}
\begin{eqnarray}
\widetilde{P}_m&\!\!\!\!=& \!\!\!\!
(c_1e^{L}+c_2e^{-L})\left[\partial_{m}(d_1e^{L}+d_2e^{-L})\right]
-\left[\partial_{m}(c_1e^{L}+c_2e^{-L})\right](d_1e^{L}+d_2e^{-L})  \nonumber   \\
&\!\!\!\!=& \!\!\!\!
\left\{
\begin{array}{l}
2\mu\gamma(c_2d_1 - c_1d_2),~~~~~~\mbox{if}~~  m=z \\
2\mu\delta(c_2d_1 - c_1d_2),~~~~~~~\!\mbox{if}~~  m=w
\end{array}
\right..
\end{eqnarray}
In this case, we obtain a concise form of gauge fields :
\begin{eqnarray}
\label{A_m_Appendix_ABCD_2}
{A_{m}}&\!\!\!\!=& \!\!\!\!
\frac{2(\lambda-\mu)}{(AD - BC)^2}
\left(
\begin{array}{cc}
\widetilde{p}_mAB - p_mCD  &  -\widetilde{p}_mA^2 + p_mC^2  \\ 
\widetilde{p}_mB^2 -p_mD^2 &  -\widetilde{p}_mAB + p_mCD
\end{array}\right),
\\
&& \!\!\!\!
(p_m,~\widetilde{p}_m)
:=
\left\{
\begin{array}{l}
(\alpha(a_2b_1 - a_1b_2),~\gamma(c_2d_1 - c_1d_2))~~~~~~
{\mbox{if}} ~~ m=z, \\ 
(\beta(a_2b_1 - a_1b_2),~\delta(c_2d_1 -c_1d_2))~~~~~~~\!
{\mbox{if}} ~~ m=w,  \\
(0,~0) ~~~~~~~~~~~~~~~~~~~~~~~~~~~~~~~~~~~~~~~~ {\mbox{if}} ~~ m=\widetilde{z},~\widetilde{w}
\end{array}
\right..    \label{p,p_tilde}
\end{eqnarray}

\subsection*{Field Strength
	~(${f}^{\prime}:=\partial_{l} f,~l=\widetilde z,~\widetilde w$ )}
\begin{eqnarray}
{F_{m l}}
&\!\!\!=& 
\!\!\! \partial_{m}A_{l}-\partial_{l}A_{m}+[A_l,A_m] ~~~ (m \neq \widetilde{z}, \widetilde{w}) \nonumber \\
&\!\!\!=&\!\!\! -\partial_{l}A_{m}   \nonumber \\
&\!\!\!=&
\!\!\! \frac{2(\lambda-\mu)}{\Delta^2}\left(
\begin{array}{cc}
U & V \\ 
W & -U
\end{array}\right),   \\
&&
\left\{
\begin{array}{l}
\Delta:=AD-BC \\
U
:=p_m\left[{B}^{\prime}D+B{D}^{\prime}-2BD({\Delta}^{\prime}/\Delta)\right]
-q_m\left[{A}^{\prime}C+A{C}^{\prime}-2AC({\Delta}^{\prime}/\Delta)\right]  \\
V
:=-2p_m\left[B{B}^{\prime}-B^{2}({\Delta}^{\prime}/\Delta)\right]
+2q_m\left[A{A}^{\prime}-A^2({\Delta}^{\prime}/\Delta)\right]  \\
W
:=2p_m\left[D{D}^{\prime}-D^2({\Delta}^{\prime}/\Delta)\right]
-2q_m\left[C{C}^{\prime}-C^2({\Delta}^{\prime}/\Delta)\right]
\end{array}
\right.  ~~~~~~~~
\end{eqnarray}
~~~~~~~~~~~~~~~~~~~where $p_m$, $\widetilde{p}_m$ are defined in \eqref{p,p_tilde}.

\subsection*{Lagrangian Density}

\begin{eqnarray}
&&{\mbox{Tr}}{F_{w\widetilde{z}}F_{z\widetilde w}}  \nonumber \\
&\!\!\!=&
\!\!\!\frac{16(\lambda-\mu)^2 \varepsilon_{0}\widetilde{\varepsilon}_{0}}{\Delta^4}
\left\{
\begin{array}{l}
4\varepsilon_{0}\widetilde{\varepsilon}_{0}\alpha \beta \gamma \delta + 
(\alpha \delta -\beta \gamma)^{2} (A{D}^{\prime}-{B}^{\prime}C) ({A}^{\prime}D-B{C}^{\prime}) 
 \nonumber  \\  
+\alpha \beta \gamma \delta [(AD-BC)
({A}^{\prime}{D}^{\prime}
-{B}^{\prime}{C}^{\prime})
+({A}^{\prime}D-B{C}^{\prime})
(A{D}^{\prime}-{B}^{\prime}C)]
\end{array}
\right\},
\nonumber \\
&&{\mbox{Tr}}{F^2_{w\widetilde{w}}}  \nonumber \\
&\!\!\!=& 
\!\!\! \frac{16(\lambda-\mu)^2 \varepsilon_{0}\widetilde{\varepsilon}_{0}}{\Delta^4}
\left\{
\begin{array}{l}
2\varepsilon_{0}\widetilde{\varepsilon}_{0}(\alpha^2\delta^2 +\beta^2 \gamma^2) + 
\nonumber  \\
+\alpha \beta \gamma \delta [(AD-BC)
({A}^{\prime}{D}^{\prime}
-{B}^{\prime}{C}^{\prime})
+({A}^{\prime}D-B{C}^{\prime})
(A{D}^{\prime}-{B}^{\prime}C)]
\end{array}
\right\},
\nonumber \\
&& ~~~~~~~~~~~~~~~~~~
\left\{
\begin{array}{l}
\varepsilon_0:=a_2b_1-a_1b_2,~~~
\widetilde{\varepsilon_0}:=c_2d_1-c_1d_2,~ \\ 
\varepsilon_1:=a_1d_1-b_1c_1,~~~
\widetilde{\varepsilon}_1:=a_2d_2-b_2c_2,~ \\
\varepsilon_2:=a_1d_2-b_1c_2,~~~
\widetilde{\varepsilon}_2:=a_2d_1-b_2c_1.   
\end{array}
\right.  \nonumber 
\end{eqnarray}
By \eqref{Complex Lagrangian density_ASD}, we have 
\begin{eqnarray}
{\mbox{Tr}}F^2 
&\!\!\!=& \!\!\!  
4({\mbox{Tr}}{F_{w\widetilde{z}}F_{z\widetilde w}}-{\mbox{Tr}}{F^2_{w\widetilde{w}}})
\nonumber \\
&\!\!\!=&
\!\!\! \frac{64(\lambda-\mu)^2(\alpha \delta -\beta \gamma)^2 \varepsilon_{0}\widetilde{\varepsilon}_{0}}{\Delta^4}
\left\{(A{D}^{\prime}-{B}^{\prime}C) ({A}^{\prime}D-B{C}^{\prime})-2\varepsilon_{0}\widetilde{\varepsilon}_{0}
\right\}
\label{Lagrangian Density_uncompleted}
\end{eqnarray}
Substituting \eqref{ABCD} into \eqref{Lagrangian Density_uncompleted}, we get the result of \eqref{Lagrangian density_cpx}.

\newpage


\end{document}